\documentclass[a4paper, 11pt, amsmath, amssymb]{article}
\pdfoutput=1
\usepackage{jheppub}

\usepackage{amsmath}
\usepackage{amssymb}

\usepackage{graphicx}
\usepackage{dcolumn}
\usepackage{bm}
\usepackage{hyperref}

\usepackage{epsfig}
\usepackage{epstopdf} 

\newcommand{\bc}{\begin{center}}
\newcommand{\ec}{\end{center}} 
\newcommand{\be}{\begin{equation}}
\newcommand{\ee}{\end{equation}}
\newcommand{\beq}[1]{\begin{equation} \label{#1}}
\newcommand{\eeq}[1]{\label{#1} \end{equation}}
\newcommand{\bea}{\begin{eqnarray}}
\newcommand{\eea}{\end{eqnarray}}
\newcommand{\beqar}{\begin{eqnarray}} 
\newcommand{\eeqar}[1]{\label{#1} \end{eqnarray}} 
\newcommand{\bit}{\begin{itemize}}
\newcommand{\eit}{\end{itemize}}
\newcommand{\bsp}{\begin{split}}
\newcommand{\esp}{\end{split}}

\newcommand{\vx}{{\bf x}}
\newcommand{\vz}{{\bf z}}

\newcommand{\lp}{\left(}
\newcommand{\rp}{\right)}

\newcommand{\bk}{\mathbf{k}}

\newcommand{\bn}{\mathbf{n}}

\newcommand{\bq}{\mathbf{q}}
\newcommand{\bx}{\mathbf{x}}

\newcommand{\bz}{\mathbf{z}}

\newcommand{\kT}{k_\perp}
\newcommand{\qT}{q_\perp}

\newcommand{\RAA}{R_{AA}}
\newcommand{\raa}{$ R_{AA} $}
\newcommand{\vtwo}{$ v_{2} $}

\newcommand{\eexp}{{\rm e}}

\newcommand{\nocontentsline}[3]{}
\newcommand{\tocless}[2]{\bgroup\let\addcontentsline=\nocontentsline#1{#2}\egroup}

\newcommand{\chisq}{$ \chi^2/d.o.f. $}
\newcommand{\amax}{$ \alpha_{max} $}
\newcommand{\etas}{$ \eta/s $}

\title{Azimuthal jet flavor tomography with CUJET2.0 of nuclear collisions at RHIC and LHC}

\author[a]{Jiechen Xu}
\author[a,b]{Alessandro Buzzatti}
\author[a,b,c,1]{Miklos Gyulassy\note{Corresponding author.}}
\affiliation[a]{Department of Physics, Columbia University,\\538 West 120th Street, New York, NY, USA}
\affiliation[b]{Nuclear Science Division, Lawrence Berkeley National Laboratory,\\1 Cyclotron Road, Berkeley, CA, USA}
\affiliation[c]{Institute for Particle and Nuclear Physics, Wigner RCP, HAS,
1121 Budapest, XII. Konkoly Thege Mikl\'{o}s \'{u}t 29, Hungary}

\emailAdd{xjc@phys.columbia.edu}
\emailAdd{buzzatti@phys.columbia.edu}
\emailAdd{gyulassy@phys.columbia.edu}

\abstract{A perturbative QCD based jet tomographic Monte Carlo model, CUJET2.0, is presented to predict jet quenching observables in relativistic heavy ion collisions at RHIC/BNL and LHC/CERN energies. This model generalizes the DGLV theory of flavor dependent radiative energy loss by including multi-scale running strong coupling effects. It generalizes CUJET1.0 by computing jet path integrations though more realistic 2+1D transverse and longitudinally expanding viscous hydrodynamical fields contrained by fits to low $p_T$ flow data. The CUJET2.0 output depends on three control parameters, $(\alpha_{max},f_E,f_M)$, corresponding to an assumed upper bound on the vacuum running coupling in the infrared and two chromo-electric and magnetic QGP screening mass scales $(f_E \mu(T), f_M \mu(T))$ where $\mu(T)$ is the 1-loop Debye mass. We compare numerical results as a function of $\alpha_{max}$ for pure and deformed HTL dynamically enhanced scattering cases corresponding to $(f_E=1,2, f_M=0)$ to data of the nuclear modification factor, $R^f_{AA}(p_T,\phi; \sqrt{s}, b)$ for jet fragment flavors $f=\pi,D, B, e$ at $\sqrt{s}=0.2-2.76$ ATeV c.m. energies per nucleon pair and with impact parameter $b=2.4, 7.5$ fm. A $\chi^2$ analysis is presented and shows that $R^\pi_{AA}$ data from RHIC and LHC are consistent with CUJET2.0 at the $\chi^2/d.o.f< 2$ level for $\alpha_{max}=0.23-0.30$. The corresponding $\hat{q}(E_{jet}, T)/T^3$ effective jet transport coefficient field of this model is computed to facilitate comparison to other jet tomographic models in the literature. The predicted elliptic asymmetry, $v_2(p_T;\sqrt{s},b)$ is, however, found to significantly underestimated relative to RHIC and LHC data. We find the $\chi^2_{v_2}$ analysis shows that $v_2$ is very sensitive to allowing even as little as 10\% variations of the path averaged $\alpha_{max}$ along in and out of reaction plane paths.}


\begin{document}

\maketitle

\flushbottom

\section{Introduction}
\label{sec:intro}

Recent experimental data of jet quenching and collective flow in relativistic heavy ion collisions at the Large Hadron Collider (LHC) and Relativistic Heavy Ion Collider (RHIC) on Pb+Pb at $\sqrt{s_{NN}}=2.76$ TeV in ALICE \cite{ALICE:2012ab,Abelev:2012di,Abelev:2012hxa}, ATLAS \cite{ATLAS:2011ah}, CMS \cite{CMS:2012aa,Chatrchyan:2012xq} detectors, and Au+Au at $\sqrt{s_{NN}}=200$ GeV in PHENIX \cite{Adare:2008qa,Adare:2010de,Adare:2010sp,Adare:2012wg}, STAR \cite{Abelev:2006db,Abelev:2009wx} detectors have provided unique new opportunities to probe the dynamical properties of deconfined QCD matter called the Quark Gluon Plasma (QGP) \cite{Gyulassy:2004zy}. The interpretation of those new data requires the development of more powerful quantitative theory to enable prediction of observables in nuclear collisions covering a wide range of energy $\sqrt{s_{NN}}=0.02-2.76$ TeV, centrality $b=1-10$ fm, transverse momentum $p_T=5-200$ GeV/c scales, and rapidity $|\Delta \eta|<10$. In addition, consistency of predictions with the all quark mass/flavor $M=0.2-4.5$ $\rm{GeV/c^2}$ is essential \cite{Abreu:2007kv}. We focus on tomographic jet probes of the strongly interacting Quark Gluon Plasma (QGP) using perturbative QCD (pQCD) based tomographic models \cite{Baier:1996kr,Zakharov:1997uu,Baier:1998kq,Wiedemann:2000za,Salgado:2003gb,Armesto:2003jh,Armesto:2004pt,Armesto:2005iq,Guo:2000nz,Wang:2001ifa,Majumder:2007ae,Arnold:2002ja,Arnold:2002zm,Arnold:2003zc,Gyulassy:1993hr,Gyulassy:1999zd,Gyulassy:2000er,Djordjevic:2003zk,WHDG,Buzzatti:2011vt}. An important parallel effort based on string theory inspired gravity dual holographic models \cite{Gubser:2006bz,Herzog:2006gh,CasalderreySolana:2006rq,Noronha:2010zc,Ficnar:2012np,Ficnar:2012yu,Ficnar:2013wba,Ficnar:2013qxa} will not be considered here.

In the pQCD framework, radiative energy loss is assumed to be the dominant dynamical mechanism along with elastic energy loss for jet-medium interactions. In the past two decades, a wide range of jet quenching models have been formulated and applied to explain or predict high transverse momentum $p_T$ measurements at RHIC and LHC. These models are based on different medium property assumptions. BDMPS-Z \cite{Baier:1996kr,Zakharov:1997uu,Baier:1998kq} and ASW \cite{Wiedemann:2000za,Salgado:2003gb,Armesto:2003jh,Armesto:2004pt,Armesto:2005iq} multiple soft scattering models are assumed to describe the medium as a series of static color scattering centers, the incoming parton is subject to Brownian motion due to multiple soft scatterings with the medium, and a constant jet transport coefficient $\hat{q}$ is presumed to characterize adequately the jet-medium interaction process; Higher twist (HT) \cite{Guo:2000nz,Wang:2001ifa,Majumder:2007ae} models formulate the medium in terms of matrix elements of gauge field operators, and the properties of the plasma are specified through the entropy density $s$; AMY \cite{Arnold:2002ja,Arnold:2002zm,Arnold:2003zc} characterizes the medium as a thermally equilibrated plasma, and describe it in the context of finite temperature field theory using the Hard Thermal Loop (HTL) rate equation approximations with all properties of the plasma specified by its local temperature T and baryonchemical potential $\mu_{B}$ fields. 

CUJET2.0 is the most recent extension of the opacity series formalism of Gyulassy-Levai-Vitev (GLV)\cite{Gyulassy:1993hr,Gyulassy:1999zd,Gyulassy:2000er,Gyulassy:2002yv,Vitev:2002pf} for applications to jet tomography of the QGP produced at RHIC and LHC. Jet tomography assumes that initial production of hard jets occurs before QGP formation and is reliably predicted via collinear factorized pQCD. The depletion or quenching of the intial rates jet fragments as a function of $(p_T, y, A, b, \sqrt{s}, M)$ can be used to probe the dynamical properties of QGP at short wavelengths under the assumption that jet medium interactions can be calculated via perturbative QCD multiple collision theory. In GLV theory, high energy jet energy loss is formulated as an expansion in the number of parton-medium scatterings and is found to be dominated by the first hard contribution in a medium in kinematic regions involving coherence of the long formation time compared to the size of the medium. If the QGP medium is well described by a quasi-parton Hard Thermal Loop plasma then the density of scattering centers $\rho$ and the Debye screening mass $\mu$ as well as the plasmon mass, $m_g$, can all be computed as a function of only the temperature T and the effective thermal coupling $\alpha_s( 4T^2)$. By relaxing the assumptions of HTL approximation, different non-perturbative models of the QGP can be tested.

The GLV theory correctly predicted in 2002 the general form of the $\sqrt{s}$ evolution of the high $p_T$ pion nuclear modification factor $R_{AA}(p_T,\eta=0;\sqrt{s}, b)= dN_{AA\rightarrow \pi}/\lp T_{AA}(b)dN_{pp\rightarrow \pi} \rp$ from SPS, RHIC to LHC energies. GLV was generalized to DGLV \cite{Djordjevic:2003zk} to include the kinematic effects due to thermal masses and extend to charm and beauty quark flavors. However, it was found that  DGLV radiative energy loss significantly underpredicted the quenching of non-photonic electrons from charm and bottom quark jets. This led to the WHDG \cite{WHDG} generalization of DGLV theory to include elastic scattering as well as  more realistic jet path length fluctuations. Those effects were found to be insufficient to solve the ``heavy quark puzzle''. This led Djordjevic\cite{Djordjevic:2007at} to develop a dynamical generalization of DGLV, replacing the Gyulassy-Wang (GW) \cite{Gyulassy:1993hr} static color electric scattering potential with dynamical color magnetic interaction included potential through the HTL weakly coupled QGP ansatz. Including the above generalizations of DGLV/WHDG approach, a powerful numerical code, CUJET1.0, was developed by Buzzatti \cite{Buzzatti:2012pe,Buzzatti:2011vt}, that finally solved the ``heavy quark puzzle'' and predicted a novel quark flavor inversion of the normal nuclear modification factor hierarchy as a unique signature of perturbative QCD based jet tomography models, which is not shared by the jet holography counterpart.

The CUJET1.0 \cite{Buzzatti:2012pe,Buzzatti:2011vt} Monte Carlo code developed
by Buzzatti implemented numerically the dynamical DGLV opacity series and featured (1) an interaction potential that could interpolate between the pure HTL dynamically screened color magnetic limit and static Debye color electric screening limit; (2) the ability to calculate high order opacity corrections to radiative energy loss up to 9th order; (3) explicit integration over jet path in diffuse nuclear geometries including Bjorken longitudinal expanding HTL QGP; (4) inclusion of fluctuating elastic energy loss; (5) the evaluation of the convolution over numerical tables of pQCD $\sqrt{s}$ dependent initial jet production spectra of all flavors; and (6) the final convolution over jet fragmentation functions and evaluation of semi-leptonic final decay into non-photonic electrons.

CUJET1.0 was found to explain for the first time \cite{Buzzatti:2011vt} the anomalous high quenching of non-photonic electrons (``heavy quark puzzle'') within a pure HTL QGP paradigm  as due to the enhanced dynamical magnetic scattering effects proposed by Djordjevic. It also predicted a novel inversion of the $\pi < D < e^- < B$ flavor ordering of $R^f_{AA}$ at high $p_T$ that has yet to be tested at RHIC and LHC. However, the ``surprising transparency'' \cite{Horowitz:2011gd} of QGP produced at LHC severely challenged the fixed coupling version of the CUJET1.0 inherited the failure of the fixed coupling approximation used in the WHDG extrapolation from RHIC to LHC.

This led to the present extension of the dynamic DGLV/WHDG theory called here CUJET2.0. This Monte-Carlo code generalizes CUJET1.0 to include multi-scale running coupling effects as well as full 2+1D transverse and longitudinal expanding medium which azimuthal tomography is very sensitive to. Both versions of CUJET were developed as part of the ongoing DOE Topical JET Collaboration Project \cite{JetCollab} with the mission to develop more quantitative jet quenching codes coupled to state of the art bulk observable constrained viscous hydrodynamic fields. First results with this code for azimuthally averaged $R^\pi_{AA}$ and the $\chi^2$ determination the jet transport coefficient $\hat{q}/T^3$ from fits to RHIC and LHC data, as well as comparison to other JET collaboration model approaches were recently reported in \cite{Burke:2013yra}. This long write up aims to document the physics and numerical details of the current CUJET2.0 model as well as present a systematic $\chi^2$ comparison not only to azimuthal averaged $R^\pi_{AA}$ but also to its elliptic azimuthal moment $v_2(p_T)$ that remains an open problem at this time. In addition, the quark flavor dependence of the above observables in this latest version is also presented.

Motivated by the high energy kinematic regions being probed presently at LHC and the complex reaction vertices involved in the jet-medium processes, in CUJET2.0 we introduce a physically motivated multi-scale running of strong coupling factors $\alpha_s(Q^2_i(x,\bk,T))$ in the DGLV opacity expansion, where $\alpha_s$ is the vacuum QCD running coupling bounded by an upper limit $\alpha_{max}$ coupling strength. In addition, we explore the sensitivity to non-perturbative color magnetic and color electric screening mass deformations by varying multiplicative screening parameters $(f_E,f_M)$. (The perturbative HTL QGP limit is $(1,0)$.)  The three control parameters $(\alpha_{max}, f_E, f_M)$ define the dynamical model in this running coupling extension of DGLV.

An especially important new feature of CUJET2.0 is its ability to adaptively read in a variety of 2+1D viscous hydrodynamic temperature $T(x,y,\tau)$ evolution grids to be used to perform jet path opacity integrations. Thus far we have used only the VISH2+1 event averaged grids available from the JET Collaboration depository. Future applications to event by event fluctuating hydro grids are planned.

This paper is organized as follows: first, we postpone many of the technical details and development of both CUJET 1.0 and 2.0 models to a series of appendices. See Appendix~\ref{sec:model} for a review of the fundamental ingredients of CUJET. In this appendix, dynamical DGLV opacity expansion, elastic energy loss, path length and energy loss fluctuation, convolution of energy loss probability distribution over initial production spectra and fragmentation functions will be discussed.

In Section~\ref{sec:development}, we discuss the choice of scales in the multi-scale running strong coupling extension of DGLV. We quantify the CUJET running coupling effects on jet energy loss in terms of a phenomenological ``abc'' power law energy loss model \cite{Betz:2013caa,oai:arXiv.org:1211.0804,BBMG2012,oai:arXiv.org:1106.4564,oai:arXiv.org:1102.5416,oai:arXiv.org:0812.4401}. We introduce an effective jet medium interaction potential which is able to interpolate color electric and magnetic screening effect, as well as extrapolate to non-HTL scenarios. 

Section \ref{sec:numerics} presents the main numerical results obtained with CUJET2.0 including pion nuclear modification factor $ R^{\pi}_{AA} $ at RHIC and LHC central and
semi-peripheral A+A collisions; $ \chi^{2}/d.o.f. $ analysis and discussions on the consistency of the model in various collision configurations; jet transport coefficient $ \hat{q}/T^3 $ and its variation with temperature and jet energy; flavor dependent suppression pattern for pion, D meson, B meson and non-photonic electron and the mass ordering in \raa; jet quenching with respect to reaction planes and single particle azimuthal anisotropy \vtwo. 

As a further exploration with CUJET2.0, in Section \ref{sec:discussion}, we study the thermalization time's effect on pion suppression factor, and non-HTL scenario's implication of non-perturbative near $T_c$ physics in the CUJET2.0 framework. Finally, we summarize our main results and conclusions, and discuss possible future works, improvements and tests on CUJET2.0 in Section \ref{sec:summary}.
\\
\section{The CUJET2.0 Framework}
\label{sec:development}

\subsection{Running strong coupling effects}
\label{sec:rc}

Recent data from the LHC showed a significantly steeper rise of $\RAA$ with $p_T$ in the range of $5-100$ $\rm GeV/c$ \cite{Abelev:2012hxa,CMS:2012aa} than predicted by fixed coupling  extrapolation from RHIC via WHDG\cite{Horowitz:2011gd}. This indicated a ``surprising transparency'' of the QGP at LHC to high energy jets as compared to bulk multiplicity scaling (by a factor $(dN^{LHC}/dy)/(dN^{RHIC}/dy) \approx 2.2$) of the QGP opacity assumed in WHDG. The fixed coupling version of CUJET1.0 also encountered the same difficulty \cite{Buzzatti:2012dy}. A generic $dE/dx$ model analysis \cite{oai:arXiv.org:1211.0804, BBMG2012} also found that effective jet medium $\kappa\propto \alpha_s^3$ coupling required to fit the slope and magnitude of central $R_{AA}(p_T)$ at both RHIC and LHC is $\sim 30\% $ less at LHC than at RHIC. 

The above problem motivated us to study whether UV running coupling effect could account for the relatively greater transparency of the QGP at LHC with CUJET2.0.

\subsubsection{Multi-scale running coupling for radiative and elastic energy loss}
\label{sec:multirc}

Earlier estimates of running coupling effects were made by Zakharov \cite{Zakharov:2008kt,Zakharov:2007pj}. In CUJET2.0 model, we follow a similar scheme using the 1-loop pQCD running coupling that is cutoff in the infrared when the coupling reaches a certain maximum saturation value $\alpha_{max}$ for $Q\le Q_{min}$: 
\be
\alpha_s \; \longrightarrow \; \alpha_s(Q^2) = \begin{cases}
\alpha_{max} & \mbox{if } Q \le Q_{min}\;, \\
\dfrac{2\;\pi}{9\;\log(Q/\Lambda_{QCD})}  & \mbox{if } Q > Q_{min}\;.
\end{cases}
\label{AlphaRun}
\ee
where the saturation scale $Q_{min}$ is fixed by $\alpha_{max}$ as $Q_{min}=\Lambda_{QCD}\;\exp\left\lbrace \dfrac{2\pi}{9\alpha_{max}}\right\rbrace\;$.

The choice of the $\alpha_{max}$ parameter is not obvious and is regarded here as a key infrared control parameter of the CUJET2.0 model. In principle, it can also depend on the local temperature field. At $T=0$, the $\alpha_{max}\approx 0.7$ estimate arose historically from the analysis of the heavy quark production in the vacuum \cite{Dokshitzer:1995ev}. However, in a QGP, suppression of $\alpha_s$ at scales $\sim T$ is also expected. Lattice QCD $q\bar{q}$ potential studies \cite{Kaczmarek:2004gv} found the effective thermal $\alpha_s(T)$ coupling decreases monotonically from $0.5$ at $T\sim 175$ MeV to $0.35$ at $T\sim 400$ MeV. More generally, the effective running coupling $\alpha_s(Q,T)$ depends on the relevant $Q$ of an observable as well as on $T$. 

In the dynamical case of CGC gluon production the generalization of fixed coupling rates in \cite{oai:arXiv.org:1009.0545,oai:arXiv.org:1106.5456} led to intricate multi-scale running coupling modifications of the fixed coupling formulae. A corresponding generalization of DGLV to running coupling remains a challenging and open problem because virtualities in radiative amplitudes depend on multiple kinematic $(x_+,\bk,\bq)$\footnote{See Appendix~\ref{app:notations} for notations and conventions.} as well as temperature dependent infrared screening scales and plasmon masses.

We explore in CUJET2.0 the running coupling effects using physics motivated ansatz for relevant virtuality scales. At leading opacity order,
%Since the DGLV opacity expansion involves various gluon radiation vertices, jet%-medium interaction vertices and thermal loops, the effective running scales mu%st depend on various processes accordingly. In CUJET, our approach consists of %letting the strong coupling run in accordance to Eq.~\ref{AlphaRun}, and 
we can identify in the fixed coupling DGLV forumla (see Eq.~\ref{fcDGLV}), three distinct scales $Q_i(i=1,2,3)$ controlling the strength of different aspects of the physics: % with which different powers of $\alpha_s(Q^2_i)$ vary:
%\footnote{Analytical calculations of running coupling corrections for inclusive gluon production are done by Horowitz and Kovchegov in \cite{oai:arXiv.org:1009.0545,oai:arXiv.org:1106.5456}, however, calculating parton energy loss would involve more complicated quark loops.}. Assuming the scale is set by the exchanged four-momentum in the relevant Feynman vertex, we have:

\begin{enumerate}
	\item Two powers $\alpha_s^2(Q^2)$ clearly originate from the jet-medium interaction vertices from the exchanged transverse momentum $\mathbf q$, and so for these we simply take $Q_1^2=\bq^2$.
	\item One power $\alpha_s(Q^2)$ originates from the radiated gluon vertex. The off-shellness in the intermediate quark propagator for one of the three amplitudes where the gluon is emitted after the scattering is
	\be
		Q_2^2=q^2-M^2=\frac{\bk^2}{x_+(1-x_+ )}+\frac{x_+  M^2}{1-x_+}+\frac{m_g^2}{x_+} \; \; .
	\label{rcscale1}
	\ee
	Where $\bk$ is the transverse momentum of the radiated gluon, $M$ is the mass of on-shell quark and $m_g$ is the plasmon mass of gluon. An ambiguity arises from other amplitudes for example if the radiated glue scatters with q instead of the quark. In the limit when $k \gg q$ and mass effects are negligible 
	\be
		Q_2^2 \approx \frac{\bk^2}{x_+ (1-x_+)} \; \; ,
	\label{rcscale2}
	\ee
	as in DGLAP radiative splitting.
	\item Running thermal couplings can arise from the Debye mass $\mu(\alpha_s(Q^2);T)$ and plasmon mass. We allow these to run with scale $Q_3^2=(2T)^2$.\footnote{Djordjevic proposed a more elaborate self-consistent equation for the thermal scale \cite{oai:arXiv.org:hep-ph/0601119,Djordjevic:2013xoa}.}
\end{enumerate}

Note in the above choices of running scales there is no explicit dependence on the jet energy, which comes instead from the kinematic limits of the $\bq$ and $\bk$ integrations. $\kT^{MAX}=x_E E$ and $\qT^{MAX}=\sqrt{4ET}$.

With the above scheme, the running coupling DGLV inclusive radiative fractional energy loss distribution at first order in opacity is then given by (the notation is as
defined in Appendix \ref{app:notations} and \ref{sec:model}):
\be\label{rcCUJETDGLV}
\begin{split}
x_E \frac{dN_g^{n=1}}{dx_E}(\bx_0,\phi) = &\; \frac{18 C_R}{\pi^2} \frac{4+n_f}{16+9n_f} \int{d\tau}\; \rho(\bz) \int{d\bk} \int{d\bq}\;\\
&\times\;\alpha_s ( \frac{\bk^2}{x_+ (1-x_+)} )\;\\
&\times\;{\frac{\alpha_s^2(\bq^2)(f_E^2-f_M^2)}{(\bq^2+f_E^2\mu^2(\vz))(\bq^2+f_M^2\mu^2(\vz))}}\\
&\times\;{\frac{-2(\bk-\bq)}{(\bk-\bq)^2+\chi^2(\bz)} \lp\frac{\bk}{\bk^2+\chi^2(\bz)} - \frac{(\bk-\bq)}{(\bk-\bq)^2+\chi^2(\bz)}\rp}\\
&\times\;{\lp1-\cos\lp\frac{(\bk-\bq)^2+\chi^2(\bz)}{2 x_+ E } \tau\rp\rp}\\
&\times\;{\lp \frac{x_E}{x_+} \rp J(x_+(x_E))}
\; \; .
\end{split}
\ee
where $C_R$ is the quadratic Casimir of the jet ($C_F=4/3$ for quark jets, $C_A=3$ for gluon jets); $ \vz=\lp x_0+\tau\cos\phi,y_0+\tau\sin\phi; \tau\rp$ is the path of the jet created at $(x_0,y_0)$ in the production plane along azimuthal angle $\phi$; $\rho(\vz)$ and $T(\vz)$ is the number density and temperature evolution profile of the medium; $\chi^2(\vz)=M^2 x_+^2+m_g^2(\vz)(1-x_+)$ controls the ``dead cone'' and Landau-Pomeranchuck-Migdal (LPM) destructive interference, squared gluon plasmon mass $ m_g^2(\vz)=f_E^2 \mu^2(\vz) / 2 $, HTL Debye mass $ \mu(\vz) = g(\vz)T(\vz)\sqrt{1+n_f/6} $, $ g(\vz)=\sqrt{4\pi\alpha\lp4T^2(\vz)\rp} $; integration limit $ 0\leqslant|\bq| \leqslant \rm{min}(|\bk|,\sqrt{4ET(\vz)}) $, $ 0\leqslant|\bk| \leqslant x_E E $. $x_+(x_E)$ and $J(x_+(x_E))$ are defined in Eq.~\eqref{x+xE} and~\eqref{x+xEJacobian} respectively.
\\

We further include running coupling effects in the elastic portion of the energy loss following the work of Peign\'{e} and Peshier \cite{Peigne:2008nd}: both powers of $\alpha_s$ in Eq.~\eqref{ELcrosssec} run with $\hat{t}$, and when integrated over $d\hat{t}$ in Eq.~\eqref{BjEloss}, we obtain
\be
\alpha_s^2 \int_{\mu^2}^{4ET} \dfrac{d\hat{t}}{\hat{t}}  \; \longrightarrow \;  \int_{\mu^2}^{4ET} \dfrac{d\hat{t}}{\hat{t}}\alpha_s^2(\hat{t})
\;\;,
\ee
\be
\alpha_s^2 \log\dfrac{4ET}{\mu^2}  \; \longrightarrow \;  \alpha_s(\mu^2) \alpha_s(4ET)\log\dfrac{4ET}{\mu^2(\alpha_s(4T^2);T)}
\;\;.
\ee
Here the limits of $\hat{t}$ are chosen according to the Bjorken computation of elastic energy losses. In CUJET model, the argument of the logarithm is modified according to Eq.~\eqref{ElasticLog}. All after, in running coupling CUJET, Eq.~\eqref{CUJETElastic} and \eqref{NumOfColl} is modified to
\be\label{rcCUJETElastic}
\begin{split}
\frac{dE(\bz)}{d\tau}= & - C_R \pi \left[ \alpha(\mu(\bz))\alpha( E(\bz) T(\bz)) \right] T(\bz)^2 \lp 1+\frac{n_f}{6} \rp \\
& \times \log \left[ \frac{4T(\bz)\sqrt{E(\bz)^2-M^2}}{\lp E(\bz)-\sqrt{E(\bz)^2-M^2}+4T(\bz)\rp\mu(\bz)} \right],
\end{split}
\ee
and
\be
\bar{N_c} = \int_{0}^{\tau_{max}} d\tau \left[ \frac{\alpha(\mu(\bz)) \alpha( E(\bz) T(\bz))}{\mu(\bz)^2} \right] \left[ \frac{18 \zeta(3)}{\pi} (4+n_f) T(\bz)^3 \right]\;\;.
\label{rcNumOfColl}
\ee
respectively. Note in the running coupling scenario, the calculation of average number of collisions involves recursively solving the $E(\bz)$ integral equation.
\\

\begin{figure*}[!t]
\centering
\includegraphics[width=0.45\textwidth]{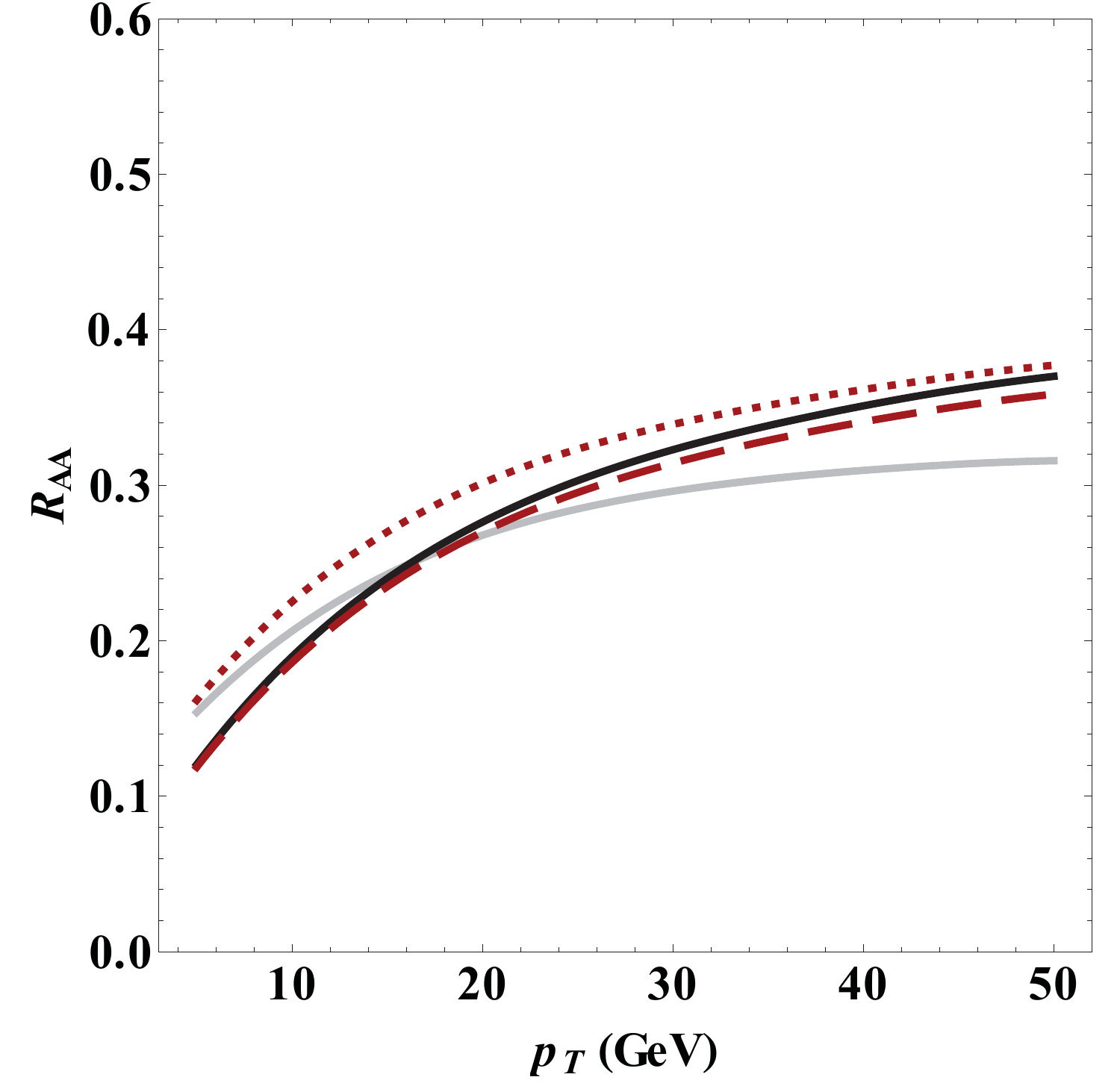}
\hspace{0.01\textwidth}
\includegraphics[width=0.45\textwidth]{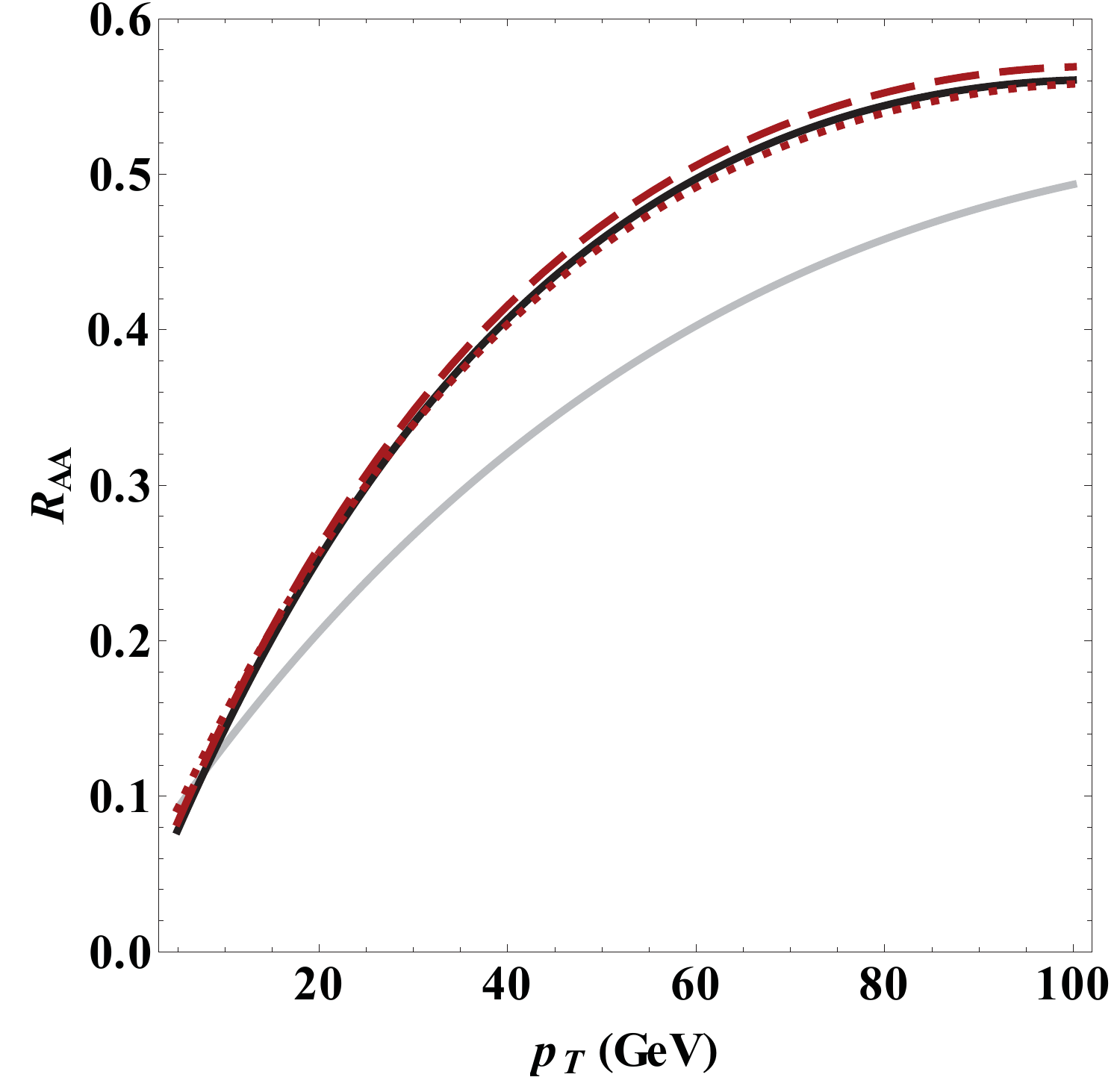}
\caption{Fixed and running coupling pion $\RAA$ results are compared side to side at RHIC (\textit{left}) and LHC (\textit{right}) in CUJET with Glauber static transverse plus Bjorken longitudinal expanding background. The gray opaque curves use a fixed coupling with $\alpha_s=0.3$, while the black curves use a running coupling with $\alpha_{max}=0.4$. The difference is notable, especially in the higher energy range available at the LHC, while RHIC results are left almost unchanged. The sensitivity to the variation of running scales $Q_i$ (cf. Eq.~\eqref{AlphaRun} and following) is measured by the red curves: on one side we decrease the value of all scales $Q_i$ by $50\%$ and lower $\alpha_{max}$ to $0.3$ (red dashed), on the other we increase all scales $Q_i$ by $25\%$ and increase at the same time $\alpha_{max}$ to $0.6$ (red dotted). $\alpha_{max}$ is constrained to fit $\RAA^{\pi ,LHC}(p_T \approx 30\;{\rm GeV})=0.35$.}
\label{fig:RAAalpha}
\end{figure*}

The choice of running scales $Q_i$ is of course subject to significant uncertainties at present. To estimate the systematic uncertainties on the nuclear modification due to the variation of running scale we increase or decrease the running scales $Q_i$ by $25$ or $50$ percent respectively, and refit the fixed reference point at $p_T=30$ GeV
by changing the free \amax~parameter. The results are shown in Fig.~\ref{fig:RAAalpha}.

With a static Glauber transverse background in CUJET, to compensate 50\% decreased running scales $Q_i$, \amax~needs to be 25\% lower; while to compensate 25\% increased running scales $Q_i$, \amax~needs to be 50\% higher. Since the radiative energy loss depends on \amax~with approximately a cubic law according to Eq.~\eqref{rcCUJETDGLV}, making the running scales $Q_i$ smaller by 50\% or greater by 25\% can result in approximately 5\% decrease or 10\% increase in \raa. The systematic error bar coming from varying running scales is therefore significantly smaller than the experimental errors at this time, indicating the relative insensitivity of CUJET to the precise choice of running scales.

\subsubsection{Generic abc model quantification of running coupling effect}
\label{sec:abc}

We quantify the impact that the running coupling has on jet quenching by using a phenomenological ``abc'' energy loss model introduced in \cite{Horowitz:2011gd, Betz:2013caa,oai:arXiv.org:1211.0804,BBMG2012,oai:arXiv.org:1106.4564,oai:arXiv.org:1102.5416,oai:arXiv.org:0812.4401}.
\be
\frac{dP}{d\tau} = - \kappa P^a \tau^b T^{2-a+b}
\; \; ,
\label{abcModel}
\ee
where $P(\tau)$ corresponds to the momentum of a massless jet passing though a plasma characterized by a local temperature $T$. The power of $T$ is constrained by simple dimensional analysis, and the index $a$ and $b$ are set by the asymptotic LPM behavior of the GLV model. In the fixed coupling case,
\be
\frac{\Delta E}{E} \propto T^3 L^2 \frac{\log (E/T)}{E} 
\; \; .
\ee
For the range of energies of interest, $\log{(E/T)}/E \sim E^{a}$, with $a\sim 1/3-1/4$.

\begin{figure*}[!t]
\bc
\includegraphics[width=0.5\textwidth]{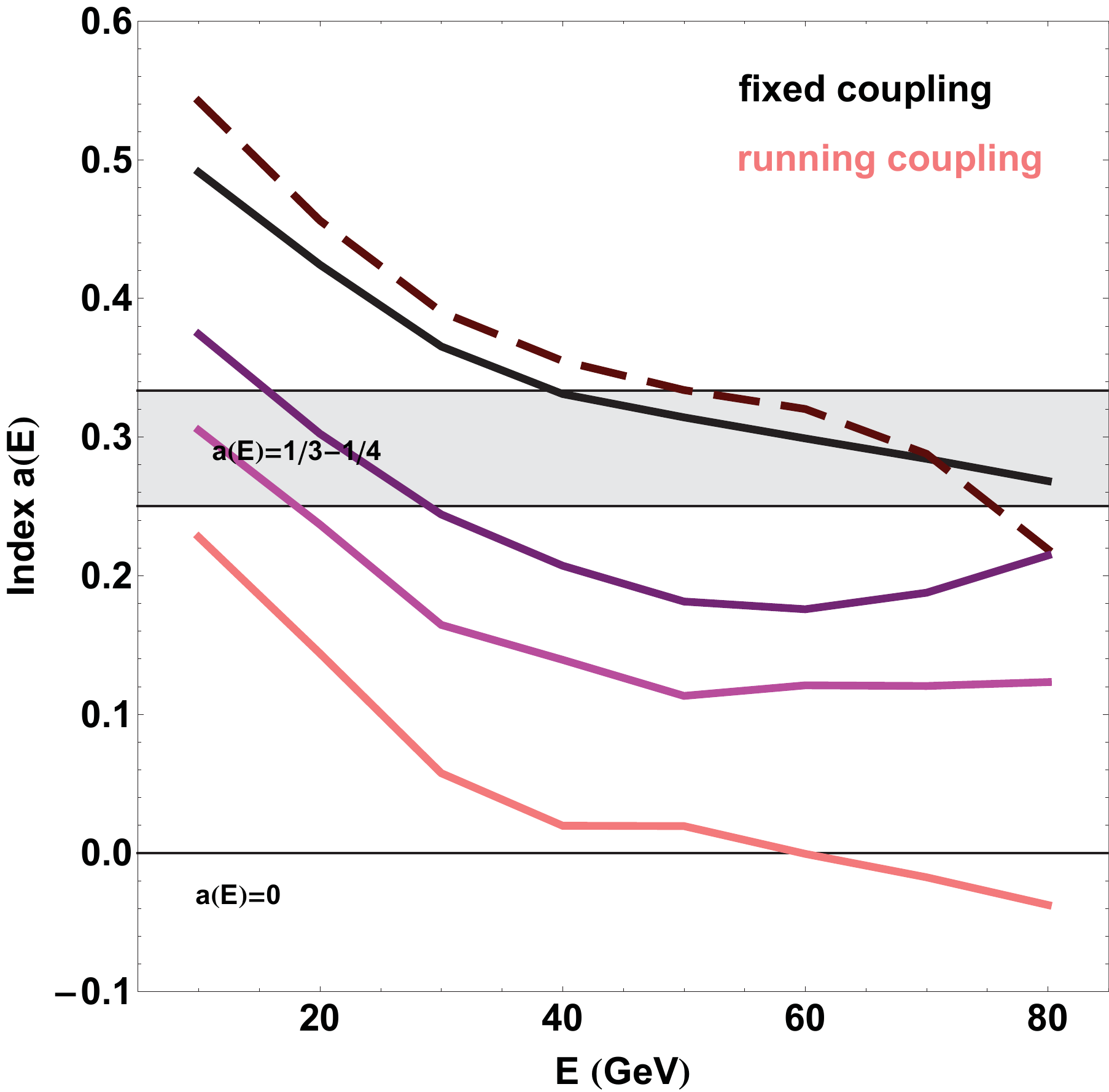}
\caption{Energy loss index $a(E)$ (cf. Eq.~\eqref{abcModel}) for different assumptions of the running coupling in CUJET: fixed effective $\alpha_s=0.3$ (black), only thermal coupling running (dashed red), only $\alpha_s^2(\bq^2)$ running (purple), only $\alpha_s^2(\bk^2/(x(1-x)))$ running (magenta), all couplings running (pink). The saturated $\alpha_{max}$ value is chosen to be equal to $0.4$, which corresponds to approximately $Q_0\sim1$ GeV. The plot shows the energy loss of a light quark ($M=0.2$ GeV) traveling from the origin of the transverse plane and through a gluonic plasma ($n_f=0$) of size $L=5$ fm, whose density profile is generated from Glauber model and resembles the medium created in a Pb+Pb $\sqrt{s_{NN}}=2.76$ TeV $b=0$ fm collision.}
\label{fig:abcIndex}
\ec
\end{figure*}

In Fig.~\ref{fig:abcIndex} we show the value of the index $a$ as a function of the jet energy $E$, for five different cases: $\alpha_s$ fixed, only $\alpha_s(4T^2)$ running, only $\alpha_s^2(\bq^2)$ running, only $\alpha_s(\bk^2/(x(1-x))$ running and finally all couplings running. In this example, we use a non-uniform density profile generated from Glauber model with Bjorken expansion that approximates the thermal medium formed in a Pb+Pb central collision at LHC. The results are insightful:
\begin{itemize}
	\item As expected, the fixed case shows $a\sim 1/3-1/4$.
	\item By introducing the thermal coupling, only the absolute value of the energy loss is affected and the energy dependence of the index remains unaltered. The scale at which the thermal $\alpha_s$ is evaluated is in fact independent of $E$. Not noticeable in this plot, at very high temperatures the reduced thermal coupling causes a stronger quenching compared to the fixed coupling case, since the smaller Debye mass diminishes the screening in the plasma. This running effect is however small: for most of the temperature ranges, $\alpha_s$ is in fact equal to the saturated value $\alpha_{max}$ (with $Q_0\sim1$ GeV, $T$ needs to be greater than $0.6$ GeV to start feeling the running effects).
	\item The couplings $\alpha_s^2(\bq^2)$ and $\alpha_s(\bk^2/(x(1-x))$ sensibly reduce the dependence of $\Delta E/E$ on $E$, and as a consequence the value of the index $a$ gets smaller and closer to $0$. The $\alpha_s^2(\bq^2)$ contribution is smaller since the $\bq$ distribution is peaked at small values of $\qT$, as opposed to the $\alpha_s(\bk^2/(x(1-x))$ contribution which is larger due to the high tails of the $\bk$ distribution.
	\item The all-running case shows almost no dependence of $\Delta E/E$ on $E$, and $a(E)\approx 0$.
\end{itemize}

According to Fig.~\ref{fig:abcIndex}, the running coupling drastically alters the jet energy dependence of the energy loss, making $\Delta E/E$ approximately independent of $E$. This naively implies less quenching at high energies and an increase in the $\RAA$ slope. Fig.~\ref{fig:RAAalpha} in fact proves our assertions. What is remarkable is the fact that the change in the slope of $\RAA$ cannot be mimic by a rescaling of the fixed coupling $\alpha_s$: this measurement constitutes a potentially clear signature of running coupling effects. 
	
\subsection{Bulk evolution profile}
\label{sec:bulk}

The evolution profile of the bulk is encoded in CUJET as local density of scattering centers $\rho(\vz)$ and local temperature $ T(\vz)$, in both the radiative energy loss Eq.~\eqref{rcCUJETDGLV} and elastic energy loss Eq.~\eqref{TGcollEq}. To gain meaningful information about the parton medium interaction mechanism from comparing predicted jet quenching observables with experimental measurements, QGP evolution profile plays an essential role, and carefully constrain it with for example bulk low $p_T$ flow data is critical. In CUJET1.0, a static transverse geometry is generated from Glauber model, and longitudinally a Bjorken expansion is applied. In CUJET2.0, this picture is replaced with more realistic fluid fields generated from 2+1D viscous hydrodynamics, and the medium has dynamical expansion both transversely and longitudinally.

\subsubsection{CUJET1.0: Glauber initial with 1+1D Bjorken expansion}
\label{sec:glauber}

In CUJET1.0, Glauber model \cite{Glauber:1970jm} is used to generate the geometry of the collision, the plasma density profile and the jet production point distribution. Time evolution of the plasma density $\rho_{QGP}$ is given by the Bjorken picture. A simple analytical expression is obtained by making few assumptions: (1) the system expands only longitudinally, along the beam direction (1+1D expanding plasma); (2) the plasma is a perfect fluid; (3) the computation is carried out in a relativistic hydrodynamical framework.

In particular, the following steps are implemented in the model calculation: firstly, Density profile of nucleus A is generated with Woods-Saxon parametrization
\be
\rho_A(r)=\frac{\cal{N}_{A}}{1+\exp((r-R)/a)}
\; \; .
\label{rhoA}
\ee
This density is normalized to mass number $A$. $R$ is the nuclear radius and $a$ represents the surface thickness. Subsequently, the thickness function of the nucleus A is defined as
\be
T_A(\bx)=\int{dz}\; \rho_A(z,\bx)
\; \; .
\ee

After acquired the corresponding thickness functions, the distribution of participants in a collision between two nuclei A and B that collide with impact parameter $\mathbf{b}$ is then given by
\be
\begin{split}
\rho_{part}(\bx,b) = & T_A(\bx) \lp 1-{\rm e}^{-\sigma_{in}T_B(\bx-\mathbf{b})} \rp  \\
                  & + T_B(\bx-\mathbf{b}) \lp 1-{\rm e}^{-\sigma_{in}T_A(\bx)} \rp \; \; \\
		 N_{part}(b) = & \int{d\bx} \; \rho_{part}(\bx,b) \; \; ,
\end{split}
\label{PartDistr}
\ee
where $\sigma_{in}$ is the inelastic nucleon-nucleon cross section. The $A, R, a, \sigma_{in}$ parameters chosen for Au+Au (RHIC) and Pb+Pb (LHC) collisions in Eq.~\eqref{rhoA}\eqref{PartDistr}\eqref{BinaryDistr} are listed in Table~\ref{wsparameters}.
\begin{table}[!t]
\centering
 \begin{tabular}{ | c || c | c | c | c |}
 \hline
  & $A$ & $R$ (fm) & $a$ (fm) & $\sigma_{in}$ (mb)\\
		\hline\hline
 Au & 197 & 6.37 & 0.535 & 42\\
		\hline
 Pb & 207 & 6.48 & 0.535  & 63 \\
		\hline
 \end{tabular}
\caption{Woods-Saxon parameters used in CUJET. $A$ is the mass number of the nucleus, $R$ is the nuclear radius, $a$ is the surface thickness and $\sigma_{in}$ is the inelastic nucleon-nucleon cross section.}
\label{wsparameters}
\end{table}

In determining the proper time $\tau$ ($=\sqrt{t^2-z^2}$) dependence of QGP density profile, which is characterized by longitudinal boost invariance, for practical applications, we concentrate on the mid-rapidity region of the collision ($y=\frac{1}{2}\log\lp\frac{t+z}{t-z}\rp=0$). The QGP density field $\rho_{QGP,0}(\bx,\tau;b)$ in the azimuthal direction is given by:
\be
\left.\rho_{QGP}(\bx,b,\tau)\right|_{y=0}=\frac{1}{\tau_0}\;\frac{\rho_{part}(\bx,b)}{N_{part}(b)}\;\left.\frac{dN_g}{dy}\right|_{y=0} f(\tau/\tau_0) \;\; ,
\label{rhoQGP}
\ee
where $dN_{g}/dy$ represents the gluon rapidity density, it is proportional to the measured charged hadrons rapidity distribution $dN_{ch}/dy$, with $dN_{g}/dy=(3/2)dN_{ch}/dy$. Thermalization time $\tau_0$ is chosen to be $\tau_0=1~\rm{fm/c}$ in CUJET1.0. We introduce $f(\tau/\tau_0)$ to characterize the pre-thermal stage and the evolution profile after the medium is fully thermalized. In absence of a clear theoretical answer to the way high energy jet couples to the medium before thermalization, we make such phenomenological assumptions:
\be
	f(\tau/\tau_0) = \begin{cases}
\tau/\tau_0 & \mbox{if } \tau\le\tau_0, \\
\tau_0/\tau & \mbox{if } \tau>\tau_0.
\end{cases}
\label{tau0}
\ee
The density ``seen'' by the jet grows linearly until thermalization time $\tau_0$ is reached, thereafter it decreases as $1/\tau$, converges to the Bjorken expansion picture. Note different parametrizations for the temporal evolution of the system exist, and by choosing the linear thermalization scheme systematic uncertainties are inevitably introduced. A detailed discussion about the choice of thermalizing schemes and associated errors can be found in Appendix~\ref{app:thermal}.

The jet production points are distributed according to the binary collision distribution, which is given by
\be
\begin{split}
\rho_{binary}(\bx,b) = & \sigma_{in} \; T_A(\bx) T_B(\bx-\mathbf{b})\;\;, \\
		 N_{binary}(b) = & \int{d\bx} \; \rho_{binary}(\bx,b) \; \; .
\end{split}
\label{BinaryDistr}
\ee

The ability of CUJET to perform a full jet path integration allows us to parametrize the evolution of the system in different ways, and we can perform comparisons with experimental measurements and draw insightful conclusions on the physics of the collision. See Appendix~\ref{app:thermal} for a comprehensive analysis of thermalization phase effects on jet quenching physics.

\subsubsection{CUJET2.0: viscous hydrodynamics}
\label{sec:hydro}

In Appendix~\ref{app:thermal}'s Fig.~\ref{PlasmaSlice}, we show a smooth temperature profile of a symmetric plasma. The shape of the region of interest takes the form of a perfect circle when the impact parameter is null, or an almond when $b\ne0$.

The reality is however different. Nuclear matter is very granular at short distances, and the nucleons are not distributed in a perfectly symmetrical way. The naive picture of a circle or an almond is an idealization of the collision geometry in most situations, and the identification of a reaction plane determined by the orientation of the impact parameter is often a hard experimental task. The average over multiple collisions might lead to a smooth temperature profile, but this is not the case on an event-by-event basis, where fluctuations of initial conditions might lead to considerably different results.

Therefore, the full three-dimensional hydrodynamic expansion may differ substantially from the Bjorken ideal 1+1D hydrodynamic evolution in CUJET1.0. A complete description of the system must include not only transverse expansion, but also viscous corrections to the perfect fluid. And furthermore, effects such as initial condition fluctuation or jet energy deposition into the medium should be considered. We hence adapt the grid of QGP fluids in CUJET2.0 to encompass more realistic 2+1D viscous hydrodynamical fields, incorporating a dynamical medium expanding both transversely and longitudinally. This is for the first time the DGLV opacity series is fully coupled to viscous hydro fields, and the combination of strongly coupled bulk dynamics and weakly coupled pQCD energy loss theory generates rather indicative results of quenching observables computed from CUJET2.0 under RHIC and LHC conditions. The numerical analysis of CUJET2.0 output in RHIC and LHC nuclear collisions will be presented in Section~\ref{sec:numerics}. 

In principle, the present CUJET2.0 framework can be coupled to virtually any complex geometries and plasma evolution profiles. A wide range of flow fields generated by external hydrodynamical code can be applied, e.g. transverse blast wave model \cite{Gyulassy:2000gk,Gyulassy:2001kr}, viscous RL hydro \cite{Luzum:2008cw,Luzum:2009sb}, VISH2+1 \cite{Song:2008si,Shen:2010uy,Renk:2010qx}, etc. At present stage, we have used only the VISH2+1 event averaged evolution profile available from the JET Collaboration depository to study the azimuthal angle and transverse momentum dependence of high-$p_T$ light to heavy flavor quenching pattern, incorporating event by event fluctuating hydro fields is a work in progress.

VISH2+1 \cite{Song:2008si,Shen:2010uy,Renk:2010qx} utilizes viscous hydrodynamics to describe the fireball evolution. In the version adapted by CUJET2.0, MC-Glauber initial conditions are used to sidestep issues related with hypothetical early non-equilibrium evolution; using Cooper-Frye algorithm \cite{Cooper:1974mv} along a hypersurface of constant temperature $T_f=120$MeV, a sharp transition from viscous fluid to free-streaming particles is generated to describe hadronic rescattering and kinetic freeze-out; the s95p-PCE (partial chemical equilibrium) equation of state (EOS) is constructed according to \cite{Huovinen:2009yb}, which matches Lattice QCD data at high temperature and recovers the hadron resonance gas at low temperature. The various input parameters are adjusted to fit final hadron spectra and elliptic flow in low transverse momentum $p_T<1.5(2.5)$ GeV/c region in \cite{Shen:2010uy}. In particular, experimental data of pion and proton spectra in 200AGeV Au+Au central collisions (0-5\% centrality, b=2.33 fm), pion, proton and charged hadron elliptic flow $v_2(p_T)$ in semi-peripheral collisions (20-30\% centrality, b=7.5 fm) are compared. With MC-Glauber initial conditions, s95p-PCE EOS and 120MeV freeze-out temperature, for a QGP with number of quarkonic flavor $n_f=2.5$, the starting time $\tau_0$ at which the system is sufficiently close to local thermal equilibrium for viscous hydrodynamics to be applicable is calculated to be $\tau_0=0.6$ fm/c\footnote{In CUJET2.0 calculations we set $\tau_0=0.6$ fm/c in the model to match the orginal hydro setting. For pre-thermal stage we use the linear scheme, systematic uncertainties resulted from the choosing different thermalization parametrizations can be found in Appendix~\ref{app:thermal}.}, and the key QGP transport parameter, shear viscosity \etas~(the ratio between shear viscosity $\eta$ and entropy density $s$), is phenomenologically extracted to be $\eta/s=0.08$.  

For initial jet production distributions, presumably binary distributions generated from corresponding viscous hydro should be used. However, in CUJET2.0 the binaries are given according to Eq.~\eqref{BinaryDistr}, because the spacing of the VISH2+1 x-y grid in CUJET2.0 is 0.5 fm while in computing final jet spectra the initial jet production points have a uniform 1.0 fm step in the radial direction. Systematic uncertainties resulting from binary distributions are therefore negligible. Nevertheless, this is a potentially interesting case if the quenching of jet occurs mostly at the edge of the medium, under which circumstance the fineness of the grid will play an important role. This topic will be explored in future studies with CUJET.

\section{CUJET2.0 Results at RHIC and LHC}
\label{sec:numerics}

In this section, CUJET2.0 is applied to RHIC Au+Au 200AGeV and LHC Pb+Pb 2.76ATeV collisions. Impact parameter b=2.4fm and 7.5fm is used to simulate 0-10\% and 10-30\% centrality respectively. For radiative energy loss, the DGLV opacity series is calculated to first order because of the computational efficiency of the Monte Carlo algorithm. The convergence of the DGLV expansion is discussed in detail in Appendix~\ref{app:conv}, from there we see, since CUJET2.0 concentrates on high energy jet suppression and averages over all possible path lengths in a realistic heavy ion collision, the first order in opacity can be regarded as a good approximation to the series. If to impose an artificial systematic uncertainty on higher order contributions, in terms of inclusive $\pi$, $D$, $B$, $e^-$ \raa, one can estimate the associated variation to be less than 15\% (cf. Appendix~\ref{app:convDGLV}).

We implement CUJET2.0 to study jet quenching observables, in particular \raa~and \vtwo~in the high transverse momentum $p_T>5$ GeV/c region where eikonal and soft approximation are applicable. RHIC inclusive neutral pion suppression factor $ R_{AA}^{\pi}(p_T=15\rm{GeV/c})= 0.3 $ at Au+Au 200AGeV 0-5\% centrality is set as a reference point to fix the maximum coupling constant \amax. We then extrapolate our calculation to $p_T=5\sim 20$ GeV/c region at RHIC, and $p_T=5\sim 100$ GeV/c region at LHC central to semi-peripheral collisions for pion, D meson, B meson and non-photonic electron \raa~and \vtwo. Rigorous \chisq~study is conducted for inclusive pion spectra, and comprehensive azimuthal tomography is applied to pion single particle anisotropy $v_2$ in A+A collisions at both RHIC and LHC.

To elucidate our theoretical predication of quenching observables, we choose to plot \raa~curves without adding error bands, and we will for completeness list the known systematic uncertainties which contribute to hadron spectra for CUJET at present stage in the next paragraph, besides the error from first order DGLV opacity calculation which has been discussed above. The uncertainties associated with elliptic flow \vtwo~can only be partially interpreted from errors in \raa~ because \vtwo~depends on more complex factors such as fluctuation, anisotropy, inhomogeneity, etc. Systematically induced variation of azimuthal anisotropy in our model is subject to future exploration.
\\

A list of systematic uncertainties in CUJET\footnote{Other theoretical and numerical systematic uncertainties may still exist, cf. for instance \cite{oai:arXiv.org:1106.3927,Brick2012}.}:
\begin{itemize}
\item Running scale variation. Increase the running scale by 50\% generates approximately 5\% enhancement of \raa, while decrease the scale by 25\% leads to about 10\% lower \raa, cf. Section~\ref{sec:rc}.

\item Kinematic limits for $\kT$ integration in Eq.~\eqref{fcDGLV} and~\eqref{rcCUJETDGLV}. Depending on the interpretation of energy fraction $x_E$ or light-cone $x_+$ in the DGLV formula, and the treatment of large angle emissions which break down the collinear approximation, fractional energy loss varies. The coupling constant needs to be tuned accordingly $\pm 10$\% at most, cf. Appendix~\ref{sec:kTlimits}.

\item Number of quarkonic flavors $n_f$ for the medium. A rescaling of coupling constant by 6\% perfectly reconciles the scenario of $n_f=0$ and $n_f=2.5$, cf Appendix~\ref{app:systematics}.

\item Fragmentation temperature $T_f$. In terms of \raa, lower $T_f$ to be below freeze-out temperature, typically around 100 MeV, by about 50\% has no significant influence. But increase it by approximately 100\% effectively generates a 20\% weaker coupling, cf. Appendix~\ref{app:systematics}.

\item Initial rapidity density $dN/dy$. In CUJET1.0, the variation of rapidity density significantly changes the magnitude of \raa, but the shape of suppression curves remains semi-stable, thus one can rescale the coupling constant to balance this effect, cf. Appendix~\ref{app:systematics}. However, the results presented in the following sections are calculated from CUJET2.0, within which framework the bulk is assumed to be properly modeled by the 2+1D viscous hydrodynamical fields, therefore variations coming from $dN/dy$ are beyond the scope of CUJET2.0.

\item Thermalization scheme. Fixing initial time $\tau_0$, divergent and free streaming pre-thermal scenario creates approximately an effective 10\% larger and 7\% smaller coupling constant comparing to linear scheme. For heavy flavor \raa, the scheme variation slightly affects the slope of the quenching pattern, however this change is minor when juxtaposed with experimental error bars, cf. Appendix~\ref{app:thermal}.

\item Energy loss fluctuations. The assumed Poisson distribution for radiative energy loss has negligible effects on \raa, but it may significantly influence \vtwo, cf. Appendix~\ref{app:fluc}.

\item Partonic pp spectra variations. Error bands from NLO and FONLL initial cross sections span 5\% at partonic \raa~level, being insignificant to a certain extent. Absolute normalizations of the spectra drop out when calculating ratios such as \raa, but steepnesses matter. Relative steepness between production spectra has influence on the calculation of pion and non-photonic electron spectrum, which is fragmented from gluons and light quarks, and charms and bottoms receptively, cf. Appendix~\ref{app:partonspc}.
\end{itemize}

\subsection{Pion nuclear modification factor}
\label{sec:pion}

Nuclear modification factor $ R_{AA} $ is a key observable at RHIC and LHC relativistic heavy-ion collisions, it describes the relative magnitude of jet quenching. \raa~is defined as the ratio of the quenched A+A spectrum to the unquenched p+p spectrum, scaled according to the number of binary collisions $N_{binary}$:
\be
R_{AA}(p_T) = \dfrac{\dfrac{d\sigma^{AA}}{dp_T}(p_T)}{N_{binary}\;\dfrac{d\sigma^{pp}}{dp_T}(p_T)}
\;\;.
\label{RAAdef}
\ee
We have suppressed the explicit dependence on rapidity $y$ and c.m. energy $\sqrt{s}$ in Eq.~\eqref{RAAdef}. Here the \raa~is understood as azimuthally averaged, and azimuthal angle $\phi$ is integrated out.

\subsubsection{Pion suppression}
\label{sec:pionRAA}

\begin{figure}[!t]
\bc
\includegraphics[width=0.85\textwidth]{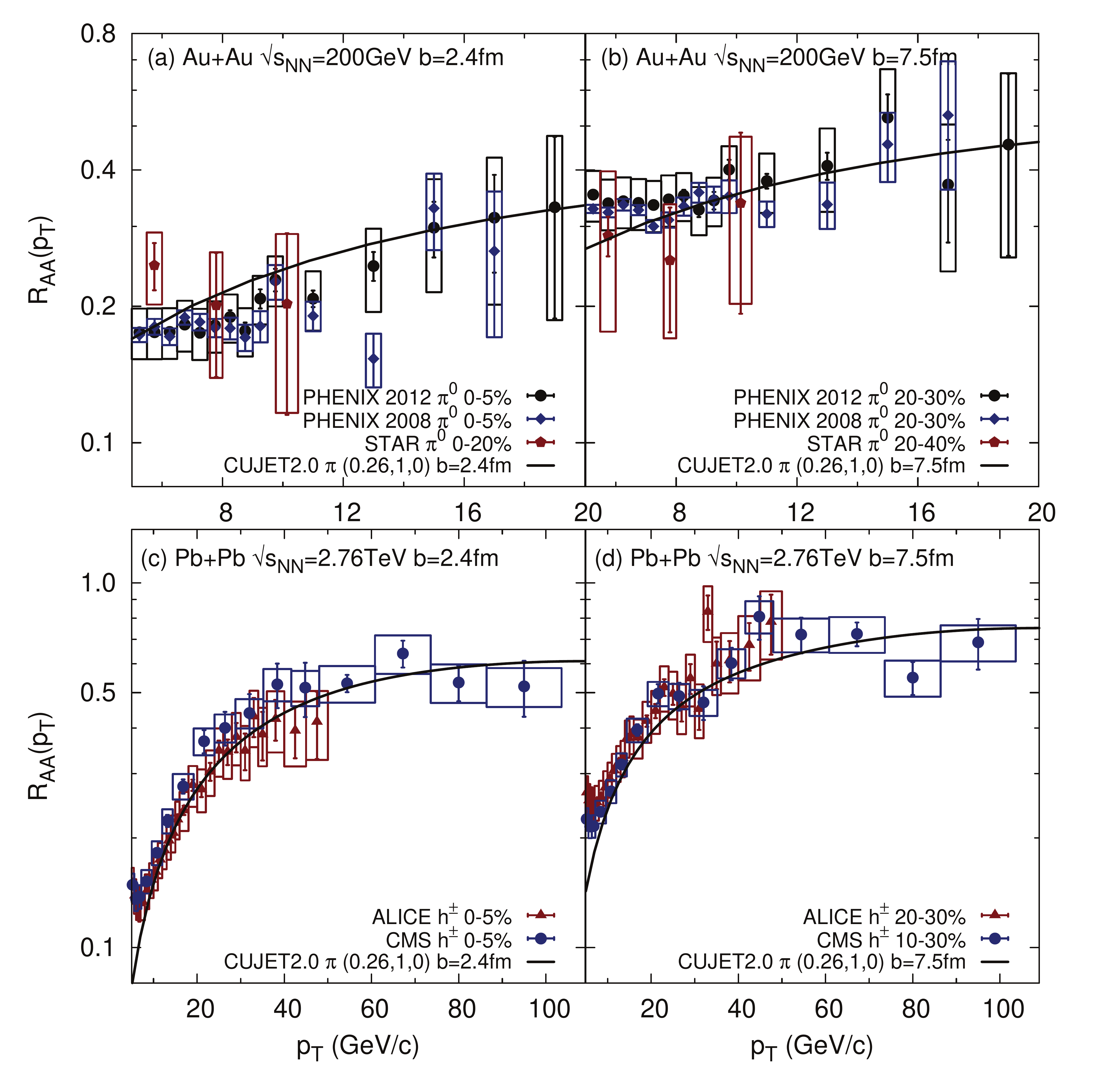}
\caption{\label{fig:RAA_pT} 
		CUJET2.0 inclusive pion nuclear modification factor $ R_{AA} $ versus transverse momentum $ p_T $, comparing with PHENIX \cite{Adare:2008qa,Adare:2012wg} and STAR \cite{Abelev:2009wx} $\pi^0 R_{AA}$ for Au+Au collisions at 200AGeV and (a) 0-5\%(PHENIX)/0-20\%(STAR), (b) 20-30\%(PHENIX)/20-40\%(STAR) centrality; with ALICE \cite{Abelev:2012hxa} and CMS \cite{CMS:2012aa} $h^\pm R_{AA}$ for Pb+Pb collisions at 2.76ATeV and (c) 0-5\%, (d) 20-30\%(ALICE)/10-30\%(CMS) centrality. The $ R_{AA} $ is calculated at leading n=1 order in opacity, and the maximum coupling constant \amax~is constrained by fitting to central RHIC Au+Au data at a reference point $ R_{AA}^{\pi}(p_T=15\rm{GeV/c})= 0.3 $, setting $ \alpha_{max}=0.26 $. For the effective potential in Eq.~\eqref{rcCUJETDGLV}, the parameters are set to be $ f_E=1,f_M=0 $, i.e. dynamical QCD medium with HTL approximation. Impact parameter b=2.4 and 7.5 fm is used in CUJET2.0 to simulate the central and semi-peripheral collisions respectively. The 2+1D viscous hydro grid is generated from VISH2+1 \cite{Song:2008si,Shen:2010uy,Renk:2010qx}, with $ \tau_0=0.6\;\rm{fm/c} $, $ \eta/s=0.08 $, 120 MeV freeze-out temperature, MC-Glauber initial conditions and Lattice QCD s95p-PCE EOS at both RHIC and LHC. Compared to CUJET1.0 \cite{Buzzatti:2012dy} which has a static transverse profile, with an transversely expanding medium in CUJET2.0, pion $ R_{AA} $ flattens out more clearly at high $ p_T $ region at LHC. The strong coupling however decreases from $ \alpha_{max}=0.4 $ in running coupling CUJET1.0 to $ \alpha_{max}=0.26 $ in CUJET2.0, indicating longer path length of jet in a transversely expanding medium overrides the reduction of density, and contribute to overall enhanced quenching.
		}
\ec
\end{figure}

We calculate in CUJET2.0 the inclusive pion \raa~at RHIC Au+Au 200AGeV and LHC Pb+Pb 2.76ATeV, central ($b=2.4$ fm) and semi-peripheral ($b=7.5$ fm) collisions, the results are illustrated in Fig.~\ref{fig:RAA_pT}.
Theoretical \raa~curves are compared side by side with corresponding experimental measurements of charged hadron suppression factor at ALICE \cite{Abelev:2012hxa} and CMS \cite{CMS:2012aa}, and neutral pion suppression factor at PHENIX \cite{Adare:2012wg}. In the radiative energy loss sector, DGLV opacity series is calculated to first order, with $f_E=1, f_M=0$ in Eq.~\eqref{rcCUJETDGLV} interpolating the pure dynamical scattering potential in the hard thermal loop scenario. Maximum coupling constant \amax~is adjusted to $ \alpha_{max}=0.26 $ to fit the \raa~reference point at RHIC central collision.

After adjusting $ \alpha_{max}=0.26 $ to match the calculated pion nuclear modification factor from CUJET2.0 with the $p_T=15\rm{GeV/c}$ reference point of experimentally measured inclusive neutral pion suppression factor from PHENIX 2012 \cite{Adare:2012wg} Au+Au 200AGeV central collisions, the rest of the \raa~curve in the range of $p_T=5 \sim 20\rm{GeV/c}$ shows reasonable compatibility with both PHENIX 2008 and 2012 data. More importantly, when move on to simulate RHIC 20-30\% centrality collisions by changing solely the impact parameter to $b=7.5$ fm and fixing all other parameters in CUJET2.0, the theoretical \raa~result demonstrates even better agreement with experimental data.

When switch to LHC after constrained all CUJET model parameters with RHIC data, fixed coupling CUJET1.0 used to encounter difficulties explaining the surprising transparency of the QGP at LHC high $p_T$ region \cite{Buzzatti:2011vt}, though this problem is eased by running coupling CUJET1.0 which has effectively reduced coupling strength at high energies, the pion \raa's steep rising and successive flattening pattern at LHC remains only partially explained \cite{Buzzatti:2012dy}. This issue is fully solved in CUJET2.0 which has a more realistic bulk evolution profile. As shown in Fig.~\ref{fig:RAA_pT}(c)(d), at both ALICE and CMS, both central and semi-peripheral collisions, the CUJET inclusive pion \raa~curves seamlessly simulate both the low $p_T$ steep rising and high $p_T$ saturating behavior of $\pi^0$ or $h^\pm$ nuclear suppression factor.

One of the most appreciable signatures of CUJET2.0 which has 2+1D viscous hydrodynamic background is the drastic reduction of the strong coupling constant compared to CUJET1.0 where transversely a static Glauber medium is assumed. In running coupling CUJET2.0 calculations, the maximum coupling strength $ \alpha_{max}$ is adjusted to $ \alpha_{max}=0.26 $ to fit inclusive pion \raa's at both RHIC and LHC, central and mid-central A+A collisions, this value is distinguishably smaller than running coupling CUJET1.0's $ \alpha_{max}=0.4 $. Compared to CUJET1.0's static Glauber bulk, CUJET2.0's average medium density is reduced because of the transversely expanding hydro fluids, and one would intuitively expect less quenching in such a medium. However, the fact that the effective strong coupling constant demands a tremendous reduction in itself to generate the same hadron suppression factor in CUJET2.0 as in CUJET1.0 indicates that there is another factor contributes more remarkably to jet quenching and in fact dominates the modification of parton shower in the expanding medium. This key contributor can only be the path length of propagating jet in QGP. The longer jet path length generated from 2+1D viscous hydro fields in CUJET2.0 plays a decisive role on parton energy loss, it overrides the effect of less quenching resulted from diminished medium density, and induces an overall more strongly quenched jet spectra in relativistic heavy ion collisions.

More comments about the small value of best fit \amax ($=0.26$) in the CUJET2.0 HTL model. We list in the beginning of Sec.~\ref{sec:pion} factors that may contribute to the systematic uncertainties of our calculation, for example, the default choice of freeze-out temperature $T_f=120$ MeV (which can in fact contribute 10\% enhancement of \amax~if increased from hadronic freeze-out to critical temperature $160$ MeV, a detailed discussion about this effect is included in Appx.~\ref{app:systematics}) and the computation of DGLV opacity series to the first order (in a small region of phase space where gluon's energy fraction $x$ and transverse momentum $k_T$ are small, calculate n=1 may lead to overestimation of radiative energy loss, detailed discussions are in Appx.~\ref{app:conv}). However, given a wide range of origins of systematic uncertainties, the absolute value of \amax~itself makes physical sense only semi-quantitatively. But the relative value of \amax, e.g. between CUJET1.0 and CUJET2.0, does enable quantitative physical statements. Significantly decreased coupling strength is observed in a transverse expanding medium (CUJET2.0) compared to the transversely static case (CUJET1.0), and we can therefore conclude the longer jet path length in 2+1D viscous hydro fluids is dominating, it overrides the reduced medium density and contribute overall more energy loss. 
 
\subsubsection{Chi square per degree of freedom}
\label{sec:chi^2}

Calculating relative variance per degree of freedom \chisq~is one of the best quantitative methods to test to what extent the theoretical \raa~results are in agreement with experimental measurements. This quantity is the average of relative variance, which is the ratio of squared difference between experimental data point and its theoretical counterpart to the quadratic sum of all theoretical and experimental statistical and systematic errors associated with that point, over all selected data. It is defined as follows:
\be
\chi^2/d.o.f.=\sum_{i=1}^{N}\Big[\dfrac{(V_{th}-V_{exp})^2}{\sum_t \sigma_t^2}\Big]_i \Big/N
\;\;.
\ee
Where $V_{th}$ is the theoretical value, $V_{exp}$ is the experimental value, $ \sum_t \sigma_t^2 $ stands for the quadratic sum over all types of errors that one chosen point has, and N is the number of data points selected. 

\begin{figure*}[!t]
\bc
\includegraphics[width=0.475\textwidth]{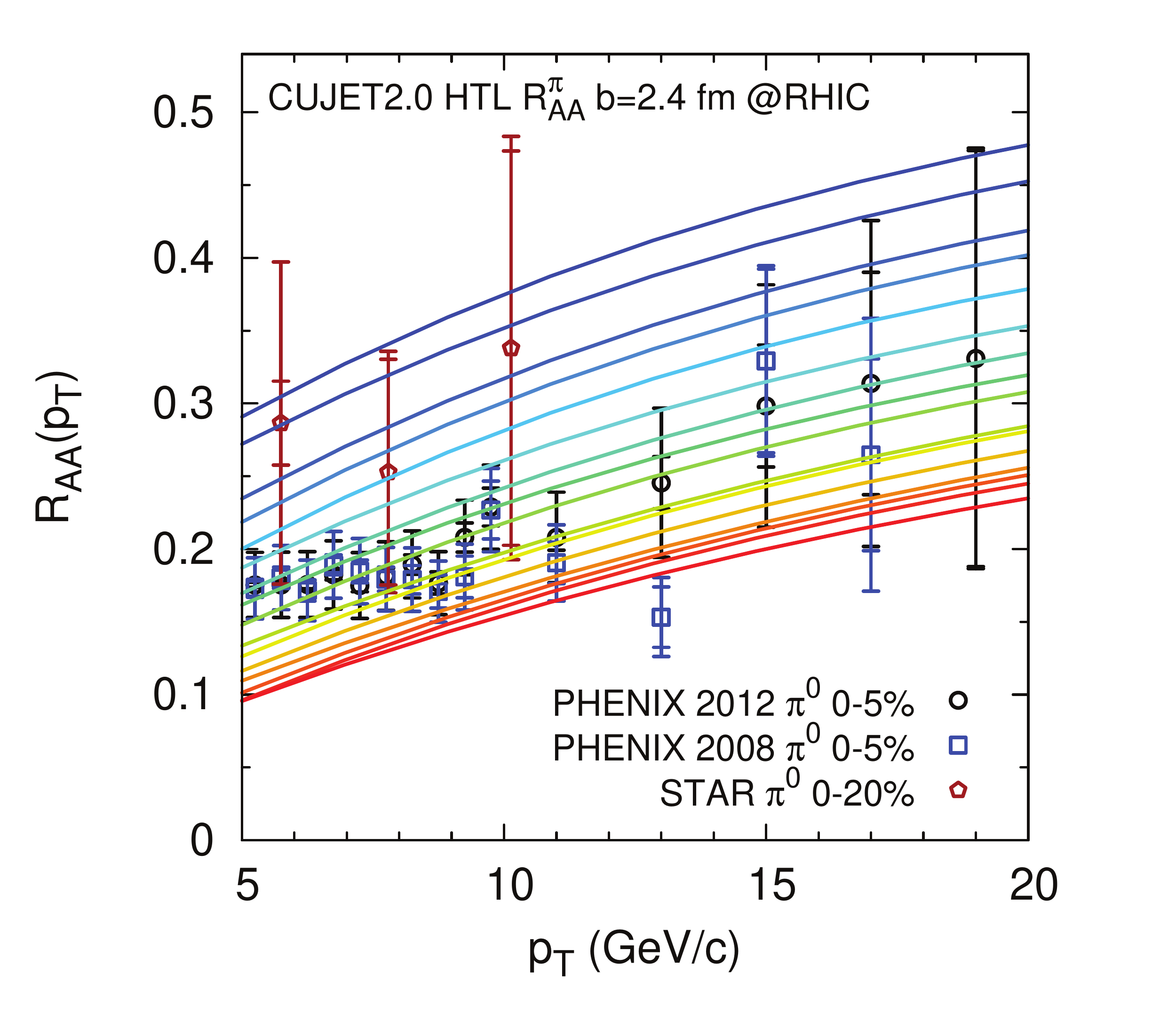}
\includegraphics[width=0.475\textwidth]{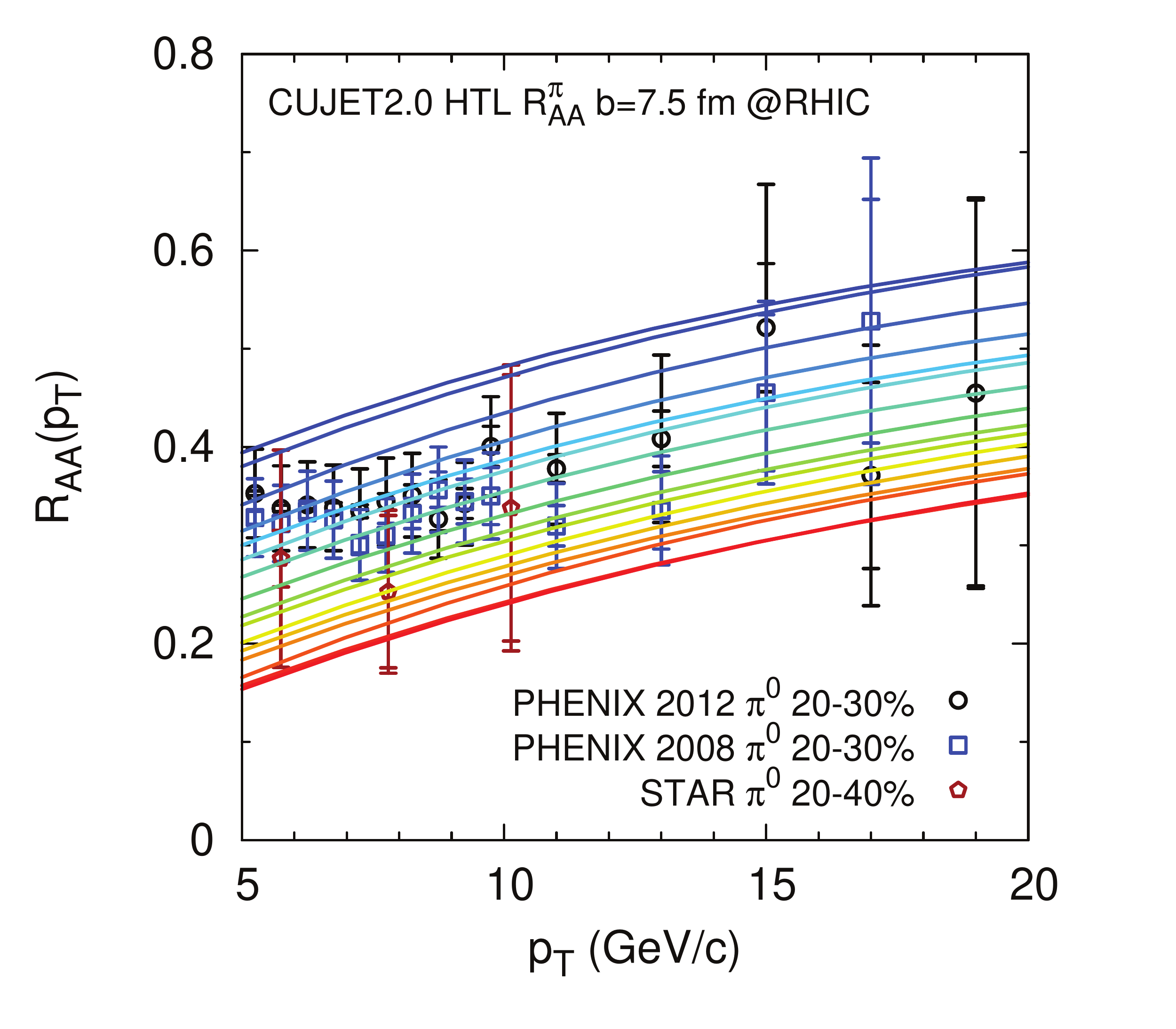}
\includegraphics[width=0.475\textwidth]{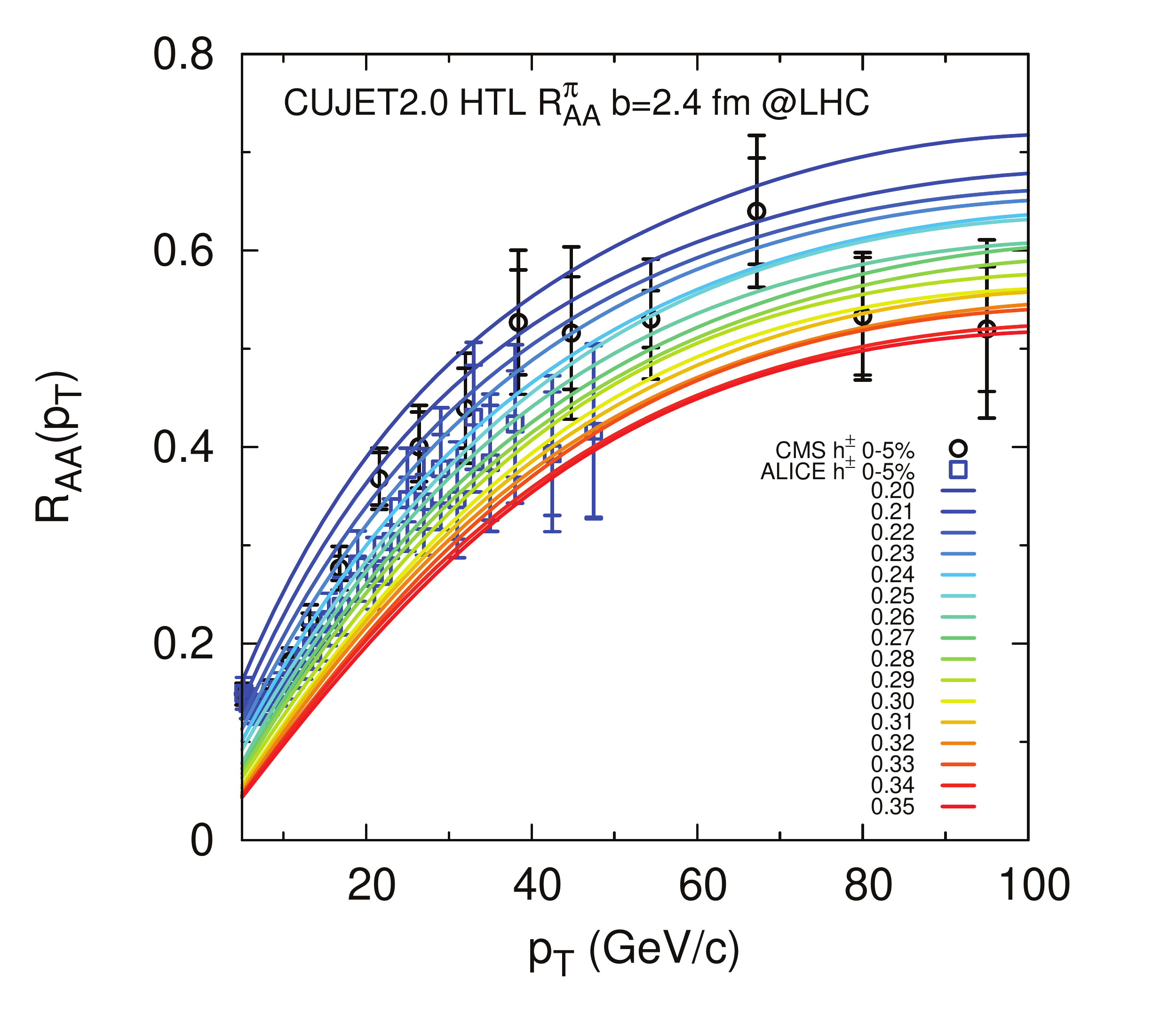}
\includegraphics[width=0.475\textwidth]{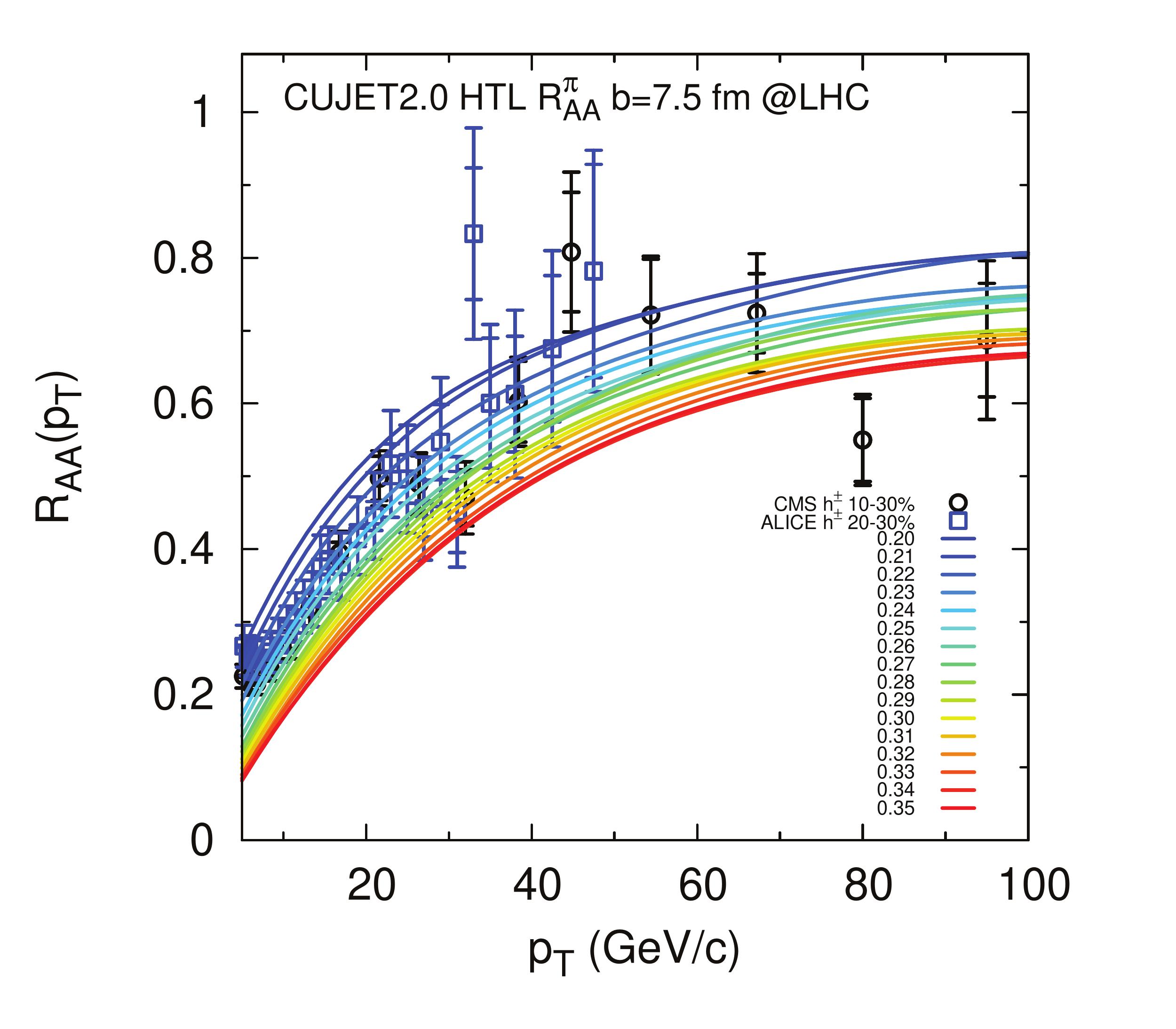}
\caption{\label{fig:Multi_Alf_HTL} 
		CUJET2.0 results for pion nuclear modification factor $ R_{AA} $ versus $ p_T $, with maximum coupling strength $ \alpha_{max}=0.20\sim0.35 $ in the dynamical HTL scenario, at RHIC Au+Au 200AGeV (top panels) and LHC Pb+Pb 2.76ATeV (bottom panels), central (b=2.4fm, left panels) and semi-peripheral (b=7.5fm, right panels) collisions. Experimental references are: PHENIX 2008 \cite{Adare:2008qa} and 2012 \cite{Adare:2012wg} Au+Au 200AGeV $\pi^0 R_{AA}$ with 0-5\% (top left) and 20-30\% (top right) centrality; STAR \cite{Abelev:2009wx} Au+Au 200AGeV $\pi^0 R_{AA}$ with 0-20\% (top left) and 20-40\% (top right) centrality;  ALICE \cite{Abelev:2012hxa} Pb+Pb 2.76ATeV $h^\pm R_{AA}$ with 0-5\% (bottom left) and 20-30\% (bottom right) centrality; CMS Pb+Pb 2.76ATeV $h^\pm R_{AA}$ with 0-5\% (bottom left) and 10-30\% (bottom right) centrality. The hydro grid being used is the same as in Fig.~\ref{fig:RAA_pT}. Despite the existence of multi-scale running coupling, the magnitude of jet quenching monotonically enhances with increasing \amax~in both central and semi-peripheral collisions at both RHIC and LHC. 
		}
\ec
\end{figure*}

We vary the maximum coupling constant \amax~in CUJET2.0 from 0.20 to 0.35 with 0.01 steps, and maintain the dynamical HTL scenario by fixing $ f_E=1, f_M=0 $, to study the most compatible CUJET2.0 one parameter (\amax) fit at RHIC and LHC, and test the consistency of the model at different $\sqrt{s_{NN}}$'s. Fig.~\ref{fig:Multi_Alf_HTL} shows the pion \raa~curves with those different \amax~values at RHIC Au+Au 200AGeV and LHC Pb+Pb 2.76ATeV central ($b=2.4$ fm) and semi-peripheral ($b=7.5$ fm) collisions. The experimental data being compared with are PHENIX 2008 \cite{Adare:2008qa}, 2012 \cite{Adare:2012wg} and STAR \cite{Abelev:2009wx} $\pi^0$ \raa~at RHIC; and ALICE \cite{Abelev:2012hxa} and CMS \cite{CMS:2012aa} $h^\pm$ \raa~at LHC.

In all four panels of Fig.~\ref{fig:Multi_Alf_HTL}, focusing on $p_T<40$ GeV region, the magnitude of inclusive hadron suppression has near uniform increment with an uniformly increasing maximum coupling \amax, with exception at relatively large \amax's where the spacing between \raa~curves becomes smaller, but the monotonic lowering of \raa~proceeds. Since the saturation scale $ Q_{min} $ for the running coupling depends solely on the maximum coupling constant \amax, i.e., $ Q_{min}= \Lambda_{QCD} \rm{Exp}\{2\pi/9 \alpha_{max}\} $. Take $ \alpha_{max}=0.20, 0.25, 0.30~\rm{and}~0.35 $ for example, the saturation scale $ Q_{min}=6.56, 3.26, 2.05~\rm{and}~1.47 \rm{GeV}$ respectively. At relatively low \amax, because of the large saturation scale, the strong coupling recovers asymptotically the fixed coupling scenario up to a relatively high energy, this explains the near uniform increment in the panels. The influence of running coupling is substantial at relative high \amax~where the minimum running scale is low, in that situation the logarithmic decay of coupling strength resulted from vacuum running shrinks the spacing of \raa's more effectively.

A significant phenomenon shows up in the bottom panels of Fig.~\ref{fig:Multi_Alf_HTL} -- the flattening pattern (slope) of \raa~in high $p_T$ ($p_T>50$ GeV) region at LHC is almost independent of the choice of \amax, this implies the relative insensitivity of \raa~saturation to the running coupling effect, and we can therefore exclude to a certain extent the influence of running on the saturation of \raa~for ultra-high energy jet. Note in \cite{Buzzatti:2012dy}, the previous calculation of multi-scale running coupling combined CUJET1.0, whose medium assumes static Glauber transverse profile plus 1+1D Bjorken longitudinal expansion, did not exhibit a clear signature of \raa~flattening. Therefore, evident \raa~saturation comes largely from the kinematics in a medium with both transverse and longitudinal expansion, which feature distinguishes CUJET2.0 from running coupling CUJET1.0. A dynamically transverse expanding medium, for instance a 2+1D viscous hydro fluid, plays a very important role in the \raa~flattening of high $p_T$ jet in ultra-relativistic heavy ion collisions, and shall receive more attention in predicting jet quenching observables in A+A collisions from pQCD energy loss models.

\begin{figure*}[!t]
\bc
\includegraphics[width=0.475\textwidth]{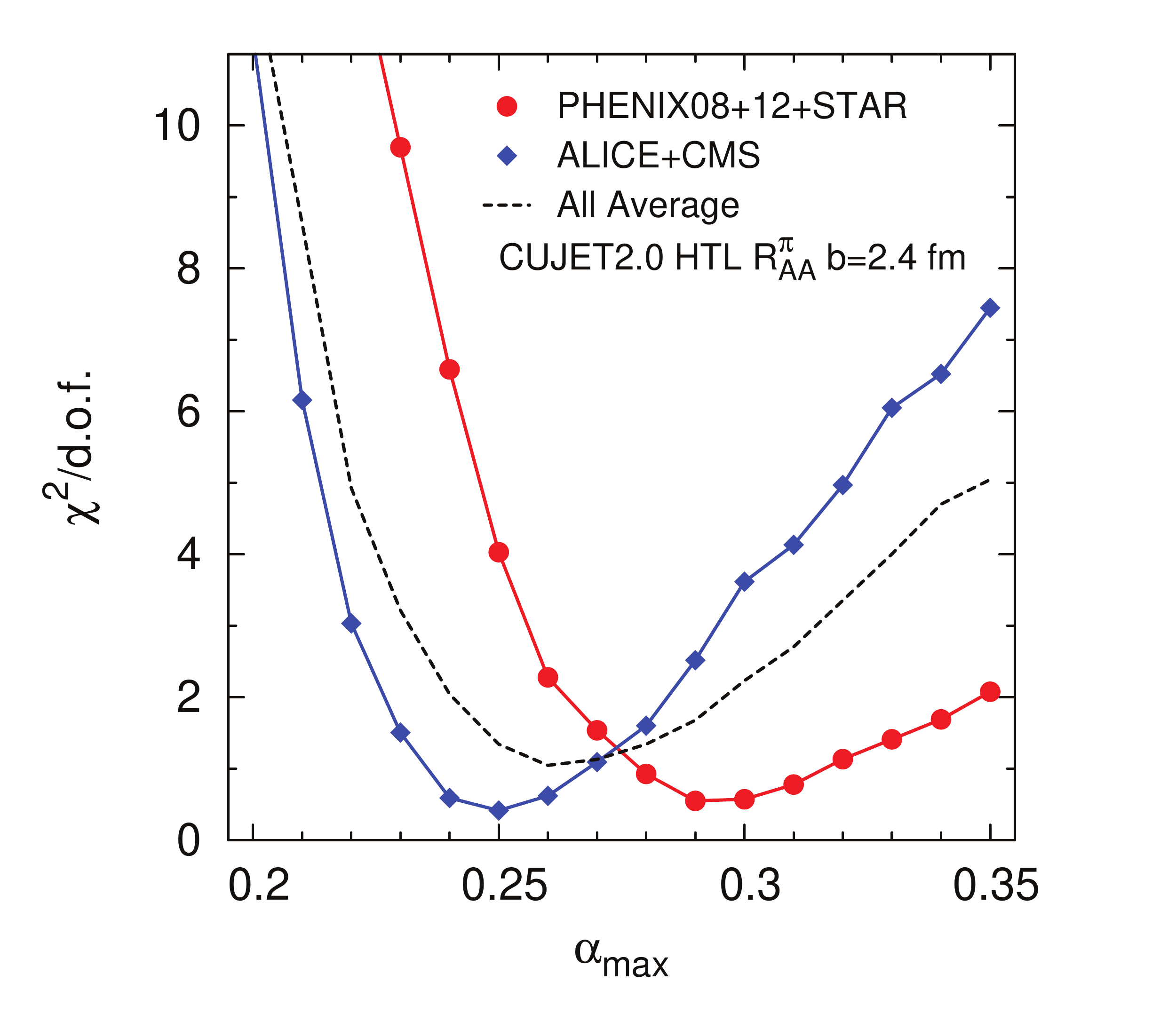}
\includegraphics[width=0.475\textwidth]{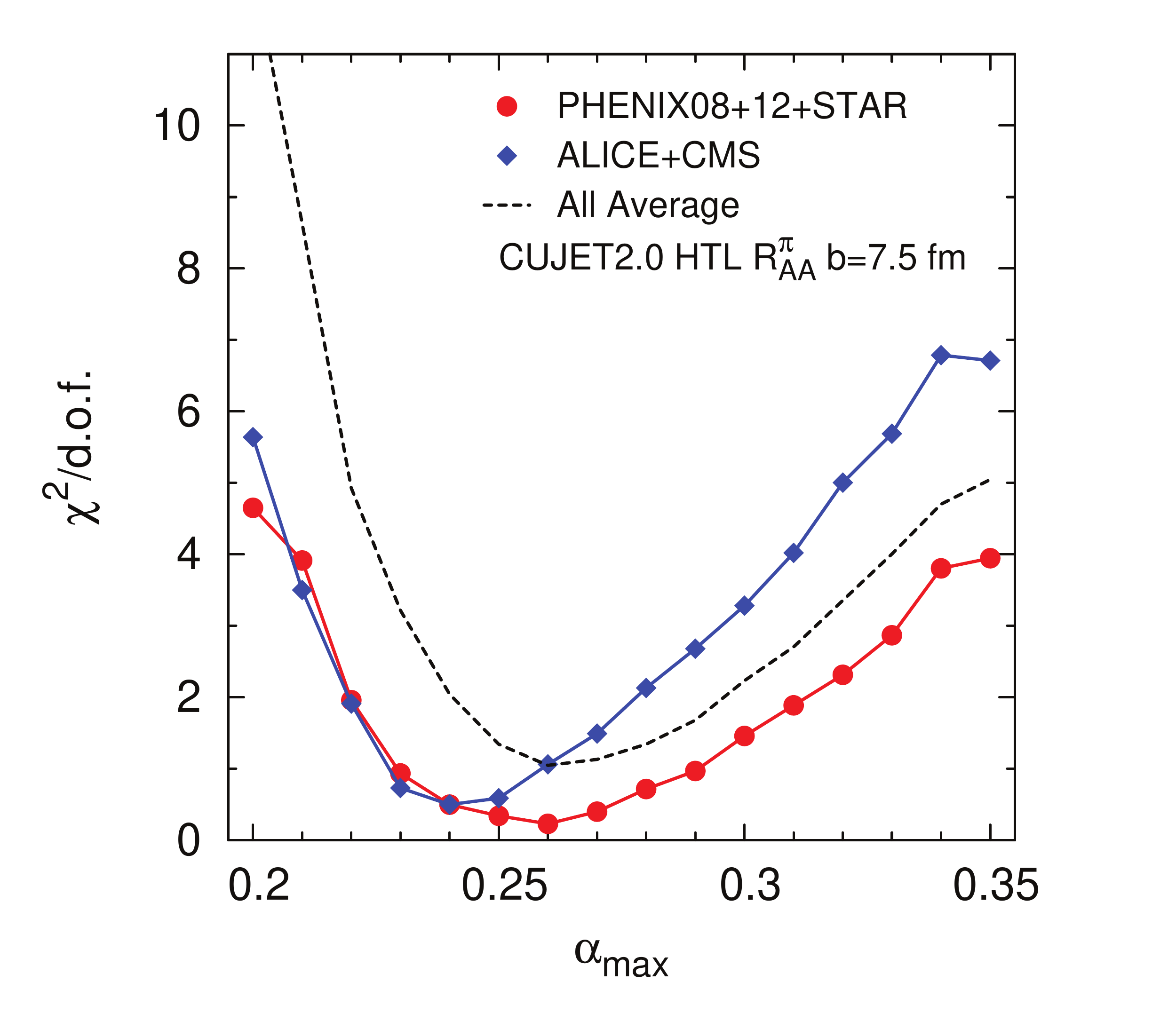}
\caption{\label{fig:Chi_square_HTL} 
		$ \chi^2/d.o.f. $ versus $ \alpha_{max} $ calculated from Fig.~\ref{fig:Multi_Alf_HTL} at RHIC (red) and LHC (blue), central ($b=2.4$ fm, left panel) and semi-peripheral ($b=7.5$ fm, right panel) collisions. The average over all four \chisq's is plotted as a reference in both panels (dashed black). Data from $ p_T>8$ GeV is used for safer preservation of DGLV's basic eikonal and soft approximations. At RHIC, PHENIX 2008, 2012 and STAR data \cite{Adare:2008qa,Adare:2012wg,Abelev:2009wx}, at LHC, ALICE and CMS data \cite{Abelev:2012hxa,CMS:2012aa} are compared respectively. The $ \alpha_{max}$ ranges for $\tilde{\chi}^2<1$ and $\tilde{\chi}^2<2$ ($\tilde{\chi}^2\equiv\chi^2/d.o.f.$) at RHIC and LHC are shown in Table~\ref{AlphaMaxChiSquare}. If allow 1.5 standard deviations per d.o.f., interpreting from the average curve, the most consistent CUJET2.0 HTL model at both RHIC and LHC has $ \alpha_{max}=0.25-0.27 $. If let average $ \chi^2/d.o.f.<2 $, then $ \alpha_{max}=0.23-0.30 $. The small value of the strong coupling constant is partially caused by the dominating longer jet path length in the transverse expanding medium, as discussed in Section~\ref{sec:pionRAA}.
		}
\ec
\end{figure*}

The essential eikonal and soft approximation in dynamical DGLV opacity expansion may break down at low $p_T$ region, hence for the purpose of \chisq~we choose experimental results in the range of $ p_T>8~\rm{GeV} $ to compare with CUJET2.0 \raa~curves. Note $ \chi^2/d.o.f.(\alpha_{max})<2 $ is an indicative signature of model consistency, and the constrained \amax~range should be independent of whether $ p_T>5~\rm{GeV} $ or $ p_T>8~\rm{GeV} $ is chosen as long as the minimum $ p_T $ is sufficient for preserving basic assumptions of the CUJET2.0 model and number of points being selected at high $ p_T $ is large enough. Hence for safer comparison we choose $ p_T>8~\rm{GeV} $, and Fig.~\ref{fig:Chi_square_HTL} shows \chisq~vs \amax~at RHIC (PHENIX08+12+STAR \cite{Adare:2008qa,Adare:2012wg,Abelev:2009wx}) and LHC (ALICE+CMS \cite{Abelev:2012hxa,CMS:2012aa}), in both central ($b=2.4$ fm) and semi-peripheral ($b=7.5$ fm) collisions.
And for better analyzing \chisq~curves in Fig.~\ref{fig:Chi_square_HTL}, we list detailed $\alpha_{max}$ ranges with $\tilde{\chi}^2<1$ and $\tilde{\chi}^2<2$ ($\tilde{\chi}^2\equiv\chi^2/d.o.f.$) at RHIC and LHC in Table~\ref{AlphaMaxChiSquare}.
\begin{table}[!t]
\centering
    \begin{tabular}{ | l | c | c | c | c | }
    \hline
    $\alpha_{max}$ & RHIC $\tilde{\chi}^2<1$ & LHC $\tilde{\chi}^2<1$ & RHIC $\tilde{\chi}^2<2$ & LHC $\tilde{\chi}^2<2$ \\ \hline
    $b=2.4$ fm & 0.28-0.32 & 0.24-0.27 & 0.26-0.35 & 0.23-0.28 \\ \hline
    $b=7.5$ fm & 0.23-0.29 & 0.23-0.25 & 0.22-0.31 & 0.22-0.27 \\
    \hline
    \end{tabular}
\caption{The ranges of $ \alpha_{max}$ with $\tilde{\chi}^2<1$ and $\tilde{\chi}^2<2$ ($\tilde{\chi}^2\equiv\chi^2/d.o.f.$) for the curves shown in Fig.~\ref{fig:Chi_square_HTL}.  If strictly limit $\tilde{\chi}^2$ to be less than 1, at $b=2.4$ fm central collisions, CUJET2.0 results for RHIC and LHC have 0.01 offset in \amax, while at $b=7.5$ fm semi-peripheral collisions, the results are perfect consistent in $\alpha_{max}=0.23-0.25$ range at RHIC and LHC. Notice also the RHIC and LHC averaged best fit \amax~value at semi-peripheral collisions is approximately 0.03 lower than central collisions, and at either centrality LHC best fit \amax~is about 0.03 lower than RHIC. If allow $\tilde{\chi}^2<2$, one can find that the intersection region of \amax~for all four collisions, i.e. RHIC Au+Au 200AGeV and LHC Pb+Pb 2.76ATeV combining $b=2.4$ fm and $b=7.5$ fm, is $\alpha_{max}=0.26-0.27$. This range of \amax~($0.26-0.27$) coincides almost perfectly with the range interpreted from $\tilde{\chi}^2<1.5$ for the average curve in Fig.~\ref{fig:Chi_square_HTL} ($\alpha_{max}=0.25-0.27$).}
\label{AlphaMaxChiSquare}
\end{table}

The combination of Fig.~\ref{fig:Chi_square_HTL} and Table~\ref{AlphaMaxChiSquare} provides quantitative information about the consistency of CUJET2.0 HTL model in various A+A collision configurations. We see that if strictly constrain $\tilde{\chi}^2$ to be less than 1, for $b=2.4$ fm central collisions, CUJET2.0 results at RHIC and LHC have 0.01 offset in \amax, and for $b=7.5$ fm semi-peripheral collisions, the results are in perfect agreements with RHIC and LHC at $\alpha_{max}=0.23-0.25$ range. We also notice the averaged best fit \amax~value at RHIC and LHC in semi-peripheral collisions is around 0.03 lower than central collisions, and at either centrality the best fit LHC \amax~is approximately 0.03 lower than RHIC. These observations will trigger useful analysis in Section~\ref{sec:v2}.

At present stage, in the CUJET2.0 HTL scenario, if restrict separate maximum $\tilde{\chi}^2$ to be 2, we find that the intersecting \amax~range for all four collisions at RHIC and LHC, i.e. Au+Au 200AGeV and Pb+Pb 2.76ATeV mix with $b=2.4$ fm and $b=7.5$ fm, is $\alpha_{max}=0.26-0.27$. This range of \amax~($0.26-0.27$) coincides almost ideally with the range interpreted from $\tilde{\chi}^2<1.5$ for the average curve in Fig.~\ref{fig:Chi_square_HTL} ($\alpha_{max}=0.25-0.27$), indicating CUJET2.0 model's rigorous consistency at varied A+A collisions, spanning a broad range of $\sqrt{s}$ and b\footnote{Note furthermore the similar curvature of the RHIC and LHC $ \chi^2/d.o.f.(\alpha_{max}) $ parabolas at both centralities, this is a circumstantial evidence of the consistency of CUJET2.0.}.

Based on all the above discussions, we conclude that from testing CUJET2.0 HTL scenario's agreement with centrality dependent neutral pion and charged hadron suppression factors at RHIC and LHC in the mid-rapidity region, the maximum coupling constant in the model is constrained to $\alpha_{max}=0.25-0.27$, in which range the averaged \chisq~is strictly less than 1.5; if allow average $ \chi^2/d.o.f.<2 $, then $ \alpha_{max}=0.23-0.30 $. As discussed in Section~\ref{sec:pionRAA}, the small $ \alpha_{max} $ value itself can be attributed to the dominating longer jet path length feature for the parton shower modification in a transversely expanding medium. Notice in the calculation of \chisq, we did not include in it any intrinsic CUJET2.0 systematic uncertainties which were discussed at the beginning of Section~\ref{sec:numerics}, this suggests the best fit \amax~region can in fact be broadened after taking those factors into account and by the mean time maintain the stringent \chisq~limit. Given the complexity of estimating the complete systematic errors in the model, we will use $\alpha_{max}=0.25-0.27$ to extract the effective jet transport coefficient in Section~\ref{sec:qhat}, and stick to $\alpha_{max}=0.26$ to extrapolate the suppression pattern of open heavy flavors and heavy flavor leptons in Section~\ref{sec:heavy}.

\subsubsection{Jet transport coefficient}
\label{sec:qhat}

\begin{figure*}[!t]
\bc
\includegraphics[width=0.9\textwidth]{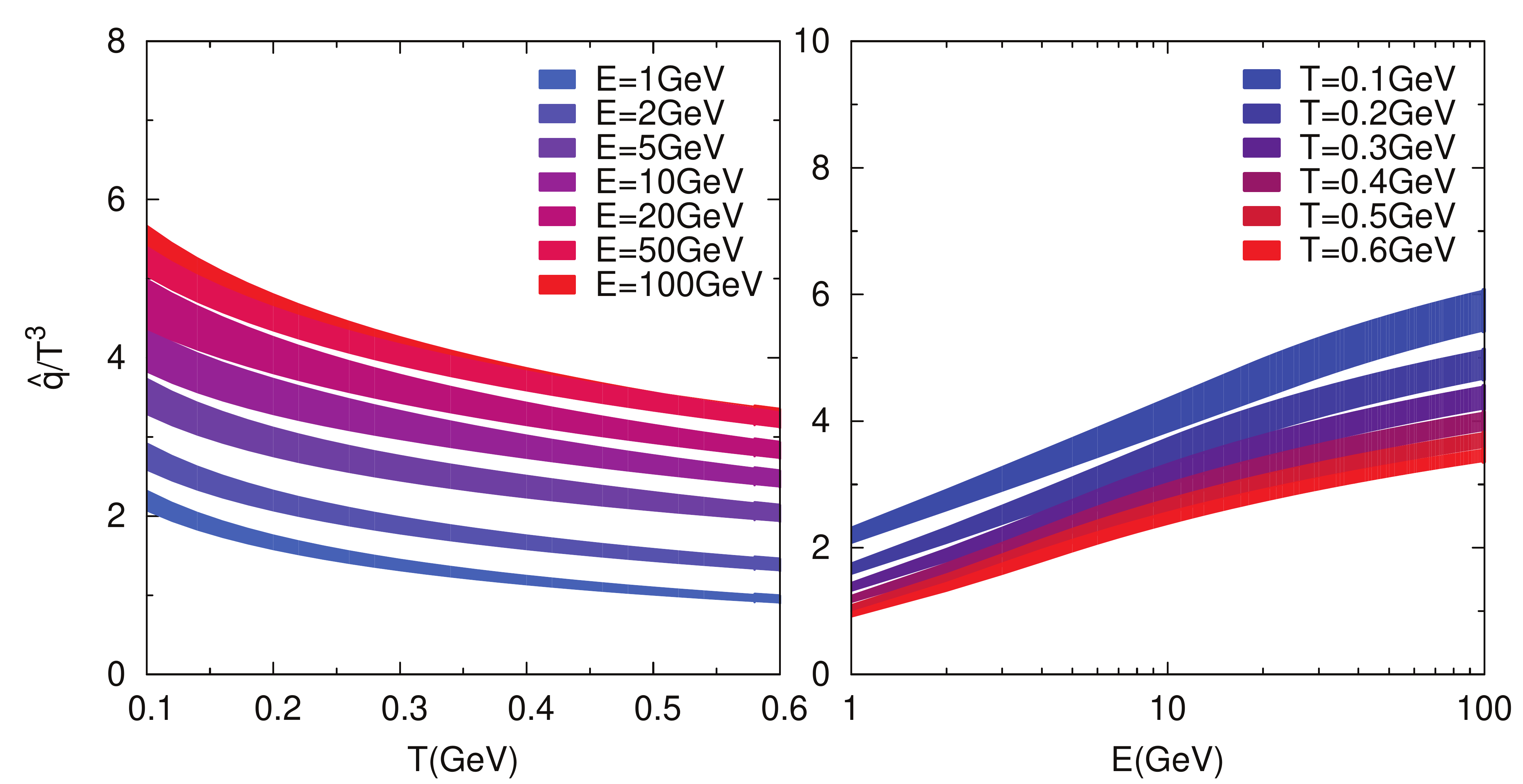}
\caption{\label{fig:qhat} 
		The absolute jet transport coefficient $ \hat{q}/T^3 $ calculated in CUJET2.0 according to Eq.~\eqref{eq:qhat1}\eqref{eq:qhat2} with parameters $ \alpha_{max}=0.25-0.27, f_E=1, f_M=0 $, which set of parameters generates consistent fits to neutral pion and charged hadron suppression factor \raa~at both RHIC and LHC both central and semi-peripheral A+A collisions. $ \hat{q}/T^3 $ versus QGP temperature T at fixed incoming jet energy E is plotted on the \textit{left} panel; the \textit{right} panel shows $ \hat{q}/T^3 $ versus E at fixed T. When E is fixed, the decrease of $ \hat{q}/T^3 $ with the increasing T follows approximately a logarithmic law, indicating $\hat{q}$ itself gains slightly slower than medium density with rising T in CUJET2.0. This feature however still suggests there is more transverse momentum transfer per mean free path at higher temperature, as intuitively expected. When T is fixed, the logarithmic E dependence of $\hat{q}/T^3$ at high energy region comes naturally from the kinematic limit of the exchanged transverse momentum. 
		} 
\ec
\end{figure*}

The suppression of hadrons at large $p_T$ is understood to be caused by scatterings of the leading parton with color charges in the near thermal QGP. This process can be characterized by the jet transport coefficient $\hat{q}$, which is defined as the squared average transverse momentum exchange per unit path length. CUJET2.0 treats thermal excitations in the assumed homogeneous QCD medium as partonic quasi-particles, and the transport parameter $\hat{q}$ in CUJET2.0 is related to the effective partonic differential cross section by the relation:
\be
\hat{q}(E,T; \alpha_{max},f_E,f_M)=\rho(T)\int_{0}^{4ET} d\bq^2 \bq^2 \dfrac{d\sigma_{\rm{eff}}}{d\bq^2}\;,
\label{eq:qhat1}
\ee
where the energy E and temperature T dependence comes in naturally from the partonic kinematics and plasma density. In CUJET2.0, $\hat{q}$ depends also on the maximum strong coupling constant \amax, as well as electric and magnetic screening mass deformation parameters $(f_E,f_M)$, all of which originate from the effective cross section of the quark-gluon process:
\be
\frac{d\sigma_{\rm{eff}}}{d\bq^2}=\frac{\alpha^2_{\rm s} (\bq^2)(f_E^2-f_M^2)}{(\bq^2{+}f^2_E \mu^2(T))(\bq^2{+}f^2_M \mu^2(T))}\;\;,
\label{eq:qhat2}
\ee
with the Debye mass $\mu(T)=T\sqrt{4 \pi \alpha_s(4T^2) (1+n_f/6)}$. Note here the temperature is non-local. Eq.~\eqref{eq:qhat2} differs from the effective scattering potential in Eq.~\eqref{rcCUJETDGLV} where the same cross section form appears but varies with local temperature $T(\vz)$. We assume $ \rho\sim2T^3 $ for an idealized uniform thermal equilibrated medium, and calculate the CUJET2.0 jet transport coefficient $\hat{q}$ according to Eq.~\eqref{eq:qhat1}\eqref{eq:qhat2}, with $ \alpha_{max}=0.25\sim0.27 $ in the HTL $f_E=1, f_M=0$ approximation, which parameters are derived in Section~\ref{sec:chi^2} through rigorous \chisq~consistency tests at various A+A collision configurations. The variations of the absolute jet transport parameter $\hat{q}/T^3$ with energy $E$ and temperature $T$ are illustrated in Fig.~\ref{fig:qhat}.

In the left panel of Fig.~\ref{fig:qhat}, it is shown for an initial quark jet with energy $E=10$ GeV, in the typical temperature range reached by RHIC for most central Au+Au collisions, i.e. $180\sim370~\rm{MeV}$, CUJET2.0 has $\hat{q}/T^3 \approx 3.8$; in the typical temperature range reached by LHC for most central Pb+Pb collisions, i.e. $300\sim470~\rm{MeV}$, CUJET2.0 has $\hat{q}/T^3 \approx 3.5$. Both values are consistent with not only LO pQCD estimates, but also the jet transport parameters extracted from HT-BW, HT-M, MARTINI and McGill-AMY models fitting to the same set of experimental hadron suppression factors at RHIC and LHC A+A central collisions \cite{Burke:2013yra}. At various initial jet energies, the reduction of $ \hat{q}/T^3 $ with rising T invariantly follows an approximate logarithmic law. Since $ \rho\sim2T^3 $, this logarithm indicates $\hat{q}$ grows slightly slower than medium density with increasing T, but there is still more elastic transverse momentum transfer per mean free path at higher temperature, as intuitively expected.

For an idealized static equilibrium QGP with fixed temperature T, as illustrated in the right panel of Fig.~\ref{fig:qhat}, the transverse momentum transfer between the quark jet and dynamical scattering centers shows logarithmic dependence on initial jet energy if the $T^3$ contribution from the medium density is factored out. This is expected from the kinematic limit of transverse momentum exchange, i.e. $(\bq^2)_{max}=4ET$ in Eq.~\eqref{eq:qhat1}. On the other hand, for quark jet with fixed initial energy E, the absolute jet transport coefficient $\hat{q}/T^3$ drops at an diminished rate when temperature grows and reaches high T region. This is expected from the Debye mass $\mu$'s temperature dependence, i.e. $\mu(T)=T\sqrt{4 \pi \alpha_s(4T^2) (1+n_f/6)}$. The thermal coupling has negligible contribution until when T is high, at that time the logarithmic decay of the coupling strength will weaken the linear increase of the Debye mass with rising temperature.

The QGP produced in heavy ion collisions is interpreted as strongly interacting medium \cite{Gyulassy:2004zy}, whose collective flow is known to be well described by relativistic hydrodynamics with a negligible shear viscosity, and effective perturbation theory may not be applicable to study interactions in a medium which is not dominated by quasi-particles. In \cite{Majumder:2007zh}, the authors derived a general expression relating the jet quenching parameter $\hat{q}$ with the shear viscosity $\eta$ of a weakly coupled QGP, and the deviation from this relation is conjectured to be a more broadly valid measure of ``strong coupling'' of the medium than considering solely the shear viscosity divide by entropy density \etas. The relation is expressed as follows:
\be
\dfrac{\eta}{s} \begin{cases} \approx 1.25T^3/\hat{q} & \mbox{for weak coupling}\;, \\
\gg 1.25T^3/\hat{q}  & \mbox{for strong coupling}\;.
\end{cases}
\label{eta/s}
\ee
In high energy region where the QCD coupling is supposed to be weak, the simplified CUJET2.0 $\hat{q}$ calculation from Eq.~\eqref{eq:qhat1} and~\eqref{eq:qhat1} shows $\hat{q}/T^3 \approx 3.7$ for typical temperatures reached by RHIC and LHC, and $1.25T^3/\hat{q}\approx4.2/4\pi$. This value is larger than the quantum limit $\eta/s=1/4\pi$, or the MC-Glauber VISH2+1's $\eta/s=0.08$ which is extracted from fitting to hadron spectra and harmonics at low $p_T$.

One way to reconcile the discrepancy is proposed in \cite{Burke:2013yra}, where the authors suggest lattice calculation indicates that the non-perturbative soft modes in the collision kernel can double the value of the NLO pQCD result for the $\hat{q}$ \cite{CaronHuot:2008ni,Panero:2013pla}. And there are also recent formal pQCD calculations showing that NLO corrections can result in more than 50\% increase in $\hat{q}$ \cite{Liou:2013qya,Blaizot:2013vha,Kang:2013raa,Iancu:2014kga,Blaizot:2014bha}. Another possibility is that the present $\hat{q}$ calculation in CUJET2.0 is over-idealized by disregarding anisotropy and heterogeneity/inhomogeneity, both factors can influence the jet-medium interaction significantly and require more careful considerations in both the bulk evolution sector and the energy loss sector.

We briefly summarize Section~\ref{sec:pion} here: CUJET2.0 inclusive pion \raa~calculated with maximum coupling constant $\alpha_{max}=0.25-0.27$ in a dynamical QCD medium with HTL scenario is strictly consistent with both RHIC Au+Au 200AGeV $\pi^0 R_{AA}$ and LHC Pb+Pb 2.76ATeV $h^\pm R_{AA}$ in both central and semi-peripheral collisions. Rigorous \chisq~calculation indicates this parameter fit has averaged \chisq~being stringently less than 1.5. And if allow average $ \chi^2/d.o.f.<2 $, then $ \alpha_{max}=0.23-0.30 $. The combined effect of multi-scale running coupling and transverse expanding medium give rise to the steep rising and subsequent flattening of inclusive hadron \raa~at LHC. The small value of \amax~itself implies that longer jet path length in a transverse expanding medium overrides the reduction of density and contribute to enhanced overall quenching. Idealized CUJET2.0 effective jet transport coefficient $\hat{q}/T^3$ is consistent with not only LO pQCD estimates, but also the $\hat{q}/T^3$'s extracted from HT-BW, HT-M, MARTINI and McGill-AMY models \cite{Burke:2013yra} by fitting to the same set of experimental hadron suppression factors at RHIC and LHC A+A central collisions.

\subsection{D meson, B meson and non-photonic electron}
\label{sec:heavy}

\begin{figure*}[!t]
\bc
\includegraphics[width=0.9\textwidth]{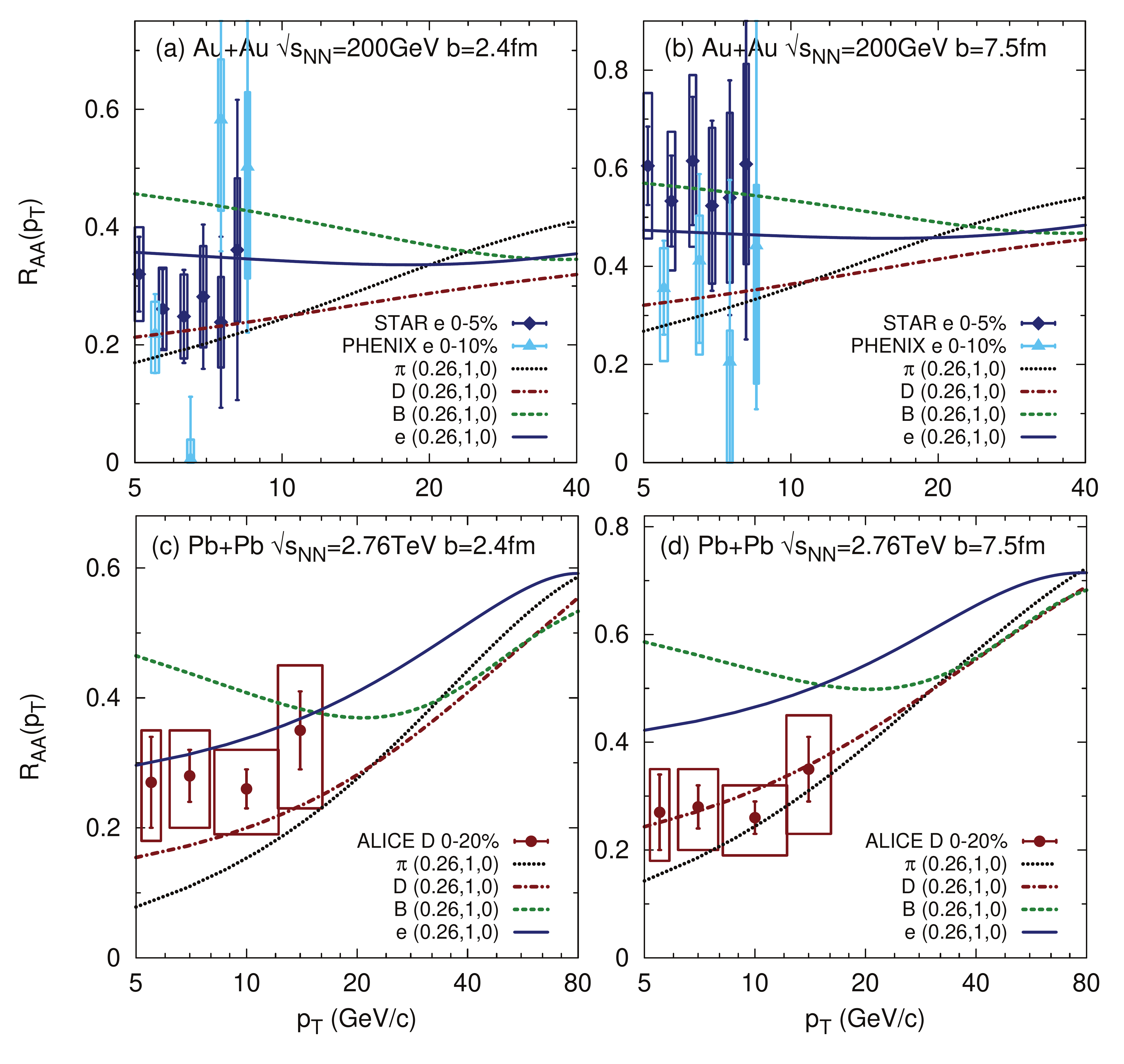}
\caption{\label{fig:Heavy_Flavor} 
		CUJET2.0 predictions of the suppression pattern of open heavy flavors and heavy flavor leptons. The mass hierarchy is illustrated as the level crossing pattern of $ R_{AA} $ versus $ p_T $ for pion (black), D meson (red), B Meson (green) and non-photonic electron (blue) fragments from quenched gluon, light, charm, bottom quark jets. Parameter $ (\alpha_{max},f_E,f_M)=(0.26,1,0) $ models a dynamical QCD medium in the HTL scenario, the bulk evolution profile being used is the same as in Fig.~\ref{fig:RAA_pT}. Au+Au $ \sqrt{s_{NN}} = 200$ GeV central and semi-peripheral collisions are simulated respectively with impact parameter (a) b=2.4fm and (b) 7.5fm at RHIC conditions; Pb+Pb $ \sqrt{s_{NN}} = 2.76$ TeV collisions are simulated respectively with (c) b=2.4fm and (d) 7.5fm at LHC conditions. The flavor dependent jet quenching results at RHIC are compared to PHENIX \cite{Adare:2010de} and STAR \cite{Abelev:2006db} non-photonic electron $ R_{AA} $ with (a) 0-5\%(STAR)/0-10\%(PHENIX), (b) 10-40\%(STAR)/20-40\%(PHENIX) centrality; results at LHC are compared to ALICE average prompt D \cite{ALICE:2012ab} $ R_{AA} $ with (c) 0-20\%. The same ALICE D $ R_{AA} $ is also plotted in panel (d) for reference since impact parameter b=7.5fm typically reproduces 10-30\% centrality. The level crossings in running coupling CUJET2.0 which has a transversely expanding medium occur at almost the same $ p_T $ as in fixed coupling CUJET1.0 which has a transversely static Glauber profile \cite{Buzzatti:2011vt}. The \raa's low $p_T$ $ B > e > D $ ordering evolves into $ e > B > D $ at high $p_T$ at LHC.
		}
\ec
\end{figure*}

We apply CUJET2.0 to study the suppression pattern of not only neutral pions or charged hadrons, but also D mesons, B mesons and non-photonic electrons at RHIC and LHC. The open heavy flavor and heavy flavor lepton nuclear modification factors calculated from CUJET2.0 $ (\alpha_{max},f_E,f_M)=(0.26,1,0) $ HTL model are shown in Fig.~\ref{fig:Heavy_Flavor}.
Inclusive non-photonic electron \raa's in central and semi-peripheral A+A collisions are compared with experimental data at PHENIX \cite{Adare:2010de} and STAR \cite{Abelev:2006db}, and D meson \raa's are compared with measurements at ALICE \cite{ALICE:2012ab}.

Both running coupling CUJET2.0 and fixed coupling CUJET1.0 \cite{Buzzatti:2011vt} predict a novel crossing pattern of $p_T$ dependent $\pi$, $D$, $B$, $e^-$ nuclear suppression factors. Compare the left panels of Fig.~\ref{fig:Heavy_Flavor} to Fig.~1 in \cite{Buzzatti:2011vt}, one finds that regardless of the inclusion of multi-scale running coupling and dynamical viscous hydro fields in CUJET2.0, and the crossings of $\pi$, $D$, $B$, $e^-$ \raa's in CUJET2.0 constantly occur at about same $p_T$ as in CUJET1.0. For example, at RHIC Au+Au 200AGeV central collisions, pion \raa~intersects D meson, non-photonic electron, B meson at $p_T \approx 9, 19, 24 $ GeV, and at LHC Pb+Pb 2.76ATeV central collisions, pion \raa~intersects D meson, B meson at $p_T \approx 23, 33$ GeV. Note all these $p_T$'s are within the range of $p_T<40$ GeV where, as discussed in Section~\ref{sec:pionRAA} and~\ref{sec:chi^2}, running coupling effect contributes significantly to a steeper rising \raa. The running coupling induced change of slope emerges from pion, D meson and B meson \raa~in a similar manner, this is as expected, because when considering the gluon radiation vertex for running coupling, the mass scale is small comparing to kinetic terms hence being dropped, cf. Eq.~\eqref{rcscale1} and~\eqref{rcscale2}. The robust $p_T$ interval of \raa~crossings in CUJET also suggests the mass ordering of $\pi$, $D$, $B$, $e^-$ suppression pattern comes intrinsically from the DGLV gluon radiation spectrum and TG elastic energy loss formula, and the bulk evolution profile has limited effect on this mass hierarchy.

Fig.~\ref{fig:Heavy_Flavor} displays an interesting crossover between pion and B meson \raa, at $p_T \simeq 25{\rm GeV}$ for RHIC and $p_T \simeq 35{\rm GeV}$ for LHC, which signature is also noticed in fixed coupling CUJET1.0 \cite{Buzzatti:2011vt}. Generally speaking, the quenched hadron (h) spectrum in an AA collision is calculated with
\be
\dfrac{E d\bar{\sigma}^{AA\rightarrow h}}{d^3 p} \equiv \dfrac{1}{N_{bin}} \dfrac{E d\sigma^{AA\rightarrow h}}{d^3 p} = \dfrac{E_i d\sigma^{pp\rightarrow q}}{d^3 p_i}\otimes P(E_i(p_i)\rightarrow E_f(p_f))\otimes D(q \rightarrow h)\;\;,
\ee
where $d\sigma^{pp\rightarrow q}/{d^3 p_i}$ is the initial partonic pp spectrum, $P$ is the energy loss probability distribution which linked to $\Delta E/E(E)$, and $D$ is the fragmentation function from parton q to hadron h. It is of great importance to study which one of the three factors plays the most critical role in maintaining the robust level crossing pattern for pion and B meson. Comparing Fig.~\ref{fig:Heavy_Flavor} with the middle and right panel of Fig.~\ref{RAAtauf}, we note that fragmentation functions alter the $p_T$ dependent quenching pattern of light and bottom quark only limitedly, i.e. $R_{AA}^{\rm light}(p_T) \sim R_{AA}^{\rm pion}(p_T)$ and $R_{AA}^{\rm bottom}(p_T) \sim R_{AA}^{\rm B}(p_T)$. At partonic level the crossing between light and bottom $R_{AA}(p_T)$ already occurred at $p_T \simeq 25{\rm GeV}$ for RHIC and $p_T \simeq 35{\rm GeV}$ for LHC, this fact suggests the near-negligible contribution of fragmentation functions to the intersection of $R_{AA}^\pi(p_T)$ and $R_{AA}^B(p_T)$. Furthermore, we notice in the bottom left panel of Fig.~\ref{RadElScan} and bottom middle panel of Fig.~\ref{PexE} that at fixed L the $\Delta E/E(E)$ for light and bottom do not intersect each other until $p_T=50 {\rm GeV}$, meaning the influence of partonic energy loss on the crossing pattern is less decisive than initial pp spectra. Indeed, we observe similar slopes of $d\sigma/dp_T(p_T)$ for light and bottom quark in the $p_T$ range of $10-15$ GeV at RHIC and $20-30$ GeV at LHC. It indicates the combined effect of partonic energy loss and initial spectra results in the crossing of $R_{AA}^\pi(p_T)$ and $R_{AA}^B(p_T)$, and among these two factors the latter is apparently more critical. A detailed study regarding the physical reason of the robust $R_{AA}(p_T)$ crossing pattern will be presented in \cite{JXMGprep}.

Notice also the quenching pattern of non-photonic electron at Au+Au 200AGeV collisions calculated from both CUJET1.0 \cite{Buzzatti:2011vt} and CUJET2.0 Fig.~\ref{fig:Heavy_Flavor}(a)(b) are in agreement with the RHIC data for central and semi-peripheral centralities. This clearly indicates the solution to ``heavy quark puzzle''\footnote{Explanations about the ``heavy quark puzzle'' can be found in \cite{Djordjevic:2013pba,Djordjevic:2013xoa,oai:arXiv.org:1209.0198,DM2012}.} is built intrinsically in the structure of CUJET energy loss framework. As discussed in Appendix~\ref{app:ElasticNumerics}, the combination of dynamical medium effect and elastic energy loss first significantly brings down the light to heavy quark energy loss ratio. Secondly, appropriately weighing path length fluctuations of initial jet production coordinates plays a pivotal role, in CUJET this is realized by cutting off the DGLV integral at a dynamical $T(\bz)|_{\tau_{max}} = T_f$ hypersurface which is parametrized by fragmentation temperature $T_f$\footnote{Also noted in Section~\ref{sec:DGLV} and Appendix~\ref{app:Thermal_Systematics}.}. The third factor is the Poisson distribution assumption for inelastic energy loss, since one has noticed Appendix~\ref{app:convDGLV} the variation of opacity can alter the gluon radiation spectrum for light and heavy quark jet differently.

Concentrating on the flavor dependent suppression pattern prediction from CUJET2.0, it is not only a revolution of \cite{Buzzatti:2011vt} with multi-scale running coupling and transverse expanding medium effect, but also a critical supplementary with comprehensive semi-peripheral A+A predications and decisive non-photonic \raa~at LHC. We can make several observations about the flavor dependent quenching scenarios in Fig.~\ref{fig:Heavy_Flavor}: firstly, \raa~for inclusive D meson and pion tangles together at low $p_T$ at both RHIC and LHC. We notice in Appendix~\ref{app:fluc} the radiative energy loss probability distribution for charm and light quark almost overlap when jet has reasonably low initial energy, the similar suppression pattern of D meson and pion in this region would suggest comparable elastic energy loss probability distribution for them, and the D meson A+A production spectrum is expected to have a steeper slope than pion at low $p_T$ (cf. also Appendix~\ref{app:partonspc}).

Secondly, CUJET2.0 predicts in Fig.~\ref{fig:Heavy_Flavor} that in low $p_T$ region the inclusive leading B meson is significantly less quenched than pion and D meson, whose $R_{AA}$ tangles together. This prediction indicates measurements of open beauty spectra at soft regime can post decisive constraints on a wide range of pQCD energy loss models. Finally, at LHC, the \raa~$ B > e > D $ mass ordering\footnote{This phenomenon is also noted in \cite{Zakharov2009} besides CUJET1.0 \cite{Buzzatti:2011vt}.} at low $ p_T $ evolves into $ e > B > D $ at $ p_T \approx 23$ GeV, and RHIC seems to have the same inversion at slightly larger $p_T$ but less discernible than LHC. This mass ordering comes from a complex interplay between total energy loss probability distribution, and initial production spectra for charm and bottom jets, and fragmentation functions in the hadronization processes. Since non-photonic electron spectrum is the combination of $ B \rightarrow e$, $ D \rightarrow e$, and $ B \rightarrow D \rightarrow e$ channels, the change in mass hierarchy can partially be attributed to a significant enhancement in the $ B \rightarrow D \rightarrow e$ channel in certain $p_T$ range, this range occurs at lower $p_T$ at LHC which has larger multiplicity density and higher temperature than RHIC, and a semi-exclusive measurement of non-photonic electron production in AA can thus be a crucial benchmark.

One final comment about the flavor dependent suppression pattern from CUJET2.0 calculations is the theoretical agreement with ALICE \cite{ALICE:2012ab} average prompt D \raa. Note impact parameter $b=2.4$ fm in CUJET typically simulates 0-10\% centrality, and $b=7.5$ fm is applicable for 10-30\% centrality, the integrated 0-20\% centrality experimental D \raa~should be within the $b=2.4$ fm and $b=7.5$ fm CUJET2.0 D meson \raa curves, and this feature shows up explicitly in Fig.~\ref{fig:Heavy_Flavor} (c) and (d). 

This brings us to a mini-summary of Section~\ref{sec:heavy}. We conclude that the robust crossing pattern of $\pi$, $D$, $B$, $e^-$ \raa's is rigorously encoded in the flavor dependent energy loss structure of DGLV opacity expansion combined with TG elastic, and a transverse expanding medium has minor effect on the mass hierarchy; solutions to the ``heavy quark puzzle'' are intrinsically integrated in the framework of CUJET; and CUJET2.0 predicts a decisively less quenched B meson \raa~which is well above D meson and pion at $5\;{\rm GeV}<p_T<15\;{\rm GeV}$, as well as a critical alternation of low $p_T$ \raa's mass ordering from $ B > e > D $  to $ e > B > D $ at $p_T\sim25$ GeV.

\subsection{Azimuthal flow}
\label{sec:v2}

Anisotropic collective flow is a key observable in relativistic heavy ion collisions, it relates directly to the formation of QGP. With a large impact parameter for A+A event, the region of interest gains an increasingly asymmetric shape. And in a strongly coupled medium, the pressure gradients due to this initial azimuthal anisotropy are effectively transferred into the collective flow of its components. The different types of collective flows are quantified in terms of Fourier components of the azimuthal angle distribution \cite{Voloshin:1994mz}:
\be
\frac{dN_h}{dyp_Tdp_T d\phi}(\sqrt{s},b)=\frac{1}{2\pi} \frac{dN_h}{dyp_Tdp_T}(\sqrt{s},b)\lp  1+2\sum_{n=1}^{\infty} v_n(y,p_T;\sqrt{s},b;h) \cos \lp n ( \phi - \Psi_n^h) \rp \rp \;\;.
\label{v2def}
\ee 
Here ${dN_h}/{dydp_T d\phi}$ represents the number of hadrons of species $h$ observed at rapidity $y$, with transverse momentum $p_T$ and azimuthal angle $\phi$. Both $\frac{dN_h}{dydp_T}$ and the Fourier coefficients $v_n$ depend on the initial rapidity density $dN_i/dy$. And $dN_i/dy$ is a function of the energy $\sqrt{s}$ and centrality b of the collision.

Generally speaking, in the transverse plane with respect to the beam axis, the collective flow generated from non-central A+A collisions is centrosymmetric if there is no fluctuation, and odd number Fourier components drop out. Thus among all collective flow harmonics, elliptic flow \vtwo~is of the most significance, and by the mean time being least sensitive to fluctuations. Therefore, a comprehensive study of single particle \vtwo~can provide critical information about the azimuthal anisotropy, as well as useful information about the jet medium interaction mechanism.

\subsubsection{Pion production with respect to reaction plane}
\label{sec:RAAinout}

\begin{figure*}[!t]
\bc
\includegraphics[width=0.475\textwidth]{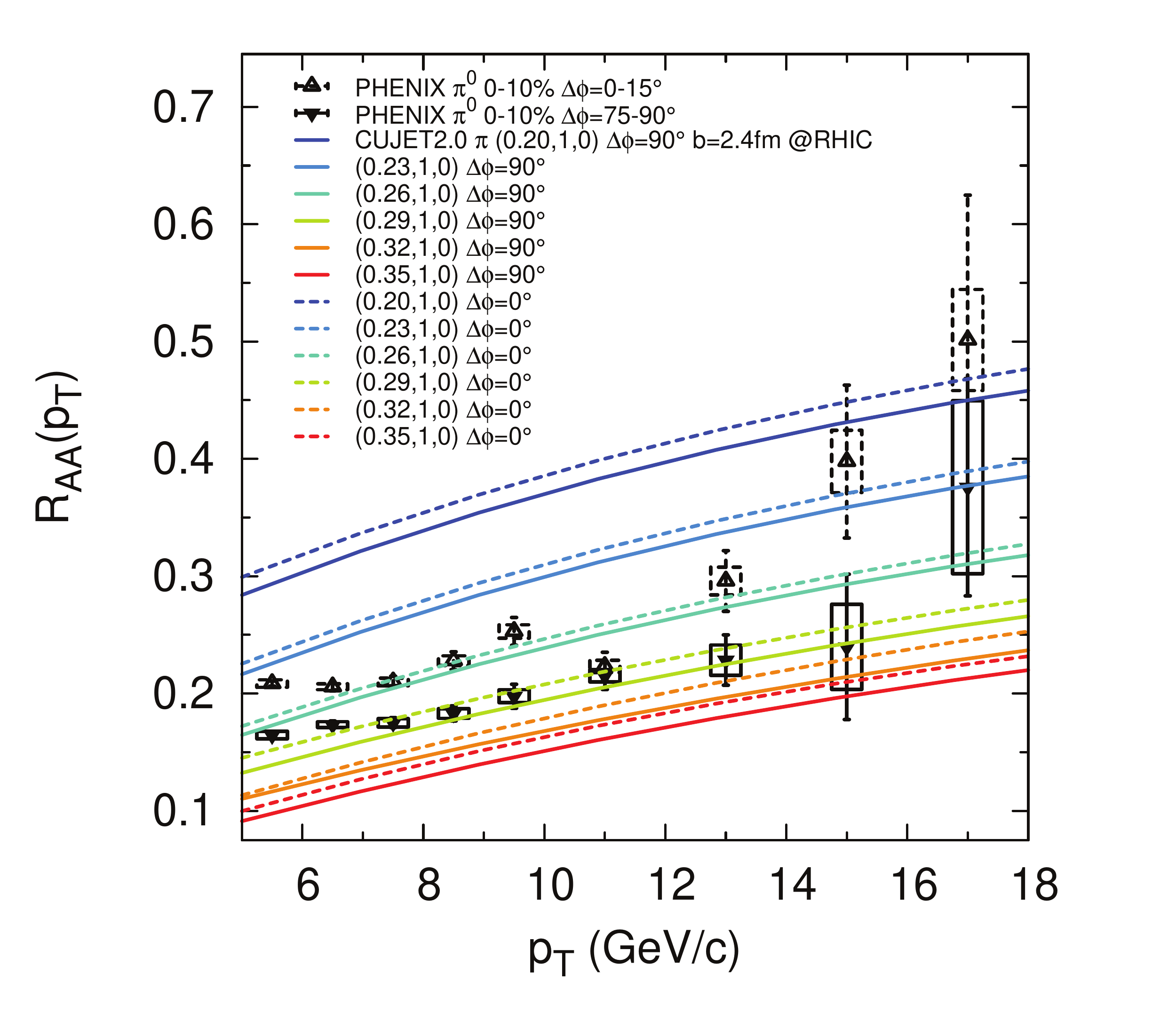}
\includegraphics[width=0.475\textwidth]{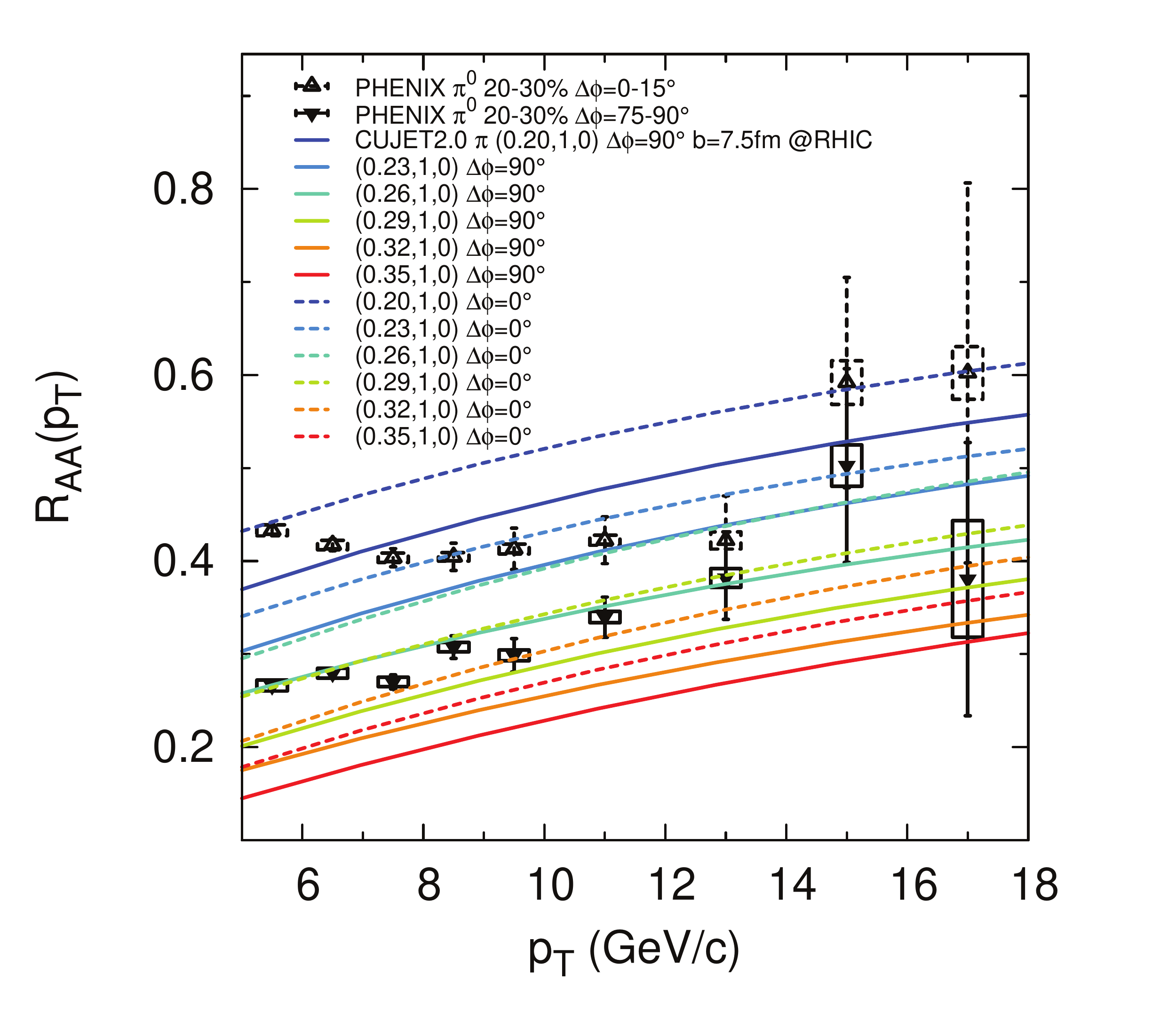}
\caption{\label{fig:RAA_in_out}
		CUJET2.0 pion $R_{AA}^{in}$ ($\Delta \phi = 0^{\circ}$, dashed curves) and $R_{AA}^{out}$ ($\Delta \phi = 90^{\circ}$, solid curves) versus $ p_T $ for Au+Au 200AGeV $b=2.4$ fm (left panel) and $b=7.5$ fm (right panel) calculated in the HTL $(f_E,f_M)=(1,0)$ scenario with maximum coupling constant \amax~varies from 0.20 to 0.35 in 0.03 steps. The bulk evolution profile being used is the same as in Fig.~\ref{fig:RAA_pT}. Theoretical results are compared with PHENIX \cite{Adare:2012wg} $\pi^0 R_{AA}$ in Au+Au collisions at $\sqrt{s_{NN}}=200$ GeV with centrality 0-10\% and reaction plane $\Delta \phi = 0-15^{\circ}$ (left panel, dashed black), 0-10\% and $\Delta \phi = 75-90^{\circ}$ (left panel, solid black), 20-30\% and $\Delta \phi = 0-15^{\circ}$ (right panel, dashed black), 20-30\% and $\Delta \phi = 75-90^{\circ}$ (right panel, solid black). In both central and semi-peripheral collisions, the $R_{AA}^{in}$ and $R_{AA}^{out}$ for $\alpha_{max}=0.26$ (the most consistent HTL fit to pion \raa's in a variety of A+A collision configurations based on rigorous \chisq~calculations in Section~\ref{sec:chi^2}) have a compatible mean value with experimental results, but they do not yield a comparable gap. Nevertheless, allow at most 10\% variations in \amax~and choose $\alpha_{max}=0.26$ $R_{AA}^{in}$ and $\alpha_{max}=0.29$ $R_{AA}^{out}$ for $b=2.4$ fm,  $\alpha_{max}=0.23$ $R_{AA}^{in}$ and $\alpha_{max}=0.26$ $R_{AA}^{out}$ for $b=7.5$ fm can generate a compatible reaction plane dependent suppression pattern for pion, and all these \amax~values fall within respective \chisq$<1$ and \chisq$<2$ entries in Table~\ref{AlphaMaxChiSquare}.
		}
\ec
\end{figure*}

Simultaneously fit particle suppression pattern and azimuthal flow asymmetry is a decisive benchmark for all jet tomography models.\footnote{Previous attempts for simultaneously fitting $R_{AA}$ and $v_{2}$ in the a-b-c model and semi-DGLV frameworks can be found in \cite{Betz:2013caa,oai:arXiv.org:1211.0804,BBMG2012,oai:arXiv.org:1106.4564,oai:arXiv.org:1102.5416,oai:arXiv.org:0812.4401,Molnar:2013eqa}.} To visualize this simultaneity more clearly, one can calculate the \raa~with respect to reaction plane. The typical choice of azimuthal angle set is the in plane $\phi=0$ and out of plane $\phi=\pi/2$, and consequent nuclear suppression factors $R_{AA}^{in}$ and $R_{AA}^{out}$ are defined as:
\be
\begin{cases}
R_{AA}^{in}(y,p_T)=\dfrac{ \frac{dN_h^{AA}}{dydp_T d\phi}|_{\phi=0} }{ N_{binary} \frac{dN_h^{pp}}{dydp_T d\phi}|_{\phi=0} } = \dfrac{ \frac{dN_h^{AA}}{dydp_T}\frac{1}{2\pi}\lp  1 + 2v_1 + 2v_2 + \cdots \rp }{ N_{binary} \frac{dN_h^{pp}}{dydp_T d\phi}|_{\phi=0} } \;\;,\\
R_{AA}^{out}(y,p_T)=\dfrac{ \frac{dN_h^{AA}}{dydp_T d\phi}|_{\phi=\frac{\pi}{2}} }{ N_{binary} \frac{dN_h^{pp}}{dydp_T d\phi}|_{\phi=\frac{\pi}{2}} } = \dfrac{ \frac{dN_h^{AA}}{dydp_T}\frac{1}{2\pi}\lp  1 - 2v_2 - \cdots \rp }{ N_{binary} \frac{dN_h^{pp}}{dydp_T d\phi}|_{\phi=\frac{\pi}{2}} } \;\;.
\end{cases}
\label{RAAinoutDef}
\ee 
The $AA, pp$ superscript and $N_{binary}$ has the same meaning as in Eq.~\eqref{RAAdef}. Since the p+p collision is presumably central, and generally no azimuthal anisotropy is expected, we have $\frac{dN_h^{pp}}{dydp_T d\phi}|_{\phi=0} = \frac{dN_h^{pp}}{dydp_T d\phi}|_{\phi=\frac{\pi}{2}}$, and $\frac{dN_h^{pp}}{dydp_T} \equiv \int_{0}^{2\pi} d\phi \frac{dN_h^{pp}}{dydp_T d\phi} = 2\pi \frac{dN_h^{pp}}{dydp_T d\phi}|_{\phi=\phi_0}$, where $\phi_0$ is an arbitrary azimuthal angle. In terms of rapidity $y$, the region we are interested in has $y \approx 0$, hence we short-write $\frac{dN_h^{pp}}{dydp_T}|_{y=0}$ as $\frac{dN_h^{pp}}{dp_T}$. Neglecting fluctuations by setting odd number harmonics to zero, in the mid-rapidity region, we get
\be
\begin{cases}
R_{AA}^{in}(p_T) \approx \dfrac{ \frac{dN_h^{AA}}{dydp_T} \lp  1 + 2v_2 + 2v_4 \cdots \rp }{ N_{binary} \frac{dN_h^{pp}}{dydp_T} } = R_{AA}^h \lp  1 + 2v_2 + 2v_4 \cdots \rp \;\;,\\
R_{AA}^{out}(p_T) = \dfrac{ \frac{dN_h^{AA}}{dydp_T} \lp  1 - 2v_2 - 2v_4 \cdots \rp }{ N_{binary} \frac{dN_h^{pp}}{dydp_T} } = R_{AA}^h \lp  1 - 2v_2 - 2v_4 \cdots \rp \;\;.
\end{cases}
\label{RAAinoutSimpleDef}
\ee 
We calculate pion's $R_{AA}^{in}$ and $R_{AA}^{out}$ in mid-rapidity region for $p_T$ up to 18 GeV/c in CUJET2.0, for Au+Au 200AGeV central and semi-peripheral collisions, and compare with corresponding PHENIX \cite{Adare:2012wg} data\footnote{In principle, better comparison with experiments can be achieved by integrating over the same $\Delta \phi$ window of measurements of $R_{AA}$~with respective to reaction planes. However, due to limited computing power, we have to stay with the faster way of computing $R_{AA}$~in-plane/out-plane, i.e. evaluating the spectra at $\phi=0,\pi/2$. The effect of window size will be explored in future works, on an event-by-event basis, it may contribute non-trivially.}. The results are shown in Fig.~\ref{fig:RAA_in_out}.

In Section~\ref{sec:chi^2}, we have constrained the maximum coupling constant $\alpha_{max}$ in the CUJET2.0 HTL scenario to be $0.25-0.27$, this range of \amax~renders the most consistent pion \raa~at RHIC Au+Au 200AGeV and LHC Pb+Pb 2.76ATeV central and semi-peripheral collisions through stringent \chisq~calculations. However, the $R_{AA}^{in}$ and $R_{AA}^{out}$ for $\alpha_{max}=0.26$ in both panels of Fig.~\ref{fig:RAA_in_out} have smaller gaps than the PHENIX measurements, indicating over-isotropized high $p_T$ single inclusive pion spectra in the model, despite mean values of them are in agreement.

\begin{figure*}[!t]
\bc
\includegraphics[width=0.475\textwidth]{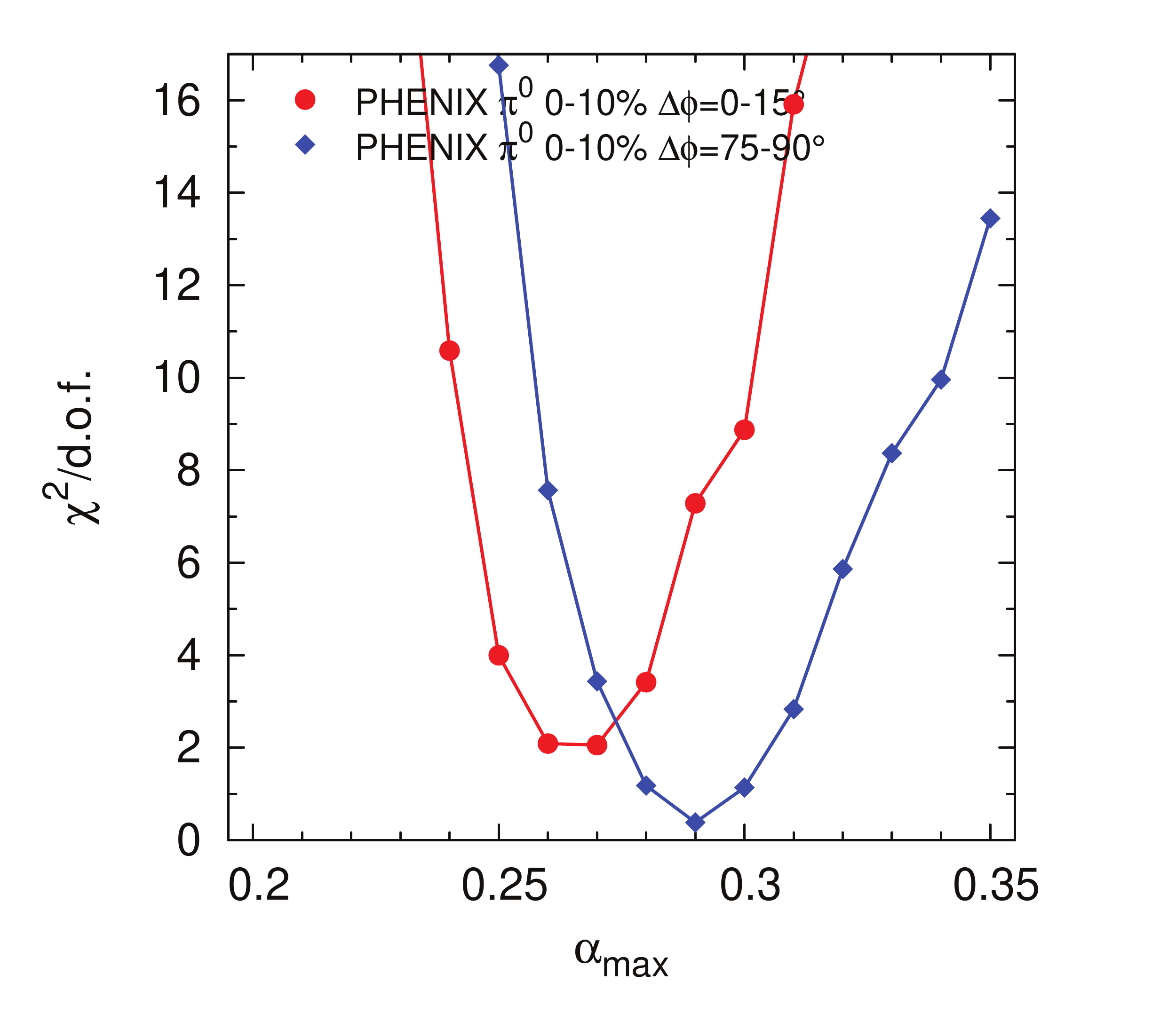}
\includegraphics[width=0.475\textwidth]{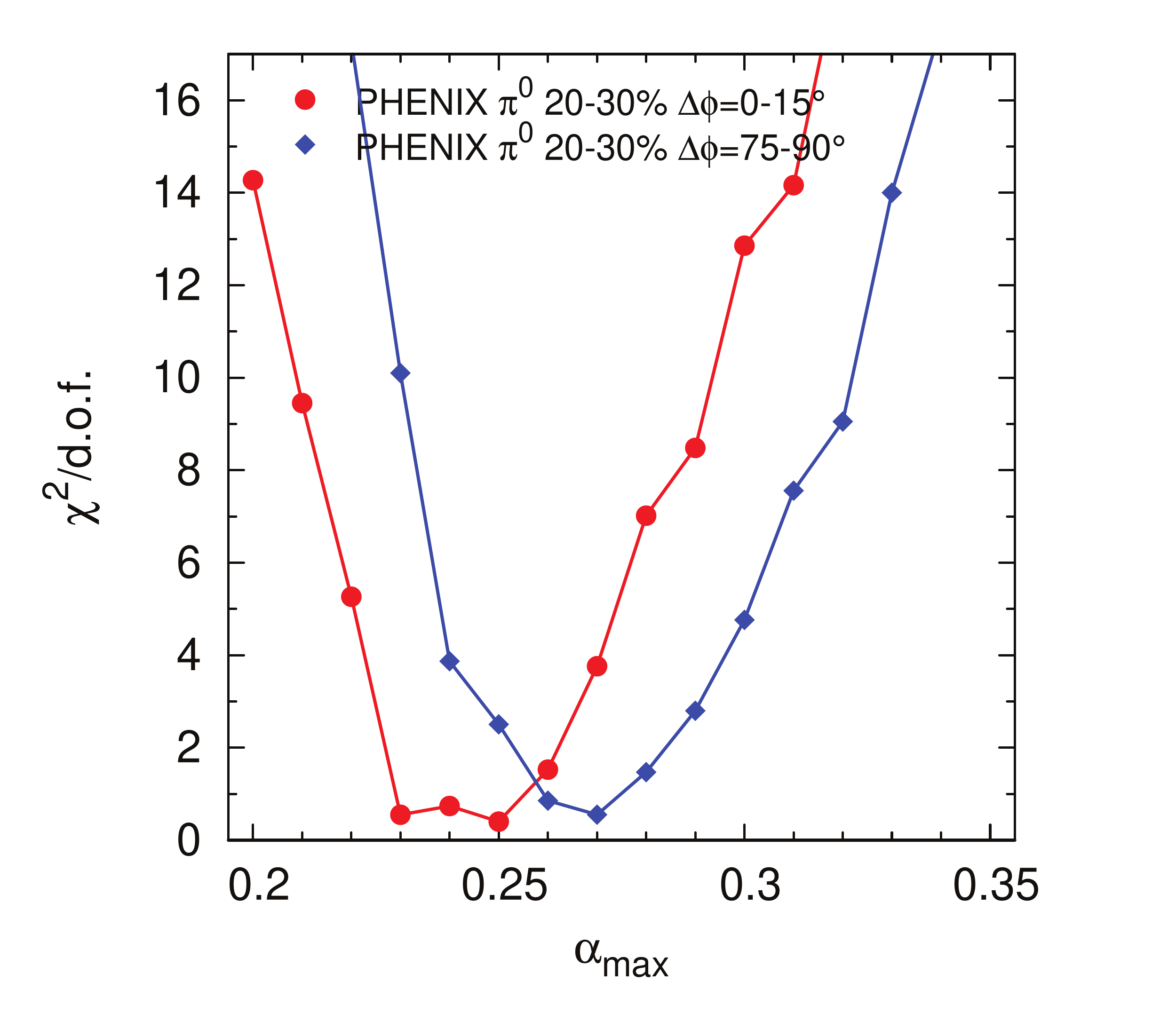}
\caption{\label{fig:ChiSqRAAinout} 
		$ \chi^2/d.o.f. $ versus $ \alpha_{max} $ calculated from Fig.~\ref{fig:RAA_in_out} at RHIC Au+Au 200AGeV central $b=2.4$ fm (left panel) and semi-peripheral $b=7.5$ fm (right panel) collisions. PHENIX \cite{Adare:2012wg} $\pi^0 R_{AA}$ with reaction plane $\Delta\phi=0-15^{\circ}$ (red) and $\Delta\phi=75-90^{\circ}$ (blue), and centrality 0-10\% (left) and 20-30\% (right) are the experimental references. For safer preservation of DGLV's basic eikonal and soft approximations, data from $ p_T>8$ GeV are used for the \chisq~calculation.
		}
\ec
\end{figure*}

Nevertheless, we note in Section~\ref{sec:chi^2} that, Fig.~\ref{fig:Chi_square_HTL} and Table~\ref{AlphaMaxChiSquare} suggest even if strictly limit \chisq~to be less than 1, \amax~maintains a non-negligible range which varies for different collisions. Using this flexibility, CUJET2.0 may create reaction plane dependent pion quenching patterns which are compatible with experiment measurements. To be rigorous, we first plot the \chisq~vs \amax~for the $R_{AA}^{in}$ and $R_{AA}^{out}$ in Figure~\ref{fig:RAA_in_out} using $p_T>8$ GeV, which $p_T$ range matches the choice in Section~\ref{sec:chi^2}, the results are shown in Figure~\ref{fig:ChiSqRAAinout}.

We find that in the left panel of Figure~\ref{fig:ChiSqRAAinout}, for $b=2.4$ fm central collisions at RHIC, in the CUJET 2.0 HTL scenario $R_{AA}^{in}$ is best fitted by $\alpha_{max}=0.26-0.27$, while $R_{AA}^{out}$ is best fitted by $\alpha_{max}=0.28-0.30$. In the right panel of Figure~\ref{fig:ChiSqRAAinout}, for $b=7.5$ fm semi-peripheral collisions at RHIC, $R_{AA}^{in}$ is best fitted by $\alpha_{max}=0.23-0.25$, while $R_{AA}^{out}$ is best fitted by $\alpha_{max}=0.26-0.27$. 

If we choose $\alpha_{max}=0.26$ $R_{AA}^{in}$ and $\alpha_{max}=0.29$ $R_{AA}^{out}$ for $b=2.4$ fm,  $\alpha_{max}=0.23$ $R_{AA}^{in}$ and $\alpha_{max}=0.26$ $R_{AA}^{out}$ for $b=7.5$ fm, the CUJET2.0 results are able to be perfectly consistent with RHIC $R_{AA}^{in}$ and $R_{AA}^{out}$ date. Since we have $R_{AA}=(R_{AA}^{in}+R_{AA}^{out})/2$, this set of \amax's effectively generates $R_{AA}$ with $\alpha_{max}=0.275$ at $b=2.4$ fm and $\alpha_{max}=0.245$ at $b=7.5$ fm. Based on Figure~\ref{fig:Chi_square_HTL} and Table~\ref{AlphaMaxChiSquare}, we see that the \chisq~for the average \raa~resulting from this \amax~sequence is: RHIC $b=2.4$ fm, $\chi^2/d.o.f.<1.5$; RHIC $b=7.5$ fm, $\chi^2/d.o.f.<1$; LHC $b=2.4$ fm, $\chi^2/d.o.f.<1.5$; LHC $b=7.5$ fm, $\chi^2/d.o.f.<1$ -- the \chisq~for average \raa~with these \amax's in the CUJET2.0 HTL scenario is strictly less than 1.5 in all four collisions. It means this modest variation in \amax~is intrinsically allowed by our model without jeopardizing its consistency for averaged hadron \raa~at a variety of collision configurations.

Notice among this set of \amax~parameters, for both $b=2.4$ fm and $b=7.5$ fm, the difference in \amax~for $R_{AA}^{in}$ and $R_{AA}^{out}$ is 0.03. And for either $R_{AA}^{in}$ or $R_{AA}^{out}$, the variance in \amax~for $b=2.4$ fm and $b=7.5$ fm is 0.03 (which surprisingly coincides with the RHIC and LHC averaged \amax~gap discussed in Section~\ref{sec:chi^2}). The maximumly 10\% \amax~deviations in $R_{AA}^{in}$ and $R_{AA}^{out}$, $b=2.4$ fm and $b=7.5$ fm imply that the anisotropic path averaged effective coupling strengths can be originated from local effects. For instance, lattice QCD simulation \cite{Bazavov:2013zha} shows that the $q\bar{q}$ potential has both distance r and temperature T dependence. The spacing r Fourier transforms into momentum Q, suggests that a complete running coupling should have both Q and T dependence, i.e. $\alpha_s \rightarrow \alpha(Q,T({\bf x}_{0\perp},\phi))$. Presently our running has only Q dependence, therefore the $T({\bf x}_{0\perp},\phi)$ dependence can be intrinsically transformed into local coordinate and azimuthal angle dependence of $\alpha_{max}$, i.e. $\alpha_{max} \rightarrow \alpha_{max}({\bf x}_{0\perp},\phi)$. On the other hand, the azimuthal sensitivity of jet quenching may strongly depend on the detailed edge geometry which varies in different hydro profiles and is yet to be studied in the CUJET2.0 framework. Both the dual running of $\alpha_s$ and the hydro sensitivity of high $p_T$ single particle $v_2$ are work in progress.

In addition, the ordering of \amax~in terms of averaged path length is noteworthy: at $\tau_0$, the length of the medium in $b=2.4$ fm and $b=7.5$ fm collisions along $\phi=0^{\circ}$ and $\phi=90^{\circ}$ direction can be approximately ordered as $7.5{\rm fm} + 0^{\circ} < 7.5{\rm fm} + 90^{\circ} \approx 2.4{\rm fm} + 0^{\circ} < 2.4{\rm fm} + 90^{\circ}$, and the best fit \amax~in corresponding situations is $0.23<0.26=0.26<0.29$. It means that within CUJET2.0, longer path length requires stronger coupling in order to predict the correct high $p_T$ single particle $v_2$.

%Meanwhile, as we have noted in Section~\ref{sec:qhat} Fig.~\ref{fig:qhat}, the decay of $\hat{q}/T^3$ with lowering E unanticipatedly reaches the ultra-soft region where presumably eikonal approximation will break down and bulk physics will emerge, but where the consequential ``shear viscosity'' is far from recovering the strong coupling limit. This suggests more careful consideration of anisotropy, nonuniformity and inhomogeneity is necessary in both the soft bulk and the hard probe sector, and high energy jet may ``see'' the ``medium'' differently from a viscous hydro fluid. All these factors combined to indicate a possibility that the variation of locally dependent strong coupling can contribute an azimuthally different path averaged \amax.

After constrained the azimuthal anisotropy in CUJET2.0 at RHIC, our next step is to test the model consistency with $R_{AA}^{in}$ and $R_{AA}^{out}$ for Pb+Pb 2.76ATeV at LHC central and semi-peripheral collisions. Because of the absence of published reaction plane dependent neutral pion or charged hadron suppression data at LHC, we turn to compare the CUJET2.0 results of \vtwo~with experimental elliptic flow measurements, which comparison is more sensitive to fluctuations from the theoretical point of view. And we will discuss this in detail in Section~\ref{sec:v2!}.

\subsubsection{Elliptic flow}
\label{sec:v2!}

\begin{figure*}[!t]
\bc
\includegraphics[width=0.475\textwidth]{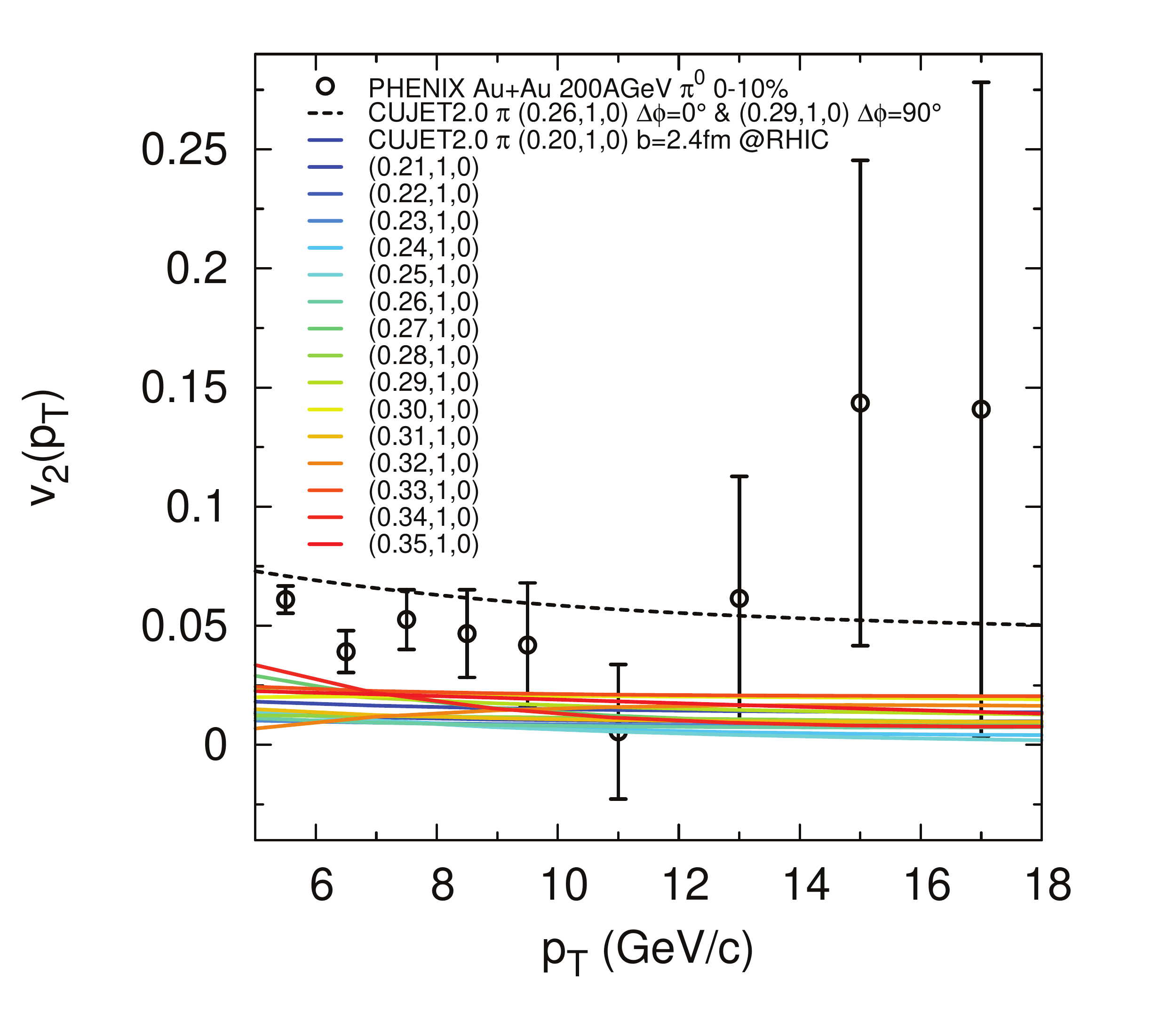}
\includegraphics[width=0.475\textwidth]{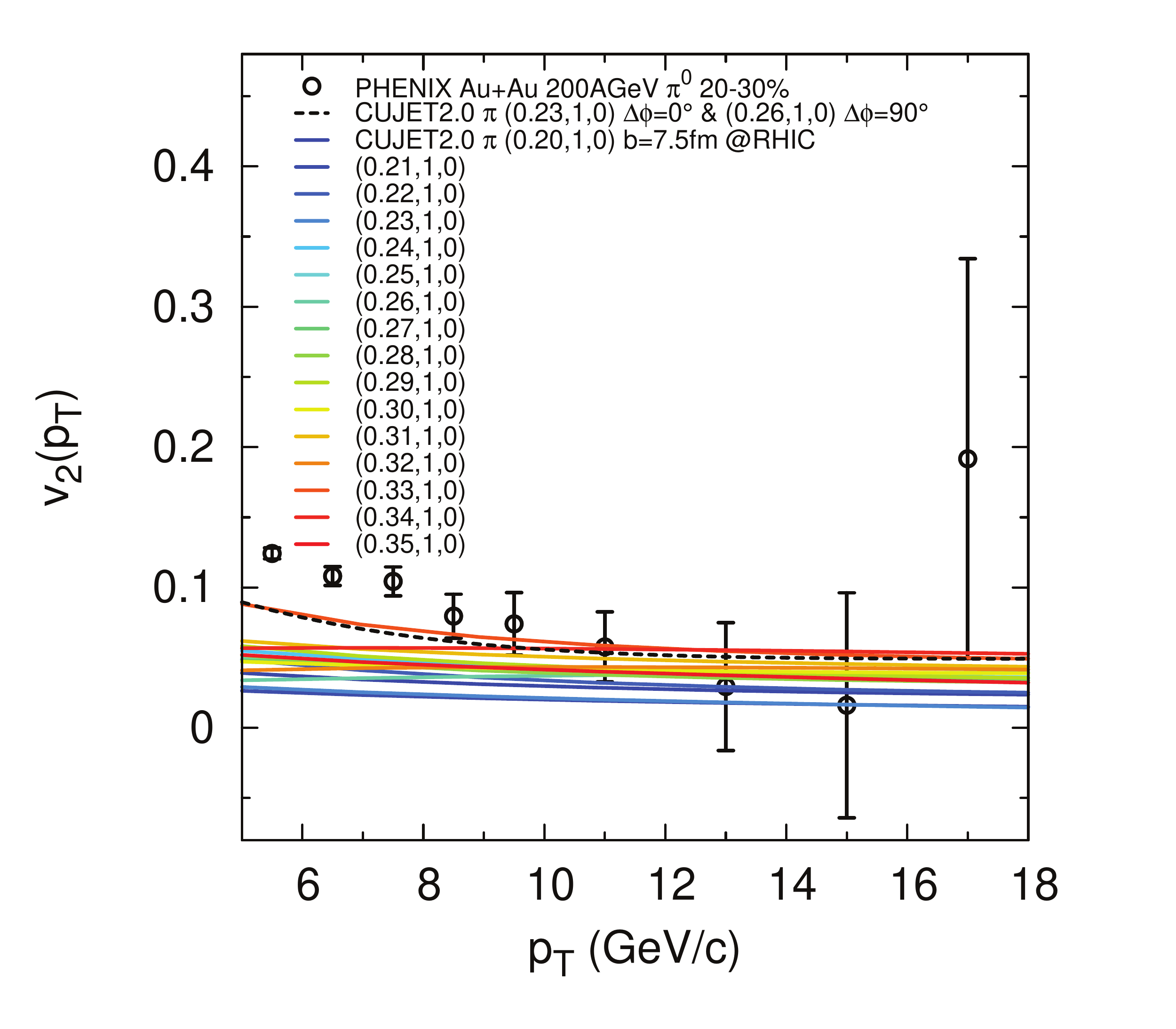}
\includegraphics[width=0.475\textwidth]{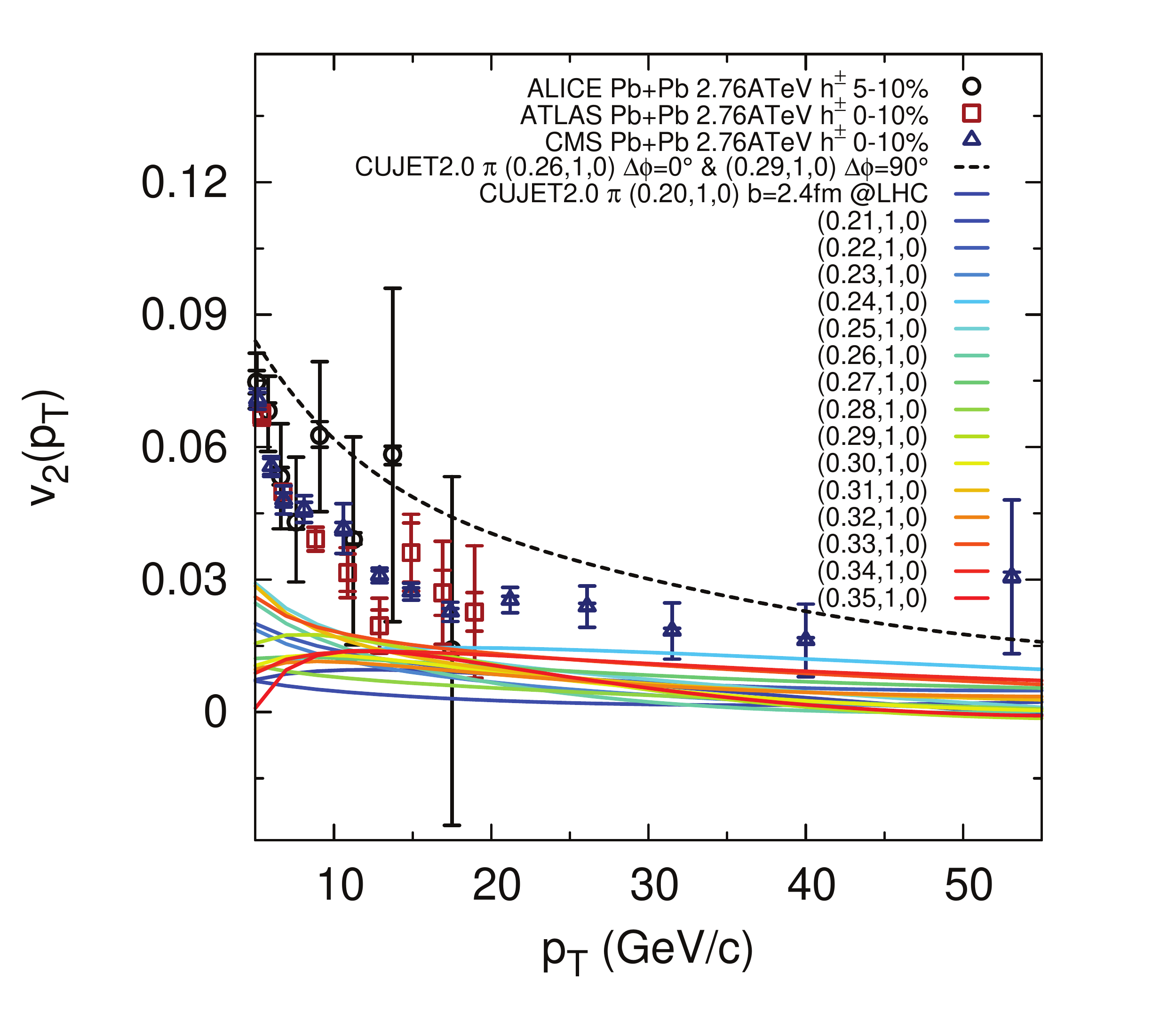}
\includegraphics[width=0.475\textwidth]{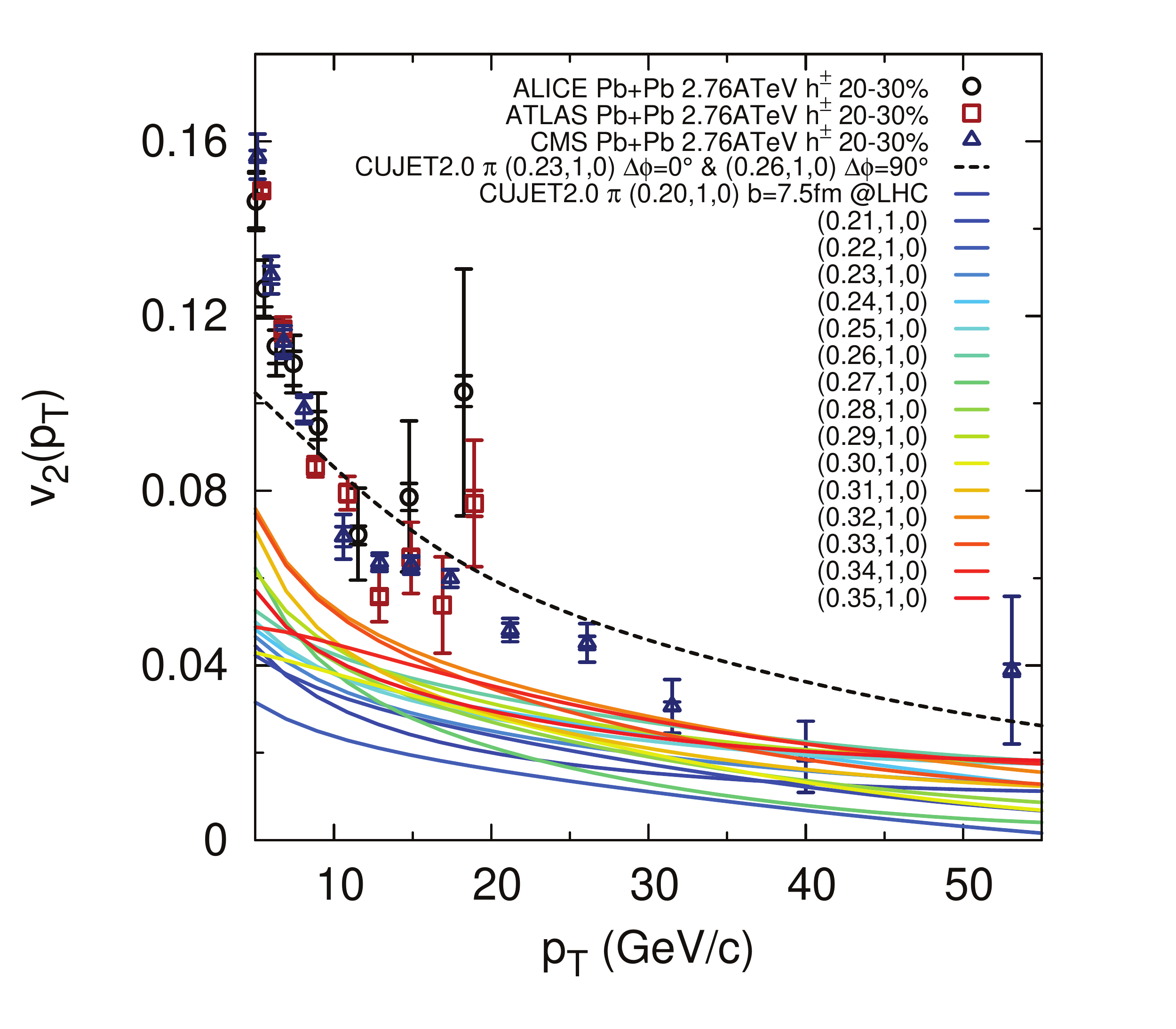}
\caption{\label{fig:v2_pT} 
		CUJET2.0 pion single particle anisotropy $ v_{2} $ versus $ p_T $ in RHIC Au+Au 200AGeV (top panels) and LHC Pb+Pb 2.76ATeV (bottom panels), central ($b=2.4$ fm, left panels) and semi-peripheral ($b=7.5$ fm, left panels) collisions. The theoretical calculations are compared with PHENIX \cite{Adare:2010sp} Au+Au 200AGeV $\pi^0 v_{2}$ with 0-10\% (top left) and 20-30\% (top right) centrality, and ALICE ($v_2\{4\}, |\eta|<0.8$) \cite{Abelev:2012di}, ATLAS ($|\eta|<1$) \cite{ATLAS:2011ah} and CMS ($|\eta|<1$) \cite{Chatrchyan:2012xq} Pb+Pb 2.76ATeV $h^{\pm} v_{2}$ with 0-10\%(ATLAS,CMS)/5-10\%(ALICE) (bottom left) and 20-30\% (bottom right) centrality. The hydro profile being used is the same as in Fig.~\ref{fig:RAA_pT}. The maximum coupling constant \amax~is varied from 0.20 to 0.35 with 0.01 steps in the CUJET2.0 HTL scenario, but as already noted in Fig.~\ref{fig:RAA_in_out}, none would be in perfect agreement with experimental data. Nevertheless, by taking into account at most 10\% \amax~azimuthal variation which is discussed in Section~\ref{sec:RAAinout}, after constraining CUJET2.0 HTL model at RHIC with $\alpha_{max}=0.26$ for $R_{AA}^{in}$ and $\alpha_{max}=0.29$ for $R_{AA}^{out}$ when $b=2.4$ fm,  $\alpha_{max}=0.23$ for $R_{AA}^{in}$ and $\alpha_{max}=0.26$ for $R_{AA}^{out}$ when $b=7.5$ fm, the CUJET2.0 theoretical $ v_{2} $'s (dashed black) are excellently consistent with LHC measurements in both central and semi-peripheral collisions. 
		}
\ec
\end{figure*}

If presuming no fluctuations in the azimuthal plane, recall Eq.~\eqref{RAAinoutSimpleDef} suggests that $R_{AA}^{in}$ and $R_{AA}^{out}$ depend solely on even harmonics. If further drop higher order components assuming they have much smaller magnitude comparing to \vtwo, we get
\be
\begin{cases}
R_{AA}^{in}(p_T) \approx R_{AA}^h \lp  1 + 2v_2 \rp \;\;,\\
R_{AA}^{out}(p_T) \approx R_{AA}^h \lp  1 - 2v_2 \rp \;\;,
\end{cases}
\ee 
in this limit. And the elliptic flow \vtwo~follows from $R_{AA}^{in}$ and $R_{AA}^{out}$ via 
\be
v_2(p_T)=\dfrac{1}{2}\dfrac{R_{AA}^{in}(p_T) - R_{AA}^{out}(p_T)}{R_{AA}^{in}(p_T) + R_{AA}^{out}(p_T)}
\;\;.
\label{vtwoapprox}
\ee 
We compute in CUJET2.0 pion \vtwo's using Eq.~\eqref{vtwoapprox} for RHIC Au+Au 200AGeV and LHC Pb+Pb 2.76ATeV, central $b=2.4$ fm and semi-peripheral $b=7.5$ fm collisions. The results are shown in Fig.~\ref{fig:v2_pT}, and corresponding ALICE \cite{Abelev:2012di}, ATLAS \cite{ATLAS:2011ah}, CMS \cite{Chatrchyan:2012xq}, and PHENIX \cite{Adare:2010sp} data are compared.

Fig.~\ref{fig:v2_pT} shows that if varying solely the maximum coupling constant \amax~from 0.20 to 0.35 with 0.01 steps in the CUJET2.0 HTL scenario, none of the theoretical curves matches the single pion \vtwo~at both Au+Au 200AGeV and Pb+Pb 2.76ATeV, central and semi-peripheral collisions. Nevertheless, as already noted in Section~\ref{sec:RAAinout}, due to the non-negligible influence that anisotropy and heterogeneity/inhomogeneity have on the jet-medium interaction, local effects can alter the CUJET2.0 framework significantly. Under present circumstances that the strong coupling running has no simultaneous energy and local temperature dependence, these effects can effectively generate azimuthally anisotropic path averaged \amax.

By choosing $\alpha_{max}=0.26$ $R_{AA}^{in}$ and $\alpha_{max}=0.29$ $R_{AA}^{out}$ for $b=2.4$ fm,  $\alpha_{max}=0.23$ $R_{AA}^{in}$ and $\alpha_{max}=0.26$ $R_{AA}^{out}$ for $b=7.5$ fm in the CUJET2.0 HTL scenario, we effectively constrained our model at RHIC with assumed azimuthal $\alpha_{max}$ anisotropy caused by possible local temperature field effects. The top panels of Fig.~\ref{fig:v2_pT} show the consequential single pion \vtwo's~at RHIC central and semi-peripheral collisions are compatible with respective experimental measurements, as expected. More importantly, if extrapolate the same CUJET2.0 framework (with azimuthal dependence of \amax) to LHC and calculate the single particle \vtwo~using the same $\alpha_{max}$ parameter set, theoretical results show even better agreements with ALICE, ATLAS and CMS data, in both central and semi-peripheral collisions (particularly the latter, cf. Table~\ref{ChiSquareTable_v2}).
\begin{table}[!t]
\centering
    \begin{tabular}{ | l | c | c | c | c | }
    \hline
    $\chi^2/d.o.f.$ ($b=7.5$ fm) & $v_2$, RHIC & $v_2$, LHC & $R_{AA}$, RHIC & $R_{AA}$, LHC \\ \hline
    $\alpha_{max}^{in} = 0.23$, $\alpha_{max}^{out} = 0.23$ & 3.72 & 43.03 & 0.93 & 0.73 \\ \hline
    $\alpha_{max}^{in} = 0.26$, $\alpha_{max}^{out} = 0.26$ & 2.06 & 24.89 & 0.23 & 1.06 \\ \hline
    $\alpha_{max}^{in} = 0.23$, $\alpha_{max}^{out} = 0.26$ & 0.50 & 4.92 & 0.42 & 0.54 \\
    \hline
    \end{tabular}
\caption{$\chi^2/d.o.f.$ for $v_2$ and azimuthally averaged $R_{AA}$ in semi-peripheral $b=7.5$ fm collisions at RHIC Au+Au 200AGeV and LHC Pb+Pb 2.76ATeV, with different choices of \amax~values for $R_{AA}^{in}$ ($\alpha_{max}^{in}$) and $R_{AA}^{out}$ ($\alpha_{max}^{out}$) in the CUJET2.0 HTL scenario. Reference curves are shown in Fig.~\ref{fig:Multi_Alf_HTL}, Fig.~\ref{fig:RAA_in_out}, and Fig.~\ref{fig:v2_pT}. The $p_T>8$ GeV range is chosen for both RHIC \raa~and LHC \raa~for safer preservation of eikonal and soft approximations, $p_T>8$ GeV and $p_T>12$ GeV range is chosen for RHIC \vtwo~and LHC \vtwo~to avoid the avalanche region. The choice of $\alpha_{max}^{in} = 0.23$, $\alpha_{max}^{out} = 0.26$ significantly reduces the \chisq~for \vtwo~at both RHIC and LHC, especially the latter one. By the mean time, this set of \amax~parameters maintains almost perfect agreement with both RHIC and LHC for azimuthally averaged \raa.}
\label{ChiSquareTable_v2}
\end{table}

The consistency, of the open heavy flavor and heavy flavor lepton's single particle azimuthal anisotropy calculated in the same CUJET2.0 framework (with azimuthal variations), with experimental measurements can shed light on the underlying physics for this azimuthal variation of path averaged coupling strength\footnote{Note by allowing azimuthal variation of the coupling strength, the prediction power of CUJET2.0 model is not jeopardized. We have shown that by fixing $\alpha_{max}$ for in-plane/out-of-plane at RHIC, the extrapolation of $R_{AA}^{in/out}(p_T)$ to LHC is in agreement with data. It means that we can use CUJET2.0 to fit RHIC at a particular centrality, then extrapolate it to predict LHC, or vice versa. Moreover, provided the built-in mass hierarchy in the CUJET2.0 model, we can extrapolate $v_{2}^{\pi}(p_T)$ to predict $v_{2}^{D,B,e^-}(p_T)$ at various centralities at both RHIC and LHC. This will be presented in \cite{JXMGprep}.}.

A mini-summary of Section~\ref{sec:v2}: by allowing the maximum coupling constant \amax~to vary by even less than 10\% in respective reaction plane at central and semi-peripheral collisions, we can gain in CUJET2.0 the simultaneous compatibility with not only measurements of $R_{AA}^{in}$ and $R_{AA}^{out}$ at RHIC, but also effective $R_{AA}^{in}$ and $R_{AA}^{out}$ (measurements of \vtwo~and averaged \raa) at LHC, while maintaining $\chi^2/d.o.f.<1.5$ for azimuthally averaged \raa~in both central and semi-peripheral collisions at RHIC and LHC.
%We postulate that anisotropy and inhomogeneity require more careful consideration in both the soft bulk and hard probe sector, and jet may ``see'' the ``medium'' differently from viscous hydro fields, therefore it is necessary to consider the azimuthal variation of path averaged coupling strength. This hypothesis is tested by extrapolating the completely constrained CUJET2.0 model at RHIC to LHC, and we observe a simultaneous agreement of theoretical \vtwo~and \raa~results with RHIC and LHC measurements in both central and semi-peripheral collisions.

\section{Further Discussion}
\label{sec:discussion}

\subsection{Thermalization time}
\label{sec:initialtime}

At very early time of relativistic heavy-ion collisions, the matter created during the collision is characterized by extremely high energy densities. While it expands, and the gluon density decreases, the strength of interaction among particles increases, facilitating the thermalization process into quark gluon plasma. The time scale over which the thermalization process takes place is generally referred to as $ \tau_0 $ and is assumed to be approximately equal to $ 0.5 \sim 1$ fm/c.

Once the plasma has reached the thermalized stage, the system can be approximated by a fluid and its evolution can be computed in the framework of relativistic hydrodynamics. The fundamental equations of energy-momentum and baryon number conservation are assumed to hold.

In CUJET2.0, the initial time $ \tau_0 $ is set to be 0.6 fm/c to match the choice of VISH2+1, which generates 2+1D viscous hydrodynamical fields as a bulk background in the parton shower modification. Additionally, the temporal evolution of the system is parametrized in such a way that the density ``seen'' by the jet grows linearly until the full thermalization is reached, and decreases as $ 1/\tau $ thereafter (cf. Eq.~\eqref{tau0} and Appendix~\ref{app:thermal}).

\begin{figure*}[!t]
\bc
\includegraphics[width=0.9\textwidth]{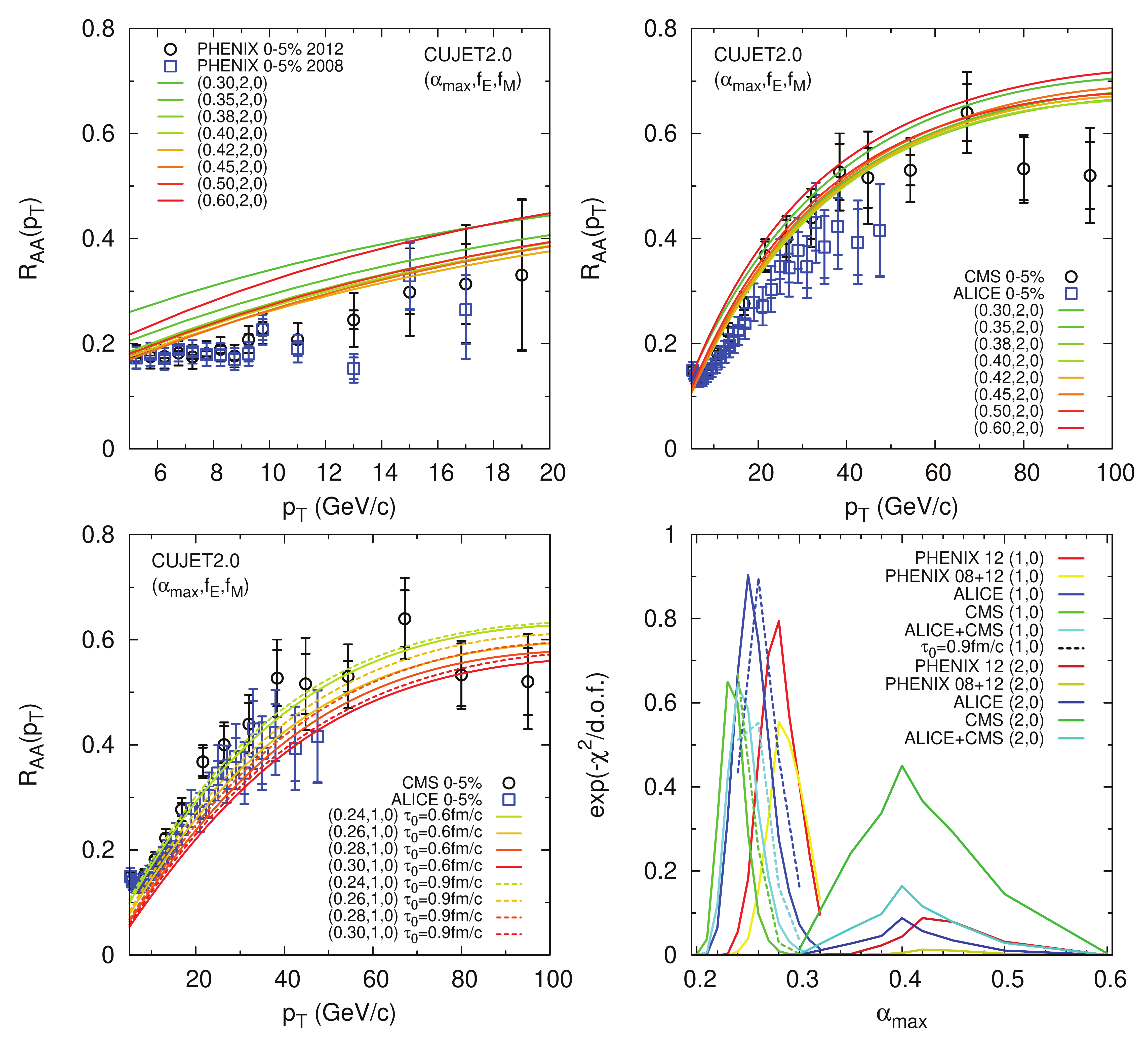}
\caption{\label{fig:Non-HTL_and_Initial-Time} 
		(top panels) CUJET2.0 pion $ R_{AA} $ with non-HTL $ f_E=2, f_M=0 $ and maximum coupling strength $ \alpha_{max}=0.30\sim0.60 $ in central $b=2.4$ fm collisions compared with PHENIX \cite{Adare:2012wg,Adare:2008qa} $\pi^0 R_{AA}$ in Au+Au collisions at 200AGeV with 0-5\% centrality (top left), and ALICE \cite{Abelev:2012hxa} and CMS \cite{CMS:2012aa} $h^\pm R_{AA}$ in Pb+Pb collisions at 2.76ATeV with 0-5\% centrality (top right). (Bottom Left) CUJET2.0 results of inclusive pion $ R_{AA} $ with initial time $ \tau_0=0.6$ fm/c (solid curves) and $0.9$ fm/c (dashed curves), scattering potential parameter $ f_E=1, f_M=0 $, and maximum coupling strength $ \alpha_{max}=0.24\sim0.30 $ in central $b=2.4$ fm collisions with LHC conditions comparing with ALICE \cite{Abelev:2012hxa} and CMS \cite{CMS:2012aa} $h^\pm R_{AA}$ in Pb+Pb collisions at 2.76ATeV with 0-5\% centrality. (Bottom Right) $ exp(-\chi^2/d.o.f.) $ calculated from the non-HTL model, the longer initial time scenario, as well as HTL models as a function of $ \alpha_{max} $ values at RHIC and LHC. Data from $ p_T>8GeV/c $ are used. All hydro evolution profiles being used are the same as in Fig.~\ref{fig:RAA_pT}, however calculations in this figure has fragmentation temperature $T_f=100$ MeV. The experimental results prefer the HTL picture, and larger initial time can improve the model consistency at RHIC and LHC.
		} 
\ec
\end{figure*}

In a recent paper by Song et al. \cite{Song:2013qma}, the authors found in VISHNU, where a UrQMD hadronic afterburner is coupled to VISH2+1, that to simultaneous fit RHIC and LHC particle production spectrum and all order collective flow harmonics, the initial time needs to be increased to 0.9 fm/c. We are therefore motivated to explore the effect of longer thermalization time, and to do so we modify the initial time $ \tau_0$ to be 0.9 fm/c at LHC conditions in CUJET2.0. The theoretical results compared with corresponding experimental data are shown in the bottom left and bottom right panel of Fig.~\ref{fig:Non-HTL_and_Initial-Time}. 

As illustrated in the bottom left panel of Fig.~\ref{fig:Non-HTL_and_Initial-Time}, with this increase in initial time, the suppression of pion at LHC slightly diminishes, which is understood by the fact that longer thermalization time in the linear scheme (cf. Appendix~\ref{app:thermal}) will create dilutely distributed scattering centers. Meanwhile, in the bottom right panel of Fig.~\ref{fig:Non-HTL_and_Initial-Time}, the best fit CUJET2.0 HTL \amax~for LHC Pb+Pb 2.76ATeV central collisions gains by 0.02 because of the prolonged $\tau_0$. Since the typical temperature range reached by LHC is higher than by RHIC, one would expect there is more reduction of density because of the $\tau_0$ growth at LHC than at RHIC. Therefore, to maintain the same quenching magnitude with $\tau_0=0.9$ GeV/c the increase of \amax~at RHIC will be smaller than LHC. This $\tau_0$ effect can hence result in a lessened gap between the best fit value of CUJET2.0 HTL \amax~at RHIC and LHC central collisions.

However, if adding the hadronic afterburner to VISH2+1 as in VISHNU, the bulk evolution profile would be significantly different, which may or may not invalidate the above argument. Therefore a thorough test with a complete VISHNU code coupled to CUJET shall be conducted before drawing any positive conclusions.

\subsection{Non-HTL scenario}
\label{sec:non-HTL}

Experimental \raa~data favors maximum running strong coupling $\alpha_{max}=0.25-0.27$ in the CUJET pQCD model with dynamical QCD medium and 1-loop HTL gluon propagator, i.e. a thermal gluon with mass $m_g=(f_E)\lp g T \sqrt{1+n_f/6}\rp/\sqrt{2}$ and $ f_E=1 $. However, lattice QCD calculation suggests this approximation breaks down in the non-perturbative region \cite{Bazavov:2013zha}. We therefore vary the HTL deformation parameters $ (f_E,f_M) $ in CUJET2.0 from HTL $ (1,0) $ to a non-HTL $ (2,0) $ to explore this effect. The consequential pion \raa~results are shown in the top left, top right, and \chisq~are shown in the bottom right panel of Fig.~\ref{fig:Non-HTL_and_Initial-Time}.

With dynamical scattering potential and doubled thermal plasmon gluon mass in CUJET2.0, a novel inversion of \raa~for pion suppression pattern appears: when increase \amax from 0.30 to 0.60, pion \raa~first decreases then increases, with the strongest quenching occurs at \amax=0.4, which is also the best fit to experimental data. This inversion is generated from a complex interplay between the reduced differential scattering cross section with an enlarged gluon mass and the variation of running coupling saturation scale $Q_{min}$ with \amax~($ Q_{min}= \Lambda_{QCD} \rm{Exp}\{2\pi/9 \alpha_{max}\} $). And at present stage there is no simple asymptotic analytical formula to track this non-HTL scenario in CUJET2.0.

On the other hand, concentrating on the \chisq~fit of this non-HTL scenario whose gluon mass is doubled, in the bottom right panel of Fig.~\ref{fig:Non-HTL_and_Initial-Time} the best fit non-HTL \amax=0.4 still has a large \chisq, implying the necessity to explore more complicated combination of HTL deformation parameters in CUJET2.0, e.g. $ f_E=2, f_M=0.6 $. In fact, a realistic non-HTL scenario would eventually have an effective running coupling and a effective scattering potential extracted from lattice QCD data on non-perturbative $q\bar{q}$ potential $V(r,T)$. And a comprehensive test of whether lattice QCD predicts the correct jet medium physics in the near $T_c$ non-perturbative region can be conducted in the CUJET framework. This is a work in progress.
\\
\section{Summary and Outlook}
\label{sec:summary}

We presented in this paper the basic features of CUJET2.0 pQCD azimuthal jet flavor tomography model, which features DGLV opacity series with multi-scale running coupling and 2+1D viscous hydrodynamical fields, as well as the TG elastic energy loss; geometry, radiative energy loss and elastic energy loss fluctuations; and the convolutions of energy loss probablity distributions with initial production spectra and fragmentation functions.

We list our main results and conclusions with CUJET2.0, derived through focusing on CUJET2.0 itself, or through cross comparison with CUJET1.0, as follows:
\begin{enumerate}

\item Rigorous \chisq~analysis indicates CUJET2.0 inclusive pion \raa~calculated with maximum coupling constant $\alpha_{max}=0.25-0.27$ in a dynamical QCD medium with HTL approximation is strictly consistent with both RHIC Au+Au 200AGeV $\pi^0 R_{AA}$ and LHC Pb+Pb 2.76ATeV $h^\pm R_{AA}$ in both central and semi-peripheral collisions at the level of average $\chi^2/d.o.f.<1.5$. If limit average $ \chi^2/d.o.f.<2 $, $ \alpha_{max}=0.23-0.30 $.

\item The effect of the multi-scale running strong coupling and the transverse expanding medium in CUJET2.0 combined to give rise to better agreement between theoretical results and LHC experimental charged hadron \raa's steep rising and subsequent flattening signatures.

\item The small value of best fit \amax~in the CUJET2.0 HTL scenario implies that longer jet path length in a transversely expanding medium overrides the reduction of density and contributes to overall enhanced quenching.

\item CUJET2.0 effective jet transport coefficient $\hat{q}/T^3$ is consistent with not only LO pQCD estimates, but also the $\hat{q}/T^3$'s extracted from HT-BW, HT-M, MARTINI and McGill-AMY models fitting to the same set of experimental hadron suppression factors at RHIC and LHC A+A central collisions \cite{Burke:2013yra}.

%\item The drop of CUJET2.0 $\hat{q}/T^3$ with decreased energy unanticipatedly extends to the ultra soft region, implying the necessity to weigh anisotropy, nonuniformity and inhomogeneity more carefully in both the soft hydro bulk and hard pQCD probe sector, and high $p_T$ jet may ``see'' the ``medium'' differently from pure viscous hydro fields.

\item The robust crossing pattern of $\pi$, $D$, $B$, $e^-$ \raa's is rigorously encoded in the flavor dependent energy loss structure of DGLV opacity expansion combined with TG elastic sector, and a transverse expanding medium has minor effect on this mass hierarchy.

\item Solutions to the ``heavy quark puzzle'' are intrinsically integrated in the framework of CUJET, through the inclusion of elastic energy loss, dynamical QCD medium effect, realistic geometry fluctuation, and energy loss fluctuations.

\item CUJET2.0 predicts a decisively less quenched B meson \raa~which is well above D meson and pion at $5\;{\rm GeV}<p_T<15\;{\rm GeV}$, as well as a critical alternation of \raa's mass ordering from $ B > e^- > D $ at $p_T<15$ GeV to $ e^- > B > D $ at $p_T\gtrsim25$ GeV.

\item We explored the effect of allowing \amax~to vary minimally in different azimuthal directions. We find that even a less than 10\% variation in the averaged coupling strength along in and out of plane paths reduces dramatically the \chisq~for the \vtwo~fit at both RHIC and LHC, and at the same time, azimuthally averaged \raa~results are consistently in agreement with experimental measurements at the level of $\chi^2/d.o.f.<1.5$. The underlying physics responsible for this local effect is subject to feature explorations.

\end{enumerate}

Future work of, improvement on and more thorough test with CUJET2.0 can be made in the following aspects:
\begin{itemize}

\item Calculate the open heavy flavor and heavy flavor lepton's single particle azimuthal anisotropy in the azimuthal dependency included CUJET2.0 framework and compare with existent experimental results to explore the underlying physics responsible for azimuthal \amax~variations.

\item Extrapolate an effective running coupling and an effective scattering potential from lattice QCD $q\bar{q}$ free energy \cite{Bazavov:2013zha} to replace the corresponding running strong coupling and differential cross section in CUJET2.0 to explore non-perturbative jet medium physics near Tc.

\item Integrate Shuryak and Liao's model \cite{Liao:2006ry,Liao:2008jg,Liao:2008dk} of near $T_c$ enhancement of jet-medium interactions in CUJET2.0 to explore non-perturbative local effect. Enhancement of coupling strength originating from non-perturbative structures, created by the color-electric jet passing a plasma of color-magnetic monopoles, dominate the near-Tc matter and could contribute to significant azimuthal variation of \amax.

\item Replace the presently assumed Poisson multiple gluon emission distribution with other radiative energy loss fluctuation patterns to explore the role that fluctuations play on jet quenching pattern and azimuthal anisotropy.

\item Consider more carefully the effects that running scale variations have on hadron spectra and collective flow harmonics. Introduce effective running coupling for each interference term in the summation of current amplitude in the DGLV opacity expansion.

\item Calculate $R_{AA}$ and $v_{2}$ in CUJET2.0 with other hydro evolution profiles such as Luzum and Romatschke hydro \cite{Luzum:2008cw,Luzum:2009sb}, VISHNU \cite{Song:2013qma} and 3+1D idea hydro.

%\item Fulfill calculation of running coupling DGLV opacity series with interpolated scattering potential to all orders in CUJET2.0, consider the convergence of the opacity expansion more carefully in more realistic QGP background.

%\item Test the CUJET2.0 calculation of flavor dependent hadron production spectra with experimental results, besides comparing ratios such as \raa.

\item Improve the structure and algorithm of the CUJET2.0 Monte-Carlo code to realize the calculation of jet-hardron correlation observables \cite{Renk:2011qf,Renk:2012cx,Renk:2011aa}. Constrain the hydro ambiguity and gain comprehensive information about the parton-medium interaction mechanism through azimuthal jet flavor tomography.

\end{itemize}

\begin{acknowledgments}
The authors thank the Nuclear Theory Group of the Nuclear Science Division at LBNL for partial support and extensive discussions with JET Project collaboration members during MG's six month sabbatical visit in 2013. JX and MG also thank the Yukawa Institute for Theoretical Physics, Kyoto University, for very useful discussions during the development of this work at the YITP workshop YITP-T-13-05 on ``New Frontiers in QCD''. MG also acknowledges partial support from the MTA Wigner RCP, Budapest during the second half of his sabbatical leave in 2014 in HU, where this work was completed. Partial support for this work under U.S. DOE Nuclear Science Grants No. DE-FG02-93ER40764 and No. DE-AC02-05CH11231 and OTKA grant NK106119 is also gratefully acknowledged. We are especially grateful to Andrej Ficnar and Xin-Nian Wang for many insightful discussions.
\end{acknowledgments}
\\

\appendix

\section{Notations and conventions}
\label{app:notations}

We adopt the following notations and conventions throughout this paper, unless otherwise footnoted:

When considering experimental observables in  an A+A collision, for example hadron suppression and/or azimuthal anisotropy, z-axis is chosen along the beam direction, and azimuthal plane refers to the plane transverse to the beam axis. In particular, we always define $p_T$ as the transverse momentum perpendicular to the beam direction. Besides $p_T$, physical quantities or concepts involving such coordinates include rapidity/pseudorapidity, bulk evolution profile, jet production distribution, etc.

When considering properties of a single jet, such as quenching in QGP, we choose the jet propagation direction as the ``z-axis'' and define the transverse plane accordingly. This coordinate system applies most importantly to the calculation of radiative and elastic energy loss. Take the scattering with a parton in a particular rapidity frame with a transverse momentum exchange q as an example, we denote $q$ as four vector $q^\mu$, $\vec{\bq}$ as $q$'s 3D space components, and $\bq$ as $\vec{\bq}$'s transverse components with respect to the z-axis, i.e. the jet propagation direction. $q \equiv q^\mu = (q_0, \vec{\bq}) = (q_0, q_z, \bq)$, and $\qT\equiv|\bq|$. Similar notations are also applied to $k^\mu$, $\epsilon^\mu$, $J^\mu$, etc.

Our discussions and calculations are within 4D Minkovski spacetime with signature $(+,-,-,-)$. The spacetime coordinates are ordered as $(x^0, x^1, x^2, x^3)=(t, z, x, y)$. We use light-cone coordinates with metric $ds^2=dx^+ dx^- - \delta_{ij} dx^i dx^j$, where $i, j = 2, 3 $, and $x^+ = x^0 + x^1, x^- = x^0 - x^1$.

\section{Review of fixed coupling DGLV}
\label{sec:model}

%Jet quenching is the collisional process between hard partons and thermal quant%a that results into a loss of energy of the original parton by both elastic and% inelastic (radiative) mechanisms, it leads to gluon radiation as the equivalen%t form of bremsstrahlung in a color plasma. Assuming factorization of scales, t%he interactions between the jet and the plasma can be isolated from the initial% and final states of the collisions and treated perturbatively. 

The DGLV opacity expansion \cite{Gyulassy:1999zd,Gyulassy:2000er,Gyulassy:2003mc,Djordjevic:2003zk} is a theory encompassing inelastic parton-medium interactions and describing gluon radiations in the pQCD framework. As in the WHDG \cite{WHDG} generalization of DGLV, CUJET1.0 and 2.0 supplement the radiative jet-medium interactions with TG \cite{Thoma:1990fm} elastic collisional energy loss in the color medium (Section~\ref{sec:elastic}).

The main computational task performed in CUJET via Monte-Carlo integration is to evaluate the number of radiated gluons per energy fraction $dN_g/dx$ for each initial jet production coordinates $(\bx_0,\hat{\bn})$\footnote{This is the 2D azimuthal plane with respect to the beam axis.}. After that the average inclusive gluon radiation distribution is calculated, fluctuations due to multiple gluon emission is computed via numerical convolution assuming uncorrelated Poisson ansatz, and the normalized radiation probability, $P_{rad}(\Delta E_{rad}, E_0; \vx_0,\hat{\bn}))$ is evaluated via fast Fourier transform including delta function end point singularities (Section~\ref{sec:erg-fluc-rad}). Normalized elastic energy loss probability, $P_{el}(\Delta E_{el}, E_0; \vx_0,\hat{\bn}))$ is also computed with Gaussian multiple collision fluctuations (Section~\ref{sec:erg-fluc-ela}). The final total energy loss probability distribution is the convolution of radiative and elastic sector, $P_{rad}\otimes P_{el}$ (Section~\ref{sec:el-prob}), it is then folded over the initial quark jet spectrum $dN_{pp}/d^2p_Td\eta$ (Section~\ref{sec:conv-initial}). Finally CUJET averages over inital jet configurations via $\int d^2\bx d\hat{n} T_A(\vx+ {\bf b}/2) T_A(\vx-{\bf b}/2)$ and fragments jets into different flavor hadrons or leptons to compare with data (Section~\ref{sec:conv-final}).

\subsection{Radiative energy loss in fixed coupling dynamical scattering DGLV}
\label{sec:DGLV}

The GLV opacity expansion model was developed by Gyulassy, Levai and Vitev \cite{Gyulassy:1999zd,Gyulassy:2000er}, built upon the foundations of the Gyulassy-Wang (GW) potential \cite{Gyulassy:1993hr}, it expresses the partonic energy loss as a series in powers of the opacity $L/\lambda$, where $L$ indicates the size of the plasma and $\lambda$ the mean free path of the parton. At $n$th order in opacity, one considers $n$ scatterings between the parton and the medium. This is often referred to as a thin plasma approximation, which is valid for small values of opacity, as opposed to the thick plasma limit where multiple soft scatterings apply.

The interaction between parton and medium is modeled according to a GW \cite{Gyulassy:1993hr} Debye screened potential with screening mass $\mu$. $\mu$ is considered a fundamental property of the plasma along with the medium density $\rho$. Both of them are expressed as functions of the local temperature T in the system. GLV model includes the power-law tail of the scattering cross section, thus scatterings with large momentum transfer (hard) are taken into account.

There are three major kinematic assumptions made in the GLV theory: soft eikonal approximation, collinear radiation and discrete scattering centers. Details of them are listed below:
\begin{itemize}
\item Eikonal approximation: both the parton energy $E$ and the emitted gluon energy $\omega$ are much larger than the transverse momentum exchanged with the medium $\qT\equiv |\bq|$: $E \gg \qT$ and $\omega \gg \qT$. Soft approximation assumes $\omega \ll E$.
\item Collinear radiation: Gluons are emitted at small angles with respect to the parent parton: $\omega \gg \kT$, where $\kT\equiv |\bk|$ represents the transverse momentum of the gluon.
\item Discrete scattering centers: the mean free path $\lambda$ is much larger than the Debye screening length $1/\mu$, $\lambda \gg 1/\mu$.
\end{itemize}
Under the soft eikonal approximation, the parent parton has sufficiently high energy such that its path is approximately straight. The gluon, which is radiated at small angles, does not carry away a significant portion of the original parton energy, and consequently the jet energy is not dynamically updated during the multiple scattering process.

The most remarkable features of the gluon radiation spectra in the GLV opacity expansion theory comprise the interference effects between production (vertex) radiation and induced radiation, and the interference effects among subsequent scatterings of the radiated gluon in the plasma (quantum cascade). These effects lead to an expression for the double-differential gluon multiplicity distribution in $x$ (fractional gluon energy) and $\kT$ (gluon transverse momentum), which is later integrated to give rise to the energy loss of the parent parton assuming no further exchange of energy with the medium takes place.

An extension of the GLV model to include massive quarks kinematic effects as well as plasmon mass for the gluons was developed by Djordjevic and Gyulassy in \cite{Djordjevic:2003zk} (DGLV). The full derivations can be found in the original papers \cite{Gyulassy:1999zd,Gyulassy:2000er,Djordjevic:2003zk}. Here we will only show the main results and provide their physical interpretations.
\\

In the soft eikonal approximation used to derive DGLV, the incoming jet, gluon and exchanged four momenta read
\be
\begin{array}{rcl}
p&=&(E,E,0)=[2E,0,0]  \;\;,\\
k&=&(\omega=x_EE,\sqrt{(x_EE)^2-\bk},\bk)=[x_+E^+,\frac{\bk^2}{x_+E^+},\bk]  \;\;,\\
q&=&(q_0,q_z,\bq)  \;\;,
\end{array}
\ee
where parenthesis and square brackets denotes Minkowski spacetime and light-cone coordinates respectively. In the above expressions we have suppressed the effective gluon plasmon mass $m_g=\mu/\sqrt{2}$ as well as the the parton mass $M$. In the static scattering center approximation $q_0\sim q_z \ll |\bq|$. The gluon fractional energy $x_E$ and fractional plus-momentum $x_+$ are related via
\be
\label{x+xE}
x_+(x_E)=\frac{1}{2}x_E\lp 1+\sqrt{1-\lp\frac{k_\perp}{x_EE}\rp^2}\rp
\; \; .
\ee
In the pure collinear limit, they coincide. Corrections need to be made for finite emission angles, which involve variations in the upper kinematic integration limit and the Jacobian of the transformation $x_+\rightarrow x_E$:
\be
\label{x+xEJacobian}
J(x_+(x_E)) \equiv\frac{dx_+}{dx_E}=\frac{1}{2}\lp1+\lp1-\lp\frac{k_\perp}{x_EE}\rp^2\rp^{-1}\rp
\; \; .
\ee
A detailed discussion about this issue can be found in Appendix~\ref{app:transmomdistr}.

The double-differential gluon multiplicity distribution in $x_+$ and $\bk$, for DGLV opacity order $n=1$, i.e. the case that the hard parton scatters one single time with weakly-coupled static quasi-particles in the deconfined thermal medium, is given by
\be
\begin{split}
\frac{dN_g^{n=1}}{dx_+ d\bk} = &\; \frac{C_R \alpha_s}{\pi^2}\frac{1}{x_+}\lp\frac{L}{\lambda_g}\rp \int{d\bq} \;{|\bar{v}(\bq)|^2}\\
&\times\;{\frac{-2(\bk-\bq)}{(\bk-\bq)^2+\chi^2} \lp\frac{\bk}{\bk^2+\chi^2}-\frac{(\bk-\bq)}{(\bk-\bq)^2+\chi^2}\rp}\\
&\times\;{\lp1-\cos\lp\frac{(\bk-\bq)^2+\chi^2}{2 x_+ E}\Delta z_1\rp\rp}
\; \; ,
\end{split}
\label{DGLV1Stat}
\ee
with the normalized modular squared scattering potential in a static QCD medium being
\be
|\bar{v}(\bq)|^2 = \frac{\mu^2}{\pi(\bq^2+\mu^2)^2}\; \; .
\label{StaticPontential}
\ee
Here $C_R$ is the quadratic Casimir of the jet ($C_F=4/3$ for quark jets, $C_A=3$ for gluon jets), $\alpha_s=g^2/4\pi$ is the strong coupling constant, and L is the jet path length. Note that the opacity is written in terms of the gluon rather than the jet mean free path, $\lambda_g$, because of a simplification in the color algebra known as ``color triviality'' \cite{Gyulassy:2000er}. $\chi^2=M^2 x_+^2+m_g^2(1-x_+)$ controls the ``dead cone'' (cf. Appendix~\ref{app:transmomdistr}) and LPM destructive interference effects due to both the finite quark current, $M$, and the thermal gluon mass $m_g=\mu(T)/\sqrt{2}$. $\Delta z_1=z_1-z_0$ represents the distance between the scattering points $z_1$ and $z_0$ (production vertex).

The DGLV all-orders result expresses an arbitrary opacity order in a closed form, for an arbitrary collision probability along the jet path. It is applicable for both coherent and incoherent geometries (cf. Appendix~\ref{app:conv}). The gluon multiplicity distribution takes the form:
\be
\begin{split}
\frac{dN_g^{n}}{dx_+ d\bk} = &\; \frac{C_R \alpha_s}{\pi^2}\frac{1}{x_+}\frac{1}{n!}\lp\frac{L}{\lambda_g}\rp^n \int\prod_{i=1}^n\lp{d\bq_i} \lp{|\bar{v}_i({\bf q}_{i})|^2 - \delta^2(\bq_i)}\rp\rp \\
&\times\;-2\;{\mathbf C}_{(1\cdots n)} \cdot \sum_{m=1}^n{\mathbf B}_{(m+1\cdots n)(m\cdots n)} \\
&\times\;{\lp \cos\lp\sum_{k=2}^m\Omega_{(k\cdots n)}\Delta z_k\rp - \cos\lp\sum_{k=1}^m\Omega_{(k\cdots n)}\Delta z_k\rp \rp}
\; \; ,
\end{split}
\label{DGLVNN}
\ee
and the normalized modular squared potential for $i$th static scattering center is
\be
|\bar{v}_i({\bf q}_{i})|^2 = \frac{\mu_i^2}{\pi(\bq_i^2+\mu_i^2)^2}
\; \; .
\label{StaticPotential0}
\ee
This interacting potential has the form of the Debye screened Gyulassy-Wang potential (cf. \cite{Gyulassy:1993hr}), and forward scattering unitarity correction $\delta^2(\bq_i)$ is subtracted from it. $\Delta z_k=z_k-z_{k-1}$ represents the distance between adjacent scattering points $z_k$ and $z_{k-1}$. The kinematic current amplitudes in Eq.~\eqref{DGLVNN} are modified versions of the Hard, Gluon-Bertsch and Cascade terms in GLV theory (cf. \cite{Gyulassy:2000er}), for finite masses:
\be
\begin{array}{lcl}
{\mathbf C}_{(1\cdots n)}&=&\dfrac{\bk-\bq_1-\cdots-\bq_n}{(\bk-\bq_1-\cdots-\bq_n)^2+\chi^2} \\
{\mathbf H}&=&\dfrac{\bk}{\bk^2+\chi^2}  \\
{\mathbf B}_{(i)}&=&{\mathbf H}-{\mathbf C}_{(i)}  \\
{\mathbf B}_{(1\cdots m)(1\cdots n)}&=&{\mathbf C}_{(1\cdots m)}-{\mathbf C}_{(1\cdots n)} \;\;,
\end{array}
\ee
with $\chi^2=M^2 x_+^2+m_g^2(1-x_+)$ and $m_g=\mu(T)/\sqrt{2}$. $\sum_2^1\equiv0$ and ${\mathbf B}_{(n+1\cdots n)(n)}\equiv{\mathbf B}_{(n)}$ is understood. The inverse of formation time $\Omega$ is given by
\be
\Omega_{m\cdots n}=\frac{(\bk-\bq_m-\cdots-\bq_n)^2+\chi^2}{2 x_+ E}
\;\;,
\ee
which regulates the LPM phase of color currents.

In principle, the opacity series should be calculated to sufficiently high order to generate the ``exact'' prediction of experimental observables. However, the numerical power required to drive the computation at such high levels of precision might prove to be insufficient to calculate more complex observables than the simple energy loss for one specific plasma setup. A limitation of this kind would indisputably hinder the capabilities of our algorithm and limit its predictive power. Thus one has to quantify the error introduced by eventually limiting the computations to lower orders in opacity, and in Appendix~\ref{app:conv} we conduct such a systematic study of the convergence of the DGLV opacity series. There we demonstrate that despite another set of observables might scale differently with the opacity, for the purpose of computing the energy loss of different quark flavors, truncating the series at first order already does not add a relevant source of systematic uncertainty. Therefore, unless otherwise stated, all the calculations presented in this paper will be at $n=1$ order.

After computed the differential gluon multiplicity distribution according to Eq.~\eqref{DGLV1Stat}, we then want to integrate over $\bk$ to get the gluon radiation spectrum $x_E dN_g / d x_E$ via
\be
\label{xEdNdxE}
x_E \frac{dN_g}{dx_E}=\int d\bk \lp x_+ \frac{dN_g}{dx_+d\bk}(\bk,x_+(x_E)) \rp \lp \frac{x_E}{x_+(x_E)} \rp J(x_+(x_E)) 
\; \; ,
\ee
where $x_+(x_E)$ and $J(x_+(x_E))$ are defined in Eq.~\eqref{x+xE} and~\eqref{x+xEJacobian} respectively. The lower integration limit for $\kT(\equiv |\bk|)$ is $\kT^{MIN}=0$. The upper kinematic limit $\kT^{MAX}$ is restricted by forward gluon emission and varies with the interpretation of $x$. We set $\kT^{MAX} = x_E E$, and leave the discussion of systematic uncertainties relevant to this kinematic boundary to Appendix~\ref{app:transmomdistr}. 

To account for generic space-time dependent plasma geometries in  Eq.~\eqref{DGLV1Stat} and~\eqref{DGLVNN}, consider a jet created at $\bx_0\equiv(x_0{\rm ,}y_0)$ pointing along $\hat{n} \equiv (\cos\phi, \sin\phi)$ direction in the transverse azimuthal plane with respect to the beam axis, we define
\be
\bz \equiv \lp \bx_0 + \hat{\bn}(\phi)\tau ; \tau \rp = \lp x_0+\tau\cos\phi, y_0+\tau\sin\phi; \tau \rp
\label{position}
\ee
as its coordinates after traveled time $\tau$. The medium seen by the jet has number density $\rho(\bz)$. Since jet travels at approximately the speed of light, in a static medium, the opacity can be expressed as
\be
\frac{L}{\lambda_g} \; \; \longrightarrow \; \; \int_0^L{d\tau}\; \rho(\bz)\sigma_{el}(\bz)
\; \; ,
\label{OpacityLocal}
\ee
and at higher orders,
\be
\frac{1}{n!}\lp \frac{L}{\lambda_g}\rp^n \; \; \longrightarrow \; \; \int_0^L{d\tau_1} \; \rho(\bz_1)\sigma_{el}(\bz_1)\; \cdots \; \int_{\tau_{n-1}}^L{d\tau_n} \; \rho(\bz_n)\sigma_{el}(\bz_n)
\; \; .
\ee
Here the elastic cross section for gluon $\sigma_{el}(\bz)$ can be expanded into gluon-quark and gluon-gluon terms, i.e.
\be
\frac{1}{\lambda_g}=\sigma_{gq}(\bz)\rho_q(\bz) + \sigma_{gg}(\bz)\rho_g(\bz) = \frac{2\pi\;\alpha_s^2}{\mu^2(\bz)}\rho_q(\bz) + \frac{9}{4}\frac{2\pi\;\alpha_s^2}{\mu^2(\bz)}\rho_g(\bz)
\; \; ,
\label{StatLambda}
\ee
where $\mu^2(\bz)=g^2 T(\bz)^2 \lp 1+n_f/6 \rp=4\pi \alpha_s T(\bz)^2 \lp 1+n_f/6 \rp$ is the squared local HTL color electric Debye screening mass in a plasma with number of flavors $n_f$ and local temperature $T(\bz)\propto\rho(\bz)^{1/3}$ along the jet path $\bz$ through the plasma. Assuming ideal gas conditions, from the boson/fermions statistics we obtain the number density of quark $\rho_q(\bz)$ and gluon $\rho_g(\bz)$ is respectively
\be\label{someRho}\begin{array}{rcl}
\rho_q(\bz) &=& \dfrac{9n_f\zeta(3)}{\pi^2} T^3(\bz) \; \; ,\\
\rho_g(\bz) &=& \dfrac{16\zeta(3)}{\pi^2} T^3(\bz)
\; \; .
\end{array}\ee
Combining Eq.~\eqref{StatLambda} and~\eqref{someRho}, we have
\be\label{LambdaPath}
\frac{1}{\lambda_g}=18\frac{\pi\;\alpha_s^2}{\mu^2(\bz)}\frac{4+n_f}{16+9n_f}\rho(\bz)= 3 \alpha_s \; T(\bz) \lp 6 \frac{\zeta(3)}{\pi^2} \frac{1+n_f/4}{1+n_f/6} \rp
\; \; ,
\ee
with $\rho(\bz)=\rho_q(\bz)+\rho_g(\bz)$.

To get the DGLV gluon radiation spectrum at the first order in opacity, combine Eq.~\eqref{DGLV1Stat}, \eqref{xEdNdxE}, \eqref{OpacityLocal}, and \eqref{LambdaPath} together, we get
\be\label{fcDGLV}
\begin{split}
x_E \frac{dN_g^{n=1}}{dx_E}(\bx_0,\phi) = &\; \frac{18 C_R \alpha_s^3}{\pi^2} \frac{4+n_f}{16+9n_f} \int{d\tau}\; \rho(\bz) \int{d\bk} \int{d\bq}\; {|\tilde{v}(\bq)|^2}\\
&\times\;{\frac{-2(\bk-\bq)}{(\bk-\bq)^2+\chi^2(\bz)} \lp\frac{\bk}{\bk^2+\chi^2(\bz)} - \frac{(\bk-\bq)}{(\bk-\bq)^2+\chi^2(\bz)}\rp}\\
&\times\;{\lp1-\cos\lp\frac{(\bk-\bq)^2+\chi^2(\bz)}{2 x_+ E } \tau\rp\rp}\\
&\times\;{\lp \frac{x_E}{x_+} \rp J(x_+(x_E))}
\; \; .
\end{split}
\ee
This is the DGLV $n=1$ kernal in fixed coupling CUJET. Recall $C_R$ is the quadratic Casimir of the jet ($C_F=4/3$ for quark jets, $C_A=3$ for gluon jets). Local $\chi^2(\bz)=M^2 x_+^2+m_g^2(\bz)(1-x_+)$, with M being the mass of the parton, gluon mass $m_g(\bz)=\mu(\bz)/\sqrt{2}$, and Debye mass
\be
\mu^2(\bz)=g^2 T(\bz)^2 \lp 1+n_f/6 \rp=4\pi \alpha_s T(\bz)^2 \lp 1+n_f/6 \rp\;\;.
\label{DebyeMass}
\ee
For a static QCD medium, $|\tilde{v}(\bq)|^2$ is defined via\footnote{Note we define $|\tilde{v}(\bq)|^2$ as the normalized squared scattering potential from now on, distinguish from $|\bar{v}(\bq)|^2$ which is defined via Eq.~\eqref{StaticPotential0}.}
\be
|\tilde{v}(\bq)|^2 = \frac{1}{(\bq^2+\mu^2)^2}
\; \; .
\label{StaticPotential}
\ee

The total energy $\Delta E$ carried away by the emitted gluons is obtained by integrating the radiation spectrum, Eq.~\eqref{fcDGLV}. Assuming no further interaction between jet and medium, this can readily been interpreted as the energy loss that the jet suffers when propagates through a hot deconfined plasma
\be
\frac{\Delta E}{E}=\int dx_E \; x_E \frac{dN_g^{n=1}}{dx_E}
\;\;.
\ee
If there are no kinematic boundaries on the $d\bq$ and $d\bk$ integrations, for a plasma with fixed size $L$ and constant gluon mean free path $\lambda_g$, a straightforward analytic computation for this first order in opacity leads to the asymptotic result
\be
\frac{\Delta E}{E} = \frac{C_R\alpha_s}{4}\frac{L^2\mu^2}{\lambda_g}\frac{1}{E}\log\frac{E}{\mu}
\;\;.
\label{asymptDE}
\ee
We immediately notice the $L^2$ dependence of the energy loss, this is the characteristic of the LPM region, which differs from the incoherent limit whose dependence on $L$ is linear.

However, as discussed previously, kinematic boundaries exist not only for the $\kT$ integration ($\kT^{MIN}=0$ and $\kT^{MAX}=x_E E$), but also for the integration of transverse momentum transfer $\qT(\equiv |\bq|)$, for which we choose $\qT^{MIN}=0$ and set the upper limit $\qT^{MAX}={\rm min}\lp\kT,\sqrt{4ET(\bz)}\rp$.

The $(\bx_0,\phi)$ dependence of $x_E dN_g^{n=1}/dx_E$ in Eq.~\eqref{fcDGLV} comes from the $\bz$ coordinates (Eq.~\eqref{position}), and strictly speaking,  
\be
x_E \frac{dN_g^{n=1}}{dx_E} = x_E \frac{dN_g^{n=1}}{dx_E}(x_E; \bx_0,\phi; M, E; \alpha_s; L, n_f)\;\;.
\ee
i.e. the radiated gluon spectrum for a quark/gluon jet with mass $M$ and energy $E(\equiv E_0)$ created at $\bx_0$ position in the transverse plane along azimuthal angle $\phi$ has explicit dependence on the strong coupling constant $\alpha_s$, jet path length $L$ and number of quarkonic flavors $n_f$\footnote{As well as other plasma parameters such as the formation time and thermalization scheme, which are discussed extensively in Section~\ref{sec:bulk}, Section~\ref{sec:initialtime} and Appendix~\ref{app:thermal}.}. In CUJET calculation, we take into account fluctuations of the geometry by cutoff the $d\tau$ integral at $\tau^{MAX}$, with $T(\bz)|_{\tau^{MAX}} = T_f$. By doing so the $L$ dependence turns into a fragmentation temperature $T_f$ dependence. We leave the discussion of the systematic uncertainties associated with varying $T_f$ and $n_f$ to Appendix~\ref{app:systematics}. There we show the variation of jet quenching spectra originating from both of them can be fully absorbed into a simple rescaling of $\alpha_s$. This makes $\alpha_s$\footnote{In the case of running coupling CUJET which is discussed in Section~\ref{sec:rc}, this parameter is the maximum strong coupling constant \amax.} the only free parameter of CUJET calculation in a static QCD medium, in which case the radiative energy loss reads
\be
\Delta E_{rad}(\bx_0,\phi;M,E_0;\alpha_s) = \int_{0}^{1} dx_E x_E \frac{dN_g^{n=1}}{dx_E}(x_E; \bx_0,\phi; M, E; \alpha_s)\;\;.
\label{DeltaErad}
\ee

However, in such a deconfined medium consisting of randomly distributed static scattering centers, the collisional energy loss is exactly zero. This contradicts recent calculations which show elastic collisional contribution is important and comparable to the radiative energy loss \cite{WHDG}. Therefore, the natural improvement on the DGLV opacity expansion is to consider a dynamically screened QCD medium. The inclusion of these dynamical effects is achieved by computing the scattering QCD diagrams in a finite temperature field theory framework, using Hard Thermal Loop resumed propagators for all gluons. The quark gluon plasma is assumed to be thermalized at temperature $T$ and has zero baryon density. Details of the computations can be found in original papers \cite{oai:arXiv.org:hep-ph/0204146,oai:arXiv.org:hep-ph/0207206,Djordjevic:2007at,Djordjevic:2008iz,Djordjevic:2009cr}.

The dynamical QCD medium brings two major corrections to the DGLV radiated gluon number distribution Eq.~\eqref{fcDGLV}: firstly, the effective dynamical mean free path for gluon $\lambda_{dyn}$, which is defined as $\lambda_{dyn}^{-1}\equiv 3\alpha_s T$, will replace its static counterpart $\lambda_g(\equiv\lambda_{stat})$ in Eq.~\eqref{DGLV1Stat}. According to Eq.~\eqref{LambdaPath}, they are related via
\be
\lambda_{dyn} = c(n_f) \lambda_{stat} =  \lp 6 \frac{\zeta(3)}{\pi^2} \frac{1+n_f/4}{1+n_f/6} \rp \lambda_{stat} \;\;,
\ee
with $n_f$ the number of effective quark flavors in equilibrium with the gluons in the plasma. However, in CUJET calculations we degenerate this mean free path effect on $\delta E /E$ by a rescaling of the effective strong coupling constant, because the coefficient $c(n_f)$ varies from $c(0)=0.73$ to $c(\infty)=1.09$, and does not contribute much to the energy loss compared to the magnetically enhanced potential which will soon be discussed.

Secondly and most importantly, the dynamical recoiling of color electric scattering centers induces an effective color magnetic screening mass which is smaller than the Debye mass, and the interaction potential
\be
|\tilde{v}({\bf q})|^2 = \frac{1}{\bq^2(\bq^2+\mu^2)}
\; \; .
\label{DynamicalPotential}
\ee
The implications of these changes are profound: the absence of the $\mu^2$ screening for soft momenta exchanges $\bq$ makes the potential diverges and the mean free path vanishes. In the limit of $\bq \rightarrow 0$, each individual Feynman diagram diverges logarithmically. These singularities however cancel out after all the contributing diagrams to the energy loss are summed over, making the gluon multiplicity finite. 

The combined effect of the enhanced cross section and reduced mean free path contributes to a remarkable increase for the magnitude of total energy loss and the ratio of heavy to light quark energy loss in the dynamical framework, systematic studies of this effect can be found in \cite{Djordjevic:2008iz} and \cite{Buzzatti:2010ck}.

In the CUJET model, motivated by lattice QCD $q\bar{q}$ potential data, we introduce an effective interaction potential with deformation parameters $(f_E,f_M)$, it reads\footnote{A similar effective potential is proposed by authors in \cite{oai:arXiv.org:1105.4359,oai:arXiv.org:1209.0198}.}
\be
|\tilde{v}({\bf q})|^2 = \frac{f_E^2-f_M^2}{(\bq^2+f_E^2\mu^2)(\bq^2+f_M^2\mu^2)}
\; \; .
\label{HybridPotential}
\ee
Here the HTL deformation parameters $(f_E,f_M)$ are used to vary the chromo-electric and chromo-magnetic screening scales relative to HTL. In principle, HTL deformations could also change $m_g(T)$. The default HTL plasma is $(1,0)$, but we also consider a deformed $(2,0)$ non-HTL plasma model which will be discussed in Section~\ref{sec:non-HTL}. $(\alpha_s\footnote{In the case of running coupling CUJET which will be discussed in Section~\ref{sec:rc}, this parameter is the maximum strong coupling \amax.}, f_E, f_M)$ are therefore our main model space control parameters.

\subsection{Elastic energy loss}
\label{sec:elastic}

The assumption that pQCD elastic energy loss is negligible compared to radiative one is questionable. In \cite{Mustafa:2003vh,Mustafa:2004dr}, the authors found that radiative and elastic average energy losses for heavy quarks were in fact comparable over a very wide kinematic range accessible at the RHIC. In \cite{WHDG}, the authors confirm these previous findings and extend them to the light quark sector, showing that elastic contributions to the total energy loss can be of the same order of magnitude of radiative ones.

It is then clear that quantitative tomographic predictions cannot ignore such large contributions to jet quenching, and elastic effects need to be included in CUJET as well.

We use Thoma-Gyulassy (TG) model \cite{Thoma:1990fm} in our calculation of the elastic energy loss. Their work was based on Bjorken's estimation of elastic energy loss in QGP (cf. appendix~\ref{app:elastic}). By using the hard thermal loop gluon propagators to provide a more natural infrared regulator, the TG computation leads to the following leading log result:
\be
\frac{dE}{dx}=-C_R \pi\alpha_s^2 T^2 \lp 1+\frac{2}{6} \rp\lp \frac{1}{v}+\frac{v^2-1}{2v^2}\log\frac{1+v}{1-v}\rp \log\lp\frac{k_{max}}{\mu}\rp
\; \; .
\label{TGcollEq}
\ee
Where $x$ is the jet path. For ultra-relativistic particles, the velocity $v$ can be approximated to $1$ and the $v$-dependent factor in parenthesis becomes approximately $1$. The integral over $k$ is infrared finite due to the Debye screening mass in the denominator, but a maximal momentum $k_{max}$ must be set in order to screen the otherwise ultraviolet divergent logarithm. Assuming that the maximal momentum transfer comes from forward scattering against target particles with average momenta $q \approx 2T$ is much smaller than the projectile momentum, the value of $k_{max}$ is $4Tp/(E-p+4T)$, with $p=\sqrt{E^2-M^2}$.

We immediately see that this model yields a result very similar to the Bjorken computation, Eq.~\eqref{bjergloss}, i.e.
\be
\frac{dE}{dx}= - C_R \pi \alpha_s^2 T^2 \lp 1+\frac{n_f}{6} \rp \log B
\; \; ,
\label{BjElasticLog}
\ee
with a different Coulomb log that reflects the more natural cutoffs which are being used now:
\be
\log\lp\frac{k_{max}}{\mu_D}\rp \equiv \log\lp\frac{4Tp}{(E-p+4T)\mu}\rp
\; \; .
\label{ElasticLog}
\ee
For the elastic energy loss sector in CUJET, assuming jet travels at the speed of light, combining Eq.~\eqref{TGcollEq},~\eqref{BjElasticLog} and~\eqref{ElasticLog}, we get the following equation to account for collisional effects:
\be
\frac{dE(\bz)}{d\tau}= - C_R \pi \alpha_s^2 T(\bz)^2 \lp 1+\frac{n_f}{6} \rp \log\lp\frac{4T(\bz)\sqrt{E(\bz)^2-M^2}}{\lp E(\bz)-\sqrt{E(\bz)^2-M^2}+4T(\bz)\rp\mu(\bz)}\rp.
\label{CUJETElastic}
\ee
Here $C_R$ is the quadratic Casimir of the jet ($C_F=4/3$ for quark jets, $C_A=3$ for gluon jets). $\bz$ and $\mu(\bz)$ are defined according to Eq.~\eqref{position} and~\eqref{DebyeMass} respectively. $T(\bz)$ is temperature profile of the medium. To compute elastic energy loss in CUJET, we solve recursively Eq.~\eqref{CUJETElastic}, with initial condition $E(\bz)|_{\tau=0}=E_0$ and evolve $\tau$ to a cutoff $\tau_{max}$ which is related to the fragmentation temperature $T_f$ via $T(\bz)|_{\tau=\tau_{max}}=T_f$, i.e.
\be
\Delta E_{el}(\bx_0,\phi;M,E_0;\alpha_s) = E(\tau;\bx_0,\phi;M,E_0;\alpha_s)\lvert_{\tau=0}^{\tau=\tau_{max}} = \int_{0}^{\tau_{max}} d\tau \frac{dE(\bz)}{d\tau}\;\;.
\label{DeltaEel}
\ee
Note similar to radiative energy loss, the $n_f$ and $T_f$ dependence of elastic energy loss is absorbed into the $\alpha_s$ degree of freedom. The recursive scheme is stopped when $E(\bz)$ drops below $M$ and returns maximum energy loss $\Delta E_{max}=E_0-M$, even if $\tau$ does not reach $\tau_{max}$. If local temperature $T(\bz)|_{\tau=\tau_0}=0$ for some $\tau_0$, the scheme will skip computing $dE/dx$ and keep evolving $\tau$.

Despite its improvement over the Bjorken result, the TG model leaves the ultraviolet region unbounded, because the classical calculation has no knowledge about the particle nature of the medium and particle recoil, which becomes important when the momentum transfer $q$ is large. The hard momentum transfer contribution is more naturally taken into account by Braaten and Thoma in \cite{Braaten:1991jj,Braaten:1991we}, but relevant analysis shows that the differences in practical applications are almost negligible.

We include here for clarity the calculation of average number of collisions $\bar{N_c}$, which plays an important role calculating the elastic energy loss fluctuations in Section~\ref{sec:erg-fluc-ela}. Recall Eq.~\eqref{StatLambda},~\eqref{someRho} and~\eqref{LambdaPath}, $\bar{N_c}$ reads
\be
\bar{N_c}=\int_{0}^{\tau_{max}} d\tau \lambda^{-1}(\bz) = \int_{0}^{\tau_{max}} d\tau \lp \frac{\alpha_s^2}{\mu(\bz)^2}\rp \lp \frac{18 \zeta(3)}{\pi} (4+n_f) T(\bz)^3 \rp\;\;.
\label{NumOfColl}
\ee

Numerical results for elastic energy loss, especially its effects on the ratio of light to heavy quark suppression magnitude, is discussed in Appendix~\ref{app:elastic}.

\subsection{Fluctuations}
\label{sec:fluc}

The radiative energy loss calculated from Eq.~\eqref{fcDGLV}\eqref{HybridPotential}\eqref{DeltaErad} and elastic energy loss calculated from Eq.~\eqref{CUJETElastic}\eqref{DeltaEel} both perform full jet path integration\footnote{This is the fixed coupling case, for running coupling CUJET, radiative energy loss is calculated from Eq.~\eqref{rcCUJETDGLV}\eqref{DeltaErad}, while elastic energy loss is calculated according to Eq.~\eqref{rcCUJETElastic}\eqref{DeltaEel}.}. The non-uniform medium's fluctuating geometry due to expanding and cooling is properly embedded in the local plasma density $\rho(\bz)$ and the cutoff fragmentation temperature $T_f$. This jet path integration provides a platform to quantify the effects of complicated heavy ion collision configurations in predicting experimental observables in the pQCD framework.

Besides the geometry, fluctuations originating from multiple gluon emissions in the radiative sector and multiple partonic collisions in the elastic sector also play important roles in jet quenching, and they may significantly influence the results of hadron multiplicity and azimuthal flow. We dedicate this section to introduce the quantification and computation of them in CUJET.

\subsubsection{Radiative energy loss fluctuation}
\label{sec:erg-fluc-rad}

The DGLV integrals, Eq.~\eqref{DGLV1Stat}, \eqref{DGLVNN}, \eqref{fcDGLV} and \eqref{rcCUJETDGLV}, are constructed starting from diagrams with only one external gluon line; multiple gluon emission can be calculated by repeating the single gluon emission kernel in an incoherent fashion.

The simplest assumption for multiple gluon emission is the Poisson ansatz, where the number of emitted gluons follows a Poisson distribution, with the mean number $\overline{N}_g$ given by the integral of the gluon emission spectrum
\be
\overline{N}_g=\int_{0}^{1} dx_E \frac{dN_g^{n=1}}{dx_E}(x_E)\;\;.
\ee
The gluon radiation can be thought of as a stochastic event, and it makes sense to speak of a probability distribution $P_{rad}(\epsilon)$ of radiating a certain amount of energy $\epsilon\equiv\Delta E_{rad}/E_0$:
\be
P_{rad}(\epsilon) = P_{r}^{null} \delta(\epsilon) + P_r(\epsilon) + P_{r}^{full} \delta(\epsilon - \epsilon_{max})\;\;,
\label{Poissonexp}
\ee
where the maximum energy loss ratio $\epsilon_{max}=1-M/E_0$. For simplicity, in the discussion below we suppress the $n=1$ superscript of $N_g^{n=1}$ and $E$ subscript of $x_E$ and write them as $N_g$ and $x$ respectively. The probability distribution Eq.~\eqref{Poissonexp} is split into three components:

The first term corresponds to the probability of zero radiation, $P_{r}^{null}=\eexp^{-\overline{N}_g}$.

The second term is given by
\be
{P_r}(\epsilon)=\sum_{n=1}^\infty {P}_n(\epsilon)
\;\;,
\label{multiGseries}
\ee
with
\be
\begin{array}{rcl}
{P}_0(\epsilon) &=& P_{null}(\epsilon)= \eexp^{-\overline{N}_g}\;\;,\\
{P}_1(\epsilon) &=& P_0\dfrac{dN_g}{dx}(x=\epsilon)\;\;,\\
\end{array}
\ee
and
\be
{P}_{n+1}(\epsilon) = \dfrac{1}{n+1}\int_0^1 dx_{n} {P}_{n}(\epsilon-x_{n}) \dfrac{dN_g}{dx}(x_n)
\;\;.
\label{Pradn}
\ee
We use fast Fourier transform techniques to solve this equation numerically. Denote $\tilde{P_i}(k)$ and ${P_i}(\epsilon)$, $\tilde{\frac{dN_g}{dk}}(k)$ and $\frac{dN_g}{dx}(x)$ as the Fourier integral pairs, i.e.
\be
\begin{array}{rcl}
\tilde{P_i}(k) &=& \int d\epsilon\;\eexp^{{\mathrm i} k \epsilon}\;{P_i}(\epsilon)\;\;,\\
\tilde{\frac{dN_g}{dk}}(k) &=& \int dx\;\eexp^{{\mathrm i} k x}\;\frac{dN_g}{dx}(x)\;\;,\\
\end{array}
\ee
we immediately get from Eq.~\eqref{multiGseries}~\eqref{Pradn} that
\be
\tilde{P_r}(k)=\sum_{n=1}^\infty \tilde{P_n}(k)
\;\;,
\label{multiGseries1}
\ee
and
\be
\tilde{P_{n}}(k) = \dfrac{1}{n!}\lp \tilde{\frac{dN_g}{dk}}(k) \rp^n P_0
\;\;.
\label{Pradn1}
\ee
Plug Eq.~\eqref{Pradn1} into Eq.~\eqref{multiGseries1}, we get
\be
\tilde{P_r}(k)=P_0\lp\exp\lp\tilde{\frac{dN_g}{dk}}(k)\rp - 1 \rp
\;\;.
\label{multiGseries2}
\ee
Fourier transform back, we have
\be
{P_r}(\epsilon)=\dfrac{\eexp^{-\overline{N}_g}}{2\pi}\int dk\;\eexp^{- {\mathrm i} k \epsilon} \lp\exp\lp\int dx\;\eexp^{{\mathrm i} k x}\;\frac{dN_g}{dx}(x)\rp - 1 \rp
\;\;.
\label{multiGseriesFinal}
\ee
Practically, the numerical evaluation of Eq.~\eqref{multiGseriesFinal} uses finite discrete ${k_i}$ and ${x_j}$ series, for example, $k_i=-1000+i\;(i=0,1,\cdots,2000)$ and $x_j=j\sigma\;(j=0,1,\cdots,\sigma^{-1};\sigma=0.0025)$, meaning the Fourier transform in the $\exp(...)$ of Eq.~\eqref{multiGseriesFinal} is in fact
\be
\int dx\;\eexp^{{\mathrm i} k x}\;\frac{dN_g}{dx}(x) \rightarrow \sum_j \eexp^{{\mathrm i} k_i x_j}\;\frac{dN_g}{dx}(x_j)~\sigma \;\;.
\ee
The $\frac{dN_g}{dx}(x)$ itself is fluctuating because of limited computing power to implement Monte-Carlo iterations. At large $|k_i|$, this fluctuation is worsened with the highly oscillating $\eexp^{{\mathrm i} k_i x_j}$, and will generate unphysical variations in ${P_r}(\epsilon)$. However, if take the $\int dk\;\eexp^{- {\mathrm i} k \epsilon}$ in Eq.~\eqref{multiGseriesFinal} into account, one sees components with larger $|k|$ will have less weight in the evaluation of ${P_r}(\epsilon)$. Therefore, we smoothfy the $\exp(...)$ in Eq.~\eqref{multiGseriesFinal} by adding a Gaussian smoother with proper width, put more weight on small $k$ Fourier components, and modify Eq.~\eqref{multiGseriesFinal} to
\be
{P_r}(\epsilon)=\dfrac{\eexp^{-\overline{N}_g}}{2\pi}\int dk\;\eexp^{- {\mathrm i} k \epsilon} \left[ \exp \left\lbrace  \sum_j \eexp^{{\mathrm i} k x_j}\;\frac{dN_g}{dx}(x_j) \:\sigma\: \eexp^{-\frac{k^2\sigma^2}{2}} \right\rbrace  - 1  \right] 
\;\;.
\label{PradnFinal}
\ee
with $x_j=j\sigma\;(j=0,1,\cdots,\sigma^{-1})$. In CUJET, ${P_r}(0)=0$, we use Eq.~\eqref{PradnFinal} to calculate ${P_r}(\epsilon)$ in the range of $0<\epsilon\le\epsilon_{max}$, as well as in the range of $\epsilon_{max}<\epsilon\le\epsilon_{leak}=1.75$ for numerical purposes.

The third and last term in Eq.~\eqref{Poissonexp} represents instead the probability of total quenching. In the soft approximation, the radiated energy $\omega$ is assumed much smaller than the initial jet energy $E$, and $x\ll1$. Consequently, the energy of the outgoing parton $E'$ is approximately equal to $E$. When the $\{x_n\}$ are integrated up to the kinematic limit $x_n=1$, a ``leakage'' error into the unphysical region ${P_r}(\epsilon>\epsilon_{max})\neq 0$ occurs, and this error is calculated in $P_{r}^{full} = \int_{\epsilon_{max}}^\infty d\epsilon\; {P_r}(\epsilon)$\footnote{In the numerical evaluation, the upper bound is $\epsilon_{leak}=1.75$ instead of infinity.}.

For the normalization of ${P_{rad}}(\epsilon)$ we keep the weight of the physical zero quenching probability $P_{null}$ unchanged, and rescale the probability distribution as follows: firstly, we calculate the norm $\mathcal{N}_{rad}$ from $\mathcal{N}_{rad}=\int_{0}^{\epsilon_{max}} {P_{rad}}(\epsilon)$. When doing this integral, the Delta functions at both boundaries are included. Secondly, we rescale the complete ${P_{rad}}(\epsilon)$ according to ${P_{rad}}(\epsilon) \rightarrow \frac{1-\eexp^{-\bar{N}_g}}{\mathcal{N}_{rad}}{P_{rad}}(\epsilon)$. And finally we replace the coefficient of $\delta(\epsilon)$ in ${P_{rad}}(\epsilon)$ with zero radiation probability, i.e. $\frac{1-\eexp^{-\bar{N}_g}}{\mathcal{N}_{rad}} P_{null} \rightarrow \eexp^{-\bar{N}_g}$. Through this procedure we maintain $\int_{0}^{\epsilon_{max}} {P_{rad}}(\epsilon)=1$, and the $\delta(\epsilon)$ at the $\epsilon=0$ boundary has weight $\eexp^{-\bar{N}_g}$. If $\bar{N}_g=0$, ${P_{rad}}(\epsilon)=\delta(\epsilon)$.

The effects of multiple gluon emission on the ratio of light to heavy quark energy loss is a topic of Appendix~\ref{app:fluc}. Note ${P_{rad}}(\epsilon)$ inherits all the jet production coordinates, parton mass and energy, and model parameter dependencies from $\frac{dN_g}{dx}$. And we write down explicitly those dependencies as:
\be
P_{rad}(\epsilon)=P_{rad}(\epsilon=\Delta E_{rad}/E_0; \bx_0, \phi; M, E_0; \alpha_s, f_E, f_M)\;\;.
\ee
	
\subsubsection{Elastic energy loss fluctuation}
\label{sec:erg-fluc-ela}

Fluctuations of the elastic energy loss around the mean were addressed in \cite{WHDG} and \cite{Wicks:2008zz}. Using a framework generally applied to diffusive processes that are characterized by a large number of soft collisions, the probability distribution to lose the collisional energy $\epsilon \equiv \Delta E_{el}/E_0$ is represented by a Gaussian centered around the average $\overline{\Delta E}_{el}$, with variance $\sigma^2=2\overline{T}\overline{\epsilon}/E_0$. Here $\overline{\epsilon} \equiv \overline{\Delta E}_{el}/E_0$, and the average elastic energy loss $\overline{\Delta E}_{el}$ is calculated according to Eq.~\eqref{DeltaEel},
\be
\overline{\Delta E}_{el} = E(\tau;\bx_0,\phi;M,E_0;\alpha_s)\lvert_{\tau=0}^{\tau=\tau_{max}}\;\;,
\ee
with $T(\bz)|_{\tau=\tau_{max}}=T_f$, and $E(\bz)$ is solved recursively from Eq.~\eqref{CUJETElastic} given $E(\bz)|_{\tau=0}=E_0$. The average temperature along the jet path is
\be
\overline{T} = \frac{1}{\tau_{max}} \int_{0}^{\tau_{max}} d\tau\; T(\bz) \;\;.
\ee
The collisional energy loss probability distribution reads
\be
P_{el}(\epsilon) = e^{-\overline{N}_c}\delta(\epsilon) + \frac{\cal N}{ \sqrt{2\pi\sigma^2}} e^{-\frac{(\epsilon-\overline{\epsilon})^2}{2\sigma^2}}
\; \; .
\label{elFluc}
\ee
The first term represents the probability of no collisions, with the average number of collisions $\overline{N}_c$ calculated according to Eq.~\eqref{NumOfColl} (or Eq.~\eqref{rcNumOfColl} in running coupling CUJET). The second term is the normalized Gaussian distribution centered around $\overline{\epsilon}$, with ${\cal N}=1-e^{-\overline{N}_c}$. The Gaussian distribution reaches unphysical regions $\epsilon<0$ and $\epsilon>\epsilon_{max}$, we absorb those ``leaks'' into the Delta function at respective boundaries, and rewrite $P_{el}(\epsilon)$ as
\be
P_{el}(\epsilon) = P_{e}^{null}\delta(\epsilon) + P_e(\epsilon) + P_{e}^{full}\delta(\epsilon-\epsilon_{max})
\; \; .
\label{Gaussianexp}
\ee
which resembles the definition of $P_{rad}(\epsilon)$ Eq.~\eqref{Poissonexp}. Here
\be
\begin{array}{rcl}
P_{e}^{null}& = & e^{-\overline{N}_c} + \int_{-\infty}^{0} d\epsilon \frac{\cal N}{ \sqrt{2\pi\sigma^2}} e^{-\frac{(\epsilon-\overline{\epsilon})^2}{2\sigma^2}}\;\;,\\
P_{e}^{full}\:& = & \int_{\epsilon_{max}}^{\infty} d\epsilon \frac{\cal N}{ \sqrt{2\pi\sigma^2}} e^{-\frac{(\epsilon-\overline{\epsilon})^2}{2\sigma^2}}\;\;,
\end{array}
\ee
and
\be
P_e(\epsilon) = \frac{\cal N}{ \sqrt{2\pi\sigma^2}} e^{-\frac{(\epsilon-\overline{\epsilon})^2}{2\sigma^2}}\;\;,
\label{PelEp}
\ee
here $0\le\epsilon\le\epsilon_{max}$. For numerical purposes we also calculate $P_e(\epsilon)$ in the range of $\epsilon_{max}\le\epsilon\le\epsilon_{leak}=1.75$ according to Eq.~\eqref{PelEp}. Note integrate $P_{el}(\epsilon)$ over $0\le\epsilon\le\epsilon_{max}$ automatically gives unity. The rearrangement of Eq.~\eqref{Gaussianexp} provides great conveniences for the convolution of radiative and elastic energy loss probability distributions, which will be studied in the following section.

Similar to Section~\ref{sec:erg-fluc-rad}, inherited from $\overline{\Delta E}_{el}$ and $\overline{N}_{c}$, the elastic energy loss probability distribution has jet production coordinates, parton mass and energy, and model parameter dependency. We write down all those dependencies for $P_{el}(\epsilon)$ as:
\be
P_{el}(\epsilon)=P_{el}(\epsilon=\Delta E_{el}/E_0; \bx_0, \phi; M, E_0; \alpha_s, f_E, f_M)\;\;.
\ee

\subsection{Convolutions}
\label{sec:convolution}

In the CUJET framework, after calculated the radiative energy loss probability distribution $P_{rad}(\epsilon)$ from Eq.~\eqref{Poissonexp} and elastic energy loss probability distribution $P_{rad}(\epsilon)$ from Eq.~\eqref{Gaussianexp}, we convolute the their contributions to get the total energy loss probability distribution $P_{tot}(\epsilon)$ (Section~\ref{sec:el-prob}). Then integrate $P_{tot}(\epsilon=\Delta E_{tot}/E_0, E_0,\bx_0,\phi; M)$ with the pQCD p+p parton (M) spectrum and binary distribution to get the quenched A+A parton (M) spectrum (Section~\ref{sec:conv-initial}). Finally, we fragment this parton spectrum to get the transverse momentum and azimuthal angle dependent production spectrum for inclusive $\pi$, $D$, $B$ and $e^-$ in A+A collisions (Section~\ref{sec:conv-final}).

\subsubsection{Total energy loss probability distribution}
\label{sec:el-prob}

To get the total energy loss probability distribution $P_{tot}(\epsilon)$, we convolute the radiative sector $P_{rad}(\epsilon)$(Eq.~\eqref{Poissonexp}) and the elastic sector $P_{el}(\epsilon)$ (cf. Eq.~\eqref{Gaussianexp}):
\be
P_{tot}(\epsilon)=\int_{0}^{\epsilon} {dx}\; P_{rad}(x) P_{el} (\epsilon-x)
\; \; .
\label{fullFluc}
\ee

Technically, in CUJET, when computing the convolution for total suppression, we keep the $\delta$ function at 0 in each sector while let $P_r(\epsilon)$ and $P_e(\epsilon)$ spread over $0\le\epsilon\le\epsilon_{leak}=1.75$, then absorb the convoluted leak to the $\delta$ function at $\epsilon_{max}$. Step by step, first we rewrite $P_{rad}(\epsilon)$ and $P_{el}(\epsilon)$ as
\bea
P_{rad}(\epsilon) = e^{-\overline{N}_g} \delta(\epsilon) + P_r(\epsilon)\; \;,\\
P_{el}(\epsilon) = e^{-\overline{N}_c} \delta(\epsilon) + P_e(\epsilon)\; \;.
\eea
with $P_r(\epsilon)$ and $P_e(\epsilon)$ calculated over $0\le\epsilon\le\epsilon_{leak}=1.75$. Then multiply them both according to Eq.~\eqref{fullFluc}, we get
\be
\begin{split}
P_{tot}(\epsilon)
& = \int_{0}^{\epsilon} {dx}\; \lp e^{-\overline{N}_g} \delta(x) + P_r(x) \rp \lp e^{-\overline{N}_c} \delta(\epsilon-x) + P_e(\epsilon-x) \rp\\
& = e^{-(\overline{N}_g + \overline{N}_c )} \delta(\epsilon) + e^{-\overline{N}_g} P_e(\epsilon) + e^{-\overline{N}_c} P_r(\epsilon) + \int_{0}^{\epsilon} {dx}\;P_r(x) P_e(\epsilon-x)
\; \; .
\end{split}
\label{PtotFull}
\ee
We define
\be
\begin{array}{rcl}
P_{t}^{null} & = & e^{-(\overline{N}_g +\overline{N}_c )}\; \;,\\
P_{t}^{full} & = & \int_{\epsilon_{max}}^{\epsilon_{leak}}{d\epsilon}\;\int_{0}^{\epsilon}{dx}\;P_r(x) P_e(\epsilon-x)\; \;,\\
P_{t}(\epsilon) & = & e^{-\overline{N}_g} P_e(\epsilon) + e^{-\overline{N}_c} P_r(\epsilon) + \int_{0}^{\epsilon} {dx} P_r(x) P_e(\epsilon-x)\; \;,
\end{array}
\label{PtotDef}
\ee
and rewrite $P_{tot}(\epsilon)$ as
\be
P_{tot}(\epsilon) = P_{t}^{null}\delta(\epsilon) + P_t(\epsilon) + P_{t}^{full}\delta(\epsilon-\epsilon_{max})\;\;,
\label{PtotFinal}
\ee
here $0\le\epsilon\le\epsilon_{max}$. The normalization of $P_{tot}(\epsilon)$ is conducted in the normal way to ensure $\int_{0}^{\epsilon_{max}} d\epsilon\:P_{tot}(\epsilon)=1$.

Strictly speaking, $P_{tot}(\epsilon)$ depends on other parameters such as parton masses which are inherited from $P_{rad}(\epsilon)$ and $P_{el}(\epsilon)$, and explicitly, 
\be
P_{tot}(\epsilon) = P_{tot}(\epsilon=\Delta E_{tot}/E_0; \bx_0, \phi; M, E_0; \alpha_s, f_E, f_M)\;\;.
\label{PtotDep}
\ee
For simplicity, throughout the paper we will suppress the ``tot'' subscript.

\subsubsection{Jet quenching spectrum}
\label{sec:conv-initial}

CUJET computes the quenched partonic AA spectrum by convolute the total energy loss probability distribution $P(\epsilon)$ calculated from Eq.~\eqref{PtotFinal} and~\eqref{PtotDef} with partonic production cross section in p+p collisions. This is a critical improvement of CUJET over its predecessor WHDG \cite{WHDG}, which assumes instead a simple and slowly varying power law distribution for the p+p spectra (spectral index approximation) and makes considerable simplifications in the computation of the nuclear modification factor. Given the sensitivity of the results to the details of the production cross sections, and the complex interplay between the latter and the energy loss mechanism, it is essential that no approximations are carried out in this delicate step of the computation.

The partonic pp spectra for CUJET are generated from pQCD calculations. For the light sector, production is based on a leading order (LO) calculation scaled by a simple K-factor and computed from the LO pQCD CTEQ5 code of X.N. Wang \cite{Wang:private}. For the heavy jet sector, both next-to-leading order \cite{Mangano:1991jk} and fixed-order plus next-to-leading-log (FONLL) \cite{Cacciari:1998it,Cacciari:2001td} computations are used. In addition to including the full NLO result \cite{Nason:1987xz,Nason:1989zy,Beenakker:1990maa}, the FONLL calculation re-sums large perturbative terms with next-to-leading logarithmic accuracy \cite{Cacciari:1993mq}. Details about the partonic spectra used in CUJET can be found in Appendix~\ref{app:partonspc}.

It is clear at this point, that the input and output of CUJET are: the model is given a parametrization of the plasma and a jet spectrum, and it returns a quenched spectrum after computing the energy loss of the jets in the medium. In this process, no approximations are made: each jet is evolved individually and its final momentum, or better momentum probability distribution, is stored along with the direction it came from (angular distribution).

This is how CUJET performs the computation of quenched partonic spectra:

\begin{enumerate}
\item The algorithm starts from a jet created at $\bx_0$ in the azimuthal plane (with respect to the beam axis) with azimuthal angle $\phi$ and mass $M$. The distribution of jets in the transverse plane in A+A collisions is given by $\rho_{binary}$ (cf. Section~\ref{sec:glauber}). The initial transverse momentum probability distribution $P_0(p_i)$\footnote{$p_i\equiv (p_T)_i$. We suppress the ``T'' (transverse) subscript in this section, and all $p_i$'s and $p_f$'s are understood as transverse momentum.} of the partons is proportional to the production cross section:
\be
P_0(p_i)\propto\frac{d\sigma^{pp\rightarrow q}}{dp_i}(p_i)
\;\; ,
\label{ppNorm}
\ee
Here $d\sigma^{pp\rightarrow q}/dp_i$ represents a generic p+p partonic production spectrum. A range of discrete transverse momenta $[p_i^{min},p_i^{max}]$ needs to be defined for the numerical computation.

\item For each initial transverse momentum $p_i$ in the range $[p_i^{min},p_i^{max}]$, CUJET computes the energy loss according to Eq.~\eqref{fcDGLV}\eqref{CUJETElastic} (or Eq.~\eqref{rcCUJETDGLV}\eqref{rcCUJETElastic} in the running coupling case). This is the most resource- and time-consuming process, where the full jet path Monte Carlo integral is evaluated over the expanding plasma and the medium-induced gluon radiation spectrum as well as elastic collisional energy loss are computed. All the dynamical properties of the plasma can be specified and their contributions to the energy loss -- radiative and/or elastic -- should be considered. Once fluctuations effects are taken into account -- Eq.~\eqref{Poissonexp},\eqref{Gaussianexp} -- the output takes the form of a distribution function which represents the probability of losing the relative energy $\epsilon$ $(\epsilon=1-E_f/E_i, E_{i,f}^2=p_{i,f}^2+M^2)$ given the initial transverse momentum $p_i$ $(p_i=\sqrt{E_i^2-M^2})$ (cf. Eq.~\eqref{PtotFinal}):
\be
P(\epsilon;p_i;\bx_0,\phi) = P_t^{null}(p_i)\delta(\epsilon) + P_t(\epsilon;p_i) + P_t^{full}(p_i)\delta(\epsilon-\epsilon_{max})
\; \; ,
\ee
with
\be
\epsilon_{max}=1-\frac{M}{\sqrt{p_i^2+M^2}}
\; \; .
\ee

\item Once all the $\{p_i\}$ in the range specified have been computed, the $\{P(\epsilon;p_i)\}$ are converted into a two-dimensional distribution map that represents the probability of a jet with initial transverse momentum $p_i$ to leave the plasma with final transverse momentum $p_f$:
\be
\begin{split}
	P(p_f,p_i) & = P(\epsilon;p_i)\frac{d\epsilon}{dp_f} \\
         & = P_t^{null}(p_i)\delta(p_f-p_i)+{P_t}(\epsilon(p_f,p_i);p_i)\frac{p_f}{E_f E_i}+P_t^{full}(p_i)\delta(p_f)
\; \; ,
\end{split}
\label{ProbMap}
\ee
with
\be
\epsilon(p_f,p_i)=1-\frac{E_f}{E_i} \;\;, \;\;\;\; E_f=\sqrt{p_f^2+M^2} \;\;, \;\;\;\; E_i=\sqrt{p_i^2+M^2}
\;\; .
\ee
The normalization is such that
\be
\int_0^{p_i}{dp_f}\;P(p_f,p_i)=1
\;\; ,
\ee
which is automatically ensured by $\int_{0}^{\epsilon_{max}} d\epsilon P(\epsilon;p_i) =1$. In Eq.~\eqref{ProbMap} we dropped the explicit dependence on the jet coordinates $\bx_0$ and $\phi$.

\item CUJET then integrates over the production spectrum, Eq.~\eqref{ppNorm}, to obtain the ``quenched'' partonic p+p spectra $\frac{d{\sigma'}^{pp \rightarrow q}}{dp_f d\phi}$:
\be
\frac{d{\sigma'}^{pp \rightarrow q}}{dp_f d\phi}(p_f;\bx_0,\phi)=\int_{p_i^{min}}^{p_i^{max}}dp_i\; P(p_f,p_i;\bx_0,\phi)\frac{d{\sigma}^{pp \rightarrow q}}{dp_i}(p_i)
\;\; .
\label{ppSpectra}
\ee

\item At last, the quenched partonic A+A spectra as a function of the observed transverse momentum $p_f(\equiv p_T)$ and azimuthal angle $\phi$ are obtained by integrating over the jet transverse distribution :
\be
\frac{d\sigma^{AA\rightarrow q}}{dp_f d\phi}(p_f;\phi)=\int{d\bx_0}\;{\rho}_{binary}(\bx_0)\;\frac{d{\sigma'}^{pp \rightarrow q}}{dp_f d\phi}(p_f;\bx_0,\phi)
\;\; .
\label{AASpectra}
\ee
\end{enumerate}

\subsubsection{Fragmentation functions}
\label{sec:conv-final}

Partonic spectra can provide useful information about jet quenching mechanism, nevertheless, comparison with data can only be carried out at the hadronic level. The quenched partonic spectra, Eq.~\eqref{AASpectra}, need to be convoluted with a set of fragmentation functions (FF's).

The process that leads to the fragmentation of partons in the medium is not theoretically well understood, especially for heavy quarks: dissociation and recombination theories \cite{ADILVITEV,Vitev:2007jj} assume that heavy D and B mesons can be formed within the plasma and lose additional energy through collisional dissociation, in a similar fashion to what has been suggested for heavy quarkonium states \cite{Wong:2004zr}. This, however, seems to contradict more recent lattice results \cite{Petreczky:2009cr}, which indicate the complete melting of open heavy flavors occurs at temperature $T\gtrsim 220$ MeV.

Since we are dealing with high $p_T$ partons, hadronization via recombination processes is suppressed compared to fragmentation. We will assume that fragmentation takes place in vacuum, on a hypersurface parametrized by $\mu(\bx,\tau_f)=\Lambda_{QCD}$, and our results do not show a particular sensitivity on the precise choice of fragmentation temperature $T_f$ (cf. Appendix~\ref{app:systematics}).

The convolution of partonic spectra over appropriate FF's takes the form
\be
\begin{split}
\frac{d\sigma^h}{dp}(p) = &  \sum_i \int_{p/p_{max}}^1 dx\; \frac{d\sigma^{i}}{dp}(\frac{p}{x})\; D^{i\rightarrow h} (x;\frac{p}{x}) \\
                        = &  \sum_i \int_{p/p_{max}}^1 dx\; \frac{1}{x} \frac{d\sigma^{i}}{d\frac{p}{x}}(\frac{p}{x})\; D^{i\rightarrow h} (x;\frac{p}{x}) \;\; .
\end{split}
\label{Fragmentation}
\ee
Here $ D^{i\rightarrow h} (y;Q)$ represents the probability that a parton $i$ fragments into a hadron $h$ which carries a fraction $y$ of the parton energy. $Q$ is the scale at which the FF is evaluated, here it is given by the energy of the parton. Eq.~\eqref{Fragmentation} is summed over all species $i$ that fragment into $h$.

For light quarks and gluons fragmenting into pions, we use leading order KKP functions \cite{KKP}. For heavy quarks fragmenting into D and B mesons ($c\rightarrow D$ and $b\rightarrow B$), we use instead the Peterson \cite{PETERSON} function with $\epsilon_c=0.06$ and $\epsilon_b=0.006$, as done also in \cite{DGVW}. While the Peterson FF does not couple well with the FONLL production cross section \cite{VOGT}, it was shown in \cite{DGVW} that similar results are produced anyway even using a more accurate fragmentation description.
Finally for the decay of the heavy mesons into non-photonic electrons ($c\rightarrow D\rightarrow e$ and $b\rightarrow B\rightarrow e$), we use the same functions as in \cite{VOGT}. The secondary decay $D\rightarrow B\rightarrow e$ is also accounted for.

\section{Convergence of DGLV opacity series}
\label{app:conv}

\subsection{Uncorrelated geometry}
\label{app:geometry}

The DGLV opacity expansion integrand in Eq.~\eqref{DGLVNN} is a function of the distance between scattering centers $\Delta z_k$. Their distribution is connected to the mean free path $\lambda$, which in the case of a non-uniform plasma is itself a function of $z$, i.e. $\lambda(z)$. To see the average over the target coordinates in a smooth background more clearly, we write
\be
\begin{split}
\frac{1}{n!} \lp \frac{L}{\lambda_g} \rp^n & \int \prod_{i=1}^n \left(d{\bf q}_{i}\, \left(|\bar{v}_i({\bf q}_{i})|^2 - \delta^2({\bf q}_{i}) \right)\right)\\
& \rightarrow \int_0^L dz_1 \cdots \int_{z_{n-1}}^L dz_n \int\prod_{i=1}^n \left(d{\bf q}_{i}\, \frac{|\bar{v}_i({\bf q}_{i})|^2 - \delta^2({\bf q}_{i}) }{\lambda(z_i)}\right)
\end{split}
\label{GeometryInt}
\ee

The $\lambda(z_i)$ dependence significantly complicates the DGLV integral, especially at higher order in opacity $n$. To study the behavior of higher order opacities more efficiently, we apply an ``uncorrelated geometry'' for quick DGLV evaluations. In this configuration, we neglect the interconnection between the location of the scattering centers and the mean free path, as well as the mutual dependence of the spacing of collisions.

The simplest medium one can study is an uncorrelated brick of uniform density, constant temperature $T$ and limited length $L$. This configuration is realized by changing Eq.~\eqref{GeometryInt} to
\be
\begin{split}
\frac{1}{n!} \lp \frac{L}{\lambda_g} \rp^n & \int \prod_{i=1}^n \left(d{\bf q}_{i}\, \left(|\bar{v}_i({\bf q}_{i})|^2 - \delta^2({\bf q}_{i}) \right)\right)\\
& \rightarrow \frac{L^n}{n!} \int_0^L dz_1 \cdots \int_0^L dz_n \bar{\rho}(z_1,\cdots,z_n) \int\prod_{i=1}^n \left(d{\bf q}_{i}\, \frac{|\bar{v}_i({\bf q}_{i})|^2 - \delta^2({\bf q}_{i}) }{\lambda(T)}\right)\;\;,
\end{split}
\label{UniBrick}
\ee
where the normalized distribution for scattering centers
\be
\bar{\rho}(z_1,\cdots,z_n)=\frac{n!}{L^n}\theta(L-z_n)\theta(z_n-z_{n-1})\cdots\theta(z_2-z_1)\theta(z_1-z_0)\;\;.
\ee
Note $z_0=0$ is the position of the production vertex. Since the 0 and L boundaries are already contained in the integration limits, one can drop either $\theta(L-z_n)$ or $\theta(z_1-z_0)$ or both in the above equation. Due to the LPM phase oscillation in Eq.~\eqref{DGLVNN}, a pure brick would easily create fluctuating gluon radiation spectra. We thus make a further generalization of the brick geometry by assuming an exponential distribution of scattering centers, i.e.
\be
\bar{\rho}(z_1,\cdots,z_n)=\prod_{l=1}^n \frac{\theta(\Delta z_l)}{L_e(n)}\eexp^{-\Delta z_l /L_e(n) }\;\;,
\label{ExpBrick}
\ee
with $\Delta z_l=z_l - z_{l-1}$. This converts the oscillating LPM phases in Eq.~\eqref{DGLVNN} into simple Lorentzian factors assuming $L$ sufficiently large,
\be
\int d\bar{\rho}\cos\lp \sum_{k=j}^{m} \omega_{(k,\cdots,n)} \Delta z_k\rp= Re \prod_{k=j}^m \frac{1}{1+i \omega_{(k,\cdots,n)} L_e(n)}\;\;.
\ee
In order to fix $L_e(n)$, we require that $\left< z_k - z_0 \right> = kL/(n+1)$. This constrains $L_e(n) = L/(n + 1)$.

In the discussion within this paper, unless otherwise stated, if referring to a ``brick'', we mean an uncorrelated brick with an exponential distribution, defined by Eq.~\eqref{UniBrick} and~\eqref{ExpBrick}.

\subsection{Convergence of DGLV}
\label{app:convDGLV}

The DGLV opacity series approach builds upon the Bertch-Gunion (GB) incoherent radiation and includes multiple coherent scatterings, interference with the production vertex radiation and gluon cascading. The LPM effect and the interplay between the cosine factors in the DGLV integral determine how fast the series converges to its asymptotic limit.

To recapitulate, the induced gluon radiation spectrum in static QCD medium at first order in opacity takes the form
\be\label{DGLVN1}
\begin{split}
\frac{dN_g^{n=1}}{dx_+ d\bk} = &\; \frac{C_R \alpha_s}{\pi^2}\frac{1}{x_+}\frac{L}{\lambda} \int{dz}\; \bar{\rho}(z)\int{d\bq} {\frac{\mu^2}{\pi(\bq^2+\mu^2)^2}}\\
&\times\;{\frac{-2(\bk-\bq)}{(\bk-\bq)^2+\chi^2} \lp\frac{\bk}{\bk^2+\chi^2} - \frac{(\bk-\bq)}{(\bk-\bq)^2+\chi^2}\rp}\\
&\times\;{\lp1-\cos\lp\frac{(\bk-\bq)^2+\chi^2}{2x_+ E}\Delta z\rp\rp}
\; \; ,
\end{split}
\ee
with $\bar{\rho}(z)$ a normalized distribution. The relevant terms in \eqref{DGLVN1} are: (1) the opacity $\frac{L}{\lambda}$; (2) the effective interaction potential $\frac{\mu^2}{\pi(\bq^2+\mu^2)^2}$; (3) the radiation antenna $\frac{\bk-\bq}{(\bk-\bq)^2+\chi^2}\cdot\left(...\right)$; (4) the LPM phases $\left[ 1-\cos \left(\frac{(\bk-\bq)^2+\chi^2}{2x_+ E}\Delta z\right)\right] $.

To understand the convergence of the DGLV opacity series quantitatively, we need to study details about the LPM phases, and the interplay between the formation time $\tau_f$ and mean free path $\lambda$. Generally speaking, if denote the size of the medium as $L$ ($L>\lambda$), the interference (coherent) effects are dominant in the region $\lambda<\tau_f<L$, whereas the Gunion-Bertsch incoherent limit (cf. Eq.~\eqref{GBspectrum}) and factorization limit is obtained in the region $\tau_f<\lambda<L$ and $\lambda<L<\tau_f$ respectively.

Formation time $\tau_f$ is the time gluon spends to become on-shell, it is approximately equal to
\be\label{tauf}
\tau_f\approx\frac{2\omega}{(\bk-\bq)^2+\chi^2}
\; \; ,
\ee
with $\omega=x_E E$ the energy of the radiated gluon. (In principle, only in the strict collinear limit $x_E$ and $x_+$ coincide. For simplicity, we set $x_+=x_E\equiv x$ in this section. And we will discuss this issue in Appendix~\ref{app:transmomdistr}.) If there are many momentum kicks from the medium within a coherence length, then $\bq\rightarrow\sum_i\bq_i$; however, for a qualitative estimate, we can assume $\bk\gg\bq$ and $\tau_f\approx2\omega/\bk^2$. In reality, the interplay between $\bk$ and $\bq$ makes the estimation of the real formation time difficult, and once the mass of a heavy quark is taken into account, the $\chi^2=M^2 x^2+m_g^2(1-x)$ factor starts playing a relevant role by reducing the formation time and by pushing the radiation back into the incoherent regime.

Mean free path $\lambda$ plays an important role in determining the effects of coherence physics. In the uncorrelated geometry assumption, the relation between $\lambda$ and the distribution of scattering centers looses, while in reality they are interconnected. Coherence effects are dominant when $\lambda < \tau_f$, and are analytically determined by the magnitude of the LPM phases $\Delta z/\tau_f$: larger phases are responsible for the oscillatory behavior characteristic of the incoherent limit, while smaller phases cause an approximate cancellation among the cosine terms, typical result of coherence physics. For instance, for $n=1$ and large formation times, the LPM term $\cos(\Delta z/\tau_f)$ approaches unity, giving rise to a neat cancellation.

\begin{figure}[!t]
\centering
%\vspace{0.25in}
\includegraphics[width=0.45\textwidth%width=1.in%height=1.in
]{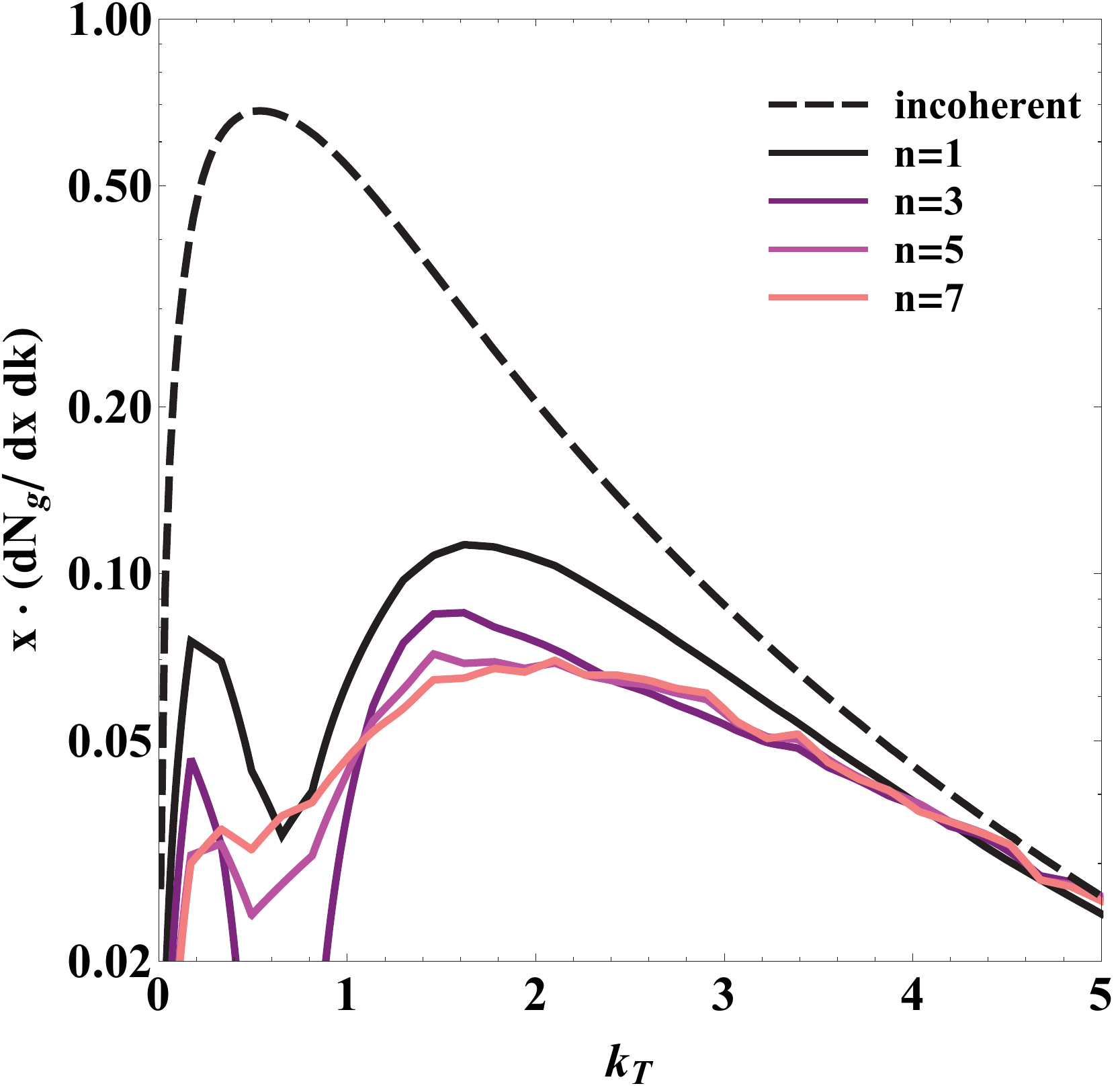}
\hspace{0.01\textwidth}
\includegraphics[width=0.45\textwidth%width=1.in%height=1.in
]{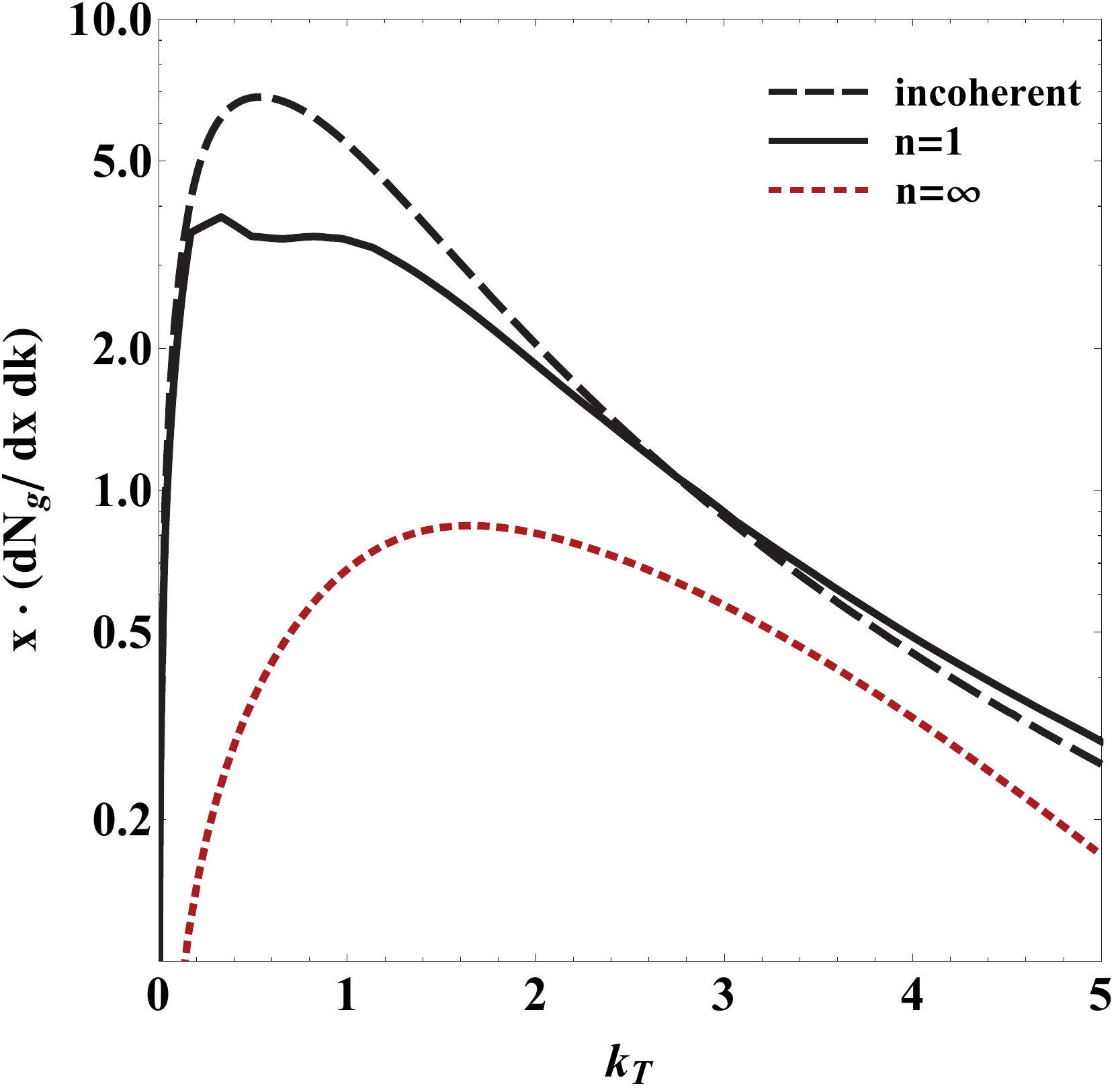}
\caption{Comparison between the DGLV $n=1$ gluon transverse momentum distribution (solid black) and the Gunion-Bertsch (GB) incoherent limit (dashed black), as well as higher order DGLV corrections added up to $n=3$ (solid purple), $n=5$ (solid magenta) and $n=7$ (solid pink), for different plasma sizes. On the left, we use a brick of size $L=5$ fm; on the right, the length $L=50$ fm. The energy of the incoming light quark ($M=0.2$ GeV) jet is $E=50$ GeV, and the radiated gluon energy $\omega=5$ GeV. Compare $n=1$ and GB, notice the suppression of the induced radiation for short path lengths due to interference with the creation radiation. Such effect vanishes in the $L\rightarrow\infty$ limit, as expected, where the average distance between the creation vertex and the scattering center becomes larger ($\overline{\Delta z}=L/2$). The higher order DGLV opacity series is shown to converge already at $n=5$, with the first order result still giving the biggest contribution to the suppression. The opacity expansion, valid at the intermediate opacities characteristic of nuclear collisions ($L=5$ fm, \textit{left}), breaks down for plasmas of the size of tens of fermi ($L=50$ fm, \textit{right}): in this case the radiation spectrum is replaced by the multiple soft scattering approximation (dashed red). Parameters used in the simulation are: $\lambda=1.16$ fm, $\mu=0.5$ GeV, $m_g=0.356$ GeV, $T=0.258$ GeV, $n_f=0$, $\alpha_s=0.3$.}
\label{BGvsNn}
\end{figure}

In order to understand how the convergence of the opacity series is related to the coherent or incoherent radiation regime, we first compare the DGLV $n=1$ result with the Gunion-Bertsch incoherent limit:
\be
\frac{dN_g}{dx_+ d\bk} = \frac{C_R \alpha_s}{\pi^2}\frac{1}{x_+} \frac{L}{\lambda} \int d\bq \frac{\mu^2}{\pi(\bq^2+\mu^2)^2} \frac{\bq^2}{(\bk^2 + \chi^2)((\bq - \bk)^2 + \chi^2)}
\;\;.
\label{GBspectrum}
\ee
At first order, the opacity series only includes interference effects between the creation and the induced radiation vertex. By plotting the gluon transverse momentum distribution for different brick sizes $L$, we demonstrate in Fig.~\ref{BGvsNn} the suppression of the induced radiation due to such interference effects.

Comparing solid and dashed black curve in Fig.~\ref{BGvsNn}, we see that this coherence effect vanishes when the size of the medium is large, as expected. To move on, we add higher order corrections to the results, shown as purple, magenta and pink curves in Fig.~\ref{BGvsNn}. Regardless of all these higher order modifications, the dominant contribution to the suppression of the induced radiation still comes from the the $n=1$ term.

In the left panel of Fig.~\ref{BGvsNn}, we observe that for $L=5$ fm, at $n=L/\lambda\approx 5$ the opacity series already converges to its asymptotic value, making further corrections negligible. This can be understood by assuming the probability of hitting a given number of scattering centers follows a Poisson distribution, its average equals the opacity, and we would expect the GLV series to peak around $n=L/\lambda$.

But this convergence is only valid for short path lengths, because the interference with formation radiation is the dominant effect, on top of which the corrections due to multiple scatterings in the medium are small. As $L$ increases, this is no longer true and the resummed result is expected to asymptotically converge to the multiple soft scattering limit, as shown in the right panel of Fig.~\ref{BGvsNn}.

The above analysis is restricted to a particular choice of $E=50$ GeV and $\omega=5$ GeV. To be general, we perform a systematic study of the properties of the series, by analyzing its convergence for several coherent and incoherent regimes, varying all relevant parameters. Our goal is to understand if there is an optimal order at which the series can be truncated for most of the practical needs, and quantify the error which is eventually made.

\begin{figure}[!t]
\centering
%\vspace{0.25in}
\includegraphics[width=0.235\textwidth%width=1.in%height=1.in
]{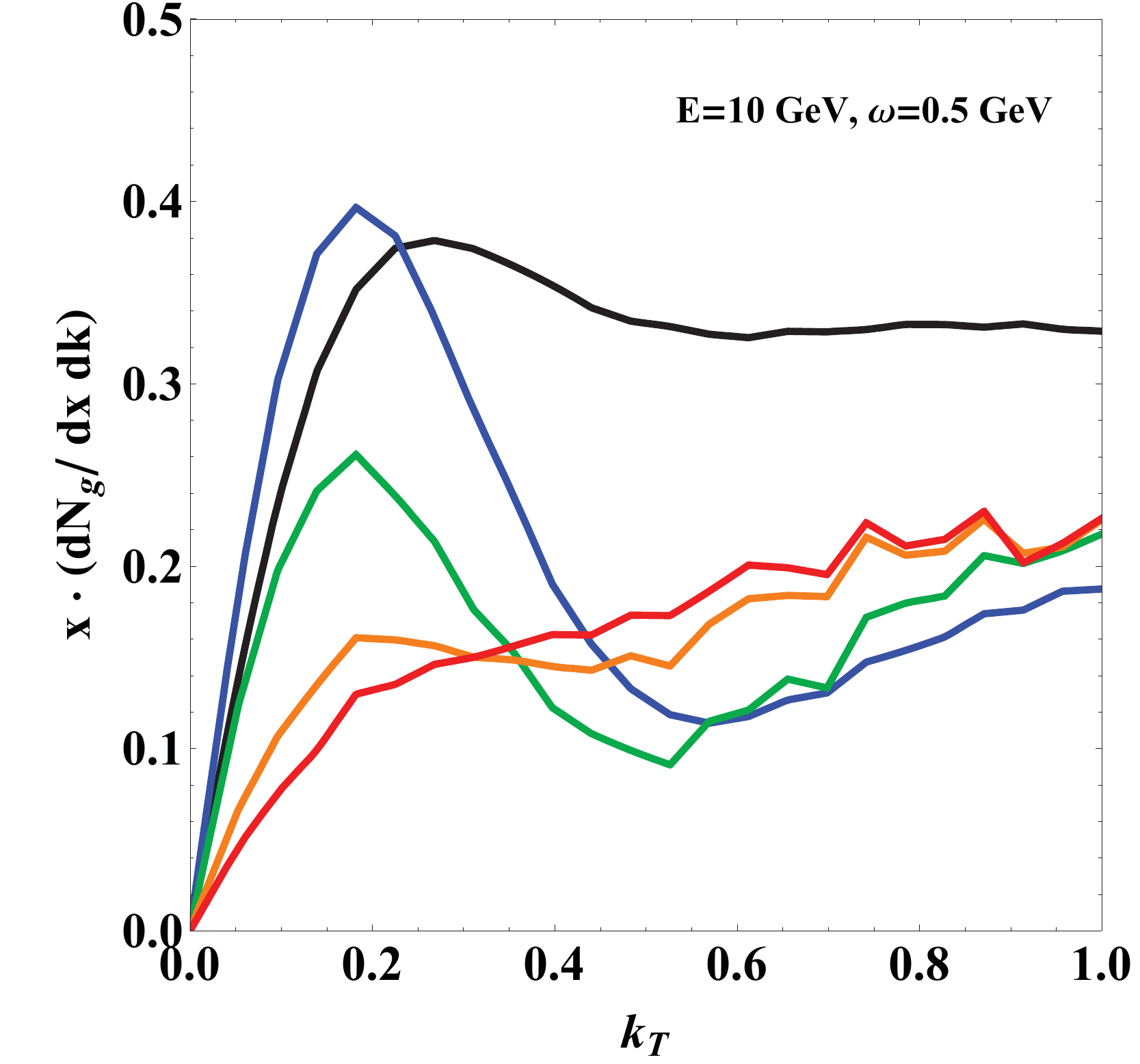}
%\hspace{0.01\textwidth}
\includegraphics[width=0.235\textwidth%width=1.in%height=1.in
]{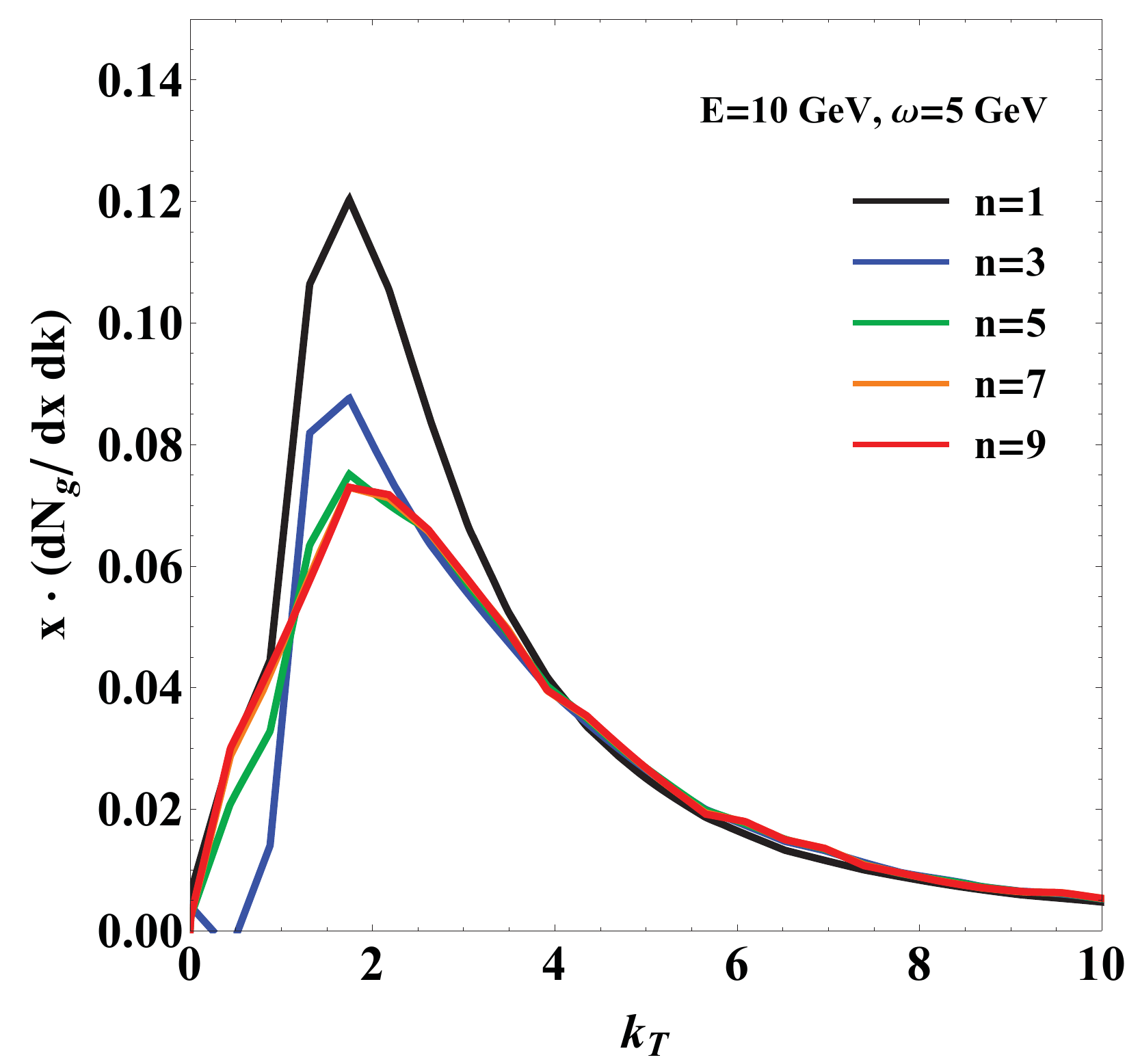}
\includegraphics[width=0.235\textwidth%width=1.in%height=1.in
]{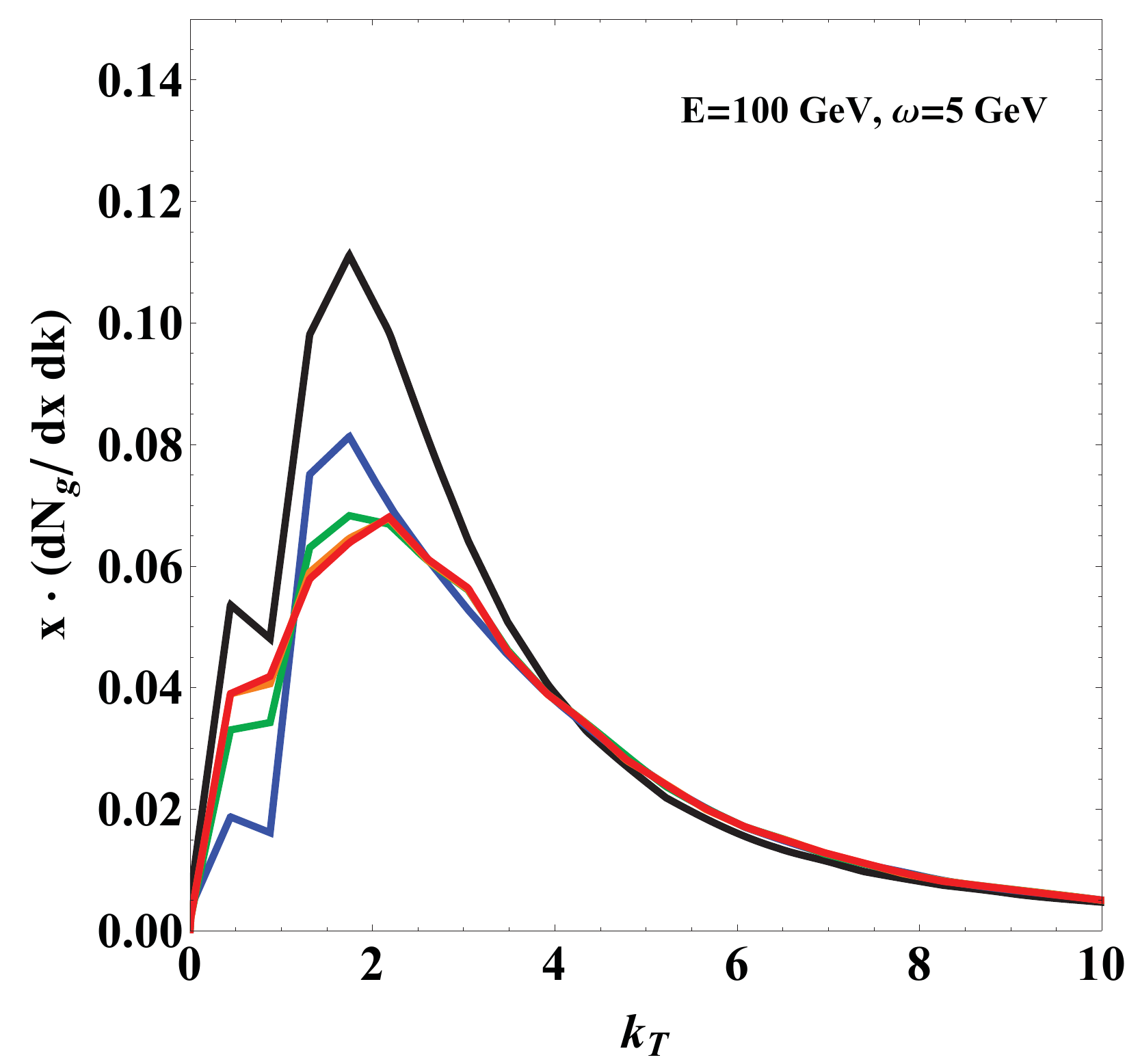}
\includegraphics[width=0.235\textwidth%width=1.in%height=1.in
]{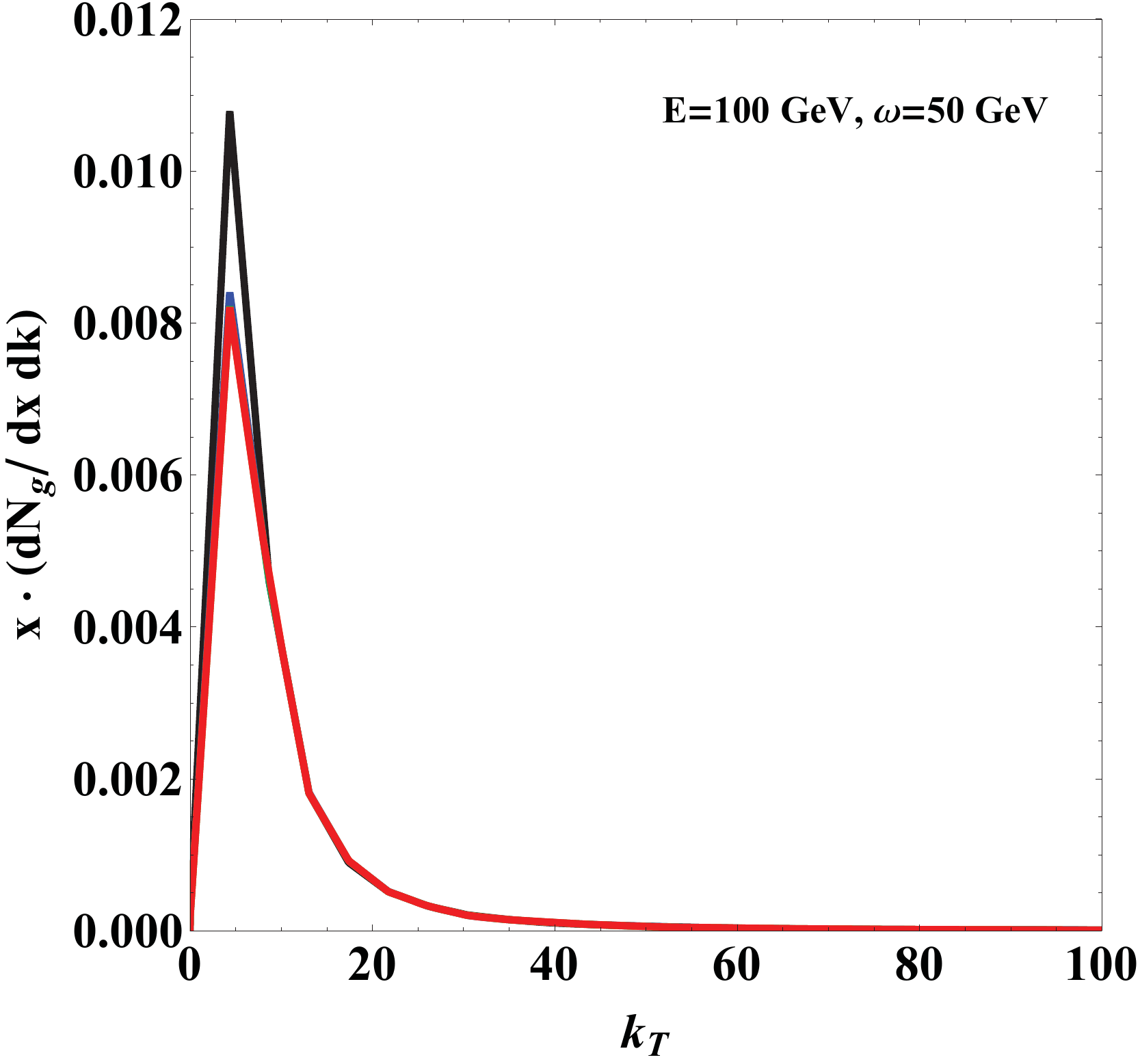}
%\hspace{0.01\textwidth}
\includegraphics[width=0.235\textwidth%width=1.in%height=1.in
]{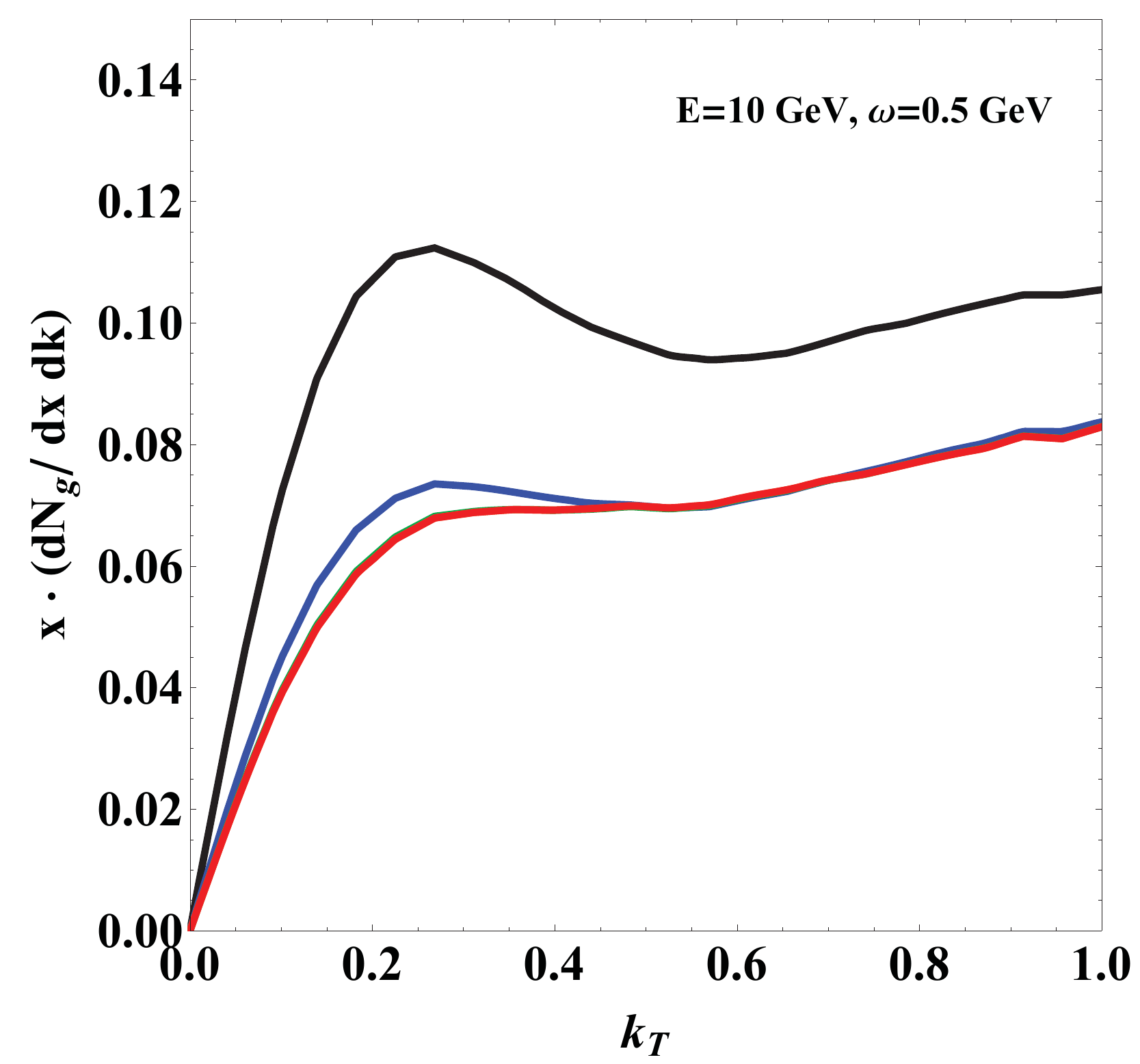}
\includegraphics[width=0.235\textwidth%width=1.in%height=1.in
]{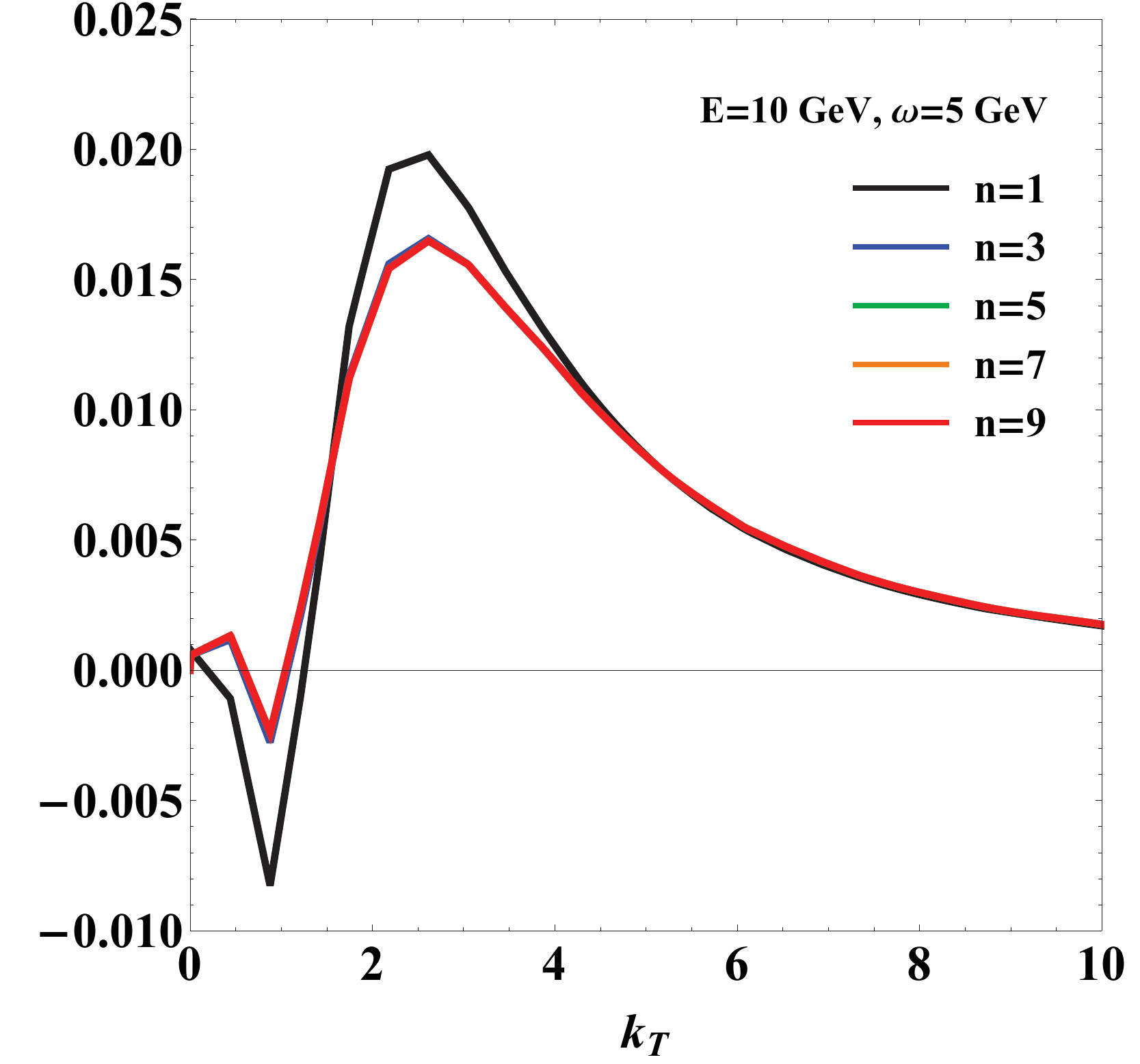}
%\hspace{0.01\textwidth}
\includegraphics[width=0.235\textwidth%width=1.in%height=1.in
]{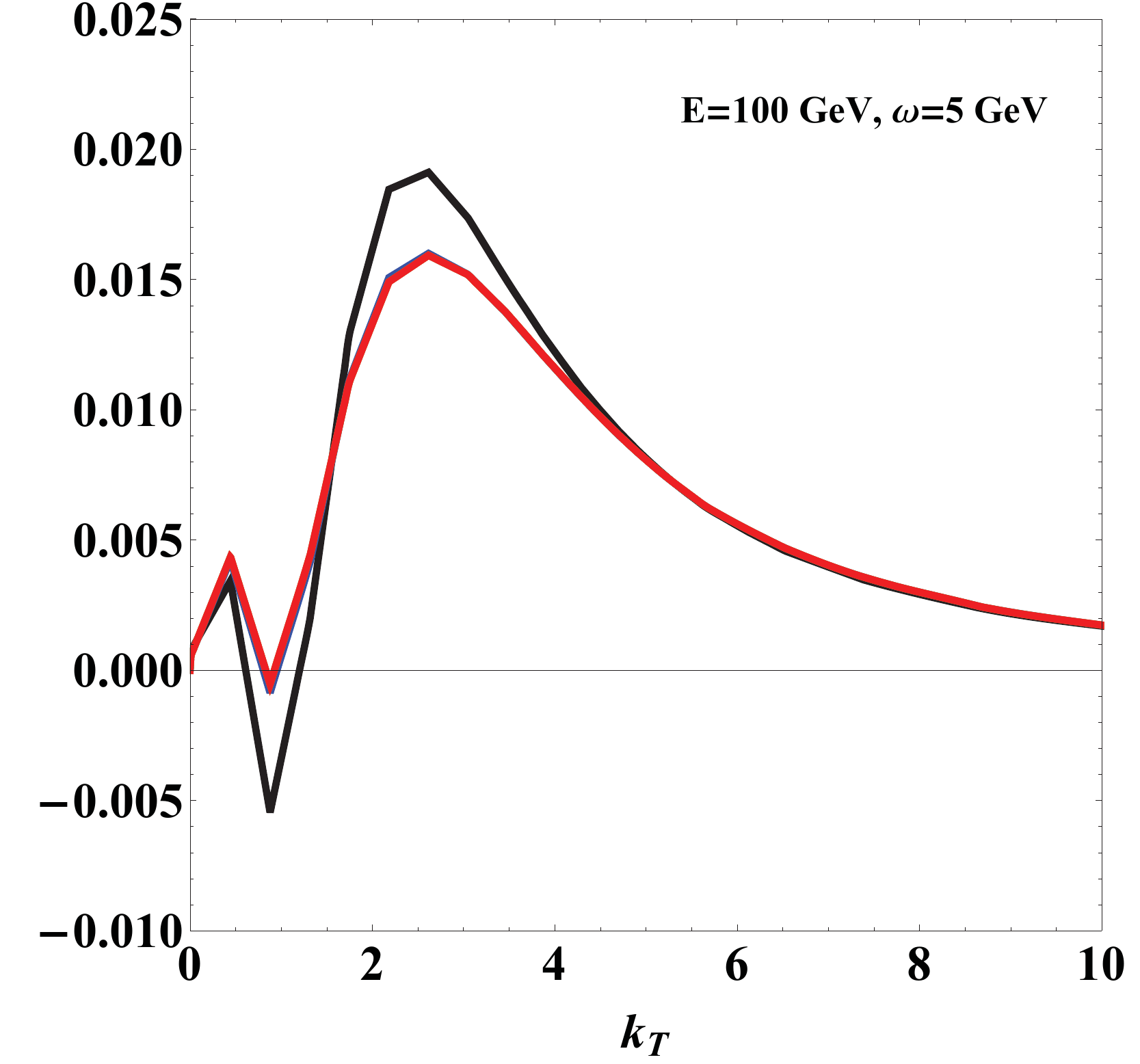}
\includegraphics[width=0.235\textwidth%width=1.in%height=1.in
]{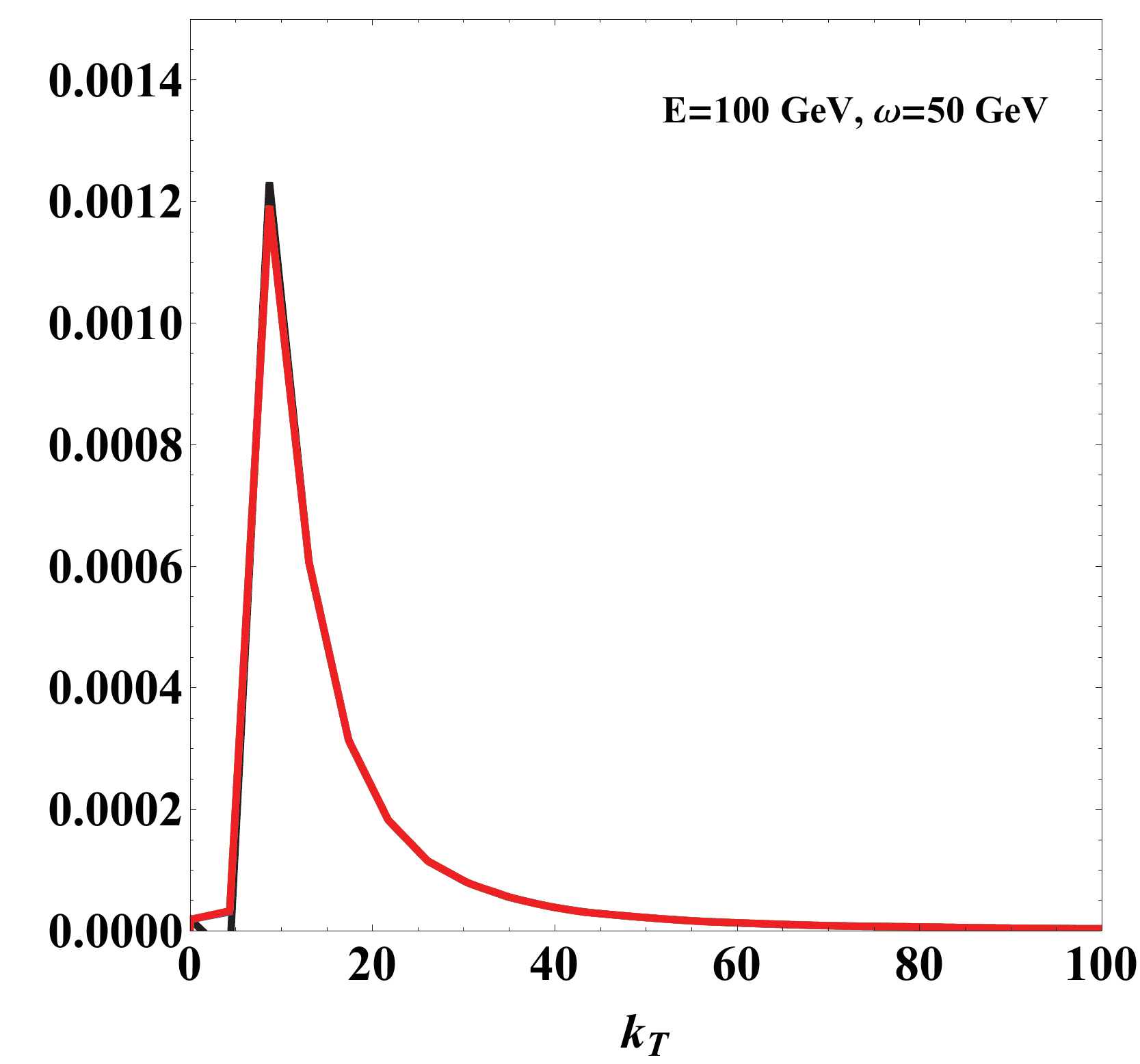}
%\hspace{0.01\textwidth}
\caption{Gluon transverse momentum distribution $xdN_g/dxd\bk$ generated by a light quark ($M=0.2$ GeV) jet traversing a brick plasma of thickness $L=5$ fm (top panels) and $L=2$ fm (bottom panels). Several orders in opacity up to $n=9$ are shown in all figures, plotted as black ($n=1$), blue ($n=3$), green ($n=5$), orange ($n=7$) and red ($n=9$) solid curves. The incoherent or coherent regime of the radiation is determined by the value of $\omega$: incoherent ($\omega=0.5$ GeV), intermediate ($\omega=5$ GeV), coherent ($\omega=50$ GeV). Note the faster convergence of the series for larger values of the gluon energy $\omega$, i.e. longer formation times, determined by the reciprocal cancellation of the oscillating LPM factors. In addition, the transverse momentum distribution depends mostly on the value of the gluon energy $\omega$, rather than the original energy of the jet $E$ (four figures in the middle). And as intuitively expected, the convergence is improved by the reduced medium size $L$. Other parameters used in the simulation are: $\lambda=1.16$ fm,  $\mu=0.5$ GeV, $m_g=0.356$ GeV, $T=0.258$ GeV, $n_f=0$, $\alpha_s=0.3$.}
\label{ExL5&L2}
\end{figure}

For different sets of ($E$, $\omega$ and $L$), we compute in Fig.~\ref{ExL5&L2} the radiation spectrum up to ninth order in opacity.

As shown in Fig.~\ref{ExL5&L2}, the coherent radiation is associated with faster convergence: the large formation time suppresses the magnitude of the LPM phases, leading to an approximate cancellation of the cosine terms in \eqref{DGLVNN}. On the other hand, the oscillatory behavior of incoherent emission results in slower convergence of the opacity series. Interestingly, the transverse momentum distribution seems to depend significantly on the gluon energy $\omega$, rather than the original jet energy $E$. Finally, the convergence is improved by the reduced medium size $L$, as expected from the assumption of Poisson distributed scatterings.

\begin{figure}[!t]
\centering
%\vspace{0.25in}
\includegraphics[width=0.45\textwidth%width=1.in%height=1.in
]{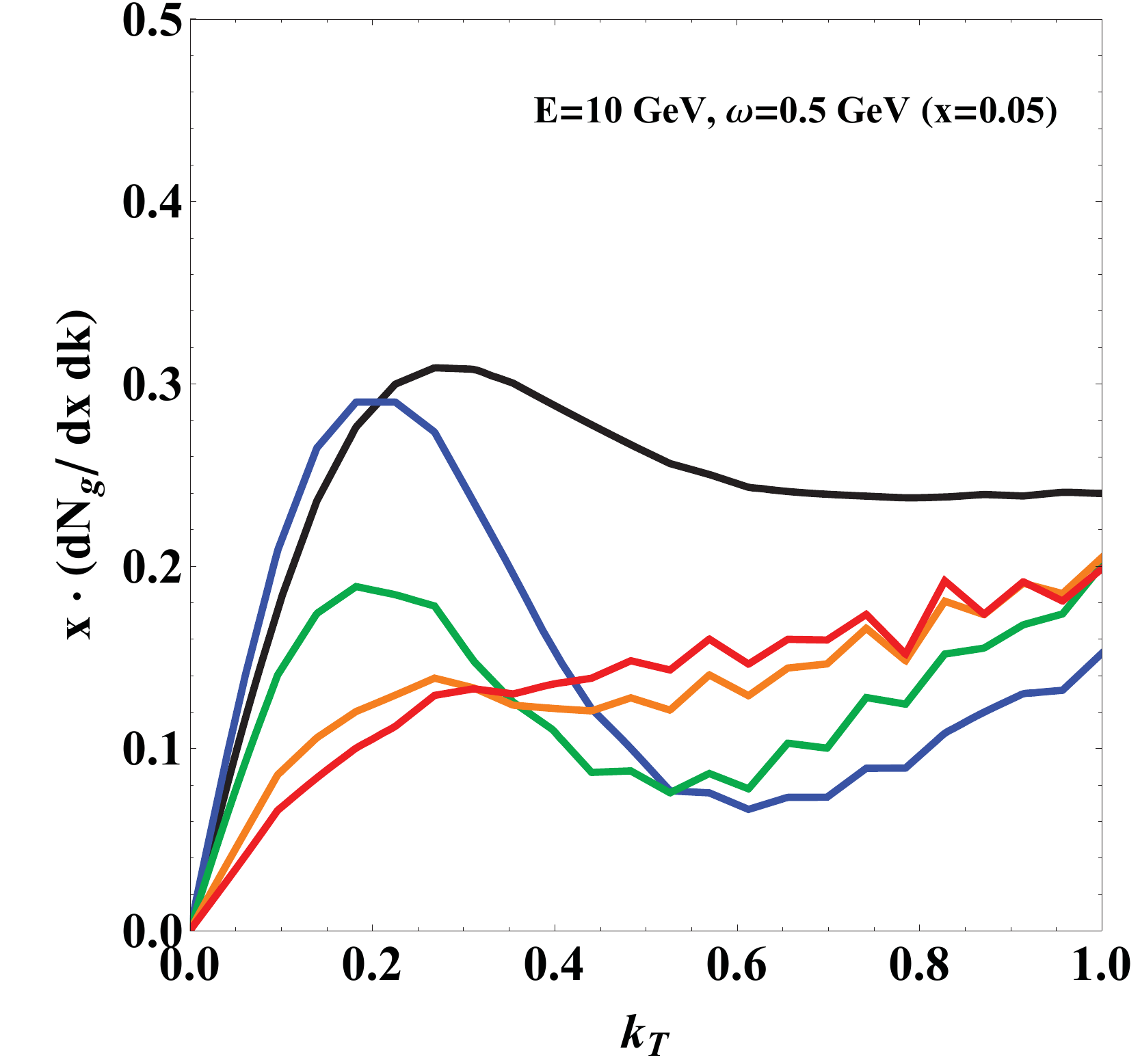}
\hspace{0.01\textwidth}
\includegraphics[width=0.45\textwidth%width=1.in%height=1.in
]{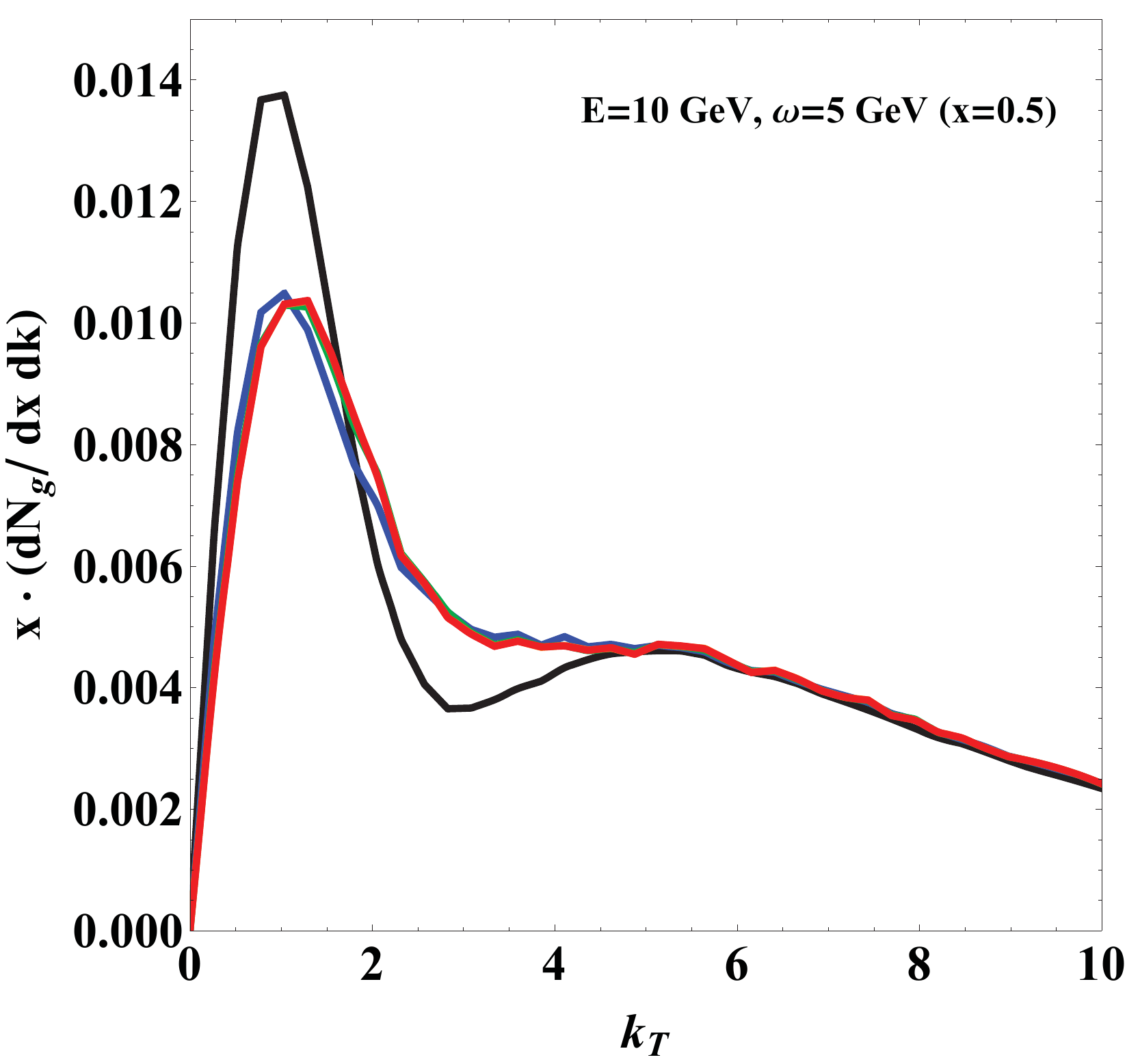}
\includegraphics[width=0.45\textwidth%width=1.in%height=1.in
]{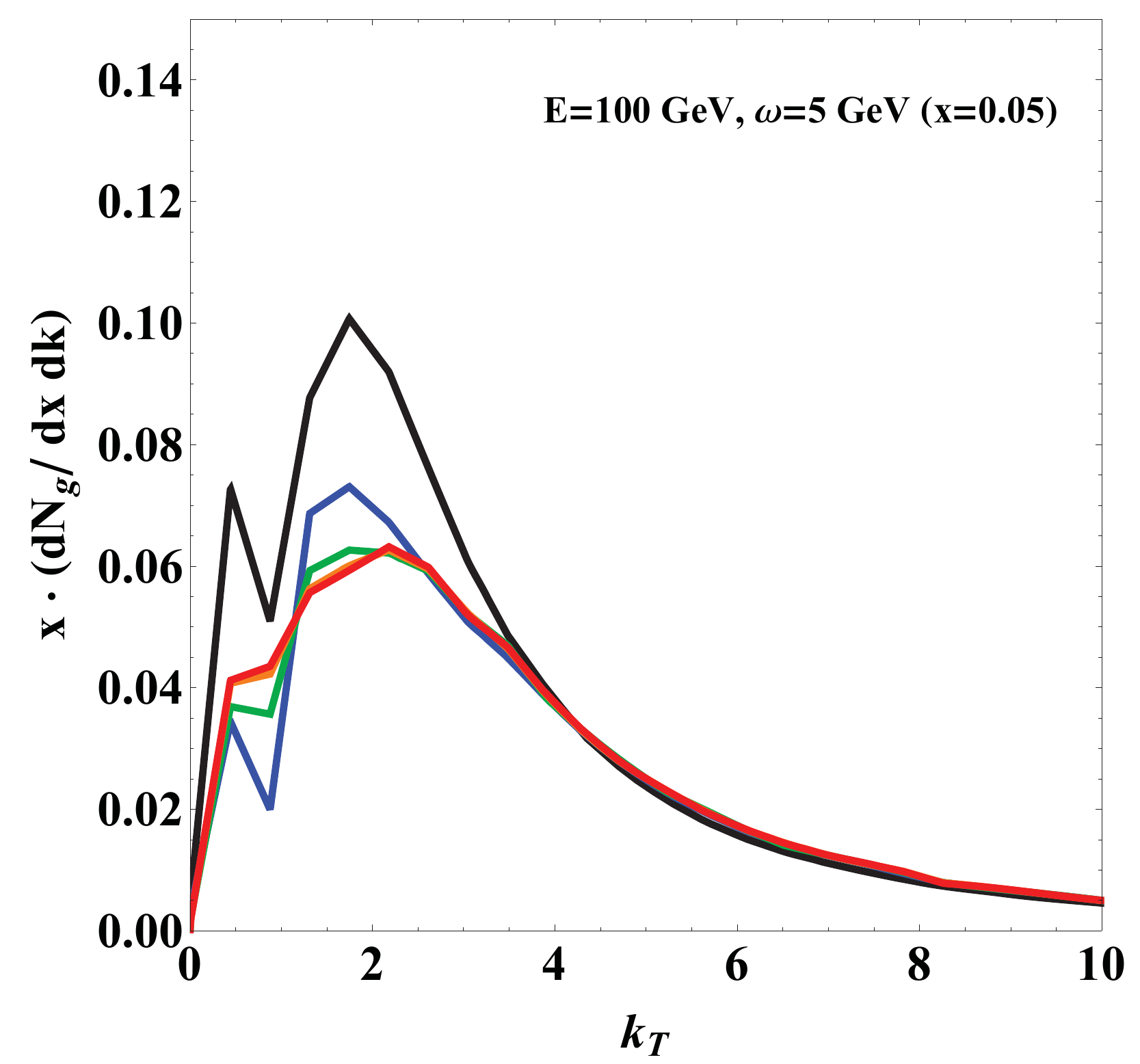}
\hspace{0.01\textwidth}
\includegraphics[width=0.45\textwidth%width=1.in%height=1.in
]{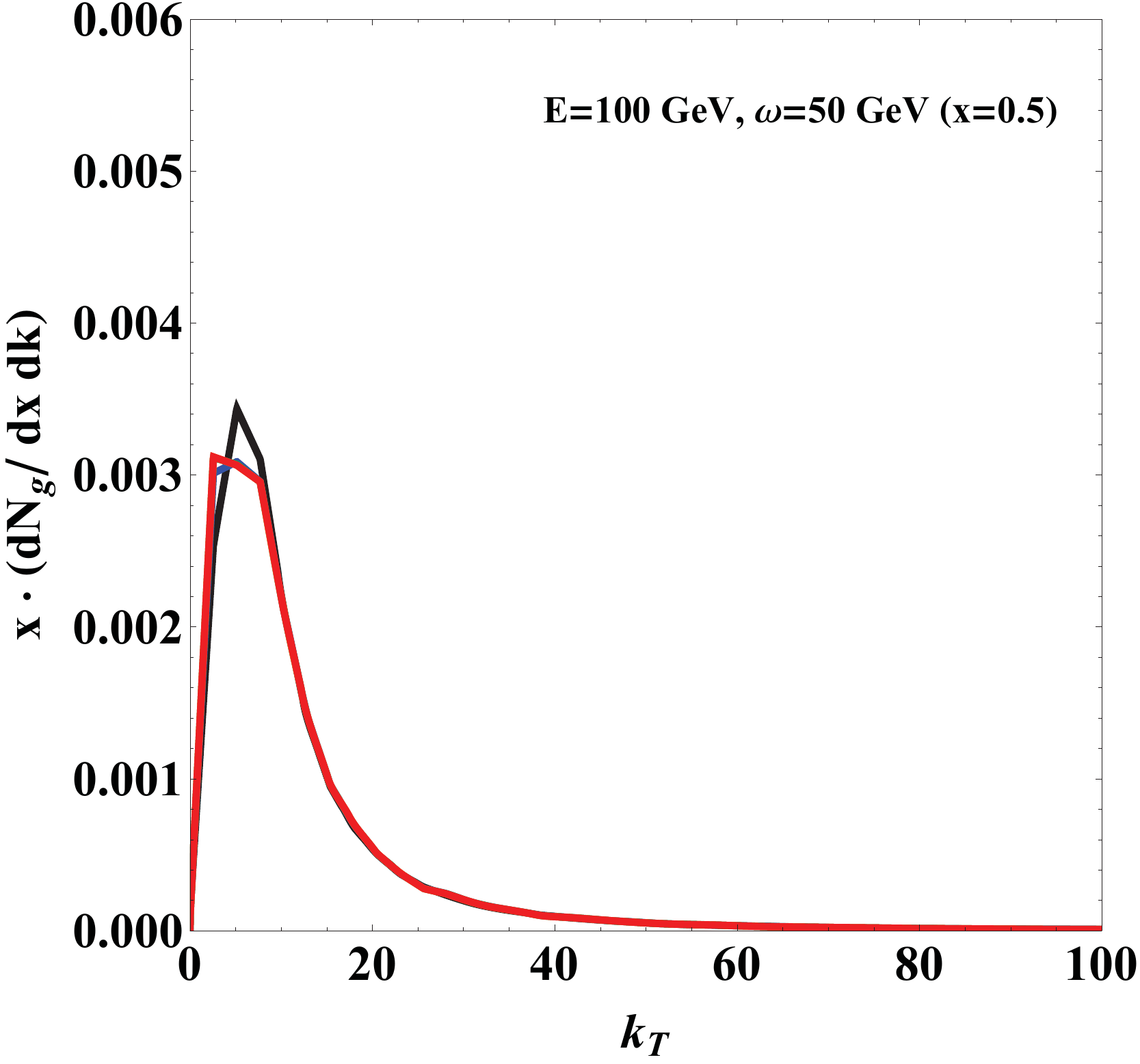}
\caption{Gluon transverse momentum distribution $xdN_g/dxd\bk$ generated by a heavy quark jet traversing a plasma of thickness $L=5$ fm. The mass of the quark $M=4.75$ GeV. All other parameters are the same as in Fig. \ref{ExL5&L2}. DGLV opacity series calculated up to $n=1$, $n=3$, $n=5$, $n=7$ and $n=9$ is plotted as black, blue, green, orange and red solid curve. The effect of the quark mass in the expression for the formation time, which intuitively would slow the convergence of the series. However, this effect is balanced by the $x$ dependence of $\chi^2=M^2 x^2+m_g^2(1-x)$: for small $x$, the results do not differ much from their light quark jet counterpart. On the contrary to light quark jet results, we observe a remarkable splitting between radiation distributions with same gluon energy $\omega$ but different heavy jet energy $E$ (top-right and bottom-left figures), due to the presence of the same $x$ dependent $\chi^2$ in the denominator of the antenna term in Eq.~\eqref{DGLVN1}, which further suppresses radiation at large $x$.}
\label{ExH5}
\end{figure}

For completeness, in Fig.~\ref{ExH5}, we show the same simulation for a heavy quark jet in a plasma of thickness $L=5$ fm: the convergence rate is almost unchanged despite the dependence of the gluon formation time on the mass of the incoming quark, manifested in the term $\chi^2=M^2 x^2+m_g^2(1-x)$.

The increase of $M$ is in fact compensated by the small value of $x$ for $\omega \ll 1$ GeV.
However, compare with the light jet results of Fig.~\ref{ExL5&L2}, now the suppression of the radiated gluon multiplicity depends jet energy $E$ as well. This is because of the presence of a $\chi^2$ term in the denominator of the DGLV radiation antenna.

Generally speaking, in a very limited phase space region where $x$ and $k_T$ are small, compute DGLV opacity series to 1st order may lead to overestimation of radiative energy loss, and hence numerically less strong coupling constant. However, since the entire phase space is integrated over in Eq.~\eqref{DGLVN1}, we conclude that except when the emission mechanism is clearly incoherent, a satisfactory result can already be obtained by truncating the expansion at third order. Furthermore, when average over all possible path lengths in a realistic nuclear collision, $2\lesssim L \lesssim 5$ fm, even the first order in opacity might be regarded as a good approximation to the series (Fig. \ref{ExL5&L2}).

\section{Transverse momentum distribution of radiated gluon}
\label{app:transmomdistr}

In this section we concentrate on the dependence of the gluon spectrum on the transverse momentum $k_\perp$, and we will see that the $k_\perp$ functions non-trivially on various aspects of the radiative jet energy loss.

In Appendix~\ref{app:conv} we observed that $\omega$ determines how fast the series converges to its asymptotic limit. However, we did not take into account the fact that the convergence appears to be faster for larger values of the transverse momentum $k_\perp$, despite the shorter formation time proportional to $1/k^2$. For instance, in Fig.~\ref{ExL5&L2}, with $E=100$ GeV, $\omega=5$ GeV and $L=5$ fm, the first order is already a good approximation for $k\geq4$ GeV, whereas between $2\leq k\leq4$ GeV the fifth order is needed; below $k=2$ GeV, only $n=7$ is a good approximation to the series. The reason can be found in the radiation antenna term of Eq.~\eqref{DGLVN1}, which determines the shape of the momentum distribution: its $1/k^3 \sim 1/k^4$ asymptotic behavior suppresses high momentum corrections and dwarfs the contribution of higher orders in opacity. This effect is very similar to what we observed for heavy quarks jets in Fig.~\ref{ExH5}, where the large contribution of $\chi^2$ in the denominator of the antenna term offsets the increased oscillatory behavior of the integral due to shorter gluon formation times.

\begin{figure}[!t]
\centering
%\vspace{0.25in}
%\includegraphics[width=0.45\textwidth%width=1.in%height=1.in
%]{kTASW,L.pdf}
%\hspace{0.01\textwidth}
\includegraphics[width=0.45\textwidth%width=1.in%height=1.in
]{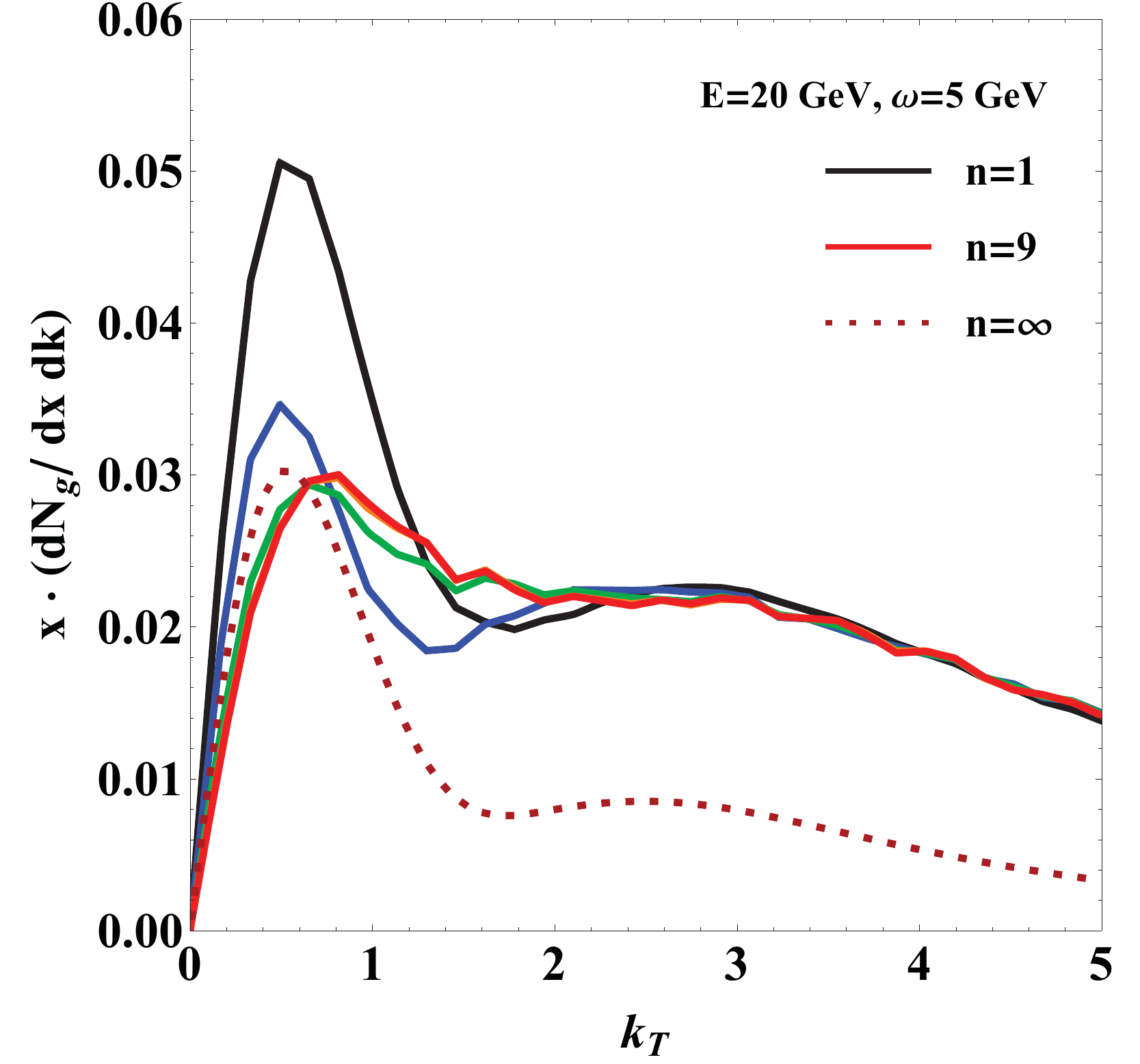}
\caption{Radiated gluon transverse momentum distribution for a heavy quark jet with energy $E=20$ GeV traversing a brick plasma of size $L=5$ fm emitting a gluon with energy $\omega=5$ GeV. The mass of the quark $M=4.75$ GeV. The DGLV opacity series calculated up to n=1 (black), 3 (blue), 5 (green), 7 (orange), 9 (red) are shown in the figure. The opacity expansion computed up to ninth order is shown to converge to the ASW multiple soft scattering limit (maroon, dashed) for small $k_\perp\lesssim\hat{q}L\approx 1$ GeV. At large $k_\perp$, differs from the ASW limit, DGLV has a robust Laudau tail. Other parameters used in the simulation are: $\lambda=1.16$ fm,  $\mu=0.5$ GeV, $m_g=0.356$ GeV, $T=0.258$ GeV, $n_f=0$, $\alpha_s=0.3$.}
\label{kTASW}
\end{figure}

The DGLV opacity expansion has the ability to interpolate between single hard scattering and multiple soft scattering limit. The latter is derived assuming the radiated gluon experiences Gaussian diffusion in the transverse momentum space: for small gluon emission angles, i.e. $k_\perp\lesssim\hat{q}L$, the momentum distribution derived from the DGLV series approaches this limit. This effect is shown in Fig.~\ref{kTASW} for a heavy quark jet.
We see that at small $k_\perp$ the series converges to the multiple soft scattering limit quickly. However for large $k_\perp$, i.e. large angle radiation which is treated poorly in the multiple soft scattering approximation, differs from ASW, the DGLV opacity expansion includes the hard power-law Landau tails of the radiation, reproduces the gluon multiplicity more accurately.

\subsection{Integration and kinematic limits}
\label{sec:kTlimits}

The hard $1/k^3 \sim 1/k^4$ tails of the DGLV distribution offer a relevant contribution to the total emitted radiation and become a source of concern once finite kinematic limits are taken into account. If the integrand in \eqref{DGLVN1} were exact, the result would vanish for unphysical values of $k_\perp$, and there is no need for worrying about integration limits. In reality, however, the model is derived assuming collinear approximation ($k_\perp \ll \omega$), therefore kinematic limits need to be imposed to enforce physicality. The integral, for consistency, should not be sensitive to the particular choice of UV $k_\perp$ cutoffs, but given the hard tails of the distribution, we will see that this is not always going to be the case.

The choice of upper bounds in the $k_\perp$ integration depends on the particular interpretation of $x$ in the expression for the gluon energy $\omega=xE$: $x$ as the fractional energy carried away by the radiated gluon ($x\equiv x_E$, $\omega=x_EE$), or $x$ as the fraction of plus-momentum in light-cone coordinates, in which case $x\equiv x_+$ and $\omega\approx x_+E^+/2$\footnote{Assuming the incoming parton four-momentum is $(E,E,\textbf{0})$, then $E_+=2E$}. In the strictly collinear limit in which the DGLV integral is derived, the two definitions coincide:
\be
\label{xplusxE}
x_+=\frac{1}{2}x_E\lp 1+\sqrt{1-\lp\frac{k_\perp}{x_EE}\rp^2}\rp
\; \; .
\ee
Equation \eqref{xplusxE} can be easily derived by writing explicitly the gluon four-momentum in Minkowski and light-cone coordinates, denoted respectively by parenthesis and square brackets:
\be
k=(x_EE,\sqrt{(x_EE)^2-\bk},\bk)=[x_+E^+,\frac{\bk^2}{x_+E^+},\bk]
\; \; .
\ee

\begin{figure}[!t]
\centering
%\vspace{0.25in}
\includegraphics[width=0.45\textwidth%width=1.in%height=1.in
]{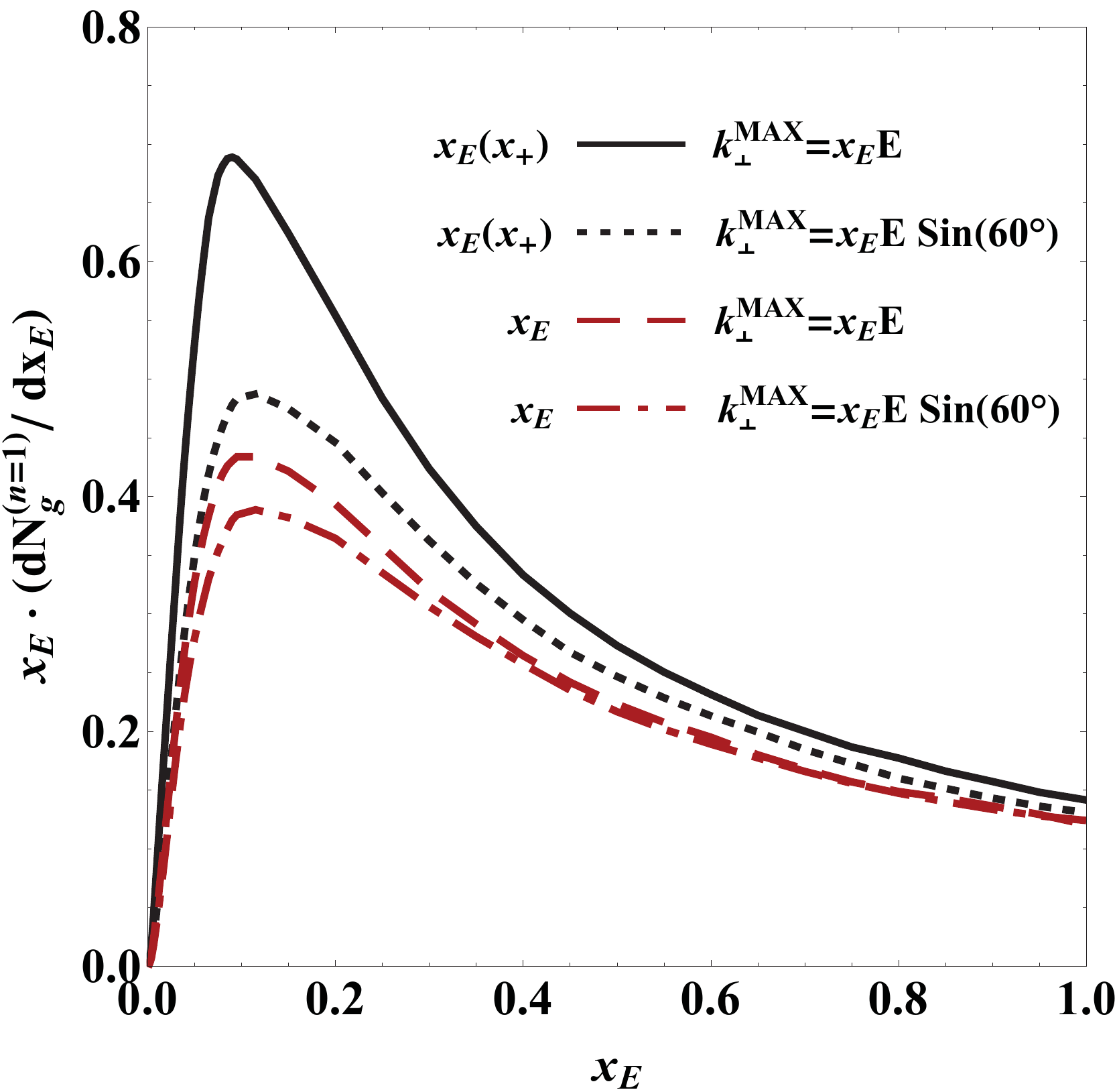}
\caption{$k_\perp$ integrated, $n=1$ gluon number distribution generated by a $E=20$ GeV light quark ($M=0.2$ GeV) jet traversing a brick plasma of thickness $L=5$ fm. The two interpretations of $x$ as gluon fractional energy ($x_E$) or gluon fractional plus-momentum ($x_+$) lead to remarkably different results, especially in the soft $x \ll 1$ region. The uncertainty due to the choice of $\theta^{MAX}$ is noticeable but less prominent. Other parameters used in the simulation are: $\lambda=1.16$ fm,  $\mu=0.5$ GeV, $m_g=0.356$ GeV, $T=0.258$ GeV, $n_f=0$, $\alpha_s=0.3$.}
\label{kTlimits}
\end{figure}

Depending on the interpretation of $x$, the upper kinematic limit on $k_\perp$ will vary: in the case of $x_+$, in order to ensure forward gluon emission we need to set $k_\perp^{MAX}=x_+ E^+$, whereas in the case of $x_E$, to keep $k_\perp$ real we must set $k_\perp^{MAX}\approx xE\sin\theta$, where $\theta$ is the angle between the radiated gluon and the propagating parton\footnote{In both cases, we neglect corrections due to the recoil of scattering centers in the medium.}. In Fig.~\ref{kTlimits}, we plot the $\kT$ integrated gluon number distribution $x\frac{dN_g}{dx}$, for both interpretations of $x$ and two different cutoff angles $\theta$; to compare apples to apples, we add the Jacobian of the transformation $x_+\rightarrow x_E$ to the $x_+$ curve and integrate up to $k_\perp^{MAX}= xE\sin\theta$:
\be
\label{Jacobian1}
x_E\frac{dN_g}{dx_E}=\int_0^{x_EE\sin(\theta)}d\bk \lp x_+\frac{dN_g}{dx_+d\bk}(\bk,x_+(x_E)) \rp \lp \frac{x_E}{x_+(x_E)}\rp J(x_E)
\; \; ,
\ee
\be
\label{Jacobian2}
J(x_E)\equiv\frac{dx_+}{dx_E}=\frac{1}{2}\lp1+\lp1-\lp\frac{k_\perp}{x_EE}\rp^2\rp^{-1}\rp
\; \; .
\ee

The differences are notable, and even more prominent in the small $x$ region, which dominates the gluon spectrum. The question of how we are going to quantify the error introduced by this systematic source of theoretical uncertainty arises immediately, and an answer will be given shortly. In the discussion above, we followed closely an in-depth analysis performed by Horowitz and Cole \cite{Horowitz:2009eb}.

\subsubsection{Systematic uncertainties}

We approach the problem of quantifying the systematic uncertainties caused by the choice of the $\kT$ integration limits in a way which will be iterated several times throughout the construction of the CUJET model. The idea is to isolate those sources of uncertainty that have a clear impact on the observables we are going to compute from other sources whose effect is hindered by the simple rescaling of a free parameter such as strong coupling constant.

In the context of the $\kT$ integration, we ask in Table~\ref{kTlimitsTABLE} what is the sensitivity of the energy loss $\Delta E/E$ to the particular choice of integration limits, provided the freedom to adjust a free parameter identified as the coupling constant $\alpha_s$.
\begin{table}[!t]
\centering
    \begin{tabular}{ | l | c | c | c |}
    \hline
    Curve & $\epsilon\equiv\frac{\Delta E}{E}$ & $\alpha_s^{\mathbf{\epsilon=0.32}}$ & $\alpha_s^{\mathbf{\epsilon=0.24}}$ \\ \hline
    $x_E(x_+)$ & \textbf{0.32} & 0.3 & 0.27 \\ \hline
    $x_E(x_+)^{\theta=60^{\circ}}$ & \textbf{0.27} & 0.32 & 0.29 \\ \hline
    $x_E$ & \textbf{0.24} & 0.33 & 0.3 \\ \hline
    $x_E^{\theta=60^{\circ}}$ & \textbf{0.23} & 0.33 & 0.30 \\
    \hline
    \end{tabular}
\caption{Fractional energy loss $\Delta E/E$ integrated from $x_EdN_g/dx_E$ for the curves shown in Fig.~\ref{kTlimits}. The results are indicated in the second column and range from $0.23$ to $0.32$. In the two rightmost columns are listed the values of the effective parameter $\alpha_s^\epsilon$ needed to obtain the energy loss specified in $\epsilon$. The free parameter $\alpha_s$ needs to be tuned at most $\pm10 \%$.}
\label{kTlimitsTABLE}
\end{table}

\begin{figure}[!t]
\centering
%\vspace{0.25in}
\includegraphics[width=0.45\textwidth%width=1.in%height=1.in
]{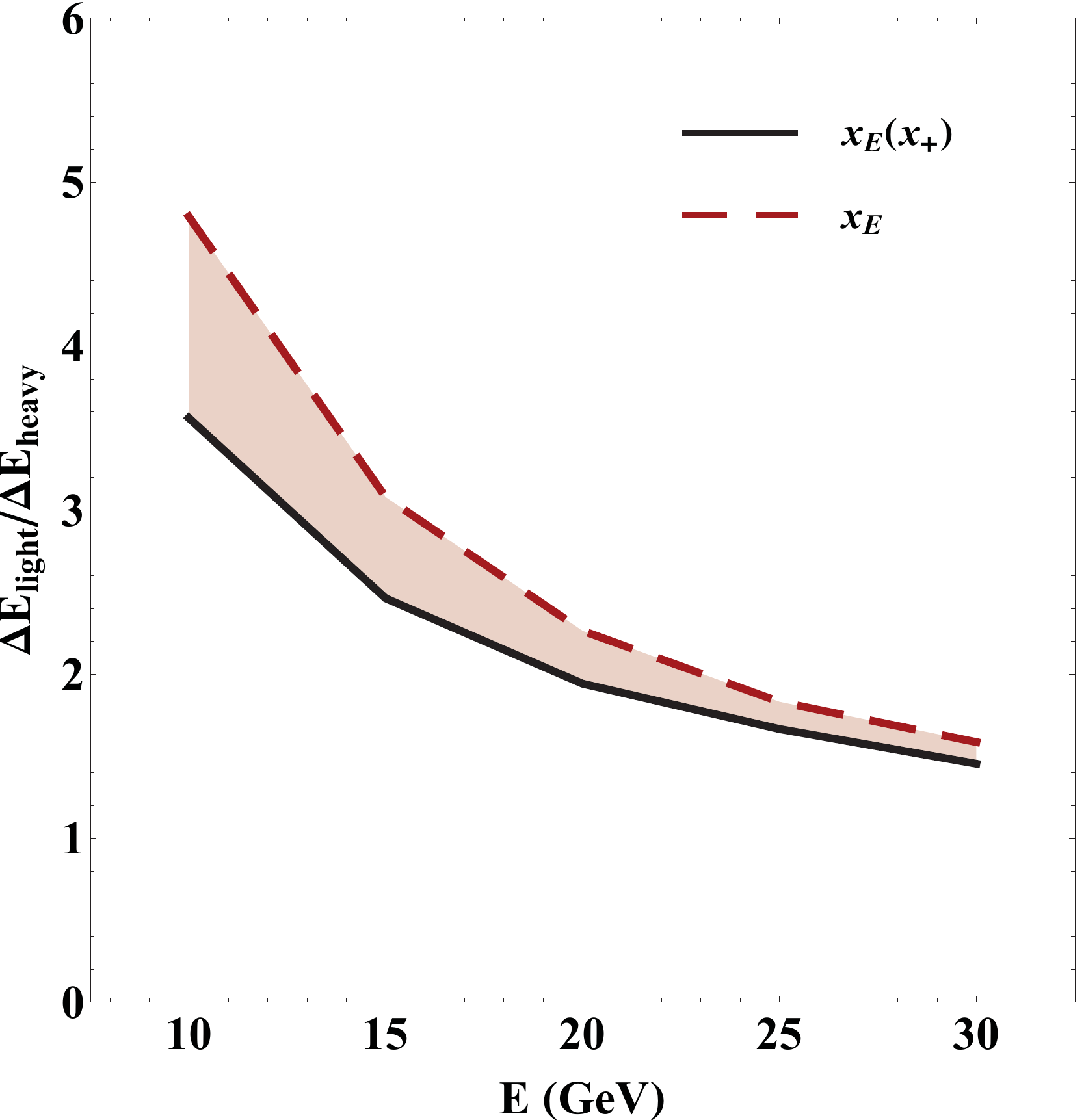}
\hspace{0.01\textwidth}
\includegraphics[width=0.45\textwidth%width=1.in%height=1.in
]{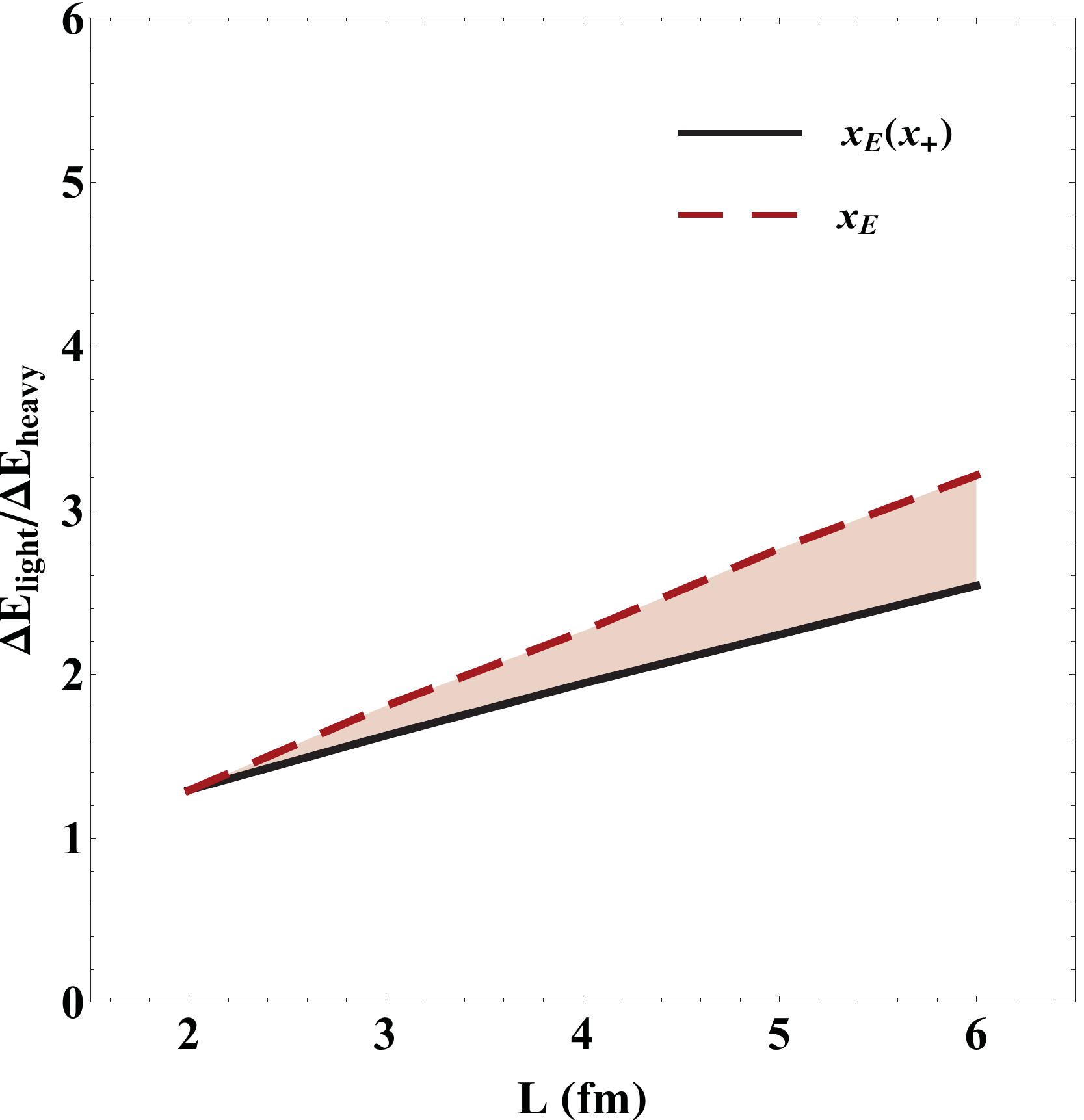}
\caption{Energy loss ratio between light ($M_l=0.2$ GeV) and heavy quark ($M_b=4.75$ GeV) jets in a brick, for different interpretations of $x$ as in Fig.~\ref{kTlimits}. Here $\alpha_s=0.3$, $L=5$ fm (\textit{left}), $E=20$ GeV (\textit{right}) and the energy loss has been computed at first order $n=1$ in opacity. An error of approximately $\sim 25\%$ is introduced for sufficiently small energies and large plasma sizes. Other parameters used in the simulation are: $\lambda=1.16$ fm,  $\mu=0.5$ GeV, $m_g=0.356$ GeV, $T=0.258$ GeV, $n_f=0$.}
\label{kTlimitsScaling}
\end{figure}

Given the interest in the ratio of light to heavy quark energy loss, we can immediately construct an error band which offers a quantitative measurement of the uncertainty generated by the choice of $k_\perp$ limits. In this way, $\alpha_s$ is factored out and the results are independent of the rescaling of the free parameter. Fig.~\ref{kTlimitsScaling} shows the scaling of $\Delta E_{light}/\Delta E_{heavy}$ with the jet energy $E$ and the plasma size $L$, for two distinct assumptions $x\equiv x_E$ and $x\equiv x_+$.

The conclusion is evident: the choice of $\kT$ limits has a relevant impact at small energies $E\le15$ GeV and long path lengths $L \ge 5$ fm. Further theoretical steps to address large angle radiation and relax the collinear approximation need to be taken. Until then, in the development of the CUJET model we adhere to the collinear derivation of GLV and interpret $x \equiv x_+$. When the gluon number distribution was needed as a differential in $x_E$, namely $x_E dN_g/dx_E$, we added the Jacobian of the transformation $x_+\rightarrow x_E$ to \eqref{DGLVN1}, and integrated $k_\perp$ up to $k_\perp^{MAX} = xE$. However, given the restricted size of such phase space, we will take our preferred assumption of $k_\perp$ limits and ignore the source of error coming from the $x$ interpretation, especially when studying results in the high energy range of LHC.

\subsection{Dead cone}

The ability to determine the quark flavor dependence of any physical observable is not only a interesting characteristic of the DLGV integral, but also an invaluable tool used to compare predictions with data. Computing the energy loss for charm and bottom quarks within the same consistent framework, in fact allows us to put additional constraints on the model and therefore gain more insights about the nature of the quark gluon plasma. In this section we want to check the effects that the parton mass has on the transverse momentum distribution of gluon radiation.

The mass term $M$ appears to have only a minor impact on the convergence of the series (Fig.~\ref{ExH5}), if nothing else by even improving it for certain combinations of $E$ and $\omega$. For very soft gluons ($x \ll 1$), the heavy quark jet radiation spectrum does not differ much from its light quark counterpart, while for large values of $x$ the radiation seems highly suppressed. The strong $x$ dependence of the magnitude and shape of $dN_g/dxd\bk$, as seen in Fig.~\ref{ExH5}, breaks the scaling with $\omega$ for typical gluon radiation from light quark jet.

\begin{figure}[!t]
\centering
%\vspace{0.25in}
\includegraphics[width=0.45\textwidth%width=1.in%height=1.in
]{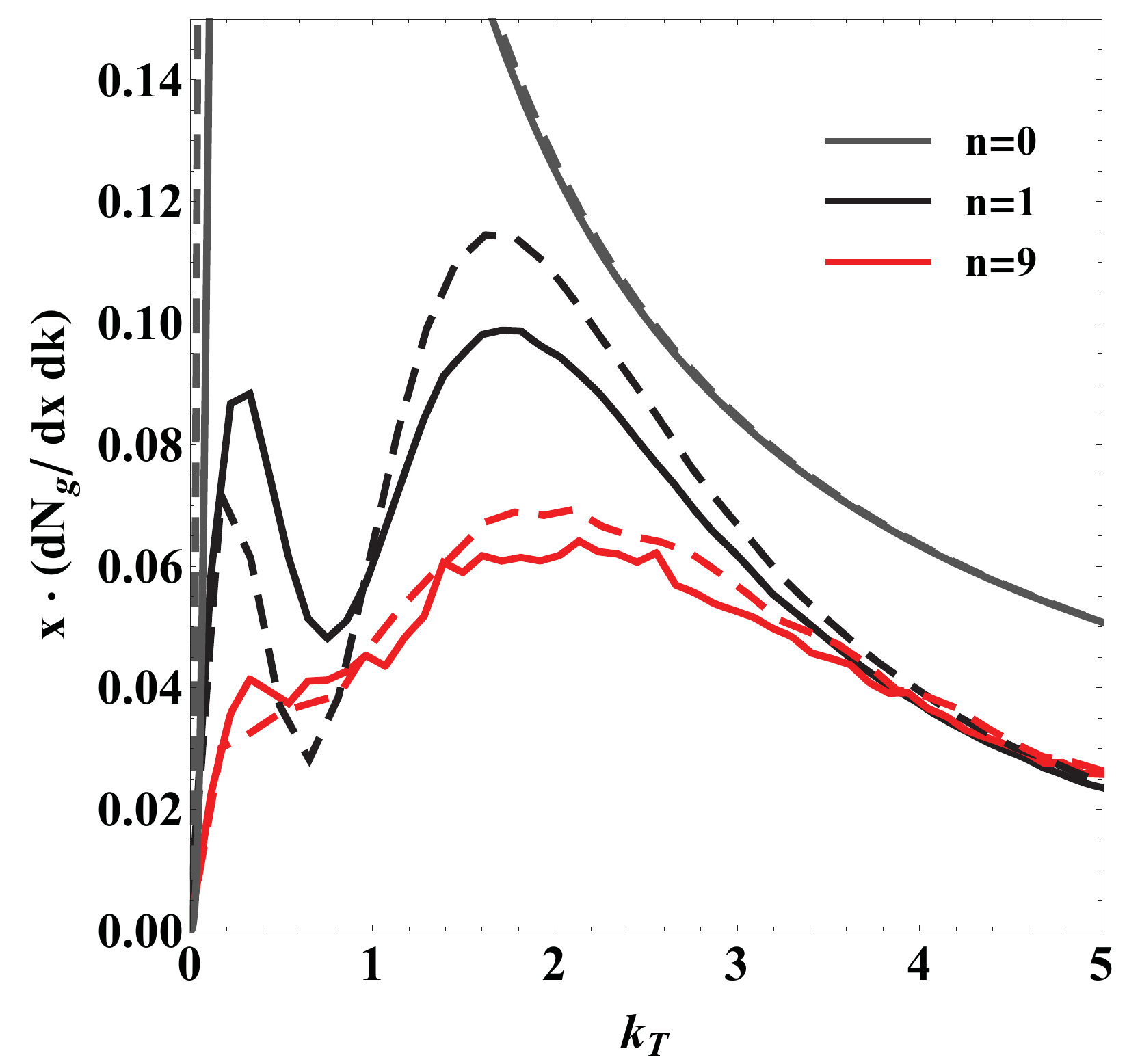}
\hspace{0.01\textwidth}
\includegraphics[width=0.45\textwidth%width=1.in%height=1.in
]{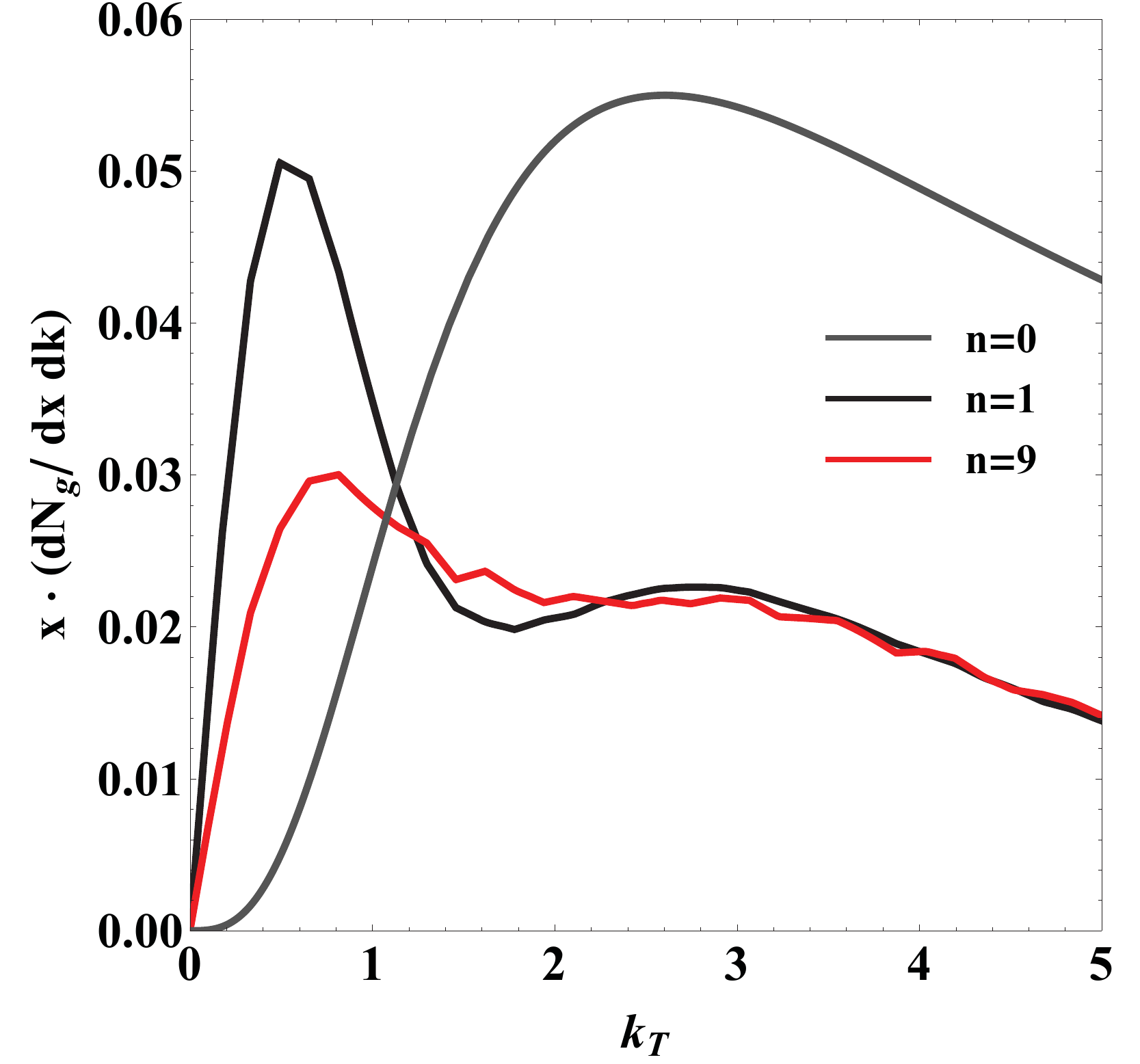}
\caption{Radiation spectrum for charm (\textit{left}) and bottom (\textit{right}) quarks traversing a brick of thickness $L=5$ fm, with $E=20$ GeV and $\omega=5$ GeV ($x=0.25$). The masses are assumed $M_c=1.2$ GeV and $M_b=4.75$ GeV. The dashed curves represent the spectrum of a light jet of mass $M_l=0.2$ GeV. Notice the similarity between the light and charm spectra, as opposed to the bottom one. The vacuum spectrum radiation is added to the plot (gray curve), showing the radiation dead cone for respective quark jets. Other parameters used in the calculation are: $\lambda=1.16$ fm,  $\mu=0.5$ GeV, $m_g=0.356$ GeV, $T=0.258$ GeV, $n_f=0$, $\alpha_s=0.3$.}
\label{deadcone}
\end{figure}

Another effect is the filling of the ``dead cone'' characteristic of the vacuum spectrum. In vacuum, the transverse momentum distribution takes the form
\be\label{vacuum}
x\frac{dN_g^0}{dxd\bk}\sim \frac{\bk^2}{(\bk^2+\chi^2)^2}
\;\;,
\ee
and the depletion of radiation takes place at angles
\be
\theta < \chi/\omega = \sqrt{M^2 x^2+m_g^2(1-x)} / (xE)
\; \; .
\ee
We compare in the right panel of Fig.~\ref{deadcone} the radiation spectrum of a heavy quark at different orders in opacity with the reference vacuum spectrum radiation, one notice immediately that the induced radiation fills in the dead cone already at first order in opacity. Since the dead cone region constitutes only a small fraction of the available phase space, the energy loss experienced by a heavy quark remains smaller than that of a light jet.

The left panel of Fig.~\ref{deadcone} shows instead a striking feature: despite its non-vanishing mass equal to $1.2$ GeV, the charm quark leads to a radiation spectrum very similar to the one of light quarks: not only the dead cone is absent and the vacuum spectrum almost divergent for $k_\perp \rightarrow 0$, but even the spectra have approximately the same shape and magnitude. This critical feature has vast phenomenological implications in the prediction of physical observables.

\section{Systematic uncertainties associated with $n_f$, $T_f$ and $dN/dy$}
\label{app:systematics}

In this section we analyze the sensitivity of CUJET to three of those parameters that govern the evolution of the medium: the number of quarkonic flavors $n_f$, the fragmentation temperature $T_f$, and the initial rapidity density $dN/dy$.\footnote{In principle, formation time and thermalization scheme also affect CUJET calculation. This issue is a topic of Appendix~\ref{app:thermal}.} As usual, eventually we will hinder their effects to the rescaling fixed or running coupling constant, and consider $\alpha_s$ or $\alpha_{max}$ the only free parameter of CUJET, which will be constrained by a specific set of experimental data, typically pion $\RAA$ at a given value of transverse momentum and center of mass collision energy. Therefore, it is of great interest to show how $\RAA$ changes for different plasma assumptions, and observe if its functional form is modified once a proper rescaling of the coupling has been performed.

\begin{figure}[!t]
%\vspace{0.25in}
\centering
\includegraphics[width=0.45\textwidth]{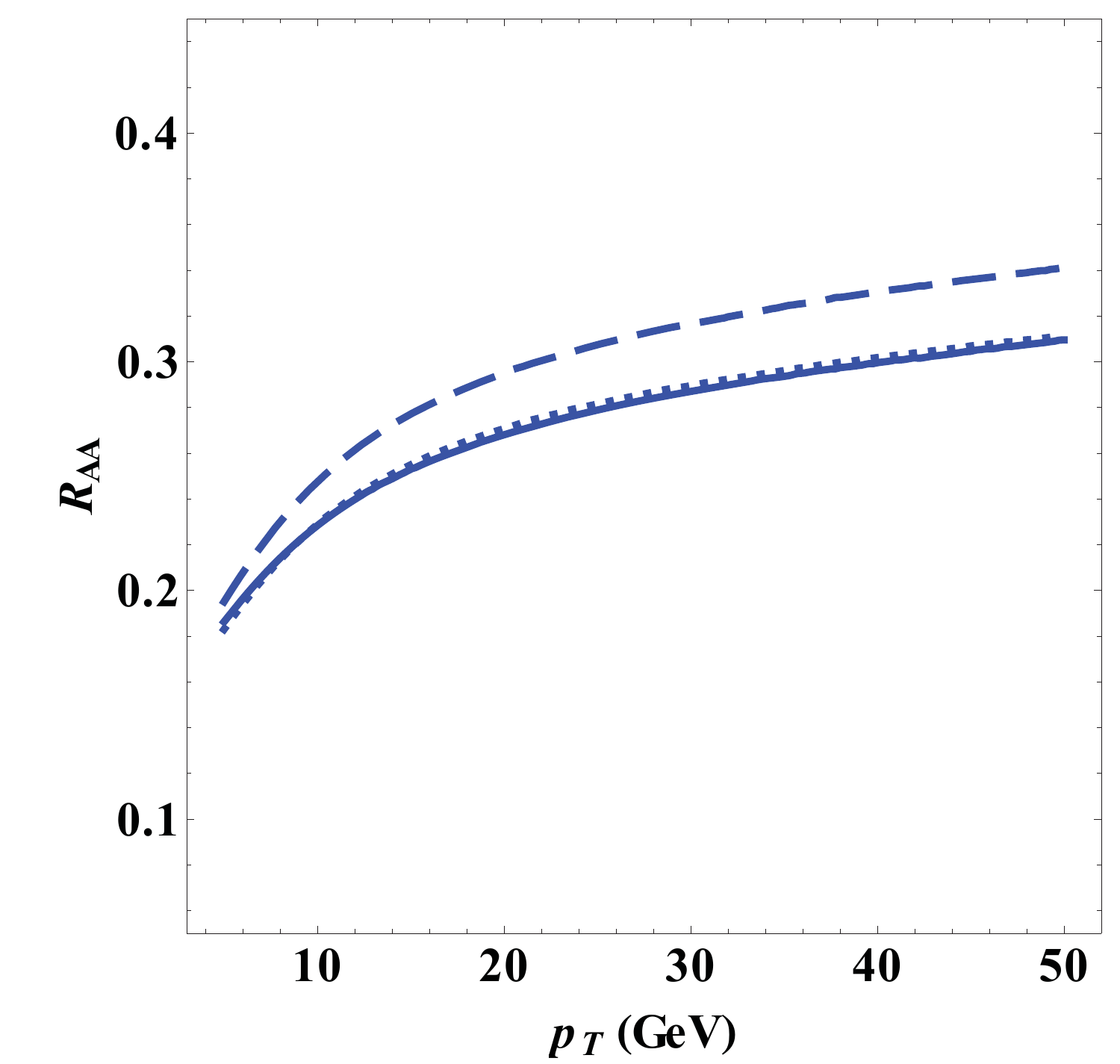}
\caption{Fixed coupling CUJET1.0 results for light quark ($M=0.2$ GeV) $\RAA$ in Au+Au 200AGeV central collisions (b=0 fm), with $n_f=0$ and $\alpha_s=0.3$ (solid line), $n_f=2.5$ and $\alpha_s=0.3$ (dashed line), $n_f=2.5$ and $\alpha_s=0.32$ (dotted line). The fragmentation temperature $T_f=100$ MeV. QGP's pre-thermal stage is linear and initial time $\tau_0=1$ fm/c, after thermalization Glauber + Bjorken evolution is assumed. The scenario of pure gluonic plasma and the scenario of equilibrated QGP with a $6\%$ increased coupling constant are indistinguishable.}
\label{Fig_RAAnf}
\end{figure}

In Fig.~\ref{Fig_RAAnf} we change the value of $n_f$ from $0$ (pure gluonic matter), to $2.5$ (mix of gluonic and quarkonic degrees of freedom in chemical equilibrium). We can easily observe that a simple rescaling of $\alpha_s$ of approximately $6\%$ leads to a perfect agreement between the two scenarios. In \cite{Zakharov:2008kt}, Zakharov reaches a similar conclusion starting from a path integral approach to the energy loss and using a running strong coupling. This simple analysis demonstrates the substantial insensitivity of CUJET to the detailed composition of the quark gluon plasma.

We now focus on the late phase of plasma evolution and measure the sensitivity of $\RAA$ to the jet hadronization temperature $T_f$.
\begin{figure}[!t]
%\vspace{0.25in}
\centering
\includegraphics[width=0.31\textwidth]{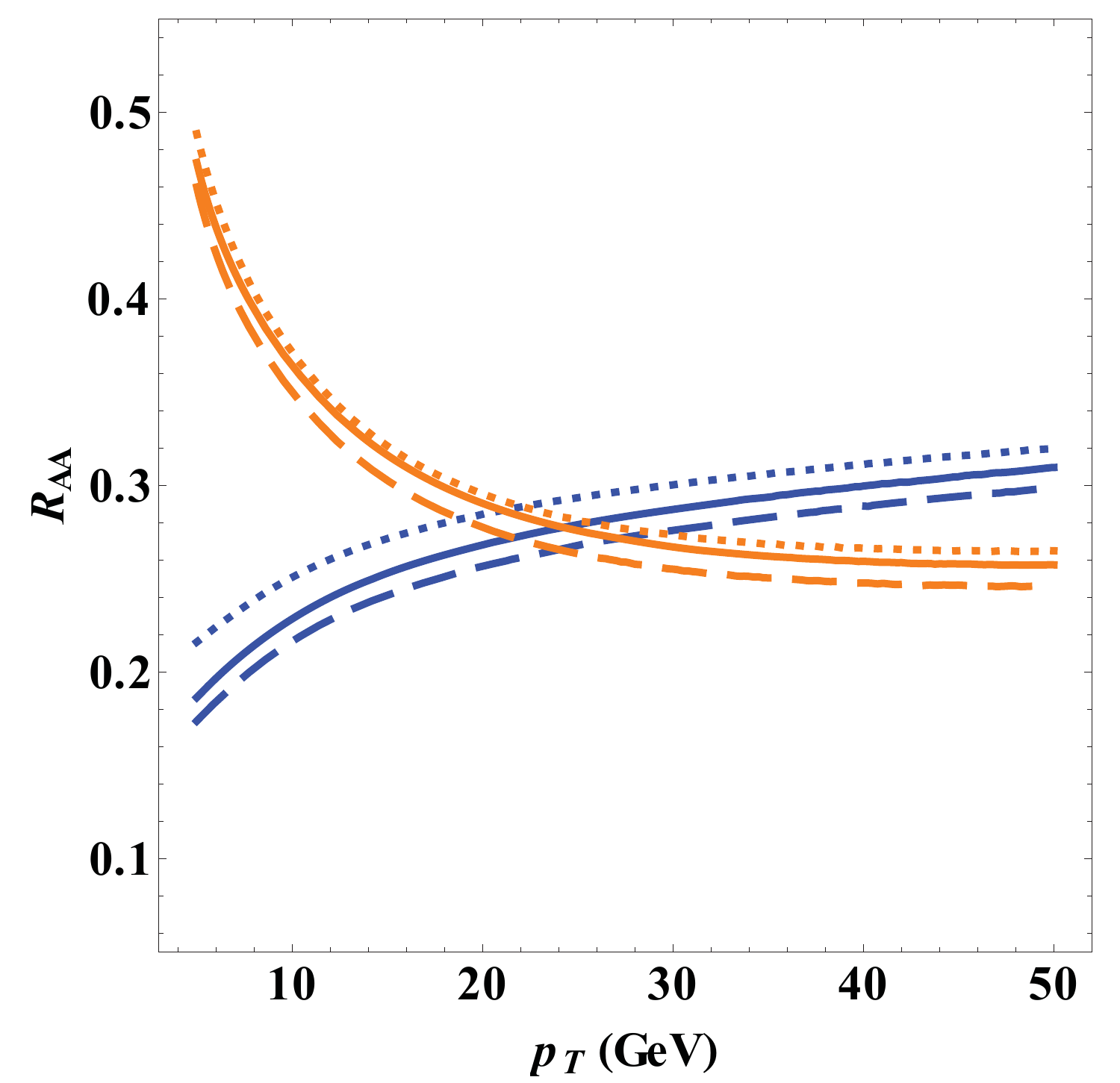}
\includegraphics[width=0.3\textwidth]{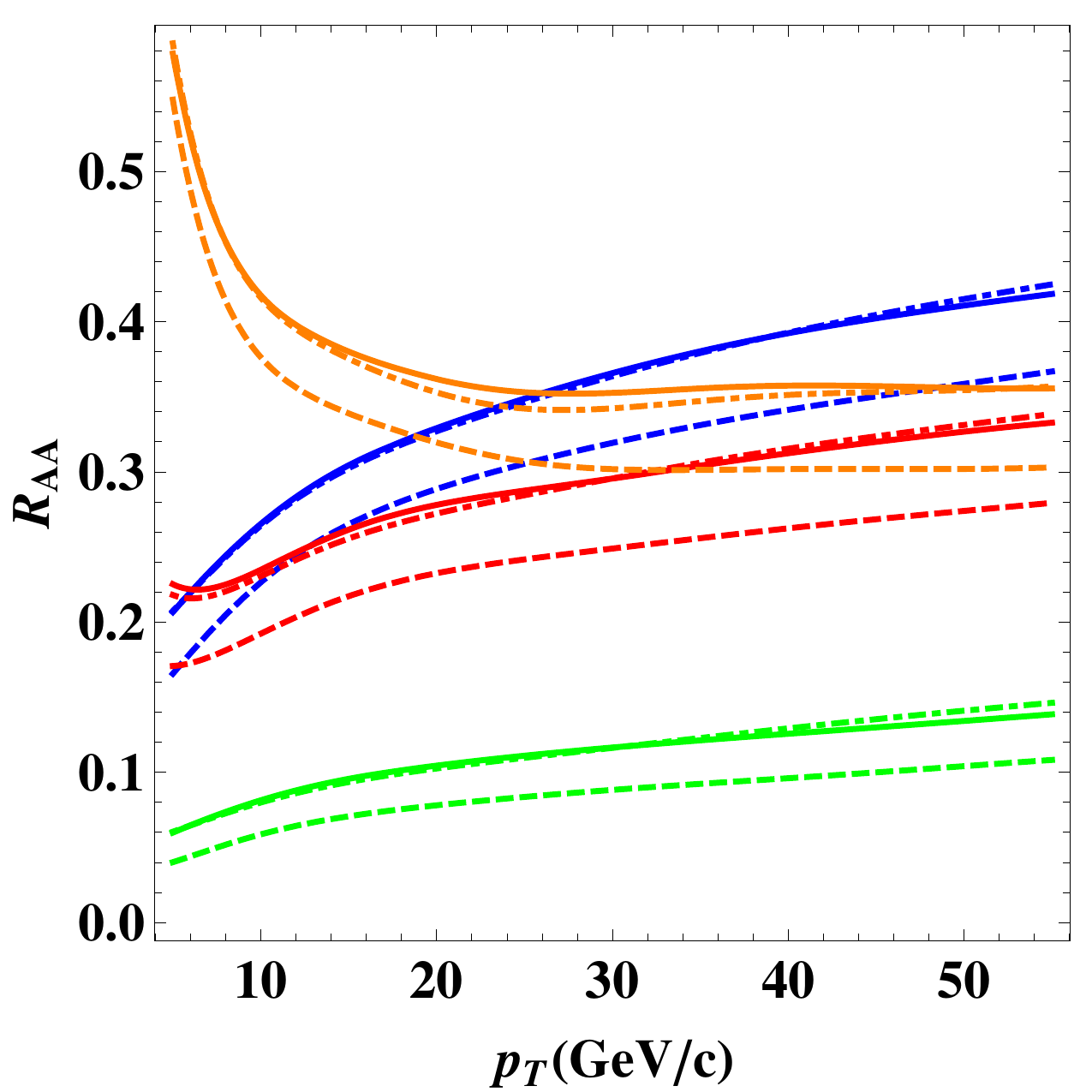}
\includegraphics[width=0.3\textwidth]{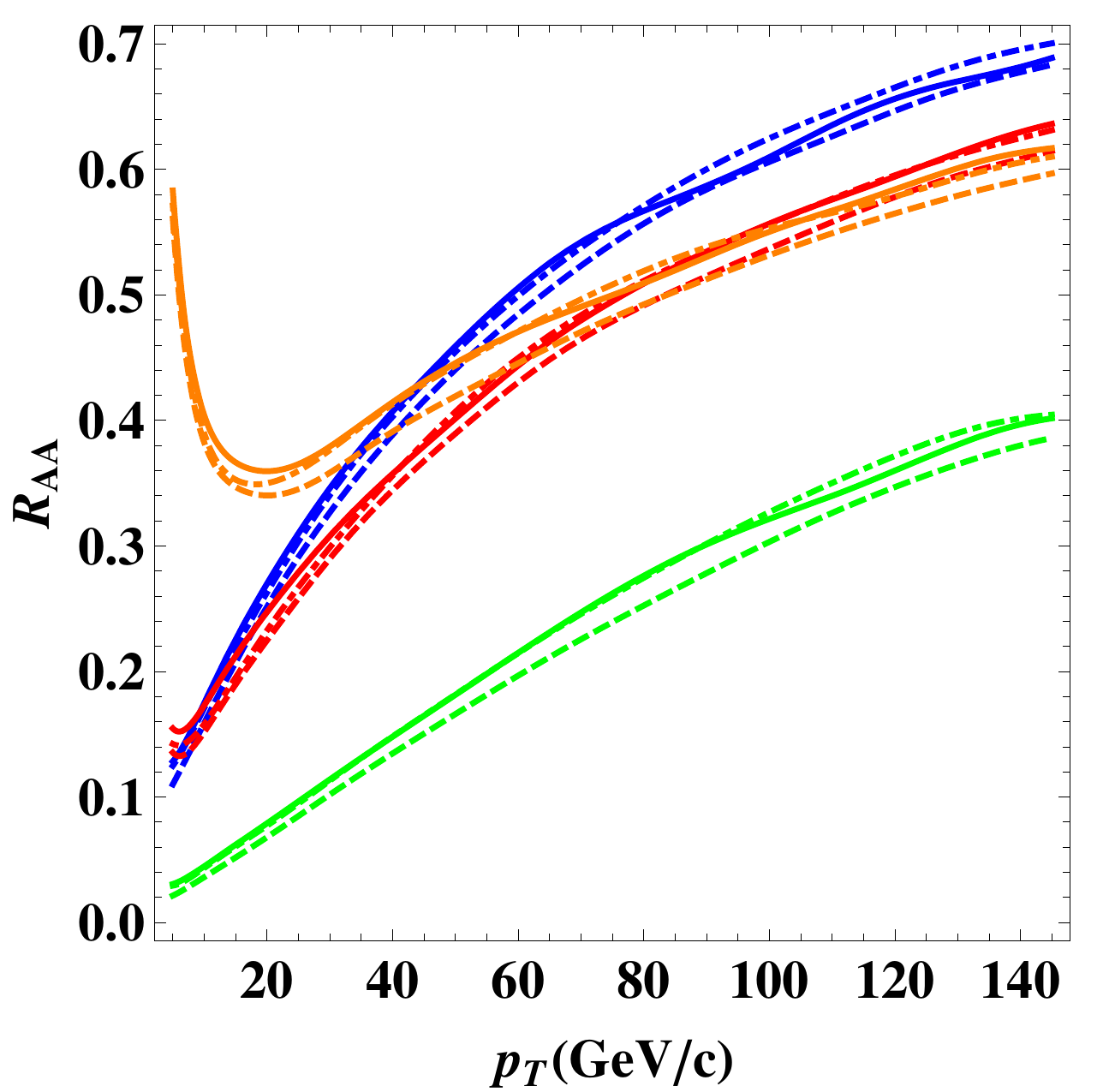}
\caption{(LEFT panel) Fixed coupling CUJET1.0 results for light quark (blue, $M=0.2$ GeV) and heavy quark (orange, $M=4.75$ GeV) $\RAA$ in Au+Au 200AGeV central collisions (b=0 fm), assuming $T_f=100$ MeV and $\alpha_s=0.3$ (solid), $T_f=50$ MeV and $\alpha_s=0.3$ (dashed), $T_f=200$ MeV and $\alpha_s=0.35$ (dotted). The number of quarkonic flavors $n_f=0$. QGP's pre-thermal stage is linear and initial time $\tau_0=1$ fm/c, after thermalization Glauber + Bjorken evolution is assumed. (MIDDLE panel) Running coupling CUJET2.0 results for light (blue, $M=0.2$ GeV), charm (red, $M=1.2$ GeV), bottom quark (orange, $M=4.75$ GeV) and gluon (green, $M=0$ GeV) $\RAA(p_T)$ in Au+Au 200AGeV central collisions (b=2.4 fm), assuming $T_f=120$ MeV and $\alpha_{max}=0.26$ (solid), $T_f=50$ MeV and $\alpha_{max}=0.26$ (dashed), $T_f=160$ MeV and $\alpha_{max}=0.28$ (dotdashed). The number of quarkonic flavors $n_f=2.5$. The bulk evolution profile is the same VISH2+1 as in Fig.~\ref{fig:RAA_pT}. The HTL deformation parameter $f_E=1$ and $f_M=0$. Note the $p_T$ dependent suppression patterns of $\pi^0$ and non-photonic $e^-$ fragmented from partonic $R_{AA}(p_T)$ with $(\alpha_{max},f_E,f_M)=(0.26,1,0)$ (solid curves) are consistent with experimental measurements of multiple collision configurations at the level of $\chi^2/d.o.f.<1.5$, cf. Sec.~\ref{sec:pion} and Sec.~\ref{sec:heavy}. At partonic level, the $R_{AA}(p_T)$ of light and bottom quark cross each other at $p_T\simeq 25$ GeV, which transverse momentum surprisingly coincides with the crossing of $R_{AA}^{\pi}(p_T)$ and $R_{AA}^{B}(p_T)$. Increase $T_f$ from hadronic freeze-out temperature $120$ MeV to critical temperature $160$ MeV requires an enhancement of \amax~from 0.26 to 0.28 for maintaining the same partonic $R_{AA}$ level, suggesting by choosing $T_f=120$ MeV the quenching effect is not significantly overestimated, and it causes less than 10\% under-prediction of \amax. (RIGHT panel) The same CUJET2.0 model applied to LHC Pb+Pb 2.76ATeV b=2.4fm. Notice again the intersection of $R_{AA}^{light}(p_T)$ and $R_{AA}^{bottom}(p_T)$ at $p_T\simeq 35$ GeV, and the relative insensitivity of $R_{AA}(p_T)$ to $T_f$ variation and \amax~rescaling.}
\label{RAAtauf}
\end{figure}
In the left panel of Fig.~\ref{RAAtauf}, the CUJET1.0 partonic nuclear modification factor is shown for light and heavy quarks, for the default $T_f=100$ MeV and $\alpha_s=0.3$ parameters, $T_f=50$ MeV and $\alpha_s=0.3$, and finally $T_f=200$ MeV and $\alpha_s$ rescaled to $0.35$. We observe that jet quenching is ``saturated'' already at $T_f=100$ MeV: even if we let the jets interact until $T$ drops to the (unphysical) value of $50$ MeV, no significant changes occur in $\RAA$. On the contrary, restricting the interaction region to $T>200$ MeV alters significantly the results and a moderate $\sim20 \%$ rescaling of the coupling constant is needed in order to reproduce the the original curve. But given the freedom to fit the coupling constant, CUJET1.0 can be regarded being insensitive to this source of theoretical uncertainty.

In the middle and right panel of Fig.~\ref{RAAtauf}, the CUJET2.0 HTL $(f_E=1,f_M=0)$ partonic $R_{AA}$ is shown for gluon and light, charm, bottom quarks at RHIC Au+Au 200AGeV and LHC Pb+Pb 2.76ATeV central collision (b=2.4fm) respectively, with $T_f=120$ MeV and $\alpha_{max}=0.26$, $T_f=160$ MeV and $\alpha_{max}=0.28$, and $T_f=50$ MeV and $\alpha_{max}=0.26$. We observe that $R_{AA}^{light}(p_T)$ and $R_{AA}^{bottom}(p_T)$ intersect at $p_T\simeq 25$ GeV for RHIC and $p_T\simeq 35$ GeV for LHC, both transverse momenta overlap with the crossing point of $R_{AA}^{\pi}(p_T)$ and $R_{B}^{bottom}(p_T)$. The physical reason for this robust level crossing has been explained semi-quantitatively in Sec.~\ref{sec:heavy}, and will be explored in more detail in \cite{JXMGprep}. Notice that $R_{AA}(p_T)$ of $\pi^0/h^{\pm}$, $D$ and non-photonic $e^-$ computed within the CUJET2.0 model of $(\alpha_{max},f_E,f_M)=(0.26,1,0)$ and $T_f=120$ MeV consistently agrees with data at average $\chi^2/d.o.f.<1.5$ level for both RHIC and LHC, the solid orange curve in the middle and right panel of Fig.~\ref{RAAtauf} is therefore the CUJET2.0 prediction for b-jet or non-prompt $J/\psi$ for Au+Au $\sqrt{s_{NN}}=200$ GeV and $\sqrt{s_{NN}}=2.76$ TeV at 0-10\% centrality respectively.

The relative insensitivity of $R_{AA}(p_T)$ to $T_f$ variation and \amax~rescaling appears again in CUJET2.0. The middle and right panel of Fig.~\ref{RAAtauf} shows that fixing \amax~and decreasing the default $T_f=120$ MeV by approximately 60\% to 50 MeV generates 10\% enhancement in the quenching of partons, and the latter magnitude is much less than the former, which suggests the ``saturation'' effect observed in CUJET1.0 occurs again in CUJET2.0. The two figures also display that a shift of \amax~from 0.26 to 0.28 can compensate the increase of $T_f$ from hadronic freeze-out temperature $120$ MeV to critical temperature $160$ MeV and thereby maintain the original partonic $R_{AA}(p_T)$. This fact indicates that choosing freeze-out rather than critical temperature does not significantly overestimated the quenching effect, and it leads to less than 10\% under-prediction of \amax.

However, regarding $T_f$, the study of experimental observables that are sensitive to the azimuthal anisotropy of the plasma and the angular distribution of jets would prove to be more insightful, because the fragmentation region generally resides in the outer shell (corona) of the plasma, and is more sensitive to the geometry of the collision, i.e. impact parameter, and the transverse expansion.

\begin{figure}[h!]
%\vspace{0.25in}
\centering
\includegraphics[width=0.45\textwidth]{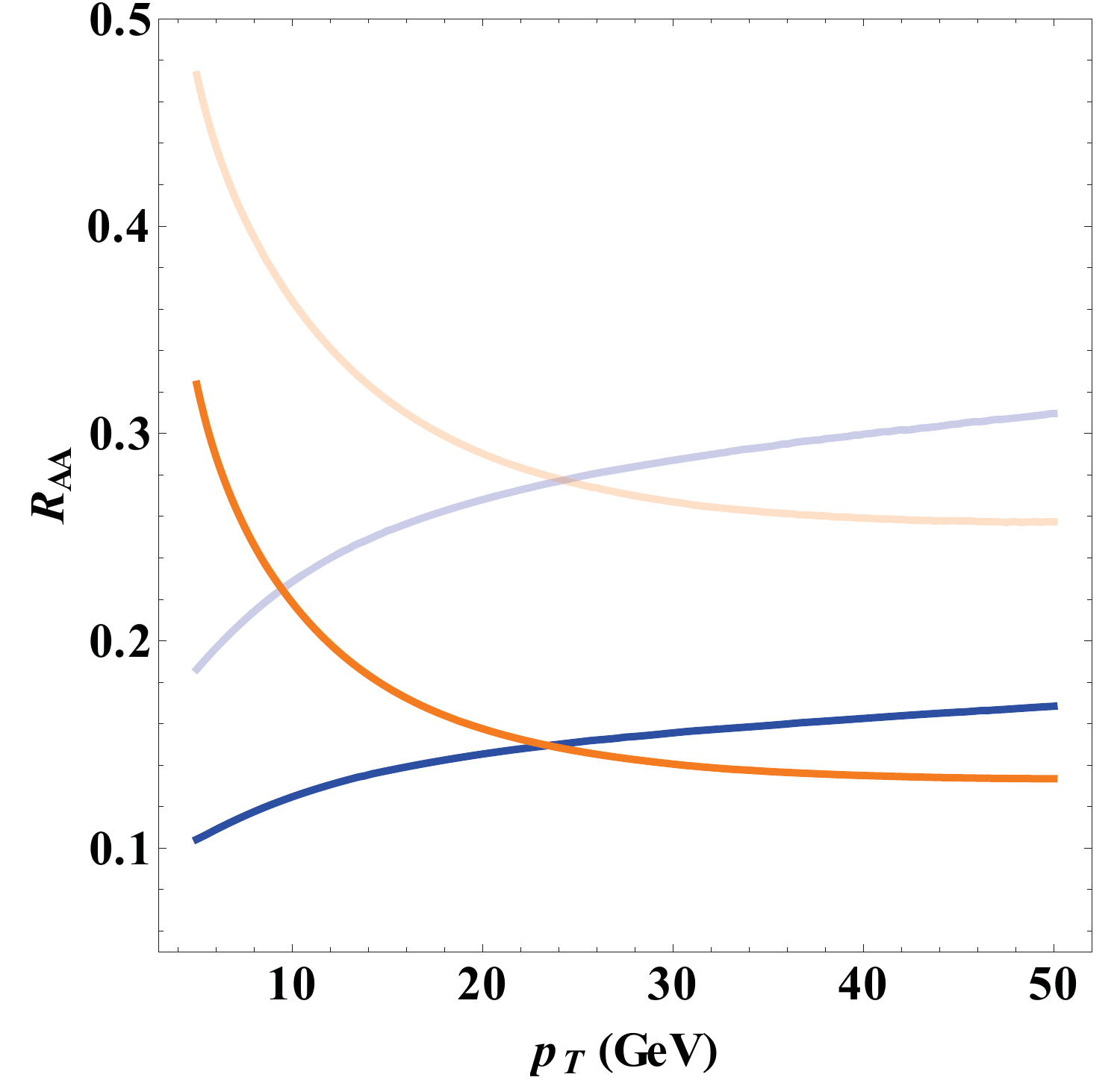}
\caption{Fixed coupling CUJET1.0 $\RAA$ for light (blue, $M=0.2$ GeV) and heavy (orange $M=4.75$ GeV) quarks in Au+Au 200AGeV central collisions. RHIC production spectra (left panel of Fig.~\ref{InitSpectra}) are used in this plot, as well as RHIC collision parameters. However, the initial rapidity density $dN_g/dy$ is increased from $1000$ (opaque lines) to $2200$ (solid lines). The increased $dN_g/dy$ is responsible for the suppression of $\RAA$. Other parameters used in the simulation are: $\alpha_s=0.3$, $n_f=0$, $T_f=100$ MeV, linear thermalization with initial time $\tau_0=1$ fm/c.}
\label{RAAdNdy}
\end{figure}

Finally, we study in CUJET1.0 the sensitivity of $\RAA$ to the initial rapidity density $dN_g/dy$. This parameter is constrained by experimental observations, it fixes the initial density and temperature of the plasma according to Eq.~\eqref{rhoQGP}. Intuitively, we expect the quenching to be higher for denser plasma, resulting in an increased suppression of $\RAA$ for collisions observed at the LHC. Our expectations are confirmed in Fig.~\ref{RAAdNdy}.

Note in CUJET2.0 the medium information has been encoded in the 2+1D viscous hydrodynamic fields which presumably fit properly the hadron production spectra and bulk harmonics in the soft region, the systematic uncertainties associated with initial rapidity density $dN/dy$ and the number of quarkonic flavor $n_f$ are therefore irrelevant in the CUJET2.0 = rcDGLV + elastic + VISH2+1 framework.

\section{Elastic energy loss}
\label{app:elastic}

\subsection{Bjorken's formula}

The first estimation for collisional energy loss in a quark gluon plasma was made by Bjorken \cite{Bjorken:1982tu}, and his work still constitutes the benchmark against which any computation of this kind should be compared. Here we briefly outline his derivation.

In the limit $E \gg k$, where $k$ the momentum of the target particle in the medium, we can approximate the quark-quark, quark-gluon and gluon-gluon elastic cross sections as
\be
\frac{d\sigma_{i,j}}{d\hat{t}}=\frac{2\pi\alpha^2}{\hat{t}^2} c_{i,j}
\; \; ,
\label{ELcrosssec}
\ee
where $c_{i,j}$ is a numerical factor equal to $4/9$, $1$, $9/4$ for $\{i,j\}=\{q,q\}$, $\{q,g\}$ or $\{g,g\}$ respectively.
The energy loss per unit length can be written as
\be
\frac{dE}{dx}=\int{d^3k}\; \rho_i(k) \Phi \int_{\hat{t}_{MIN}}^{\hat{t}_{MAX}}{d\hat{t}} \frac{d\sigma_{i,j}}{d\hat{t}}\cdot (E-E')
\; \; .
\label{BjEloss}
\ee
Here $E-E'$ represents the energy lost in the collision, $\rho_i(k)$ is the quark or gluon number density, and $\Phi$ is the flux factor that accounts for the relative orientation of the target and projectile.
Defining $\theta$ as the angle between the incoming parton and the target,
\be\begin{array}{rcl}
E-E' &=& -\dfrac{\hat{t}}{2k(1-\cos \theta)} \\
\Phi &=& 1 - \cos \theta
\; \; .
\end{array}\ee
Integrating \eqref{BjEloss} over $d\hat{t}$, we obtain
\be
\frac{dE}{dx}=\int{d^3k} \; \rho_i(k) \lp -\frac{\pi\alpha^2}{k}c_{i,j} \log B \rp
\; \; ,
\ee
where $B$ is defined by the integration limits $\hat{t}_{MAX}$ and $\hat{t}_{MIN}$. If assuming $B$ is independent of $k$ for simplicity, we can set $\hat{t}_{MAX}\approx 2\left<k \right>E \approx 4TE$ and $\hat{t}_{MIN}=\mu^2$, with $\mu$ being the Debye screening mass of the plasma.

If we further write the quark and gluon number densities as
\be\begin{array}{rcl}
\rho_q(k) &=& \dfrac{12 n_f}{(2\pi)^3} \dfrac{1}{\rm{e}^{\beta k}+1}  \\
\rho_g(k) &=& \dfrac{16}{(2\pi)^3} \dfrac{1}{\rm{e}^{\beta k}-1}
\; \; ,
\end{array}\ee
we can perform the last integration over all momenta $d^3k$ and finally get to the Bjorken energy loss formula
\be
\frac{dE}{dx}= -\pi C_R \frac{\alpha^2}{\beta^2} \lp 1+\frac{n_f}{6} \rp \log B
\; \; .
\label{bjergloss}
\ee
In order to derive this short analytic result, several approximations were made in the way that infrared and ultraviolet divergences are regulated, i.e. $\hat{t}_{MIN}$ and $\hat{t}_{MAX}$. Such divergences 
%typical of the $1/t^2$ matrix element in the elastic cross section Eq.~\eqref{ELcrosssec}, 
are physically related to the absence of collective medium effects (soft scattering) and recoil (hard scattering) in the derivation of the theory.

\subsection{Numerical effects}
\label{app:ElasticNumerics}

\begin{figure}[!t]
\centering
\includegraphics[width=0.3\textwidth%width=1.in%height=1.in
]{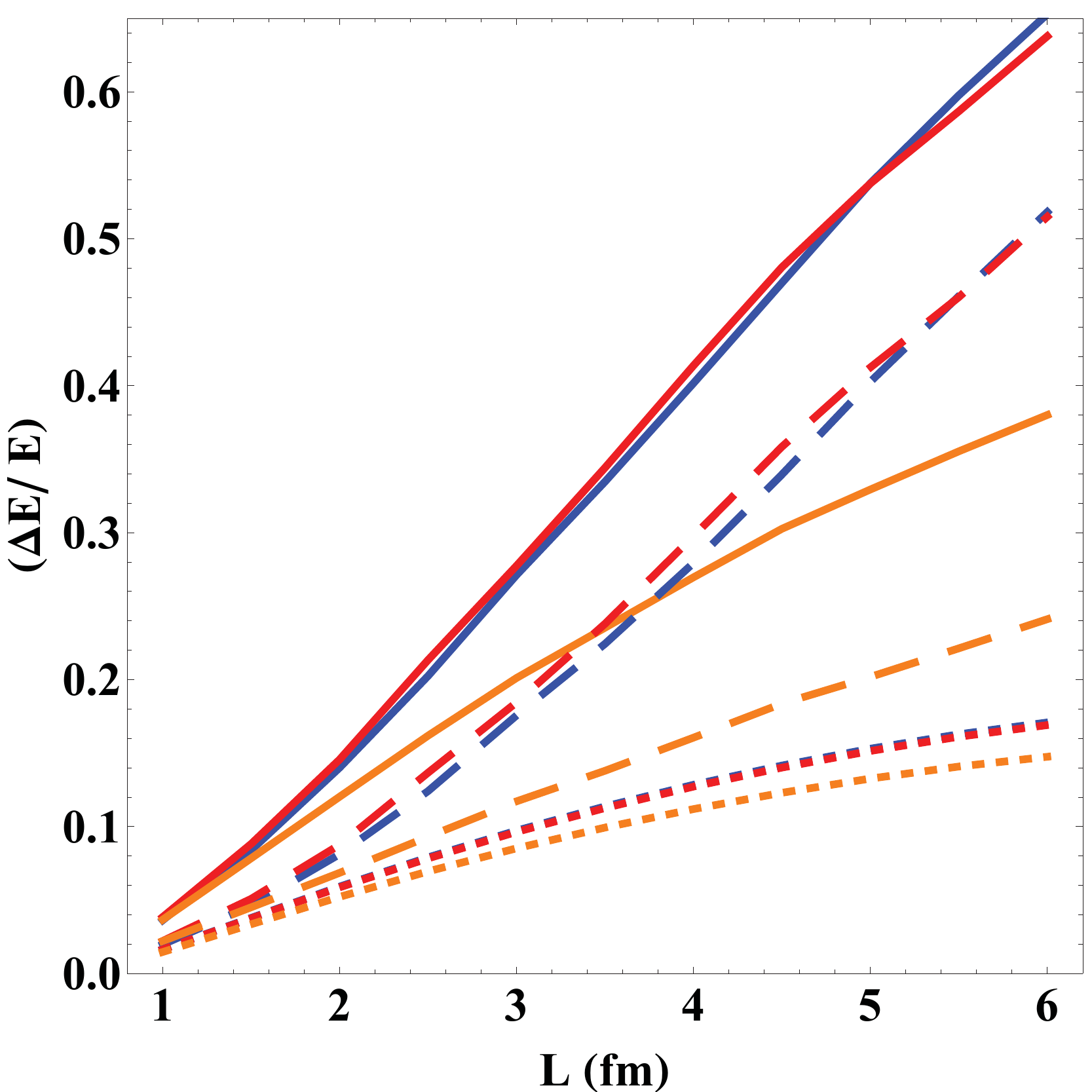}
\includegraphics[width=0.3\textwidth%width=1.in%height=1.in
]{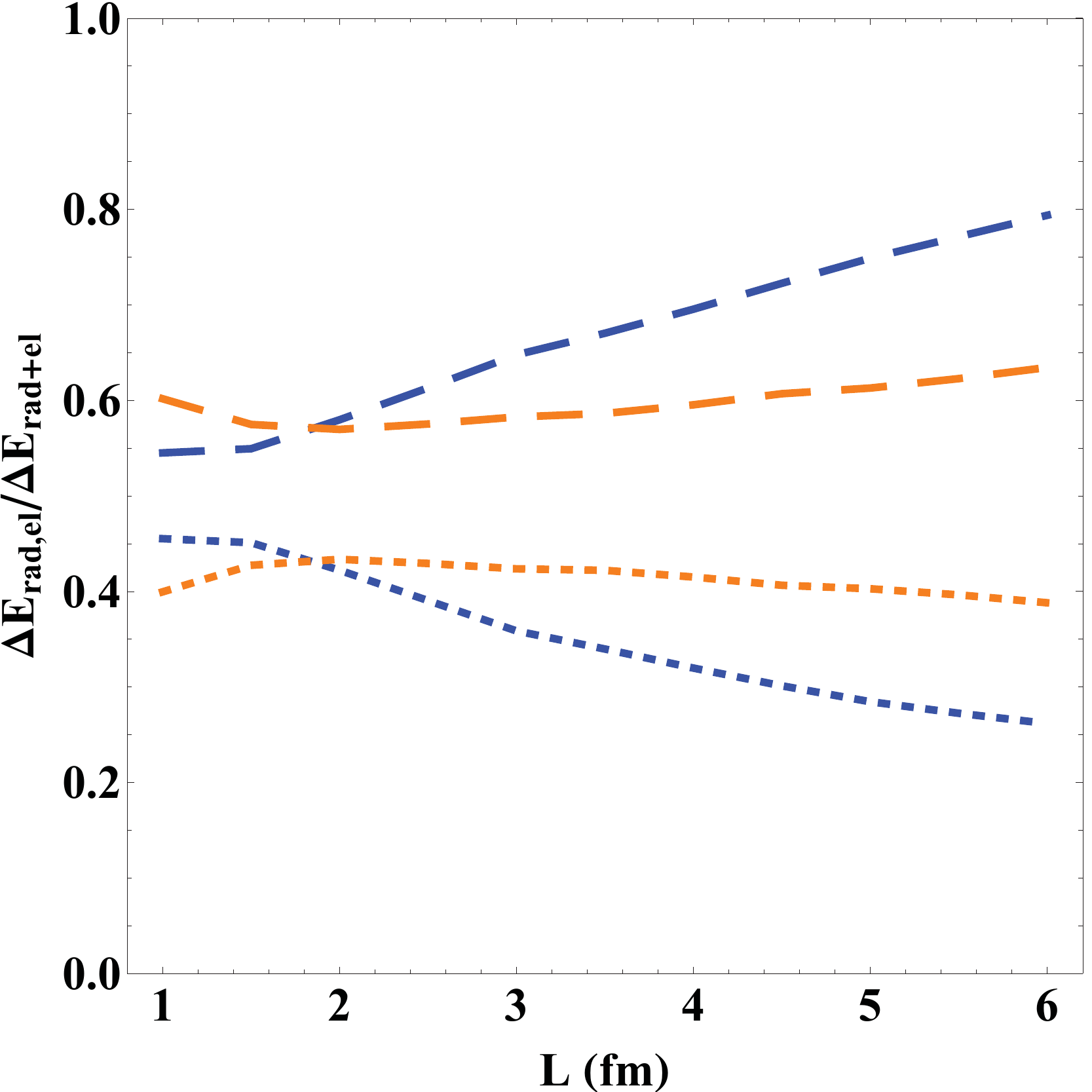}
\includegraphics[width=0.3\textwidth%width=1.in%height=1.in
]{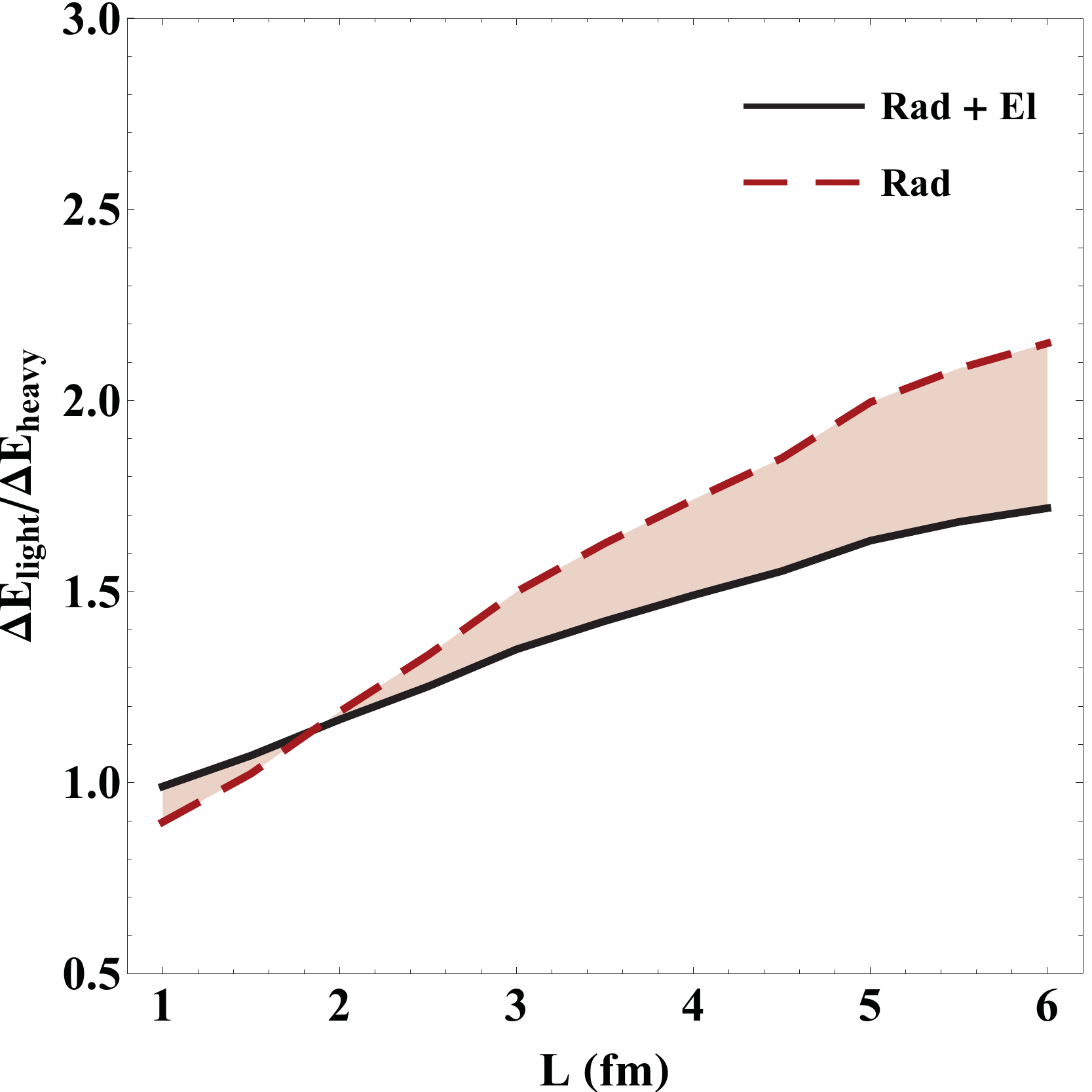}
%\hspace{0.01\textwidth}
\includegraphics[width=0.3\textwidth%width=1.in%height=1.in
]{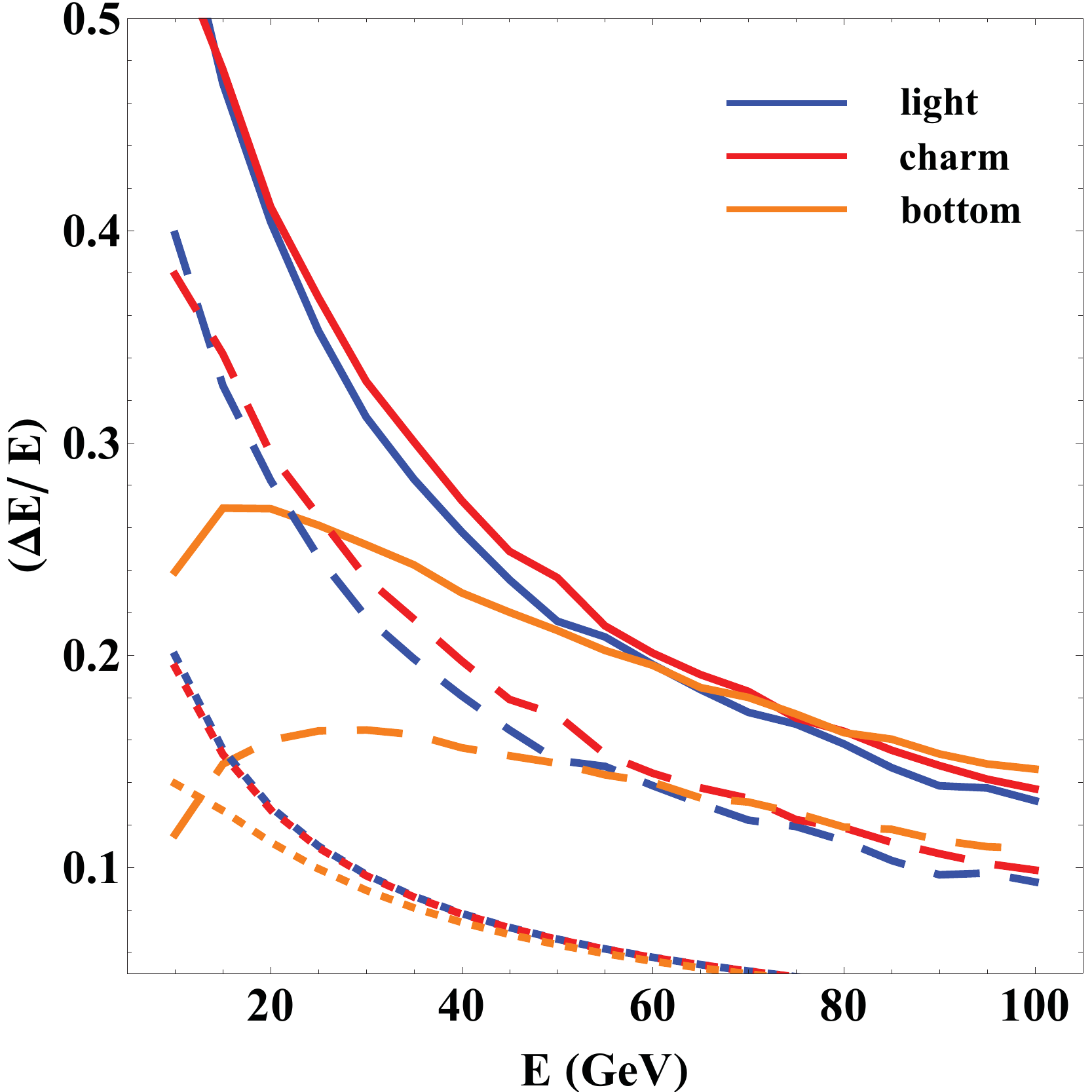}
\includegraphics[width=0.3\textwidth%width=1.in%height=1.in
]{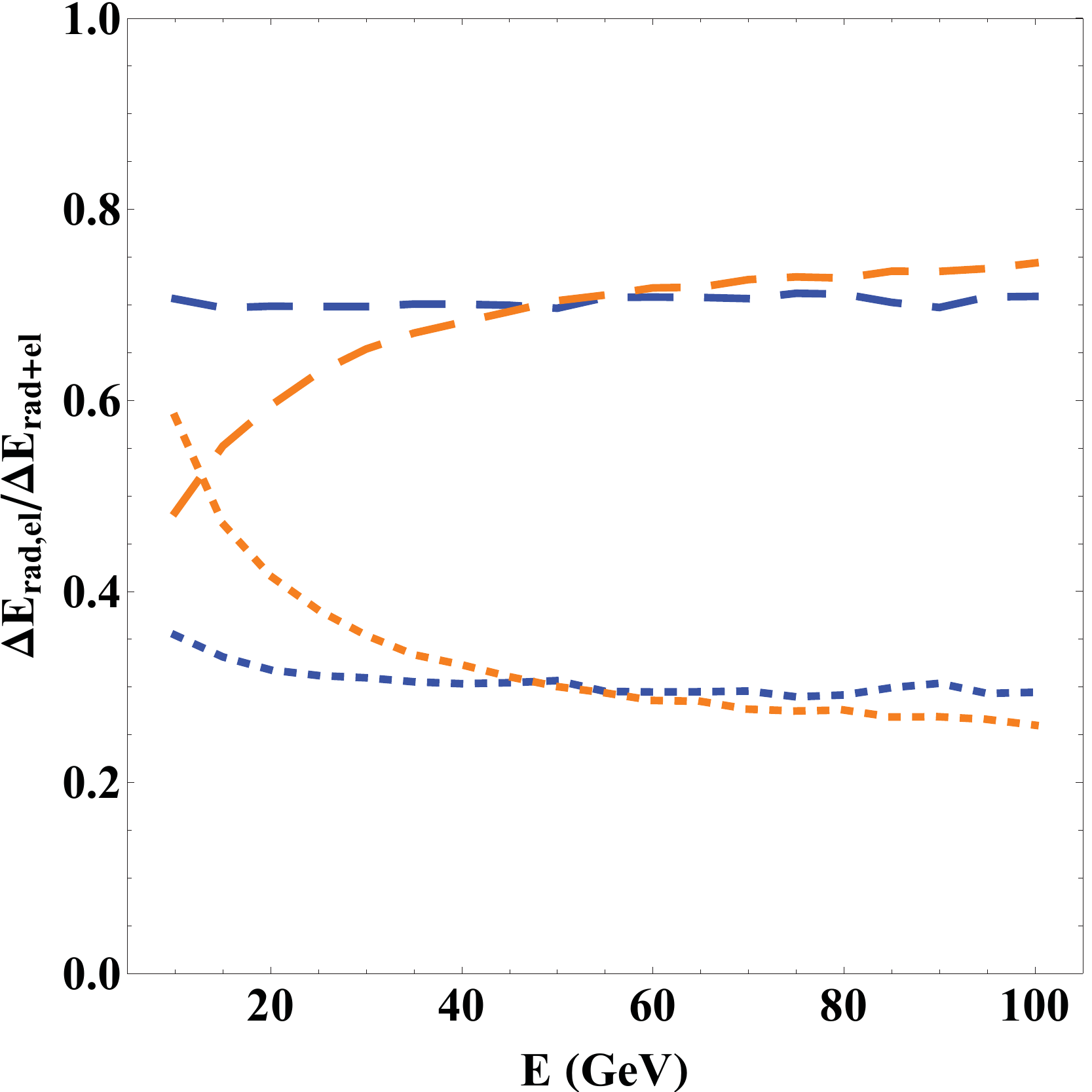}
\includegraphics[width=0.3\textwidth%width=1.in%height=1.in
]{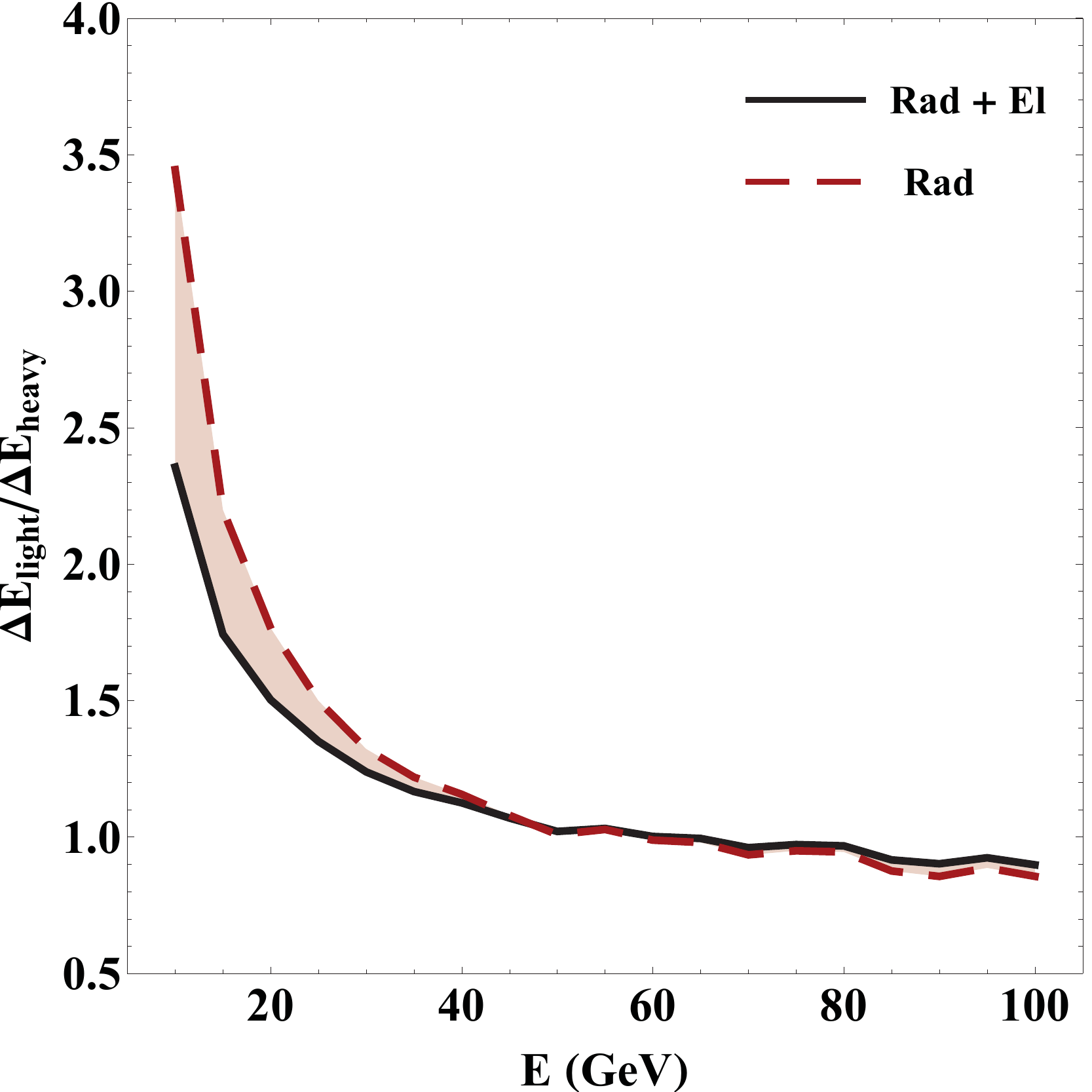}
\caption{(LEFT panels) Radiative (dashed), elastic (dotted) and total (solid) energy loss for light (blue, $M=0.2$ GeV), charm (red, $M=1.2$ GeV) and bottom (orange, $M=4.75$ GeV) quarks in a dynamical brick plasma of size $L$. The plasma is thermalized at temperature $T=0.258$ GeV and characterized by only gluonic degrees of freedom ($n_f=0$). Poisson fluctuations for the radiative sector and Gaussian fluctuations for the elastic sector are taken into account. The total energy loss is calculated from the convolution of both sectors. TOP: The quark jet energy is set to $E=20$ GeV. BOTTOM: $L=4$ fm. (MIDDLE panels) Ratio $\Delta E_{rad}/\Delta E_{rad+el}$ (dashed lines) and $\Delta E_{el}/\Delta E_{rad+el}$ (dotted lines), for light (blue) and bottom (orange) quarks. $\Delta E_{rad+el}$ denotes the total energy loss. The ratios are computed from the results in the left panels. The dominant contribution to the total energy loss comes from inelastic collisions. (RIGHT panels) Light to bottom quark energy loss ratio, for radiative only (dashed) and total (solid) energy loss. The curves are obtained from the same data plotted in the left panels. Other parameters used in the calculations are: $\lambda=1.16$ fm,  $\mu=0.5$ GeV, $m_g=0.356$ GeV, $\alpha_s=0.3$.}
\label{RadElScan}
\end{figure}

In the left panels of Fig.~\ref{RadElScan} we observe a gain in $\Delta E/E$ after elastic collisions are taken into account. We notice three main features: (1) the energy loss is increased by up to $20 \%$; (2) elastic losses almost do not distinguish between light and heavy quarks; (3) For sufficiently large $L$, $\Delta E/E$ shows signs of saturation, indicating complete quenching of the jets.

The partial contribution of radiative and elastic losses to the total $\Delta E$ is given in the middle panels of Fig.~\ref{RadElScan}, assuming a dynamical medium. Here we immediately appreciate the difference between light and bottom quarks: while the relative elastic contribution diminishes with $L$ and is approximately constant with $E$ in the case of light partons, the exact opposite behavior is observed for heavy quarks. This has a remarkable impact on the phenomenology: the ratio $\Delta E_{light}/\Delta E_{heavy}$, shown in the right panels of Fig.~\ref{RadElScan}, drops by almost $25 \%$ in the large $L$ and small $E$ regions.

The inclusion of dynamical effects first, and elastic collisions later, has brought the light to heavy quark energy loss ratio down from a factor of more than $2$x to about $1.5$x, in the range of energies $E\sim10-30$ GeV and path lengths $L\sim4-6$ fm. These improvements constitute a promising step toward closing the gap between theoretical models and experimental data, which is shown at RHIC as a surprising similarity between the quenching of light and heavy quark jets.

\section{Radiative energy loss probability distribution and fluctuations}
\label{app:fluc}

\begin{figure}[!t]
\centering
%\vspace{0.25in}
\includegraphics[width=0.3\textwidth%width=1.in%height=1.in
]{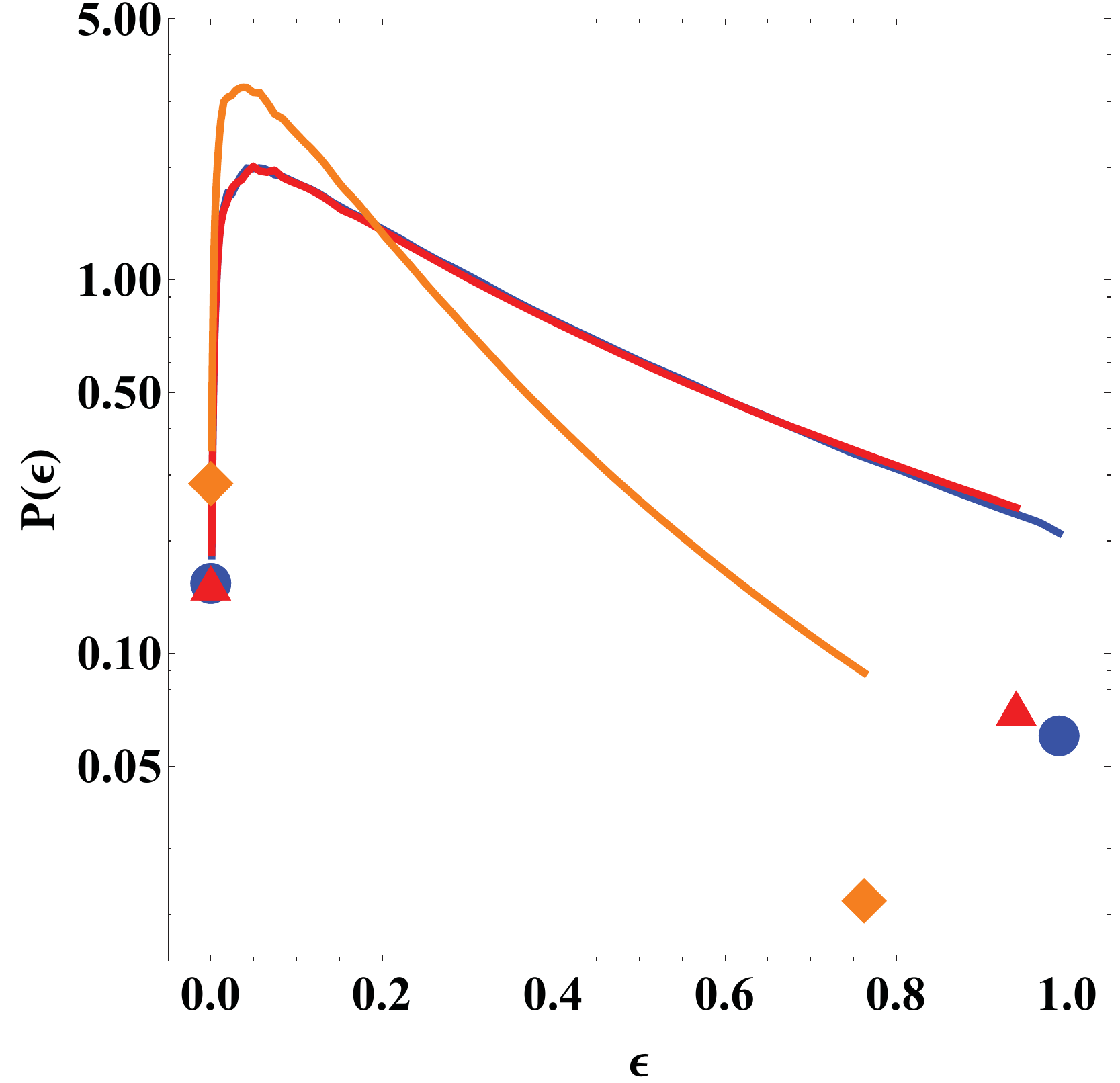}
\includegraphics[width=0.3\textwidth%width=1.in%height=1.in
]{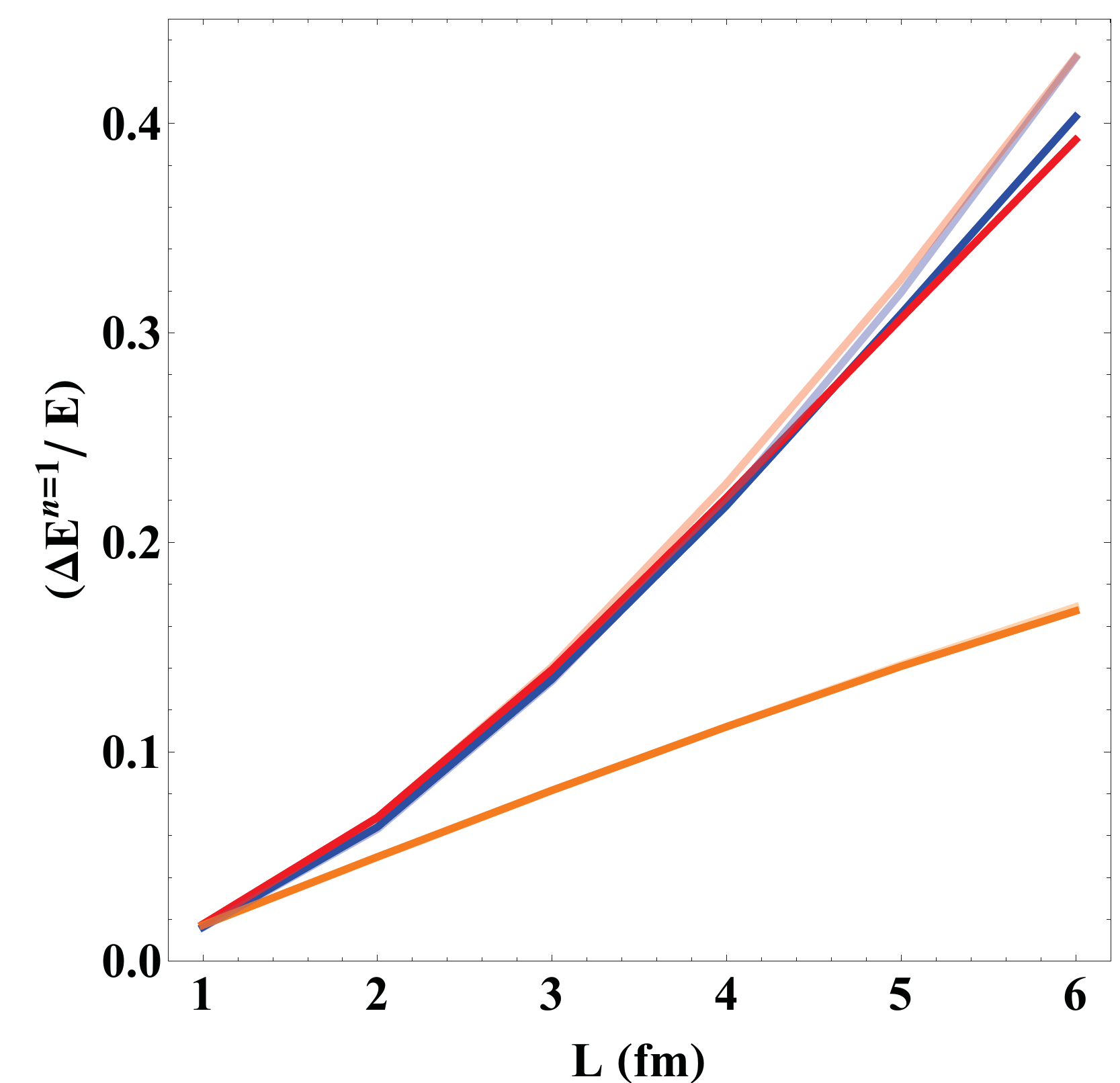}
\includegraphics[width=0.3\textwidth%width=1.in%height=1.in
]{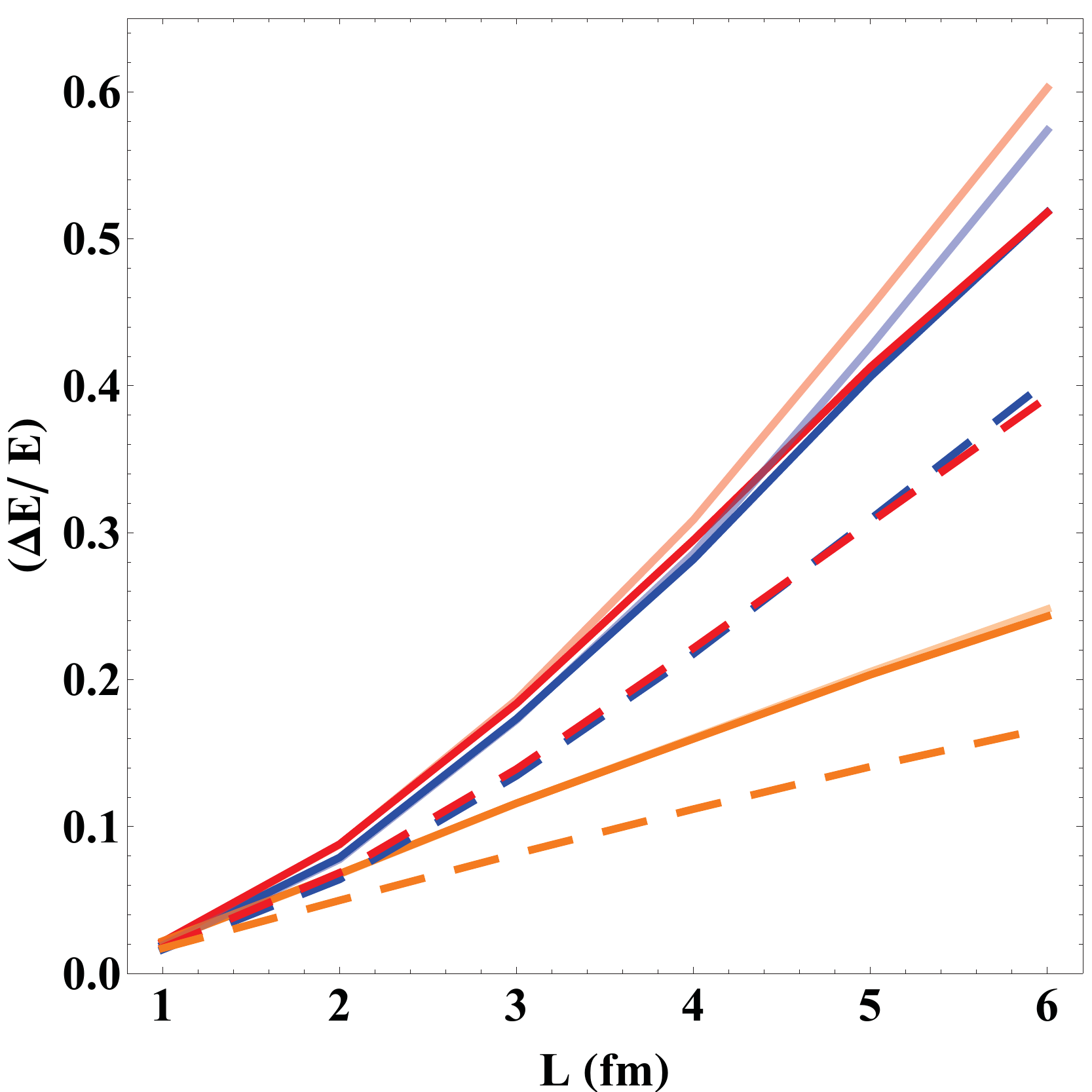}
%\hspace{0.01\textwidth}
\includegraphics[width=0.3\textwidth%width=1.in%height=1.in
]{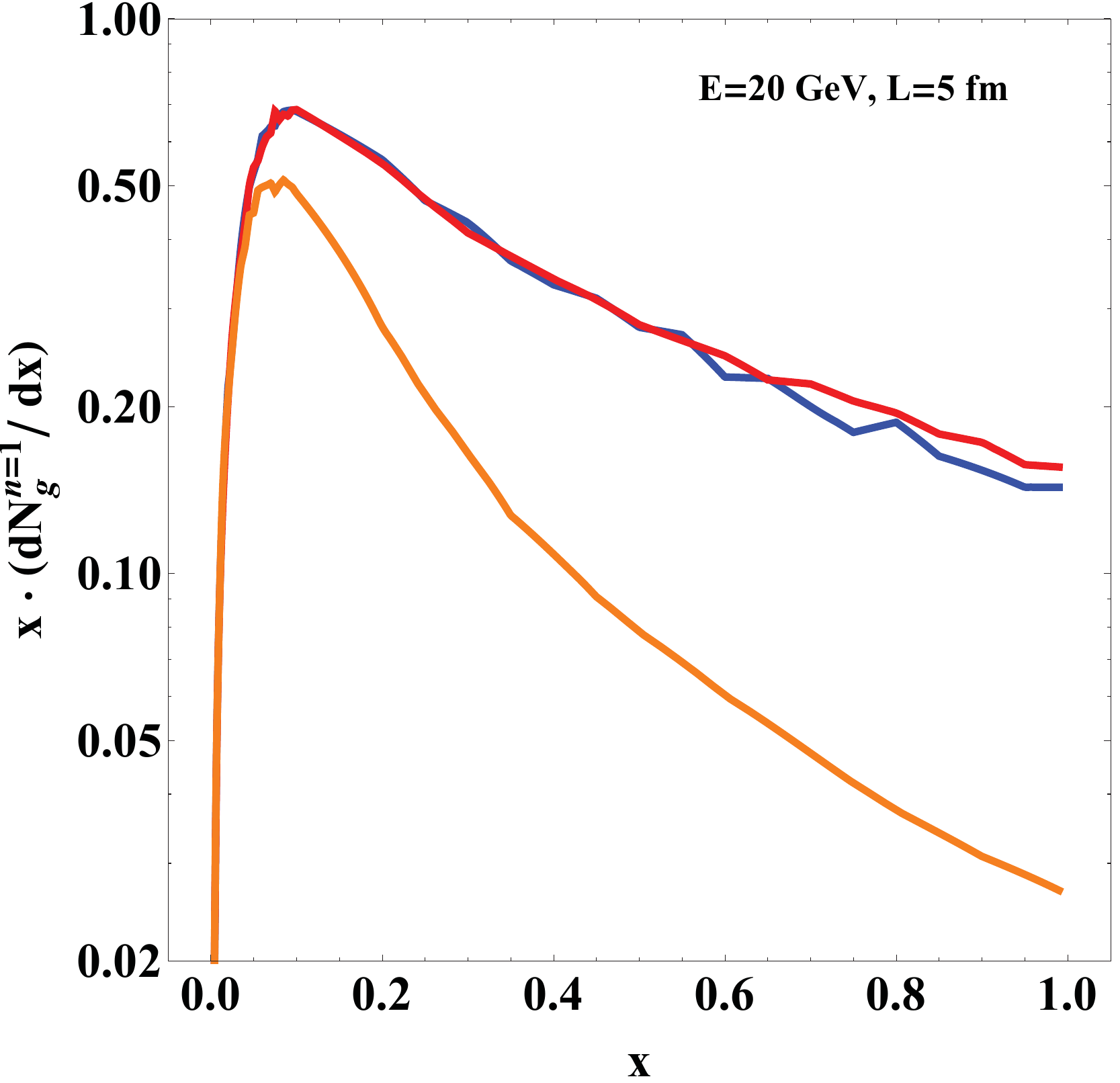}
\includegraphics[width=0.3\textwidth%width=1.in%height=1.in
]{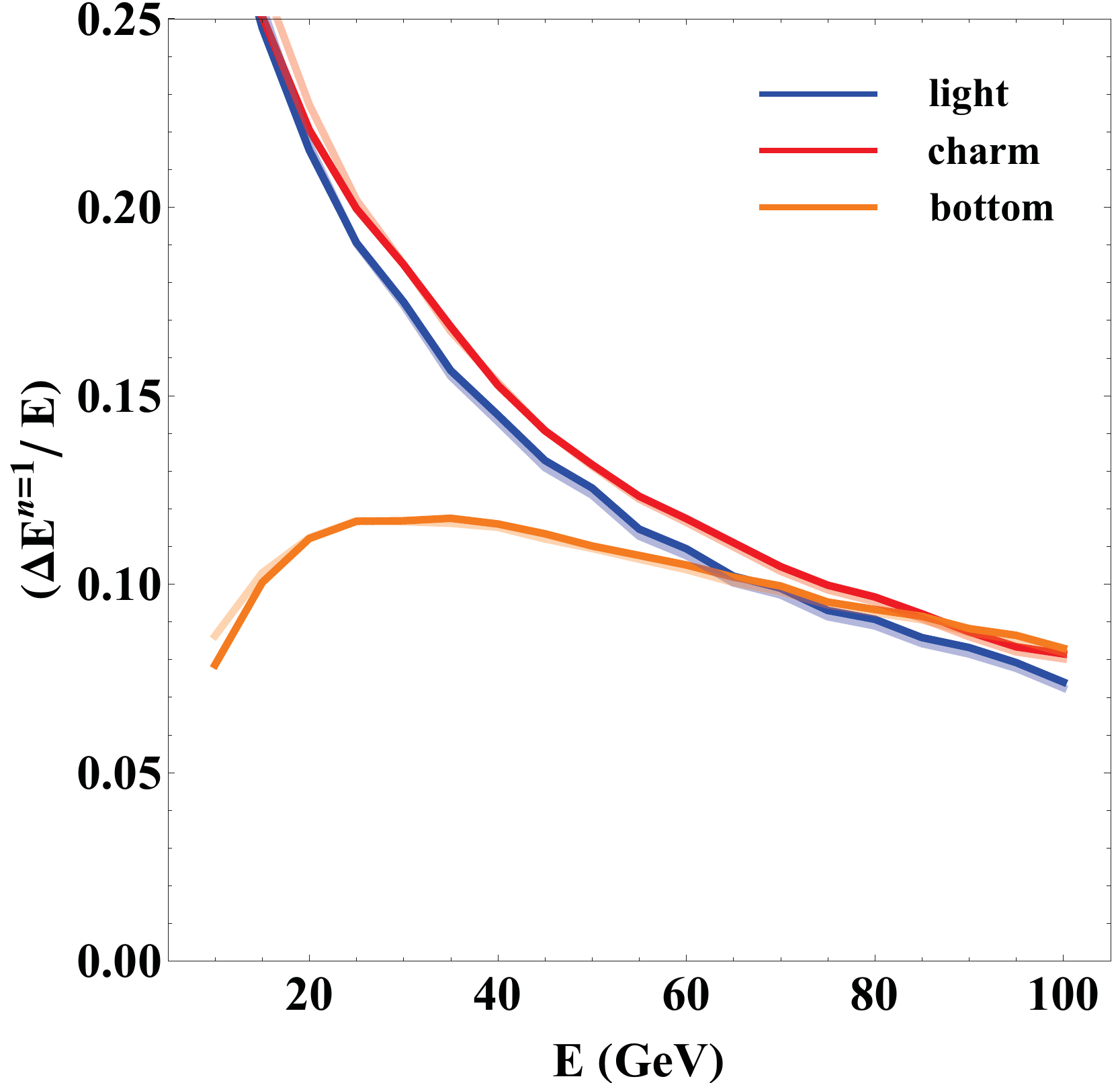}
\includegraphics[width=0.3\textwidth%width=1.in%height=1.in
]{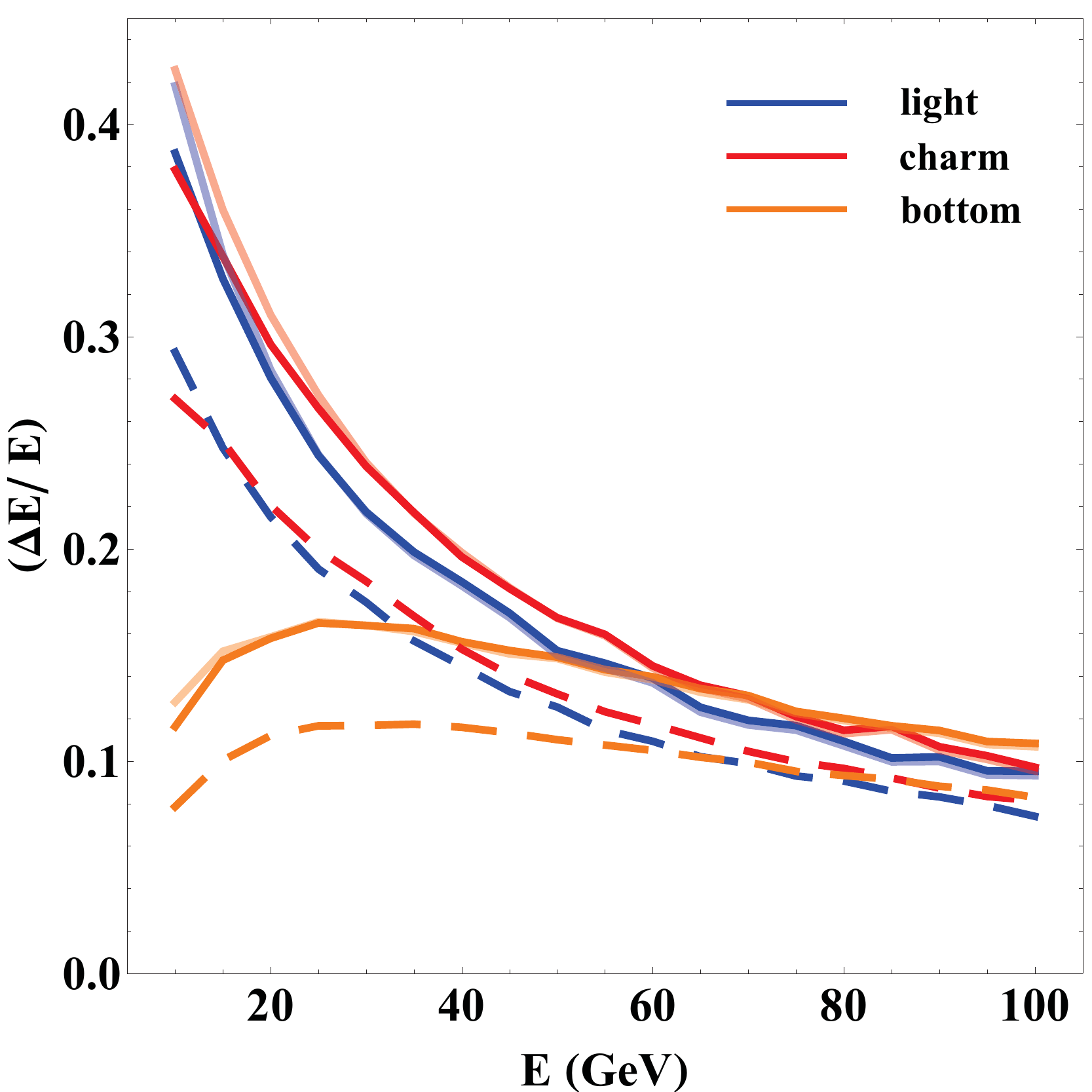}
\caption{(LEFT panels) Top: Normalized radiative energy probability distribution $P(\epsilon)$ for light (blue, $M=0.2$ GeV), charm (red, $M=1.2$ GeV) and bottom (orange, $M=4.75$ GeV) quark jet. The initial energy of the jet is $E=20$ GeV and the size of the brick is $L=5$ fm. All results are computed at $n=1$ in the opacity series. The markers on the left represent the probability of zero gluon emission ($\epsilon=0$, no energy loss), whereas the markers on the right represent the probability of complete quenching ($\epsilon=1-M/E$). Notice again how bottom quarks consistently lose less energy than light ones. Bottom: The gluon spectrum $xdN_g/dx$ used to compute the distribution on the top via Eq.~\eqref{Pradn} is shown for reference. Other parameters used in the calculation are: $\lambda=1.16$ fm,  $\mu=0.5$ GeV, $m_g=0.356$ GeV, $T=0.258$ GeV, $n_f=0$, $\alpha_s=0.3$. (MIDDLE panels) Radiative energy loss $\Delta E/E$ for light (blue), charm (red) and bottom (orange) quark jet traversing a dynamical QCD brick medium of thickness $L$, with (solid lines) or without (opaque lines) the inclusion of fluctuation effects. DGLV is computed at first order in opacity. The former are obtained by integrating $\epsilon P(\epsilon)$, the latter by integrating $xdN/dx$. We show the dependence of $\Delta E/E$ on $L$ fixing $E=20$ GeV (\textit{top}), and on $E$ fixing $L=4$ fm (\textit{bottom}). Other model parameters used in the simulation are the same as those in the left panels. (RIGHT panels) Total energy loss $\Delta E/E$ for light (blue), charm (red) and bottom (orange) quark jet, computed in the same configuration as the middle panels. Solid lines are results with fluctuation effects, while opaque curves represent the same dynamical computation without fluctuation effects. Dashed curves represent the results in the middle panels.}
\label{PexE}
\end{figure}

We dedicate this section to discuss the flavor dependent energy loss probability distribution and fluctuation effect. The energy loss is computed by integrating $\int d\epsilon~\epsilon~P(\epsilon)$ over the range $[0,\epsilon^{MAX}]$. In the top left panel of Fig.~\ref{PexE} we show the radiative energy loss probability distribution $P(\epsilon)$ for different quark flavors that propagate in a plasma of size $L$.
The integrated radiative energy loss $\Delta E/E$ is shown in the middle panels of Fig.~\ref{PexE}, alongside a comparison with the same quantity obtained without the inclusion of fluctuation effects, i.e. obtained by simply integrating the gluon spectrum $\int dx~xdN/dx$. We see typical energy loss probability distribution peaks at small $\epsilon$ for all flavors, and light and charm have almost negligible difference in terms of $P(\epsilon)$, both of them lose more energy than the bottom quark. The inclusion of fluctuation effects appears to alter only minimally the result.

The total energy loss of the jet as a function of $E$ and $L$ is presented in right panels of Fig.~\ref{PexE}. The same features observed in the context of DGLV are also present in the dynamical scenario, from the coherence physics that determines the quadratic or linear $L$ dependence of $\Delta E/E$, to the similarity between light and charm quark jets across a broad range of energies and path lengths.

\section{Partonic pp and AA spectra}
\label{app:partonspc}

In the section we will first show the pp spectra being used in CUJET at RHIC and LHC energies -- for light sector, LO CTEQ5 production spectra \cite{Wang:private}; for heavy sector, both numerical NLO and FONLL initial cross sections \cite{VOGT}, and we will estimate the error band associated with this source of systematic uncertainty.

\begin{figure}[!t]
\centering
\includegraphics[width=0.45\textwidth]{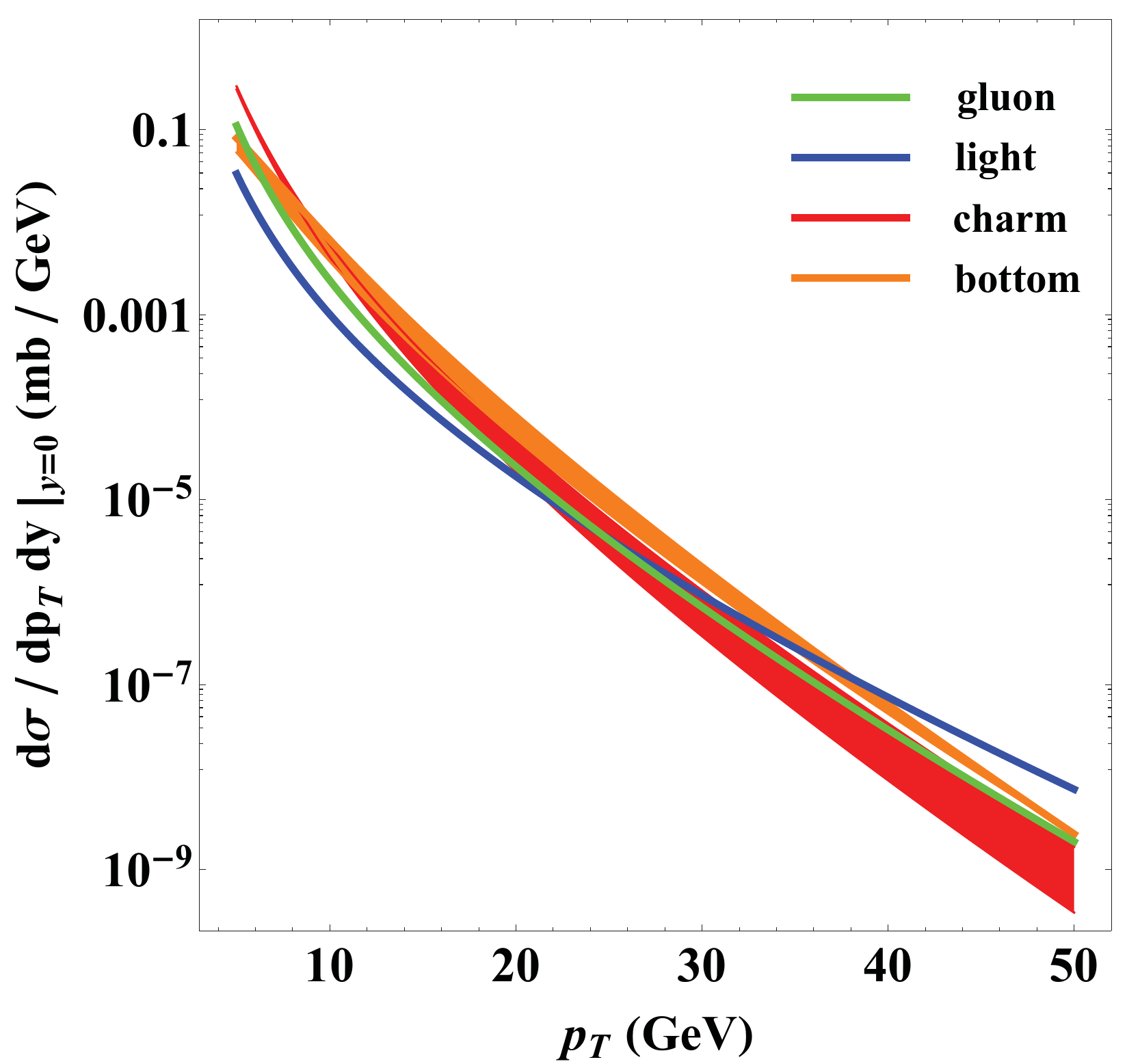}
\hspace{0.01\textwidth}
\includegraphics[width=0.45\textwidth]{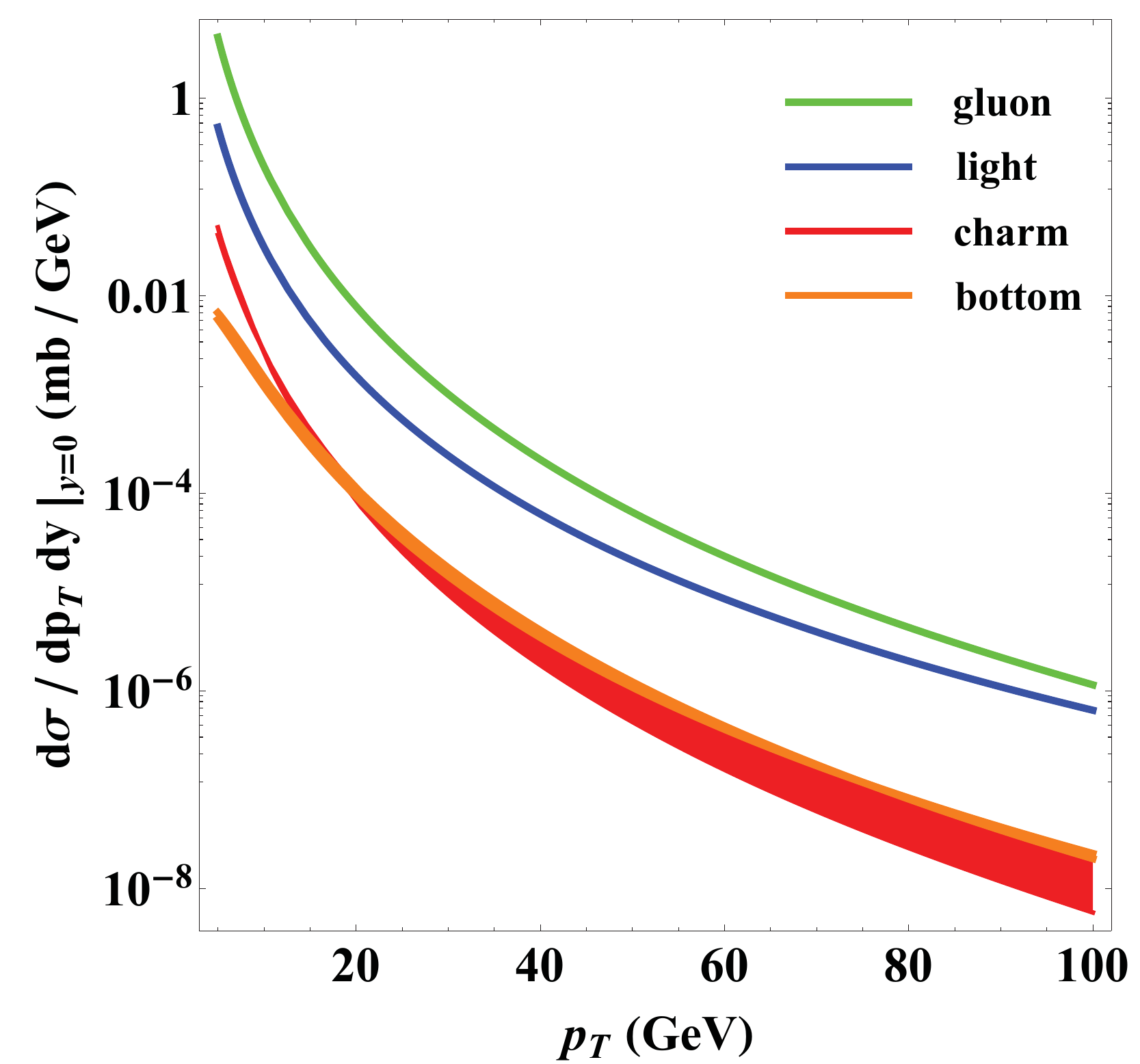}
\caption{pQCD p+p production spectra at $\sqrt{s_{NN}}=200$ GeV (RHIC, \textit{left}) and $\sqrt{s_{NN}}=2.76$ TeV (LHC, \textit{right}). Notice how steeper the RHIC spectra are compared to LHC ones. The light spectra are computed from the LO pQCD CTEQ5 code provided in \cite{Wang:private}. Numerical computations of the NLO and FONLL initial cross sections for the heavy sector are provided in \cite{VOGT}.}
\label{InitSpectra}
\end{figure}

In Fig.~\ref{InitSpectra}, we illustrate the initial partonic production spectra for gluon, light, charm and bottom quark at RHIC and LHC energies.

To compare spectrum variations from RHIC to LHC, in CUJET1.0, we use separate initial rapidity density $dN_g/dy$ and production spectra for RHIC and LHC. The theoretical curves are shown in \ref{RAAdNdySpectra}, superimposed on Fig.~\ref{RAAdNdy}:
\begin{figure}[!t]
%\vspace{0.25in}
\centering
\includegraphics[width=0.45\textwidth]{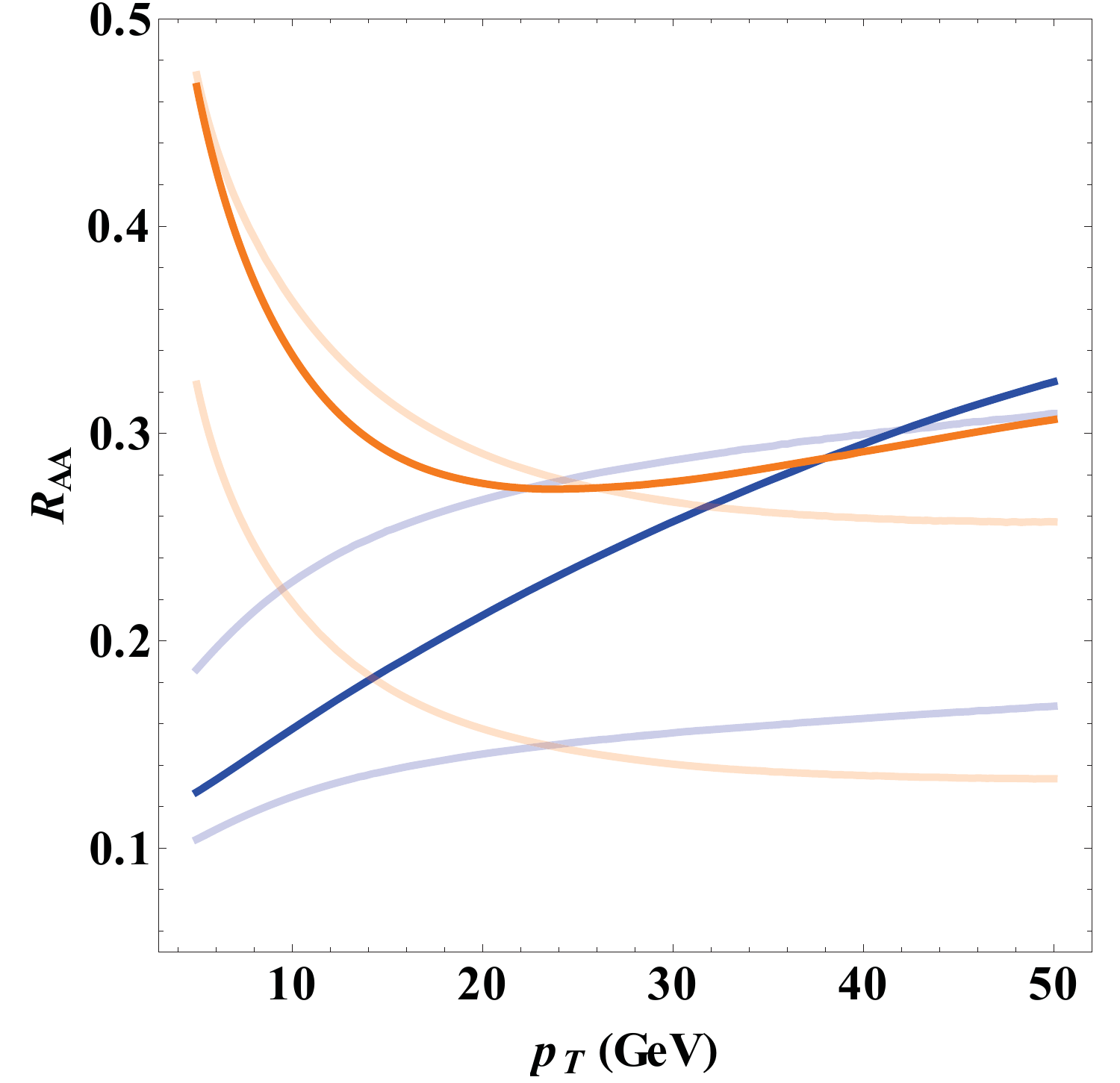}
\caption{LHC production spectra are used in conjunction with $dN_g/dy=2200$ to show the sensitivity of partonic $\RAA$ to the steepness of the p+p partonic cross sections in fixed coupling CUJET1.0. The results for light (blue, $M=0.2$ GeV) and heavy (orange, $M=4.75$ GeV) quark \raa~are presented as solid curves. Both of them are superimposed on the plot of Fig.~\ref{RAAdNdy} (opaque curves), where either a combination of RHIC spectra with RHIC initial rapidity density $dN_g/dy=1000$ (upper opaque curves) or RHIC spectra with LHC $dN_g/dy=2200$ (lower opaque curves) is used. Other parameters used in the simulation are: $\alpha_s=0.3$, $n_f=0$, $T_f=100$ MeV, linear thermalization with initial time $\tau_0=1$ fm/c.}
\label{RAAdNdySpectra}
\end{figure} 
The impact on $\RAA$ is large, and two separate effects can be noticed: (1) softer LHC spectra cause a vertical lift in $\RAA$ that completely counters the suppression generated by the increased density; (2) pion $\RAA$ rises faster with $p_T$, due to the particular shape of the light quark spectra at LHC.

\begin{figure}[!t]
\centering
\includegraphics[width=0.45\textwidth]{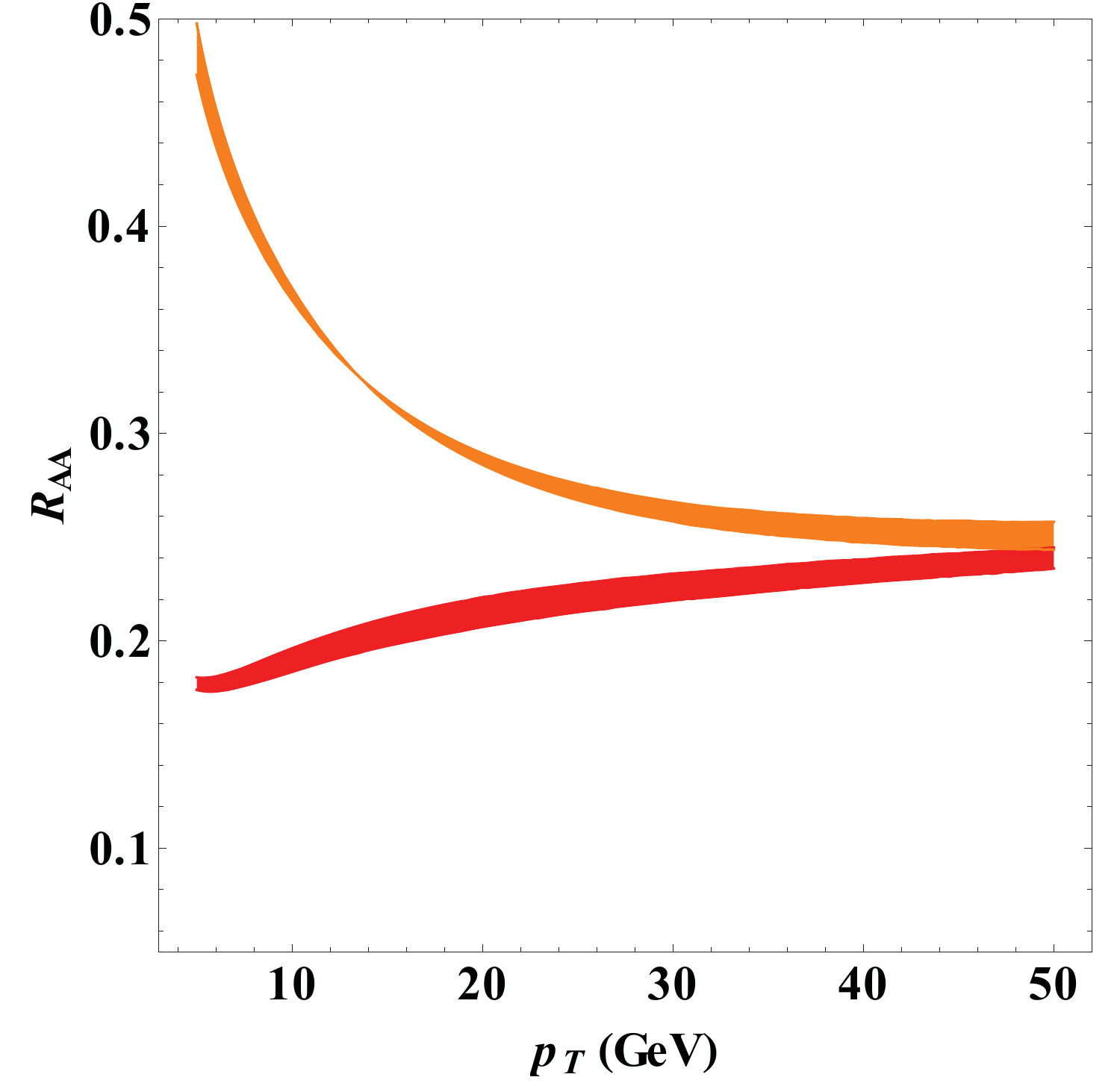}
\hspace{0.01\textwidth}
\includegraphics[width=0.45\textwidth]{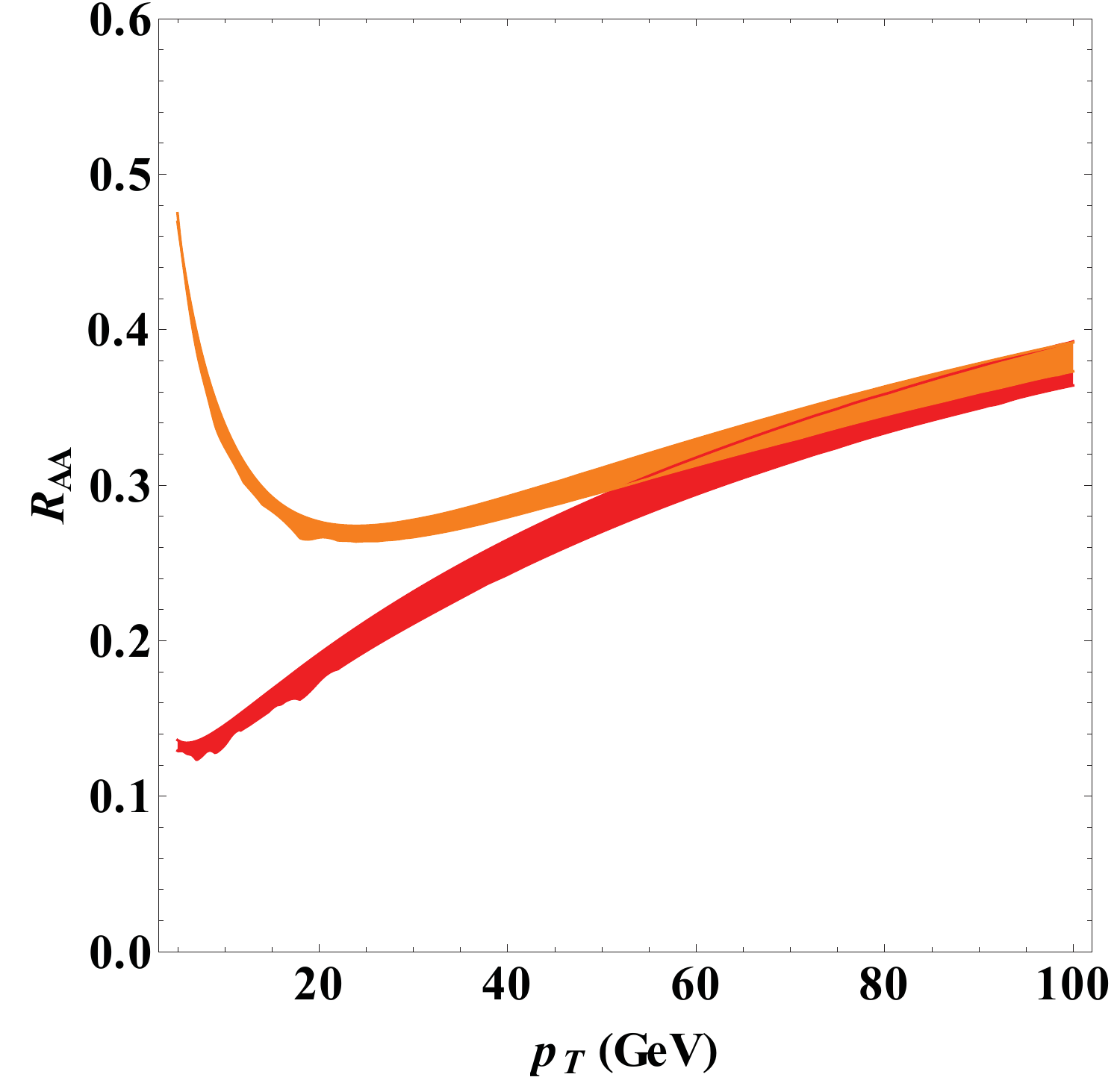}
\caption{Fixed coupling CUJET1.0 calculation of charm (red, $M=1.2$ GeV) and bottom (orange, $M=4.75$ GeV) quark $\RAA$, at RHIC Au+Au 200AGeV (\textit{left}) and LHC Pb+Pb 2.76ATeV (\textit{right}) central collisions. Only the uncertainty in the slope of the spectra matters, since the uncertainty in the absolute normalization is canceled when the $\RAA$ ratio is taken. Other parameters used in the simulation are: $\alpha_s=0.3$, $n_f=0$, $T_f=100$ MeV, linear thermalization with initial time $\tau_0=1$ fm/c.}
\label{RAAnlofonll}
\end{figure}

The uncertainties that arise from the choice of NLO or FONLL schemes for heavy quark initial spectra are shown in Fig.~\ref{RAAnlofonll}.
The error bands shown in the Figure are relatively small, and we estimate that to be 5\% in \raa~at most. At the partonic level, in fact, any uncertainty in the normalization of the production spectra is factored out: $\RAA$ is only sensitive to changes in the slope.

Depending on what physical observables we are interested to compute, different features of the partonic spectra may or may not assume a relevant role. Since $\RAA$ is defined as a ratio of particles yields, the absolute value of the cross section matters little and the normalization drops out in the definition of the observable. What influences the computation is rather the slope of the cross section, as well as the relative normalization between different flavors.

\begin{figure}[!t]
\centering
\includegraphics[width=0.45\textwidth]{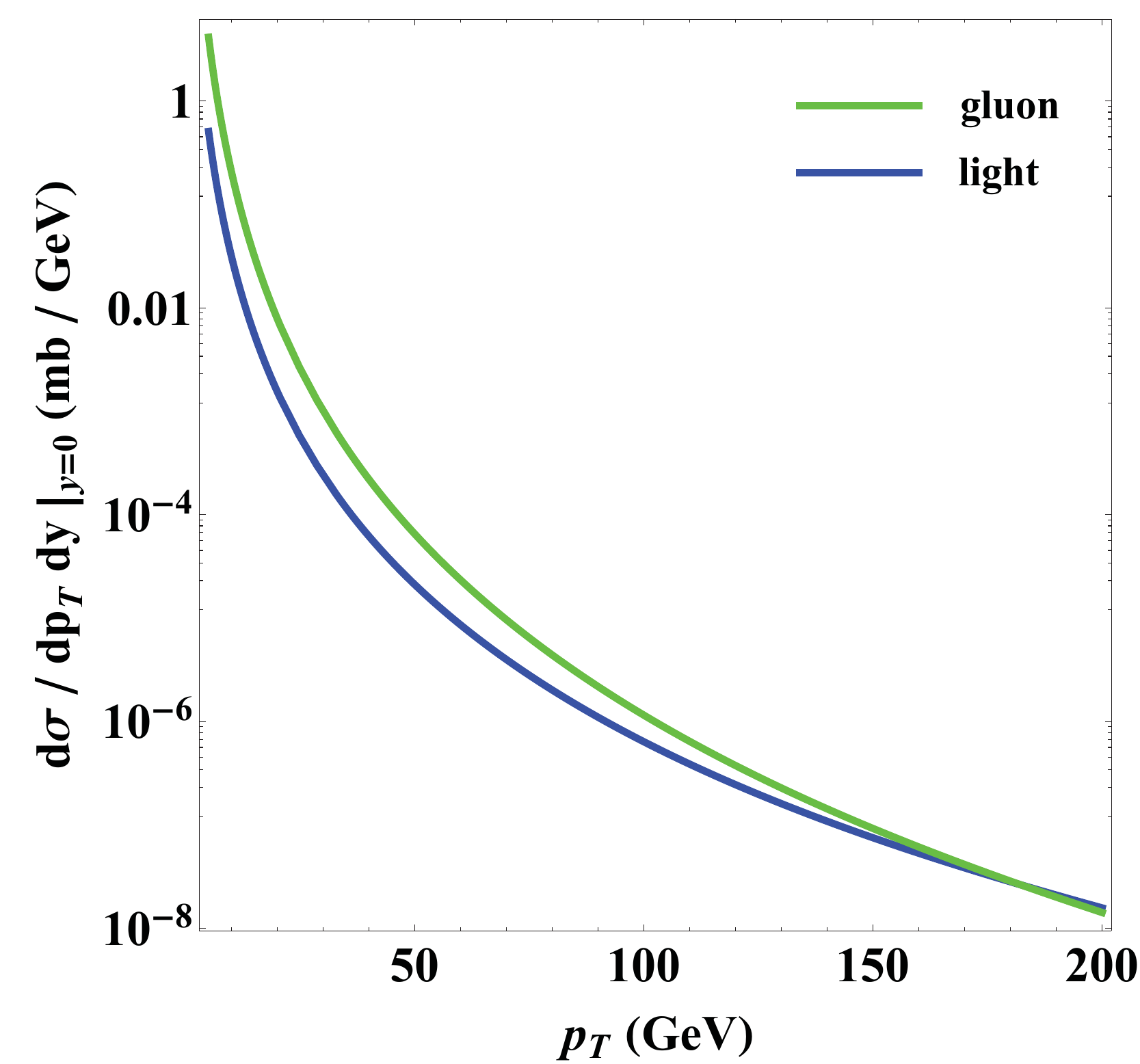}
%\hspace{0.01\textwidth}
\includegraphics[width=0.45\textwidth]{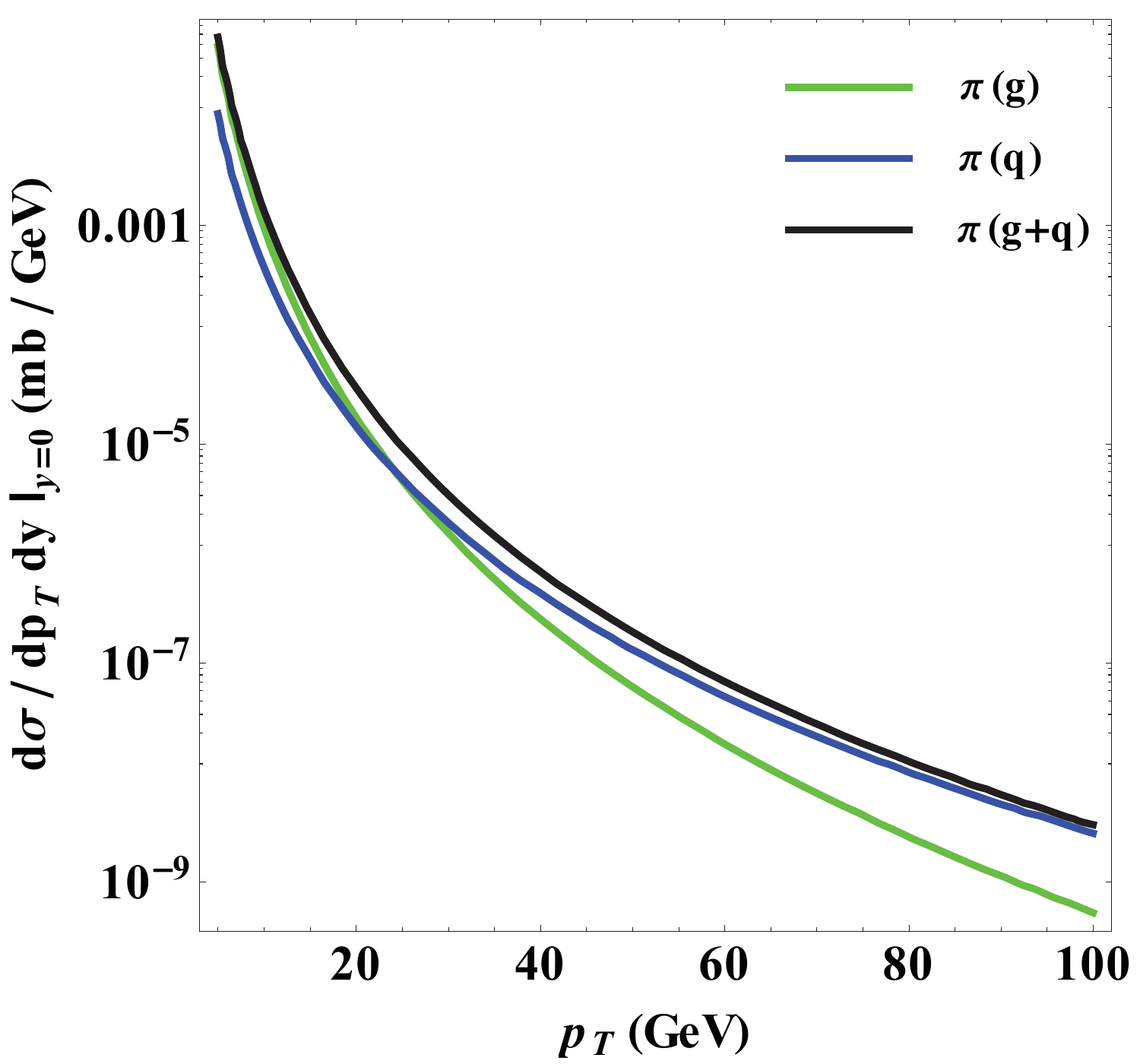}
\caption{\textit{left}: p+p gluon and light quark production spectra, same as Fig.~\ref{InitSpectra}. \textit{right}: p+p pion spectra from gluon only contribution (green), quark only contribution (blue), and total gluon plus quark contribution (black), assuming no ``cold'' nuclear effects. The pion spectra are computed using KKP fragmentation functions.}
\label{PionYield}
\end{figure}

An insightful example comes from the pion yield in p+p events at LHC, Fig.~\ref{PionYield}, which is computed by convoluting the production spectra of quarks and gluons with the appropriate fragmentation functions (more details in Section~\ref{sec:conv-final}).
We can make two observations: (1) Since gluons and light quark contributions are summed together to get the pion yield, the relative normalization between the two matters. The absolute normalization, on the other hand, drops out once the nuclear modification factor ratio is taken. (2) Despite the high production of gluons at low $p_T$, the gluon distribution is much steeper than the quark one. As a consequence, once fragmentation is taken into account, the gluonic contribution to the total number of pions produced sinks below the quarkonic one already at $p_T \gtrsim 25$ GeV. It is then reasonable to expect $R_{AA}$ to depend on the light quark sector only for sufficiently high transverse momentum.

Another example where only the relative steepness between production spectra matters is given by the comparison between the (unphysical) partonic yields of light and charm quarks. In previous appendices we saw that light and charm quarks approximately lose the same amount of energy when they propagate through a deconfined medium, however the production spectrum of charm quarks is much steeper than the one of light quarks (cf. Fig.~\ref{InitSpectra}). The immediate consequence is that the partonic yield for charm quark is suppressed in AA collisions compared to the other, regardless of the separate normalization of the production spectra.

A similar effect applies when we compare RHIC (steeper) and LHC (flatter) spectra: the expected energy loss increase at LHC with respect to RHIC due to higher densities and temperatures, which itself would drive the particle yields down, is going to be partly compensated by the flatter production cross-sections, which in turn drive the yields up.

Finally, a comment on heavy jet quenching: the measurement of heavy flavors has often been limited to the experimental analysis of non-photonic electrons, produced mainly in the secondary decays $c \rightarrow D \rightarrow e$ and  $b \rightarrow B \rightarrow e$ (where $D$ and $B$ refer to the D and B meson respectively). In such case, the relative norm between charm and bottom spectra plays a critical role, in the same way that both gluon and light quarks contribute to pion $R_{AA}$. Unfortunately, the uncertainties are more significant in the heavy flavor scenario than in the pion scenario, therefore a direct measurement of the intermediate mesons would undoubtedly provide a much cleaner and insightful measurement to be compared to CUJET predictions.

\section{Thermalization schemes}
\label{app:thermal}

\subsection{Pre-thermal stages}

In Section~\ref{sec:glauber}, we briefly mentioned that the linear thermalization scheme for the plasma (cf. Eq.~\eqref{rhoQGP}\eqref{tau0}) is a phenomenological assumption, because of the absence of a clear theoretical answer to the way high energy jets couple to the system before thermalization. Different temporal evolution profiles exist and hence induce systematic uncertainties.

\begin{figure}[!t]
%\vspace{0.25in}
\centering
\includegraphics[width=0.45\textwidth]{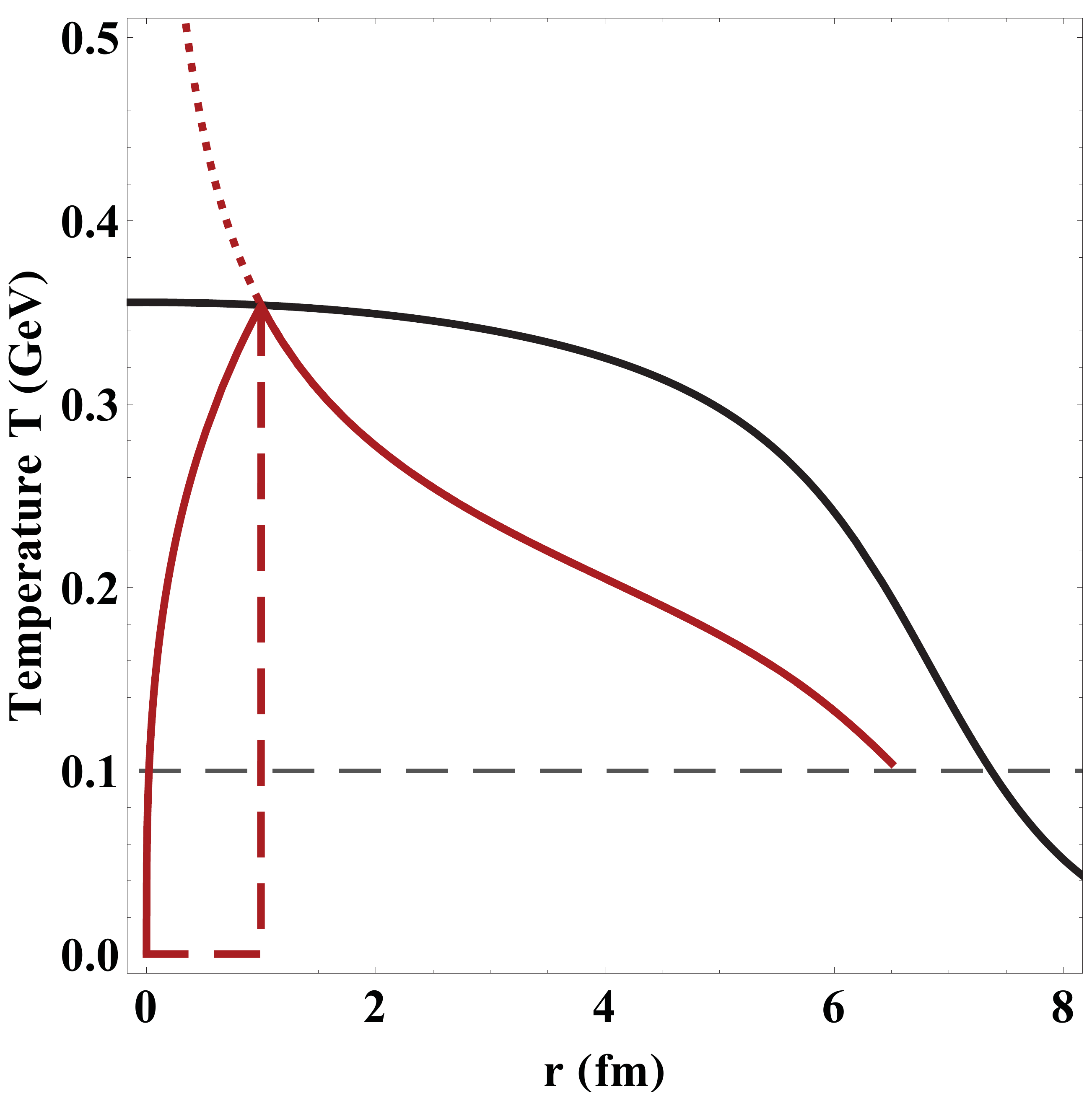}
\caption{Temperature profile of the QGP in a central ($b=0$) collision at RHIC energies. The density is constrained by the observed $dN/dy=1000$. The black curve represents the temperature at constant $\tau_0=1$ fm/c for a radial section of plasma. The red curves represents the $1/\tau^{1/3}$ temperature probed by a quark that is created at $r=0$ and propagates outward along $z\equiv r$ (with the solid, dotted and dashed curves representing the linear, divergent and free streaming cases respectively). The dashed black $T\approx100$ MeV line corresponds to the fragmentation temperature of the jet.}
\label{PlasmaSlice}
\end{figure}

The ability of CUJET to perform a full jet path integration allows us to parametrize the evolution of the system in different ways, and we can therefore draw insightful conclusions on the physics of the collision. In the discussions followed, we characterize the pre-thermal stage and the evolution profile after the medium been fully thermalized by varying $f(\tau/\tau_0)$ in Eq.~\eqref{rhoQGP} using three different methods:
\begin{enumerate}
	\item The plasma takes a proper time $\tau_0$ to thermalize, and the density ``seen'' by the jet grows linearly until thermalization is reached. The density decreases as $1/\tau$ thereafter. Referring to Eq.~\eqref{tau0},
	  \be
		f(\tau/\tau_0) = \begin{cases}
      \tau/\tau_0 & \mbox{if } \tau\le\tau_0\;, \\
      \tau_0/\tau & \mbox{if } \tau>\tau_0\;.
      \end{cases}
   	\ee
	\item The jet ``sees'' a divergent density at $\tau=0$ that decreases with $1/\tau$ (instant thermalization):
	  \be
		f(\tau/\tau_0) = \frac{\tau_0}{\tau}\;.
	  \ee
	\item The jet doesn't couple with the medium until the plasma has thermalized (free streaming):
	  \be
		f(\tau/\tau_0) = \begin{cases}
      0 & \mbox{if } \tau\le\tau_0\;, \\
      \tau_0/\tau & \mbox{if } \tau>\tau_0\;.
      \end{cases}
	  \ee
\end{enumerate}
Note in all three schemes $f(\tau/\tau_0) = \tau_0/\tau$ after thermalization time $\tau_0$, which recovers the Bjorken idea 1+1D hydro profile, i.e. the choice for CUJET1.0 bulk evolution. Therefore, strictly speaking, our discussion about the temporal evolution parametrization here is applicable only to CUJET1.0 which has Glauber + Bjorken profile. However, since the variation of pre-thermal stage dominates the deformation of medium profile, the systematic uncertainty analysis here is partially applicable to CUJET2.0 as well. In CUJET1.0, our standard choice for $\tau_0$ is $\tau_0=1$ fm/c. We name the three schemes listed above as ``linear'', ``divergent'' and ``free streaming'' respectively, and illustrate the time evolution of QGP temperature in these schemes in Fig.~\ref{PlasmaSlice}.

\subsection{Systematic uncertainties}
\label{app:Thermal_Systematics}

\begin{figure}[!t]
\centering
\includegraphics[width=0.45\textwidth]{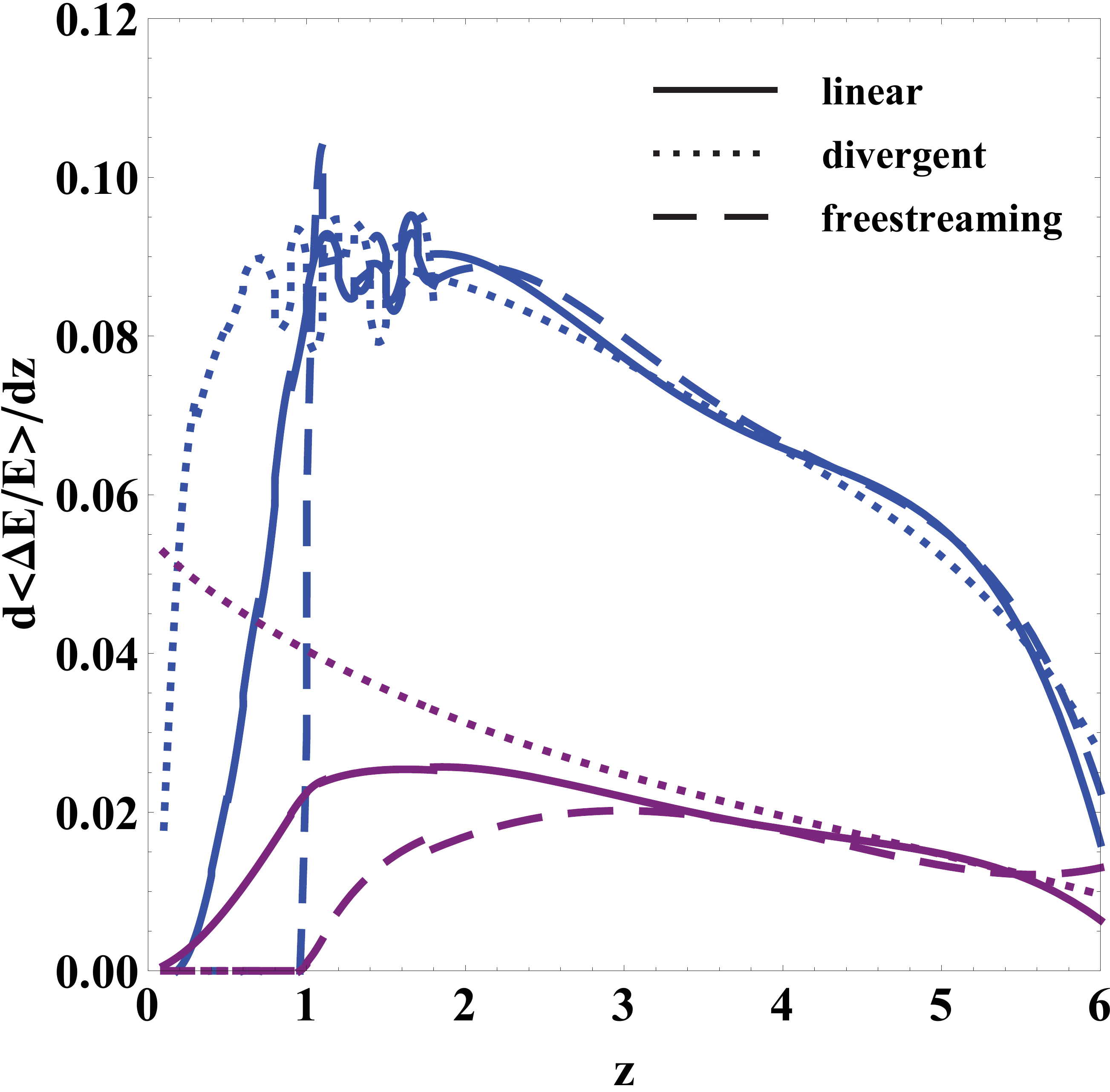}
%\hspace{0.01\textwidth}
\includegraphics[width=0.45\textwidth]{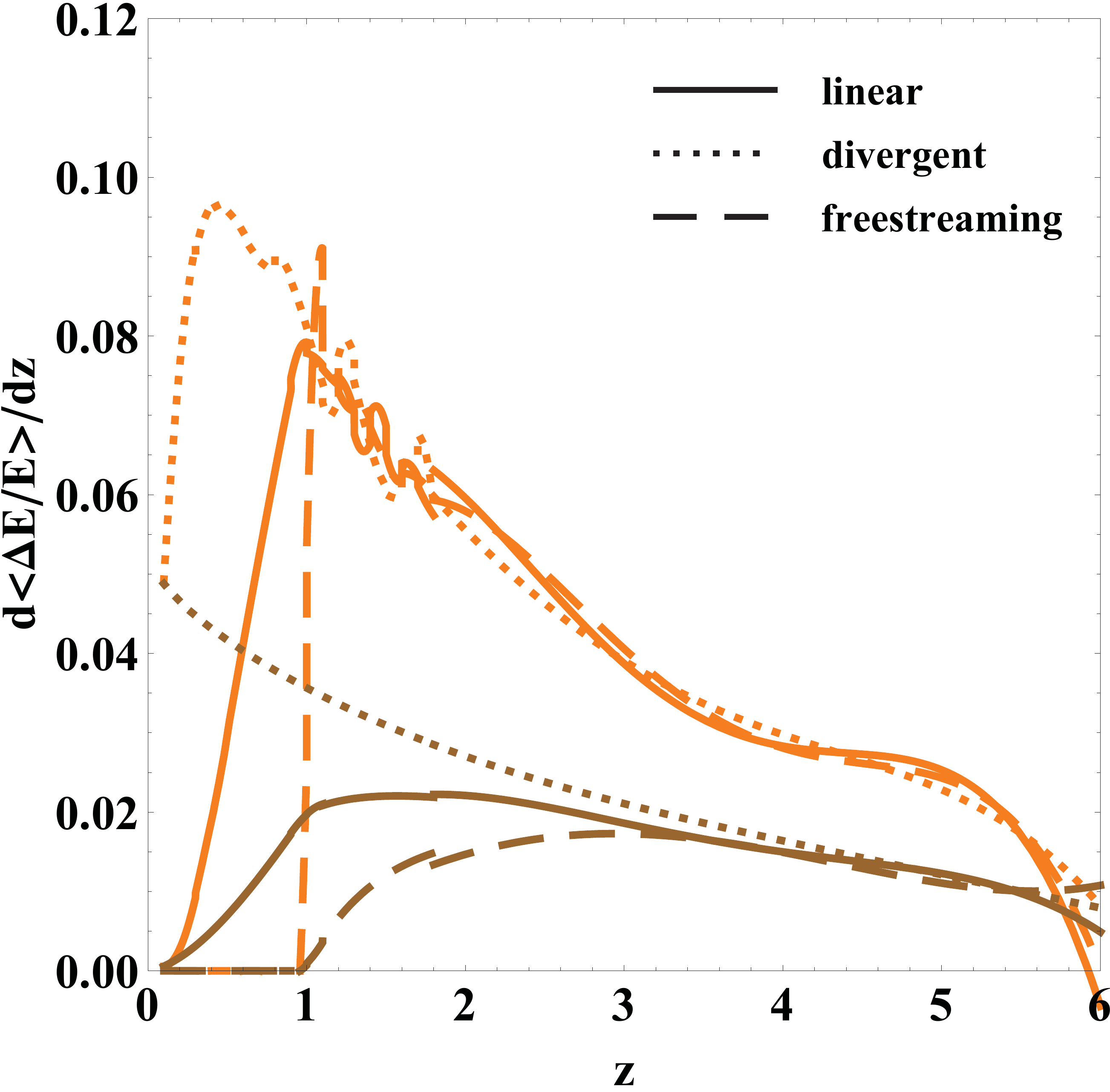}
\caption{Differential $d<\Delta E/E>/dz$ for light (\textit{left}) and heavy (\textit{right}) quarks, in a QGP defined by $dN/dy=1000$, $\tau_0=1$ fm/c and $n_f=0$. The initial energy of the quarks is $20$ GeV. Blue and orange colors refer to radiative losses, whereas purple and brown to elastic ones. Notice how quickly $d<\Delta E/E>/dz$ drops for heavy quarks compared to light jets. LPM interference effects are responsible for the finite value of the energy loss at very short $z$ in the divergent plasma scenario. Results are calculated within the framework of fixed coupling CUJET1.0 with $\alpha_s$=0.3.}
\label{dDedxQ}
\end{figure}

We show in Fig.~\ref{dDedxQ} how differently light and heavy quarks lose energy due to elastic and inelastic collisions are during different early stages of the plasma longitudinal expansion. For jets produced in central Au+Au events, the differential $d<\Delta E/E>/dz$ indicates the fractional energy loss during the first fm's of the jet evolution. Heavy quarks lose a larger percentage of their energy via radiative processes and its radiative energy loss rate follows the medium thermalization.

\begin{figure}[!t]
%\vspace{0.25in}
\centering
\includegraphics[width=0.45\textwidth]{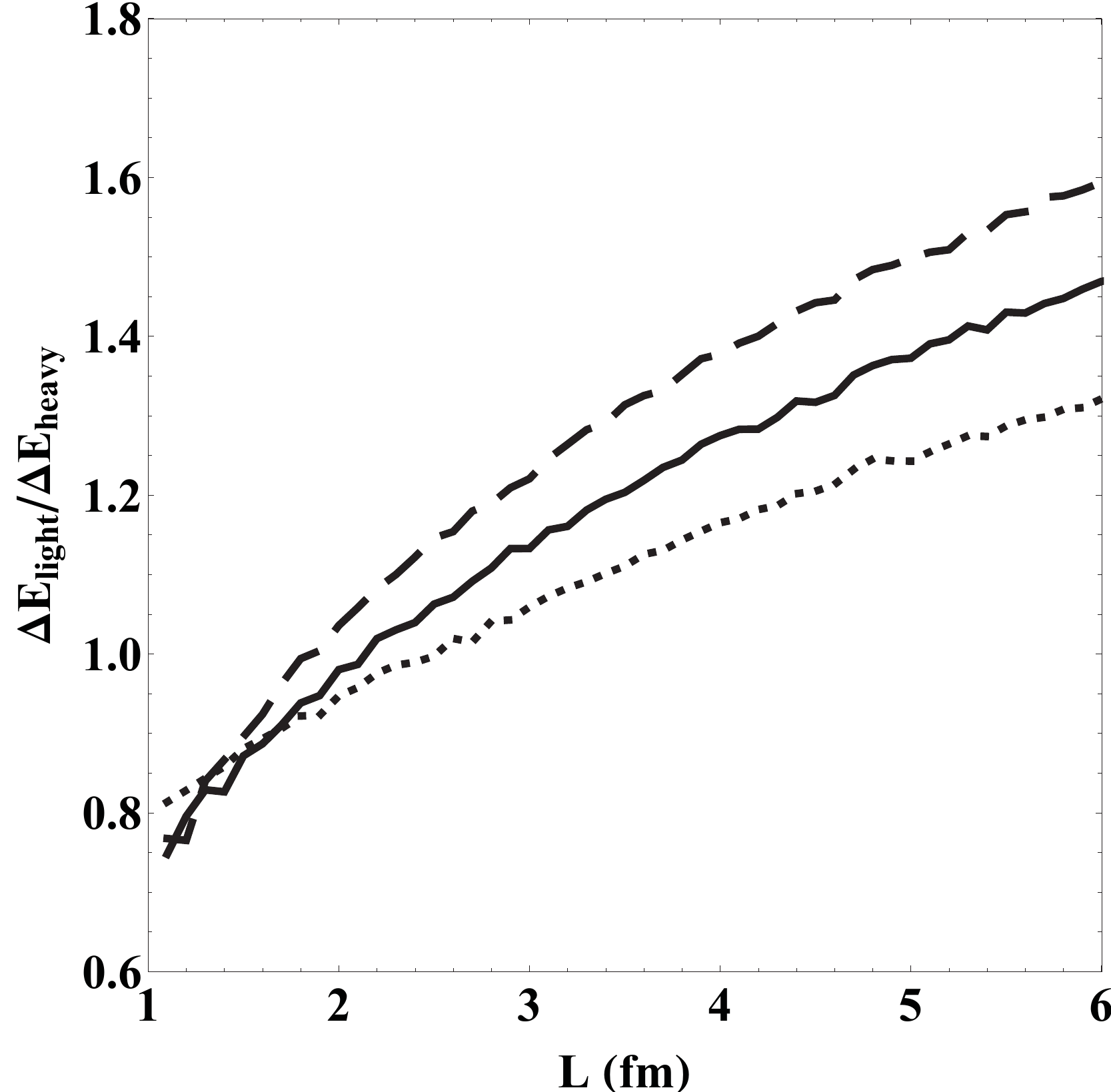}
\caption{Energy loss ratio $\Delta E_{light} / \Delta E_{heavy}$ as a function of $L$ between light and bottom quarks, for the three linear (solid), divergent (dotted) and free streaming (dashed) initial conditions. The energy loss is obtained by integrating the curves in Fig.~\ref{dDedxQ} up to $z=L$. For sufficiently long path lengths, the relative difference between the three approximations reaches approximately 10\%.}
\label{ratioLDF}
\end{figure}

The mass-dependent jet behavior observed in Fig.~\ref{dDedxQ} could be used as a phenomenological indicator of the thermalization mechanism. For different parametrizations of $f(\tau/\tau_0)$, one could expect a different relative yield between light and heavy quark jets. We can in fact expect that once the free parameters of the model ($\alpha_s$ or $\alpha_{max}$) are fixed by a comparison of the light sector with data, each assumptions of $f(\tau/\tau_0)$ will yield a different result for the heavy sector. This fact is portrayed in Fig.~\ref{ratioLDF}, where the ratio $\Delta E_{light}/\Delta E_{bottom}$ is given as a function of $L$ for all possible temporal envelopes.

\begin{figure}[!t]
\centering
\includegraphics[width=0.45\textwidth]{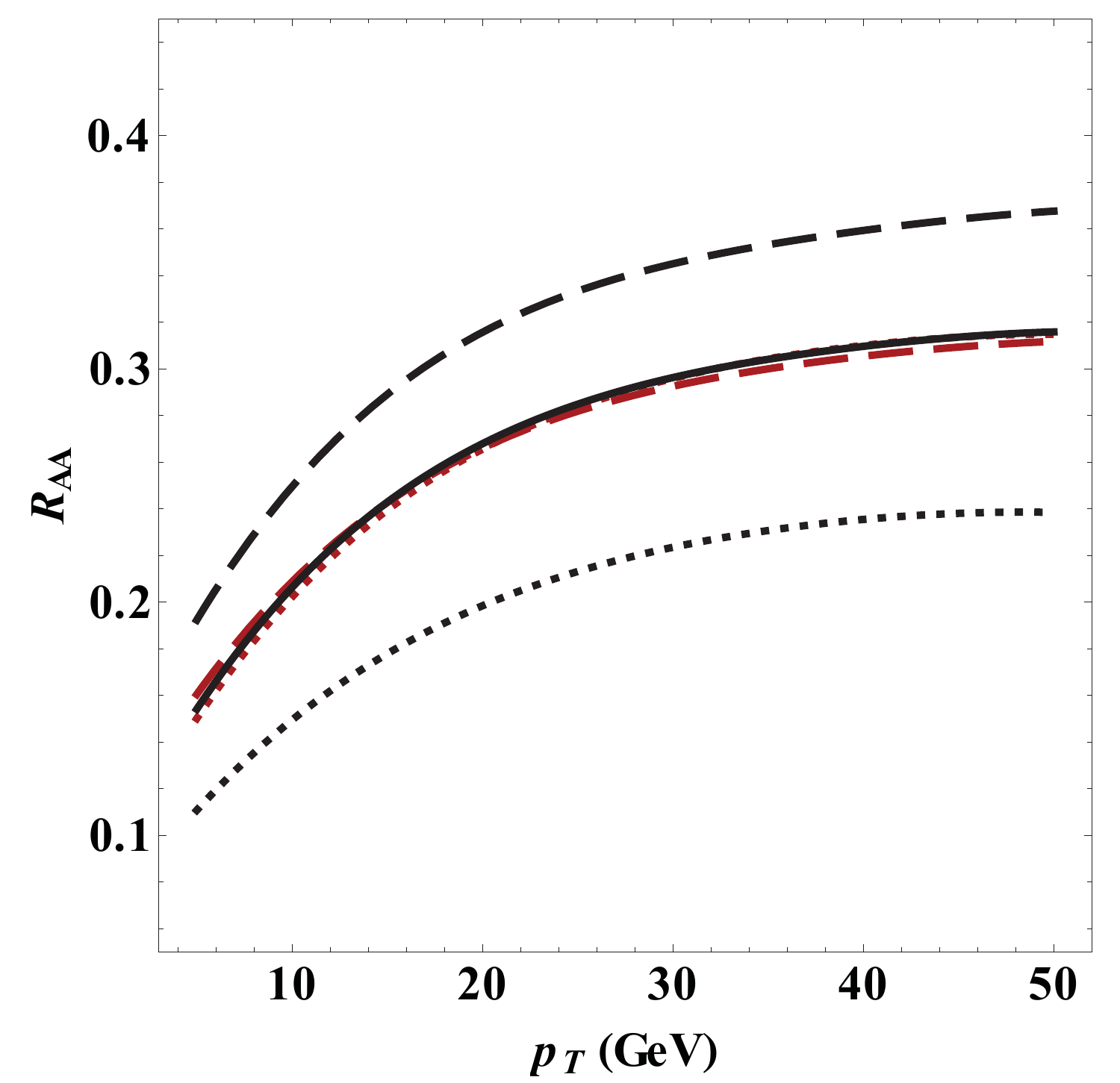}
%\hspace{0.01\textwidth}
\includegraphics[width=0.45\textwidth]{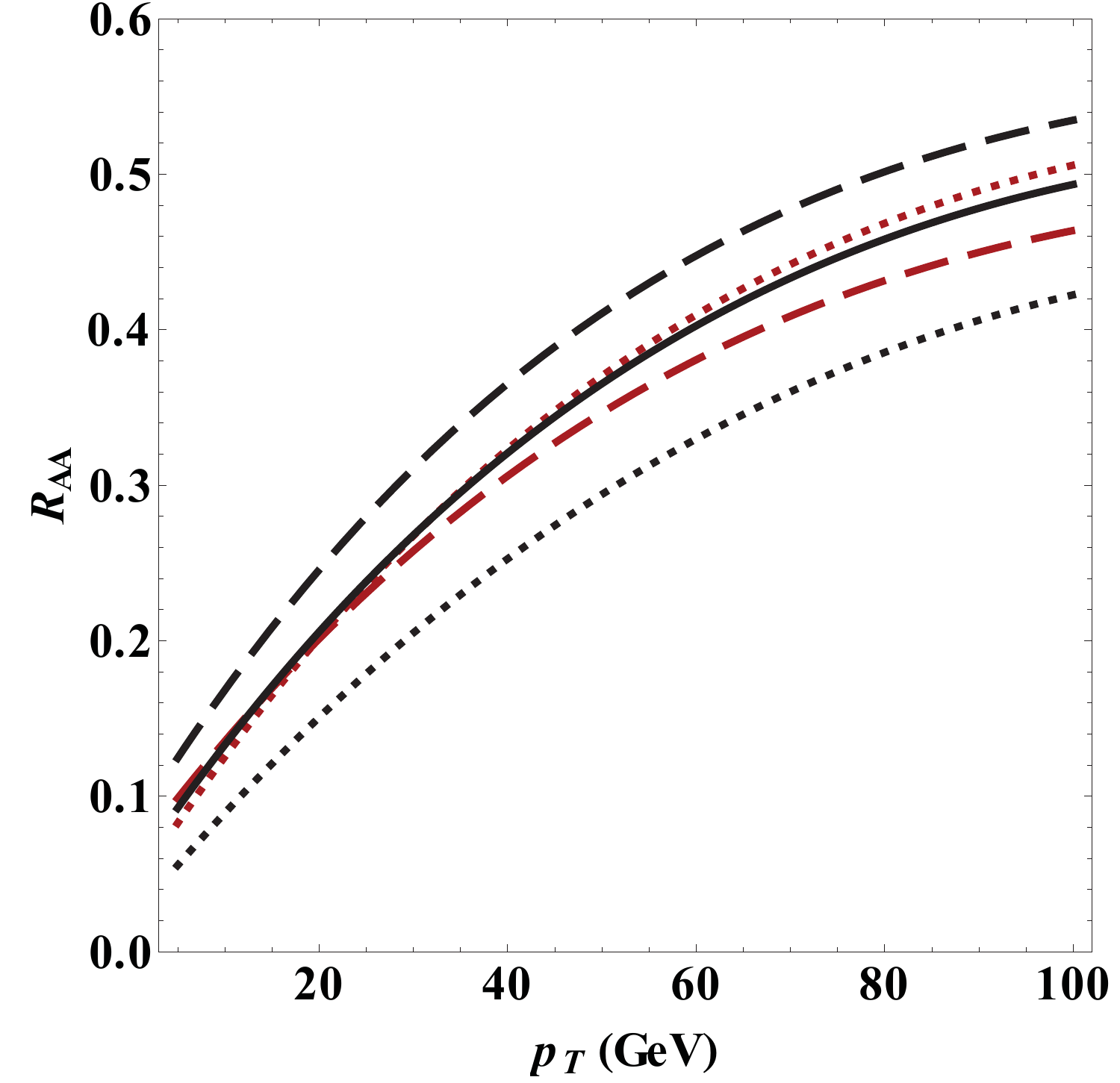}
\caption{Pion $\RAA$ for three distinct plasma thermalization scenarios, with and without rescaling of the coupling constant: linear with $\alpha_s=0.3$ (solid black); divergent with $\alpha_s=0.3$ (dotted black) or $\alpha_s=0.27$ (dotted red); free streaming with $\alpha_s=0.3$ (dashed black) or $\alpha_s=0.32$ (dashed red). The coupling constant is rescaled to fit $\RAA^{\pi ,RHIC}(p_T = 10\;{\rm GeV/c})=0.2$ (\textit{left}), and the constrained extrapolation to LHC is shown on the \textit{right}. Results are calculated within the framework of fixed coupling CUJET1.0.}
\label{RAAtau0}
\end{figure}

Next, we study the hadron suppression factor's sensitivity to the thermalization phase of the plasma. The results from CUJET1.0 calculations for RHIC Au+Au 200AGeV and LHC Pb+Pb 2.76ATeV central collisions are shown in Fig.~\ref{RAAtau0}.

\begin{figure}[!t]
\centering
\includegraphics[width=0.45\textwidth]{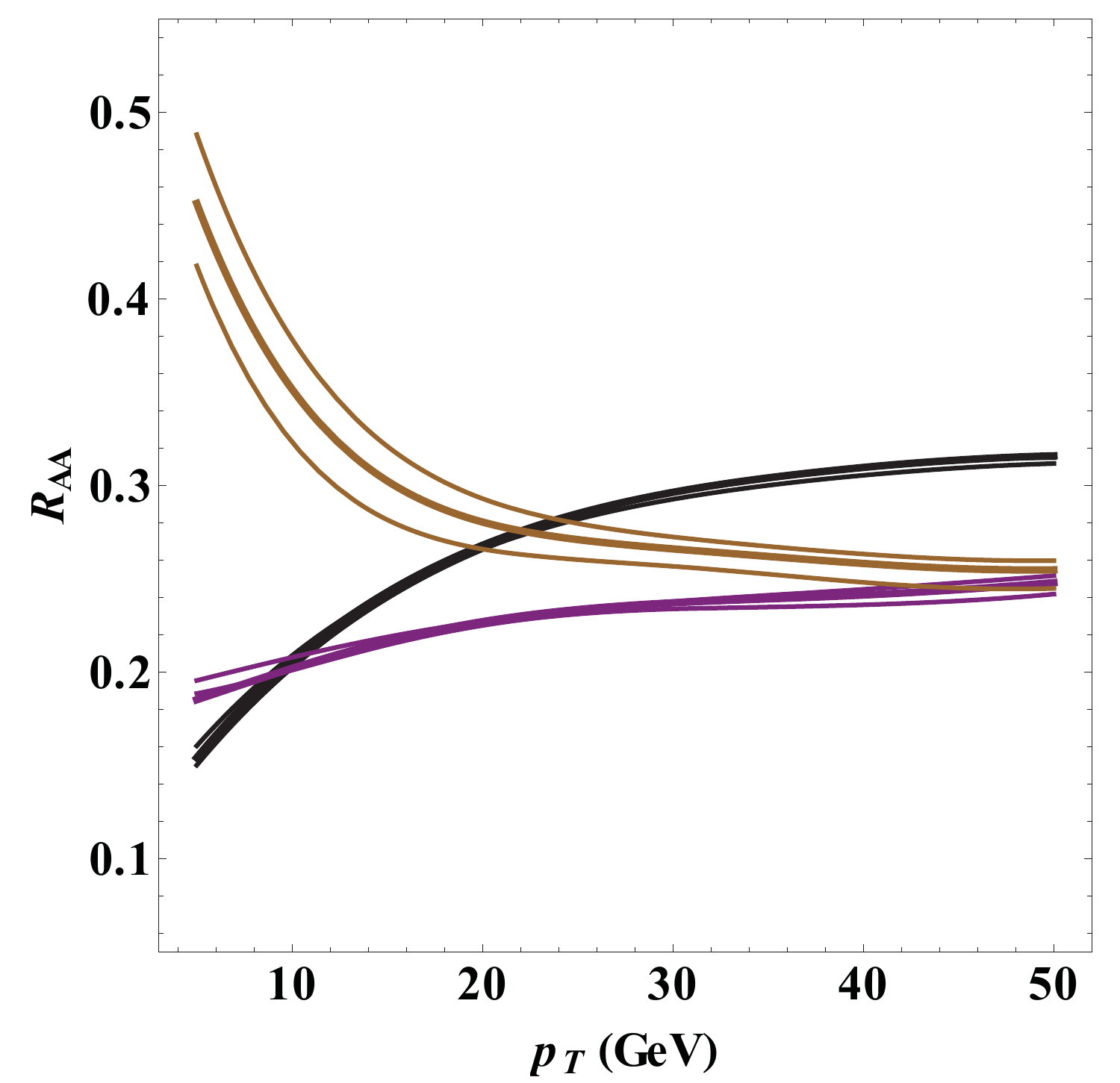}
%\hspace{0.01\textwidth}
\includegraphics[width=0.45\textwidth]{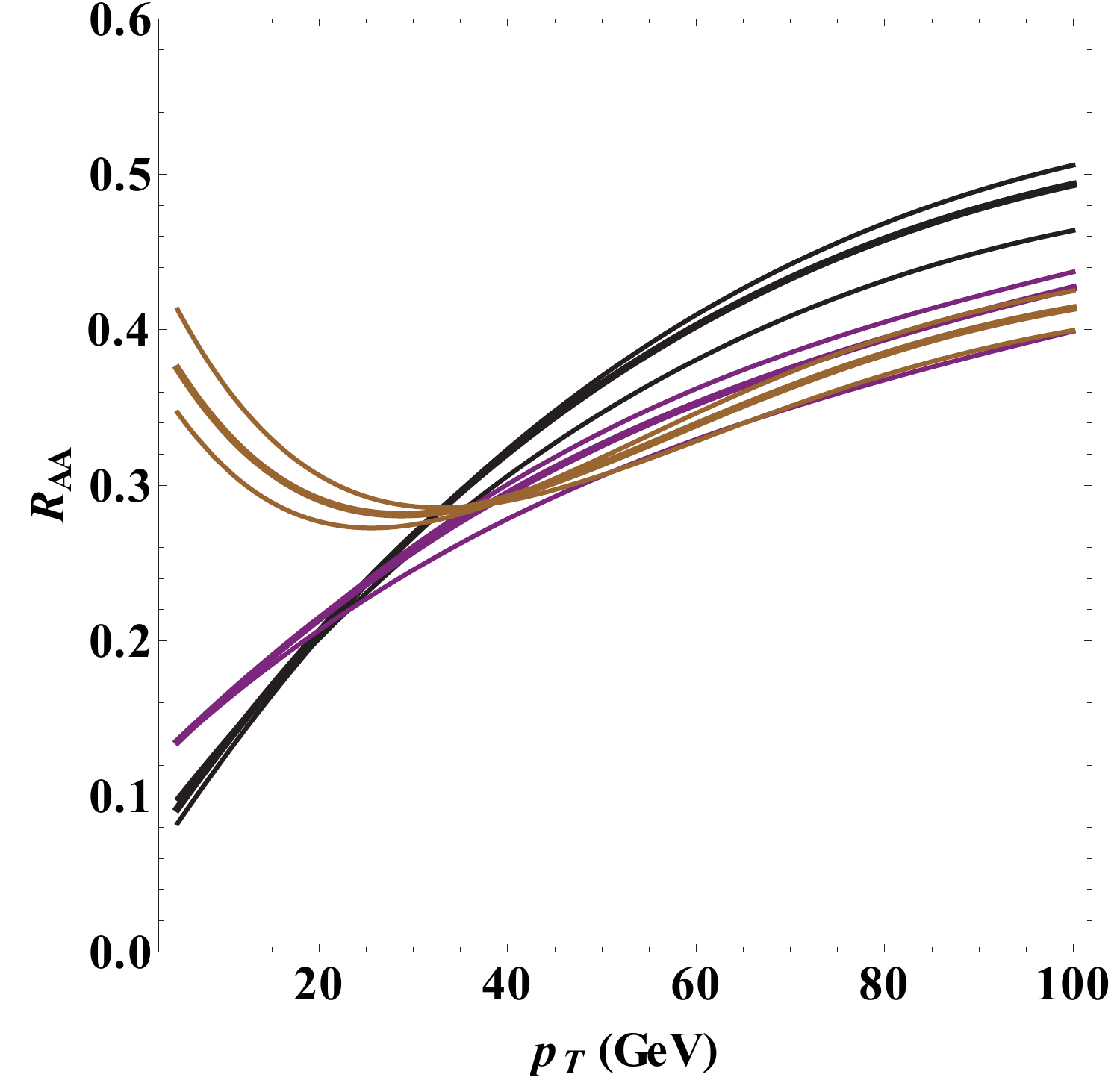}
\caption{Flavor $\RAA$ at RHIC (\textit{left}) and LHC (\textit{right}). In black the pions, in purple the D mesons, in brown the B mesons. Thick lines correspond to the linear thermalization model with $\alpha_s=0.3$, thin lines represent both the divergent and free streaming models with $\alpha_s=0.27{\rm ,\;}0.32$ respectively. Results are calculated within the framework of fixed coupling CUJET1.0.}
\label{RAAtau0total}
\end{figure}

We see a great sensitivity of pion \raa~to the pre-thermalization phase of the evolution in Fig.~\ref{RAAtau0}, which however can be counter-balanced by an adequate rescaling of the coupling constant $\alpha_s$, i.e. varying $\alpha_s$ in divergent or free streaming scheme down or up by 10\% recovers the linear scenario. If we constrain $\alpha_s$ to fit a specific $p_T$ point of pion $\RAA$ at RHIC initial conditions, in Fig.~\ref{RAAtau0} \textit{left}, we observe a complete overlap -- or ``degeneracy'' -- among the linear, divergent and free streaming scenarios. The constrained fit extrapolated to LHC energies in Fig.~\ref{RAAtau0} \textit{right}, shows on the other hand a moderate ``splitting'' at high $p_T$ among the same curves. Although the difference is too small to be measured experimentally, in theory this effect can be studied to discriminate among pre-thermalization phenomenological models.

The same effect is visible in Fig.~\ref{RAAtau0total}, where pions, D and B meson $\RAA$ is plotted assuming RHIC (\textit{left}) and LHC (\textit{right}) initial conditions. The curves are constrained by the same RHIC fit of Fig.~\ref{RAAtau0} \textit{left}. We observe a moderate ``splitting'' of B meson $\RAA$ across all $p_T$, which is a signature of the differences between the light and heavy quark quenching mechanism during the early evolution of the plasma (cf. Fig.~\ref{dDedxQ}). The splitting in the nuclear modification factor is less than $10\%$ in Fig.~\ref{RAAtau0total}, and it is difficult experimentally resolve this splitting in the near future. Nevertheless, what we have observed here is a clear indication of the importance of making simultaneous constrained fits to as many ``orthogonal'' observables as possible, and it implies that the flavor dependent quenching pattern and single particle azimuthal anisotropy are key observables of interest for refining the phase space of CUJET pQCD tomographic model.

\bibliographystyle{JHEP}
\bibliography{CUJET2.0}

\providecommand{\href}[2]{#2}\begingroup\raggedright\begin{thebibliography}{100}

\bibitem{ALICE:2012ab}
{\bf ALICE Collaboration} Collaboration, B.~Abelev et~al., {\it {Suppression of
  high transverse momentum D mesons in central Pb-Pb collisions at
  $\sqrt{s_{NN}}=2.76$ TeV}},  {\em JHEP} {\bf 1209} (2012) 112,
  [\href{http://xxx.lanl.gov/abs/1203.2160}{{\tt arXiv:1203.2160}}].

\bibitem{Abelev:2012di}
{\bf ALICE Collaboration} Collaboration, B.~Abelev et~al., {\it {Anisotropic
  flow of charged hadrons, pions and (anti-)protons measured at high transverse
  momentum in Pb-Pb collisions at $\sqrt{s_{NN}}$=2.76 TeV}},  {\em Phys.Lett.}
  {\bf B719} (2013) 18--28, [\href{http://xxx.lanl.gov/abs/1205.5761}{{\tt
  arXiv:1205.5761}}].

\bibitem{Abelev:2012hxa}
{\bf ALICE Collaboration} Collaboration, B.~Abelev et~al., {\it {Centrality
  Dependence of Charged Particle Production at Large Transverse Momentum in
  Pb--Pb Collisions at $\sqrt{s_{\rm{NN}}} = 2.76$ TeV}},  {\em Phys.Lett.}
  {\bf B720} (2013) 52--62, [\href{http://xxx.lanl.gov/abs/1208.2711}{{\tt
  arXiv:1208.2711}}].

\bibitem{ATLAS:2011ah}
{\bf ATLAS Collaboration} Collaboration, G.~Aad et~al., {\it {Measurement of
  the pseudorapidity and transverse momentum dependence of the elliptic flow of
  charged particles in lead-lead collisions at $\sqrt{s_{NN}}=2.76$ TeV with
  the ATLAS detector}},  {\em Phys.Lett.} {\bf B707} (2012) 330--348,
  [\href{http://xxx.lanl.gov/abs/1108.6018}{{\tt arXiv:1108.6018}}].

\bibitem{CMS:2012aa}
{\bf CMS Collaboration} Collaboration, S.~Chatrchyan et~al., {\it {Study of
  high-pT charged particle suppression in PbPb compared to $pp$ collisions at
  $\sqrt{s_{NN}}=2.76$ TeV}},  {\em Eur.Phys.J.} {\bf C72} (2012) 1945,
  [\href{http://xxx.lanl.gov/abs/1202.2554}{{\tt arXiv:1202.2554}}].

\bibitem{Chatrchyan:2012xq}
{\bf CMS Collaboration} Collaboration, S.~Chatrchyan et~al., {\it {Azimuthal
  anisotropy of charged particles at high transverse momenta in PbPb collisions
  at $\sqrt{s_{NN}}=2.76$ TeV}},  {\em Phys.Rev.Lett.} {\bf 109} (2012) 022301,
  [\href{http://xxx.lanl.gov/abs/1204.1850}{{\tt arXiv:1204.1850}}].

\bibitem{Adare:2008qa}
{\bf PHENIX Collaboration} Collaboration, A.~Adare et~al., {\it {Suppression
  pattern of neutral pions at high transverse momentum in Au + Au collisions at
  $\sqrt{s_{NN}}$ = 200 GeV and constraints on medium transport coefficients}},
   {\em Phys.Rev.Lett.} {\bf 101} (2008) 232301,
  [\href{http://xxx.lanl.gov/abs/0801.4020}{{\tt arXiv:0801.4020}}].

\bibitem{Adare:2010de}
{\bf PHENIX Collaboration} Collaboration, A.~Adare et~al., {\it {Heavy Quark
  Production in $p+p$ and Energy Loss and Flow of Heavy Quarks in Au+Au
  Collisions at $\sqrt{s_{NN}}=200$ GeV}},  {\em Phys.Rev.} {\bf C84} (2011)
  044905, [\href{http://xxx.lanl.gov/abs/1005.1627}{{\tt arXiv:1005.1627}}].

\bibitem{Adare:2010sp}
{\bf PHENIX Collaboration} Collaboration, A.~Adare et~al., {\it {Azimuthal
  anisotropy of neutral pion production in Au+Au collisions at $\sqrt{s_{NN}}$
  = 200 GeV: Path-length dependence of jet quenching and the role of initial
  geometry}},  {\em Phys.Rev.Lett.} {\bf 105} (2010) 142301,
  [\href{http://xxx.lanl.gov/abs/1006.3740}{{\tt arXiv:1006.3740}}].

\bibitem{Adare:2012wg}
{\bf PHENIX Collaboration} Collaboration, A.~Adare et~al., {\it {Neutral pion
  production with respect to centrality and reaction plane in Au$+$Au
  collisions at $\sqrt{s_{NN}}$=200 GeV}},  {\em Phys.Rev.} {\bf C87} (2013)
  034911, [\href{http://xxx.lanl.gov/abs/1208.2254}{{\tt arXiv:1208.2254}}].

\bibitem{Abelev:2006db}
{\bf STAR Collaboration} Collaboration, B.~Abelev et~al., {\it {Erratum:
  Transverse momentum and centrality dependence of high-$p_T$ non-photonic
  electron suppression in Au+Au collisions at $\sqrt{s_{NN}} = 200$\,GeV}},
  {\em Phys.Rev.Lett.} {\bf 98} (2007) 192301,
  [\href{http://xxx.lanl.gov/abs/nucl-ex/0607012}{{\tt nucl-ex/0607012}}].

\bibitem{Abelev:2009wx}
{\bf STAR Collaboration} Collaboration, B.~Abelev et~al., {\it {Neutral Pion
  Production in Au+Au Collisions at $\sqrt{s_{NN}} = 200$\,GeV}},  {\em
  Phys.Rev.} {\bf C80} (2009) 044905,
  [\href{http://xxx.lanl.gov/abs/0907.2721}{{\tt arXiv:0907.2721}}].

\bibitem{Gyulassy:2004zy}
M.~Gyulassy and L.~McLerran, {\it {New forms of QCD matter discovered at
  RHIC}},  {\em Nucl.Phys.} {\bf A750} (2005) 30--63,
  [\href{http://xxx.lanl.gov/abs/nucl-th/0405013}{{\tt nucl-th/0405013}}].

\bibitem{Abreu:2007kv}
N.~Armesto, N.~Borghini, S.~Jeon, U.~Wiedemann, S.~Abreu, et~al., {\it {Heavy
  Ion Collisions at the LHC - Last Call for Predictions}},  {\em J.Phys.} {\bf
  G35} (2008) 054001, [\href{http://xxx.lanl.gov/abs/0711.0974}{{\tt
  arXiv:0711.0974}}].

\bibitem{Baier:1996kr}
R.~Baier, Y.~L. Dokshitzer, A.~H. Mueller, S.~Peigne, and D.~Schiff, {\it
  {Radiative energy loss of high-energy quarks and gluons in a finite volume
  quark - gluon plasma}},  {\em Nucl.Phys.} {\bf B483} (1997) 291--320,
  [\href{http://xxx.lanl.gov/abs/hep-ph/9607355}{{\tt hep-ph/9607355}}].

\bibitem{Zakharov:1997uu}
B.~Zakharov, {\it {Radiative energy loss of high-energy quarks in finite size
  nuclear matter and quark - gluon plasma}},  {\em JETP Lett.} {\bf 65} (1997)
  615--620, [\href{http://xxx.lanl.gov/abs/hep-ph/9704255}{{\tt
  hep-ph/9704255}}].

\bibitem{Baier:1998kq}
R.~Baier, Y.~L. Dokshitzer, A.~H. Mueller, and D.~Schiff, {\it {Medium induced
  radiative energy loss: Equivalence between the BDMPS and Zakharov
  formalisms}},  {\em Nucl.Phys.} {\bf B531} (1998) 403--425,
  [\href{http://xxx.lanl.gov/abs/hep-ph/9804212}{{\tt hep-ph/9804212}}].

\bibitem{Wiedemann:2000za}
U.~A. Wiedemann, {\it {Gluon radiation off hard quarks in a nuclear
  environment: Opacity expansion}},  {\em Nucl.Phys.} {\bf B588} (2000)
  303--344, [\href{http://xxx.lanl.gov/abs/hep-ph/0005129}{{\tt
  hep-ph/0005129}}].

\bibitem{Salgado:2003gb}
C.~A. Salgado and U.~A. Wiedemann, {\it {Calculating quenching weights}},  {\em
  Phys.Rev.} {\bf D68} (2003) 014008,
  [\href{http://xxx.lanl.gov/abs/hep-ph/0302184}{{\tt hep-ph/0302184}}].

\bibitem{Armesto:2003jh}
N.~Armesto, C.~A. Salgado, and U.~A. Wiedemann, {\it {Medium induced gluon
  radiation off massive quarks fills the dead cone}},  {\em Phys.Rev.} {\bf
  D69} (2004) 114003, [\href{http://xxx.lanl.gov/abs/hep-ph/0312106}{{\tt
  hep-ph/0312106}}].

\bibitem{Armesto:2004pt}
N.~Armesto, C.~A. Salgado, and U.~A. Wiedemann, {\it {Measuring the collective
  flow with jets}},  {\em Phys.Rev.Lett.} {\bf 93} (2004) 242301,
  [\href{http://xxx.lanl.gov/abs/hep-ph/0405301}{{\tt hep-ph/0405301}}].

\bibitem{Armesto:2005iq}
N.~Armesto, A.~Dainese, C.~A. Salgado, and U.~A. Wiedemann, {\it {Testing the
  color charge and mass dependence of parton energy loss with heavy-to-light
  ratios at RHIC and CERN LHC}},  {\em Phys.Rev.} {\bf D71} (2005) 054027,
  [\href{http://xxx.lanl.gov/abs/hep-ph/0501225}{{\tt hep-ph/0501225}}].

\bibitem{Guo:2000nz}
X.-F. Guo and X.-N. Wang, {\it {Multiple scattering, parton energy loss and
  modified fragmentation functions in deeply inelastic e A scattering}},  {\em
  Phys.Rev.Lett.} {\bf 85} (2000) 3591--3594,
  [\href{http://xxx.lanl.gov/abs/hep-ph/0005044}{{\tt hep-ph/0005044}}].

\bibitem{Wang:2001ifa}
X.-N. Wang and X.-F. Guo, {\it {Multiple parton scattering in nuclei: Parton
  energy loss}},  {\em Nucl.Phys.} {\bf A696} (2001) 788--832,
  [\href{http://xxx.lanl.gov/abs/hep-ph/0102230}{{\tt hep-ph/0102230}}].

\bibitem{Majumder:2007ae}
A.~Majumder, C.~Nonaka, and S.~Bass, {\it {Jet modification in three
  dimensional fluid dynamics at next-to-leading twist}},  {\em Phys.Rev.} {\bf
  C76} (2007) 041902, [\href{http://xxx.lanl.gov/abs/nucl-th/0703019}{{\tt
  nucl-th/0703019}}].

\bibitem{Arnold:2002ja}
P.~B. Arnold, G.~D. Moore, and L.~G. Yaffe, {\it {Photon and gluon emission in
  relativistic plasmas}},  {\em JHEP} {\bf 0206} (2002) 030,
  [\href{http://xxx.lanl.gov/abs/hep-ph/0204343}{{\tt hep-ph/0204343}}].

\bibitem{Arnold:2002zm}
P.~B. Arnold, G.~D. Moore, and L.~G. Yaffe, {\it {Effective kinetic theory for
  high temperature gauge theories}},  {\em JHEP} {\bf 0301} (2003) 030,
  [\href{http://xxx.lanl.gov/abs/hep-ph/0209353}{{\tt hep-ph/0209353}}].

\bibitem{Arnold:2003zc}
P.~B. Arnold, G.~D. Moore, and L.~G. Yaffe, {\it {Transport coefficients in
  high temperature gauge theories. 2. Beyond leading log}},  {\em JHEP} {\bf
  0305} (2003) 051, [\href{http://xxx.lanl.gov/abs/hep-ph/0302165}{{\tt
  hep-ph/0302165}}].

\bibitem{Gyulassy:1993hr}
M.~Gyulassy and X.-N. Wang, {\it {Multiple collisions and induced gluon
  Bremsstrahlung in QCD}},  {\em Nucl.Phys.} {\bf B420} (1994) 583--614,
  [\href{http://xxx.lanl.gov/abs/nucl-th/9306003}{{\tt nucl-th/9306003}}].

\bibitem{Gyulassy:1999zd}
M.~Gyulassy, P.~Levai, and I.~Vitev, {\it {Jet quenching in thin quark gluon
  plasmas. 1. Formalism}},  {\em Nucl.Phys.} {\bf B571} (2000) 197--233,
  [\href{http://xxx.lanl.gov/abs/hep-ph/9907461}{{\tt hep-ph/9907461}}].

\bibitem{Gyulassy:2000er}
M.~Gyulassy, P.~Levai, and I.~Vitev, {\it {Reaction operator approach to
  nonAbelian energy loss}},  {\em Nucl.Phys.} {\bf B594} (2001) 371--419,
  [\href{http://xxx.lanl.gov/abs/nucl-th/0006010}{{\tt nucl-th/0006010}}].

\bibitem{Djordjevic:2003zk}
M.~Djordjevic and M.~Gyulassy, {\it {Heavy quark radiative energy loss in QCD
  matter}},  {\em Nucl.Phys.} {\bf A733} (2004) 265--298,
  [\href{http://xxx.lanl.gov/abs/nucl-th/0310076}{{\tt nucl-th/0310076}}].

\bibitem{WHDG}
S.~Wicks, W.~Horowitz, M.~Djordjevic, and M.~Gyulassy, {\it {Elastic,
  inelastic, and path length fluctuations in jet tomography}},  {\em
  Nucl.Phys.} {\bf A784} (2007) 426--442,
  [\href{http://xxx.lanl.gov/abs/nucl-th/0512076}{{\tt nucl-th/0512076}}].

\bibitem{Buzzatti:2011vt}
A.~Buzzatti and M.~Gyulassy, {\it {Jet Flavor Tomography of Quark Gluon Plasmas
  at RHIC and LHC}},  {\em Phys.Rev.Lett.} {\bf 108} (2012) 022301,
  [\href{http://xxx.lanl.gov/abs/1106.3061}{{\tt arXiv:1106.3061}}].

\bibitem{Gubser:2006bz}
S.~S. Gubser, {\it {Drag force in AdS/CFT}},  {\em Phys.Rev.} {\bf D74} (2006)
  126005, [\href{http://xxx.lanl.gov/abs/hep-th/0605182}{{\tt
  hep-th/0605182}}].

\bibitem{Herzog:2006gh}
C.~Herzog, A.~Karch, P.~Kovtun, C.~Kozcaz, and L.~Yaffe, {\it {Energy loss of a
  heavy quark moving through N=4 supersymmetric Yang-Mills plasma}},  {\em
  JHEP} {\bf 0607} (2006) 013,
  [\href{http://xxx.lanl.gov/abs/hep-th/0605158}{{\tt hep-th/0605158}}].

\bibitem{CasalderreySolana:2006rq}
J.~Casalderrey-Solana and D.~Teaney, {\it {Heavy quark diffusion in strongly
  coupled N=4 Yang-Mills}},  {\em Phys.Rev.} {\bf D74} (2006) 085012,
  [\href{http://xxx.lanl.gov/abs/hep-ph/0605199}{{\tt hep-ph/0605199}}].

\bibitem{Noronha:2010zc}
J.~Noronha, M.~Gyulassy, and G.~Torrieri, {\it {Conformal Holography of Bulk
  Elliptic Flow and Heavy Quark Quenching in Relativistic Heavy Ion
  Collisions}},  {\em Phys.Rev.} {\bf C82} (2010) 054903,
  [\href{http://xxx.lanl.gov/abs/1009.2286}{{\tt arXiv:1009.2286}}].

\bibitem{Ficnar:2012np}
A.~Ficnar, {\it {AdS/CFT Energy Loss in Time-Dependent String Configurations}},
   {\em Phys.Rev.} {\bf D86} (2012) 046010,
  [\href{http://xxx.lanl.gov/abs/1201.1780}{{\tt arXiv:1201.1780}}].

\bibitem{Ficnar:2012yu}
A.~Ficnar, J.~Noronha, and M.~Gyulassy, {\it {Falling Strings and Light Quark
  Jet Quenching at LHC}},  {\em Nucl. Phys.} {\bf A910-911} (2013) 252--255,
  [\href{http://xxx.lanl.gov/abs/1208.0305}{{\tt arXiv:1208.0305}}].

\bibitem{Ficnar:2013wba}
A.~Ficnar and S.~S. Gubser, {\it {Finite momentum at string endpoints}},  {\em
  Phys.Rev.} {\bf D89} (2014) 026002,
  [\href{http://xxx.lanl.gov/abs/1306.6648}{{\tt arXiv:1306.6648}}].

\bibitem{Ficnar:2013qxa}
A.~Ficnar, S.~S. Gubser, and M.~Gyulassy, {\it {Shooting String Holography of
  Jet Quenching at RHIC and LHC}},
  \href{http://xxx.lanl.gov/abs/1311.6160}{{\tt arXiv:1311.6160}}.

\bibitem{Gyulassy:2002yv}
M.~Gyulassy, P.~Levai, and I.~Vitev, {\it {Reaction operator approach to
  multiple elastic scatterings}},  {\em Phys.Rev.} {\bf D66} (2002) 014005,
  [\href{http://xxx.lanl.gov/abs/nucl-th/0201078}{{\tt nucl-th/0201078}}].

\bibitem{Vitev:2002pf}
I.~Vitev and M.~Gyulassy, {\it {High $p_{T}$ tomography of $d$ + Au and Au+Au
  at SPS, RHIC, and LHC}},  {\em Phys.Rev.Lett.} {\bf 89} (2002) 252301,
  [\href{http://xxx.lanl.gov/abs/hep-ph/0209161}{{\tt hep-ph/0209161}}].

\bibitem{Djordjevic:2007at}
M.~Djordjevic and U.~Heinz, {\it {Radiative heavy quark energy loss in a
  dynamical QCD medium}},  {\em Phys.Rev.} {\bf C77} (2008) 024905,
  [\href{http://xxx.lanl.gov/abs/0705.3439}{{\tt arXiv:0705.3439}}].

\bibitem{Buzzatti:2012pe}
A.~Buzzatti and M.~Gyulassy, {\it {An overview of the CUJET model: Jet Flavor
  Tomography applied at RHIC and LHC}},
  \href{http://xxx.lanl.gov/abs/1207.6020}{{\tt arXiv:1207.6020}}.

\bibitem{Horowitz:2011gd}
W.~Horowitz and M.~Gyulassy, {\it {The Surprising Transparency of the sQGP at
  LHC}},  {\em Nucl.Phys.} {\bf A872} (2011) 265--285,
  [\href{http://xxx.lanl.gov/abs/1104.4958}{{\tt arXiv:1104.4958}}].

\bibitem{JetCollab}
``{Topical Collaboration on Jet and Electromagnetic Tomography of Extreme
  Phases of Matter in Heavy-ion Collisions}.'' \url{http://jet.lbl.gov/}.

\bibitem{Burke:2013yra}
K.~M. Burke, A.~Buzzatti, N.~Chang, C.~Gale, M.~Gyulassy, et~al., {\it
  {Extracting jet transport coefficient from jet quenching at RHIC and LHC}},
  \href{http://xxx.lanl.gov/abs/1312.5003}{{\tt arXiv:1312.5003}}.

\bibitem{Betz:2013caa}
B.~Betz and M.~Gyulassy, {\it {Azimuthal Jet Tomography of Quark Gluon Plasmas
  at RHIC and LHC}},  \href{http://xxx.lanl.gov/abs/1305.6458}{{\tt
  arXiv:1305.6458}}.

\bibitem{oai:arXiv.org:1211.0804}
B.~Betz and M.~Gyulassy, {\it {Quantifying a Possibly Reduced Jet-Medium
  Coupling of the sQGP at the LHC}},  {\em Nucl.Phys.A904-905} {\bf 2013}
  (2013) 717c--720c, [\href{http://xxx.lanl.gov/abs/1211.0804}{{\tt
  arXiv:1211.0804}}].

\bibitem{BBMG2012}
B.~Betz and M.~Gyulassy, {\it {Examining a reduced jet-medium coupling in Pb+Pb
  collisions at the Large Hadron Collider}},  {\em Phys.Rev.} {\bf C86} (2012)
  024903, [\href{http://xxx.lanl.gov/abs/1201.0281}{{\tt arXiv:1201.0281}}].

\bibitem{oai:arXiv.org:1106.4564}
B.~Betz, M.~Gyulassy, and G.~Torrieri, {\it {Sensitivity of Azimuthal Jet
  Tomography to Early Time Energy-Loss at RHIC and LHC}},  {\em J.Phys.} {\bf
  G38} (2011) 124153, [\href{http://xxx.lanl.gov/abs/1106.4564}{{\tt
  arXiv:1106.4564}}].

\bibitem{oai:arXiv.org:1102.5416}
B.~Betz, M.~Gyulassy, and G.~Torrieri, {\it {Fourier Harmonics of High-pT
  Particles Probing the Fluctuating Intitial Condition Geometries in Heavy-Ion
  Collisions}},  {\em Phys.Rev.} {\bf C84} (2011) 024913,
  [\href{http://xxx.lanl.gov/abs/1102.5416}{{\tt arXiv:1102.5416}}].

\bibitem{oai:arXiv.org:0812.4401}
B.~Betz, J.~Noronha, G.~Torrieri, M.~Gyulassy, I.~Mishustin, et~al., {\it
  {Universality of the Diffusion Wake from Stopped and Punch-Through Jets in
  Heavy-Ion Collisions}},  {\em Phys.Rev.} {\bf C79} (2009) 034902,
  [\href{http://xxx.lanl.gov/abs/0812.4401}{{\tt arXiv:0812.4401}}].

\bibitem{Buzzatti:2012dy}
A.~Buzzatti and M.~Gyulassy, {\it {A running coupling explanation of the
  surprising transparency of the QGP at LHC}},  {\em Nucl.Phys.A904-905} {\bf
  2013} (2013) 779c--782c, [\href{http://xxx.lanl.gov/abs/1210.6417}{{\tt
  arXiv:1210.6417}}].

\bibitem{Zakharov:2008kt}
B.~Zakharov, {\it {Jet quenching with running coupling including radiative and
  collisional energy losses}},  {\em JETP Lett.} {\bf 88} (2008) 781--786,
  [\href{http://xxx.lanl.gov/abs/0811.0445}{{\tt arXiv:0811.0445}}].

\bibitem{Zakharov:2007pj}
B.~Zakharov, {\it {Parton energy loss in an expanding quark-gluon plasma:
  Radiative versus collisional}},  {\em JETP Lett.} {\bf 86} (2007) 444--450,
  [\href{http://xxx.lanl.gov/abs/0708.0816}{{\tt arXiv:0708.0816}}].

\bibitem{Dokshitzer:1995ev}
Y.~L. Dokshitzer, V.~A. Khoze, and S.~Troian, {\it {Specific features of heavy
  quark production. LPHD approach to heavy particle spectra}},  {\em Phys.Rev.}
  {\bf D53} (1996) 89--119, [\href{http://xxx.lanl.gov/abs/hep-ph/9506425}{{\tt
  hep-ph/9506425}}].

\bibitem{Kaczmarek:2004gv}
O.~Kaczmarek, F.~Karsch, F.~Zantow, and P.~Petreczky, {\it {Static quark
  anti-quark free energy and the running coupling at finite temperature}},
  {\em Phys.Rev.} {\bf D70} (2004) 074505,
  [\href{http://xxx.lanl.gov/abs/hep-lat/0406036}{{\tt hep-lat/0406036}}].

\bibitem{oai:arXiv.org:1009.0545}
W.~Horowitz and Y.~V. Kovchegov, {\it {Running Coupling Corrections to High
  Energy Inclusive Gluon Production}},  {\em Nucl.Phys.} {\bf A849} (2011)
  72--97, [\href{http://xxx.lanl.gov/abs/1009.0545}{{\tt arXiv:1009.0545}}].

\bibitem{oai:arXiv.org:1106.5456}
W.~Horowitz and Y.~Kovchegov, {\it {Running coupling corrections to inclusive
  gluon production}},  {\em J.Phys.} {\bf G38} (2011) 124064,
  [\href{http://xxx.lanl.gov/abs/1106.5456}{{\tt arXiv:1106.5456}}].

\bibitem{oai:arXiv.org:hep-ph/0601119}
A.~Peshier, {\it {Running coupling and screening in the (s)QGP}},
  \href{http://xxx.lanl.gov/abs/hep-ph/0601119}{{\tt hep-ph/0601119}}.

\bibitem{Djordjevic:2013xoa}
M.~Djordjevic and M.~Djordjevic, {\it {LHC jet suppression of light and heavy
  flavor observables}},  \href{http://xxx.lanl.gov/abs/1307.4098}{{\tt
  arXiv:1307.4098}}.

\bibitem{Peigne:2008nd}
S.~Peigne and A.~Peshier, {\it {Collisional energy loss of a fast heavy quark
  in a quark-gluon plasma}},  {\em Phys.Rev.} {\bf D77} (2008) 114017,
  [\href{http://xxx.lanl.gov/abs/0802.4364}{{\tt arXiv:0802.4364}}].

\bibitem{Glauber:1970jm}
R.~Glauber and G.~Matthiae, {\it {High-energy scattering of protons by
  nuclei}},  {\em Nucl.Phys.} {\bf B21} (1970) 135--157.

\bibitem{Gyulassy:2000gk}
M.~Gyulassy, I.~Vitev, and X.~Wang, {\it {High p(T) azimuthal asymmetry in
  noncentral A+A at RHIC}},  {\em Phys.Rev.Lett.} {\bf 86} (2001) 2537--2540,
  [\href{http://xxx.lanl.gov/abs/nucl-th/0012092}{{\tt nucl-th/0012092}}].

\bibitem{Gyulassy:2001kr}
M.~Gyulassy, I.~Vitev, X.-N. Wang, and P.~Huovinen, {\it {Transverse expansion
  and high p(T) azimuthal asymmetry at RHIC}},  {\em Phys.Lett.} {\bf B526}
  (2002) 301--308, [\href{http://xxx.lanl.gov/abs/nucl-th/0109063}{{\tt
  nucl-th/0109063}}].

\bibitem{Luzum:2008cw}
M.~Luzum and P.~Romatschke, {\it {Conformal Relativistic Viscous Hydrodynamics:
  Applications to RHIC results at s(NN)**(1/2) = 200-GeV}},  {\em Phys.Rev.}
  {\bf C78} (2008) 034915, [\href{http://xxx.lanl.gov/abs/0804.4015}{{\tt
  arXiv:0804.4015}}].

\bibitem{Luzum:2009sb}
M.~Luzum and P.~Romatschke, {\it {Viscous Hydrodynamic Predictions for Nuclear
  Collisions at the LHC}},  {\em Phys.Rev.Lett.} {\bf 103} (2009) 262302,
  [\href{http://xxx.lanl.gov/abs/0901.4588}{{\tt arXiv:0901.4588}}].

\bibitem{Song:2008si}
H.~Song and U.~W. Heinz, {\it {Multiplicity scaling in ideal and viscous
  hydrodynamics}},  {\em Phys.Rev.} {\bf C78} (2008) 024902,
  [\href{http://xxx.lanl.gov/abs/0805.1756}{{\tt arXiv:0805.1756}}].

\bibitem{Shen:2010uy}
C.~Shen, U.~Heinz, P.~Huovinen, and H.~Song, {\it {Systematic parameter study
  of hadron spectra and elliptic flow from viscous hydrodynamic simulations of
  Au+Au collisions at $\sqrt{s_{NN}}=200$ GeV}},  {\em Phys.Rev.} {\bf C82}
  (2010) 054904, [\href{http://xxx.lanl.gov/abs/1010.1856}{{\tt
  arXiv:1010.1856}}].

\bibitem{Renk:2010qx}
T.~Renk, H.~Holopainen, U.~Heinz, and C.~Shen, {\it {A Systematic comparison of
  jet quenching in different fluid-dynamical models}},  {\em Phys.Rev.} {\bf
  C83} (2011) 014910, [\href{http://xxx.lanl.gov/abs/1010.1635}{{\tt
  arXiv:1010.1635}}].

\bibitem{Cooper:1974mv}
F.~Cooper and G.~Frye, {\it {Comment on the Single Particle Distribution in the
  Hydrodynamic and Statistical Thermodynamic Models of Multiparticle
  Production}},  {\em Phys.Rev.} {\bf D10} (1974) 186.

\bibitem{Huovinen:2009yb}
P.~Huovinen and P.~Petreczky, {\it {QCD Equation of State and Hadron Resonance
  Gas}},  {\em Nucl.Phys.} {\bf A837} (2010) 26--53,
  [\href{http://xxx.lanl.gov/abs/0912.2541}{{\tt arXiv:0912.2541}}].

\bibitem{oai:arXiv.org:1106.3927}
K.~Dusling, F.~Gelis, and R.~Venugopalan, {\it {The initial spectrum of
  fluctuations in the little bang}},  {\em Nucl.Phys.} {\bf A872} (2011)
  161--195, [\href{http://xxx.lanl.gov/abs/1106.3927}{{\tt arXiv:1106.3927}}].

\bibitem{Brick2012}
N.~Armesto, B.~Cole, C.~Gale, W.~A. Horowitz, P.~Jacobs, et~al., {\it
  {Comparison of Jet Quenching Formalisms for a Quark-Gluon Plasma 'Brick'}},
  {\em Phys.Rev.} {\bf C86} (2012) 064904,
  [\href{http://xxx.lanl.gov/abs/1106.1106}{{\tt arXiv:1106.1106}}].

\bibitem{Majumder:2007zh}
A.~Majumder, B.~Muller, and X.-N. Wang, {\it {Small shear viscosity of a
  quark-gluon plasma implies strong jet quenching}},  {\em Phys.Rev.Lett.} {\bf
  99} (2007) 192301, [\href{http://xxx.lanl.gov/abs/hep-ph/0703082}{{\tt
  hep-ph/0703082}}].

\bibitem{CaronHuot:2008ni}
S.~Caron-Huot, {\it {O(g) plasma effects in jet quenching}},  {\em Phys.Rev.}
  {\bf D79} (2009) 065039, [\href{http://xxx.lanl.gov/abs/0811.1603}{{\tt
  arXiv:0811.1603}}].

\bibitem{Panero:2013pla}
M.~Panero, K.~Rummukainen, and A.~Schäfer, {\it {A lattice study of the jet
  quenching parameter}},  \href{http://xxx.lanl.gov/abs/1307.5850}{{\tt
  arXiv:1307.5850}}.

\bibitem{Liou:2013qya}
T.~Liou, A.~Mueller, and B.~Wu, {\it {Radiative $p_\bot$-broadening of
  high-energy quarks and gluons in QCD matter}},  {\em Nucl.Phys.} {\bf A916}
  (2013) 102--125, [\href{http://xxx.lanl.gov/abs/1304.7677}{{\tt
  arXiv:1304.7677}}].

\bibitem{Blaizot:2013vha}
J.-P. Blaizot, F.~Dominguez, E.~Iancu, and Y.~Mehtar-Tani, {\it {Probabilistic
  picture for medium-induced jet evolution}},
  \href{http://xxx.lanl.gov/abs/1311.5823}{{\tt arXiv:1311.5823}}.

\bibitem{Kang:2013raa}
Z.-B. Kang, E.~Wang, X.-N. Wang, and H.~Xing, {\it {Next-to-Leading QCD
  Factorization for Semi-Inclusive Deep Inelastic Scattering at Twist-4}},
  {\em Phys.Rev.Lett.} {\bf 112} (2014) 102001,
  [\href{http://xxx.lanl.gov/abs/1310.6759}{{\tt arXiv:1310.6759}}].

\bibitem{Iancu:2014kga}
E.~Iancu, {\it {The non-linear evolution of jet quenching}},
  \href{http://xxx.lanl.gov/abs/1403.1996}{{\tt arXiv:1403.1996}}.

\bibitem{Blaizot:2014bha}
J.-P. Blaizot and Y.~Mehtar-Tani, {\it {Renormalization of the jet-quenching
  parameter}},  \href{http://xxx.lanl.gov/abs/1403.2323}{{\tt
  arXiv:1403.2323}}.

\bibitem{JXMGprep}
J.~Xu and M.~Gyulassy {\em In preparation}.

\bibitem{Djordjevic:2013pba}
M.~Djordjevic, {\it {Heavy flavor puzzle at LHC: a serendipitous interplay of
  jet suppression and fragmentation}},  {\em Phys.Rev.Lett.} {\bf 112} (2014)
  042302, [\href{http://xxx.lanl.gov/abs/1307.4702}{{\tt arXiv:1307.4702}}].

\bibitem{oai:arXiv.org:1209.0198}
M.~Djordjevic, {\it {Heavy flavor suppression in a dynamical QCD medium with
  finite magnetic mass}},  \href{http://xxx.lanl.gov/abs/1209.0198}{{\tt
  arXiv:1209.0198}}.

\bibitem{DM2012}
M.~Djordjevic, {\it {Jet suppression of pions and single electrons at Au+Au
  collisions at RHIC}},  {\em Phys.Rev.} {\bf C85} (2012) 034904,
  [\href{http://xxx.lanl.gov/abs/1105.6082}{{\tt arXiv:1105.6082}}].

\bibitem{Zakharov2009}
P.~Aurenche and B.~Zakharov, {\it {Anomalous mass dependence of radiative quark
  energy loss in a finite-size quark-gluon plasma}},  {\em JETP Lett.} {\bf 90}
  (2009) 237--243, [\href{http://xxx.lanl.gov/abs/0907.1918}{{\tt
  arXiv:0907.1918}}].

\bibitem{Voloshin:1994mz}
S.~Voloshin and Y.~Zhang, {\it {Flow study in relativistic nuclear collisions
  by Fourier expansion of Azimuthal particle distributions}},  {\em Z.Phys.}
  {\bf C70} (1996) 665--672,
  [\href{http://xxx.lanl.gov/abs/hep-ph/9407282}{{\tt hep-ph/9407282}}].

\bibitem{Molnar:2013eqa}
D.~Molnar and D.~Sun, {\it {High-pT suppression and elliptic flow from
  radiative energy loss with realistic bulk medium expansion}},
  \href{http://xxx.lanl.gov/abs/1305.1046}{{\tt arXiv:1305.1046}}.

\bibitem{Bazavov:2013zha}
A.~Bazavov and P.~Petreczky, {\it {Static meson correlators in 2+1 flavor QCD
  at non-zero temperature}},  {\em Eur.Phys.J.} {\bf A49} (2013) 85,
  [\href{http://xxx.lanl.gov/abs/1303.5500}{{\tt arXiv:1303.5500}}].

\bibitem{Song:2013qma}
H.~Song, S.~Bass, and U.~W. Heinz, {\it {Spectra and elliptic flow for
  identified hadrons in 2.76 A TeV Pb+Pb collisions}},
  \href{http://xxx.lanl.gov/abs/1311.0157}{{\tt arXiv:1311.0157}}.

\bibitem{Liao:2006ry}
J.~Liao and E.~Shuryak, {\it {Strongly coupled plasma with electric and
  magnetic charges}},  {\em Phys.Rev.} {\bf C75} (2007) 054907,
  [\href{http://xxx.lanl.gov/abs/hep-ph/0611131}{{\tt hep-ph/0611131}}].

\bibitem{Liao:2008jg}
J.~Liao and E.~Shuryak, {\it {Magnetic Component of Quark-Gluon Plasma is also
  a Liquid!}},  {\em Phys.Rev.Lett.} {\bf 101} (2008) 162302,
  [\href{http://xxx.lanl.gov/abs/0804.0255}{{\tt arXiv:0804.0255}}].

\bibitem{Liao:2008dk}
J.~Liao and E.~Shuryak, {\it {Angular Dependence of Jet Quenching Indicates Its
  Strong Enhancement Near the QCD Phase Transition}},  {\em Phys.Rev.Lett.}
  {\bf 102} (2009) 202302, [\href{http://xxx.lanl.gov/abs/0810.4116}{{\tt
  arXiv:0810.4116}}].

\bibitem{Renk:2011qf}
T.~Renk, {\it {Using Hard Dihadron Correlations to constrain Elastic Energy
  Loss}},  {\em Phys.Rev.} {\bf C84} (2011) 067902,
  [\href{http://xxx.lanl.gov/abs/1110.2313}{{\tt arXiv:1110.2313}}].

\bibitem{Renk:2012cx}
T.~Renk, {\it {On the sensitivity of the dijet asymmetry to the physics of jet
  quenching}},  {\em Phys.Rev.} {\bf C85} (2012) 064908,
  [\href{http://xxx.lanl.gov/abs/1202.4579}{{\tt arXiv:1202.4579}}].

\bibitem{Renk:2011aa}
T.~Renk, {\it {Constraining the Physics of Jet Quenching}},  {\em Phys.Rev.}
  {\bf C85} (2012) 044903, [\href{http://xxx.lanl.gov/abs/1112.2503}{{\tt
  arXiv:1112.2503}}].

\bibitem{Gyulassy:2003mc}
M.~Gyulassy, I.~Vitev, X.-N. Wang, and B.-W. Zhang, {\it {Jet quenching and
  radiative energy loss in dense nuclear matter}},
  \href{http://xxx.lanl.gov/abs/nucl-th/0302077}{{\tt nucl-th/0302077}}.

\bibitem{Thoma:1990fm}
M.~H. Thoma and M.~Gyulassy, {\it {Quark damping and energy loss in the high
  temperature QCD}},  {\em Nucl.Phys.} {\bf B351} (1991) 491--506.

\bibitem{oai:arXiv.org:hep-ph/0204146}
P.~Aurenche, F.~Gelis, and H.~Zaraket, {\it {A Simple sum rule for the thermal
  gluon spectral function and applications}},  {\em JHEP} {\bf 0205} (2002)
  043, [\href{http://xxx.lanl.gov/abs/hep-ph/0204146}{{\tt hep-ph/0204146}}].

\bibitem{oai:arXiv.org:hep-ph/0207206}
B.~Zakharov, {\it {Coherent final state interaction in jet production in
  nucleus-nucleus collisions}},  {\em JETP Lett.} {\bf 76} (2002) 201--205,
  [\href{http://xxx.lanl.gov/abs/hep-ph/0207206}{{\tt hep-ph/0207206}}].

\bibitem{Djordjevic:2008iz}
M.~Djordjevic and U.~W. Heinz, {\it {Radiative energy loss in a finite
  dynamical QCD medium}},  {\em Phys.Rev.Lett.} {\bf 101} (2008) 022302,
  [\href{http://xxx.lanl.gov/abs/0802.1230}{{\tt arXiv:0802.1230}}].

\bibitem{Djordjevic:2009cr}
M.~Djordjevic, {\it {Theoretical formalism of radiative jet energy loss in a
  finite size dynamical QCD medium}},  {\em Phys.Rev.} {\bf C80} (2009) 064909,
  [\href{http://xxx.lanl.gov/abs/0903.4591}{{\tt arXiv:0903.4591}}].

\bibitem{Buzzatti:2010ck}
A.~Buzzatti and M.~Gyulassy, {\it {Dynamical magnetic enhancement of light and
  heavy quark jet quenching at RHIC}},  {\em Nucl.Phys.} {\bf A855} (2011)
  307--310, [\href{http://xxx.lanl.gov/abs/1012.0614}{{\tt arXiv:1012.0614}}].

\bibitem{oai:arXiv.org:1105.4359}
M.~Djordjevic and M.~Djordjevic, {\it {Generalization of radiative jet energy
  loss to non-zero magnetic mass}},  {\em Phys.Lett.} {\bf B709} (2012)
  229--233, [\href{http://xxx.lanl.gov/abs/1105.4359}{{\tt arXiv:1105.4359}}].

\bibitem{Mustafa:2003vh}
M.~G. Mustafa and M.~H. Thoma, {\it {Quenching of hadron spectra due to the
  collisional energy loss of partons in the quark gluon plasma}},  {\em Acta
  Phys.Hung.} {\bf A22} (2005) 93--102,
  [\href{http://xxx.lanl.gov/abs/hep-ph/0311168}{{\tt hep-ph/0311168}}].

\bibitem{Mustafa:2004dr}
M.~G. Mustafa, {\it {Energy loss of charm quarks in the quark-gluon plasma:
  Collisional versus radiative}},  {\em Phys.Rev.} {\bf C72} (2005) 014905,
  [\href{http://xxx.lanl.gov/abs/hep-ph/0412402}{{\tt hep-ph/0412402}}].

\bibitem{Braaten:1991jj}
E.~Braaten and M.~H. Thoma, {\it {Energy loss of a heavy fermion in a hot
  plasma}},  {\em Phys.Rev.} {\bf D44} (1991) 1298--1310.

\bibitem{Braaten:1991we}
E.~Braaten and M.~H. Thoma, {\it {Energy loss of a heavy quark in the quark -
  gluon plasma}},  {\em Phys.Rev.} {\bf D44} (1991) 2625--2630.

\bibitem{Wicks:2008zz}
S.~Wicks, {\it {Fluctuations with small numbers: Developing the perturbative
  paradigm for jet physics in the QGP at RHIC and LHC}}, .

\bibitem{Wang:private}
X.-N. Wang {\em Private communication}.

\bibitem{Mangano:1991jk}
M.~L. Mangano, P.~Nason, and G.~Ridolfi, {\it {Heavy quark correlations in
  hadron collisions at next-to-leading order}},  {\em Nucl.Phys.} {\bf B373}
  (1992) 295--345.

\bibitem{Cacciari:1998it}
M.~Cacciari, M.~Greco, and P.~Nason, {\it {The P(T) spectrum in heavy flavor
  hadroproduction}},  {\em JHEP} {\bf 9805} (1998) 007,
  [\href{http://xxx.lanl.gov/abs/hep-ph/9803400}{{\tt hep-ph/9803400}}].

\bibitem{Cacciari:2001td}
M.~Cacciari, S.~Frixione, and P.~Nason, {\it {The p(T) spectrum in heavy flavor
  photoproduction}},  {\em JHEP} {\bf 0103} (2001) 006,
  [\href{http://xxx.lanl.gov/abs/hep-ph/0102134}{{\tt hep-ph/0102134}}].

\bibitem{Nason:1987xz}
P.~Nason, S.~Dawson, and R.~K. Ellis, {\it {The Total Cross-Section for the
  Production of Heavy Quarks in Hadronic Collisions}},  {\em Nucl.Phys.} {\bf
  B303} (1988) 607.

\bibitem{Nason:1989zy}
P.~Nason, S.~Dawson, and R.~K. Ellis, {\it {The One Particle Inclusive
  Differential Cross-Section for Heavy Quark Production in Hadronic
  Collisions}},  {\em Nucl.Phys.} {\bf B327} (1989) 49--92.

\bibitem{Beenakker:1990maa}
W.~Beenakker, W.~van Neerven, R.~Meng, G.~Schuler, and J.~Smith, {\it {QCD
  corrections to heavy quark production in hadron hadron collisions}},  {\em
  Nucl.Phys.} {\bf B351} (1991) 507--560.

\bibitem{Cacciari:1993mq}
M.~Cacciari and M.~Greco, {\it {Large $p_{T}$ hadroproduction of heavy
  quarks}},  {\em Nucl.Phys.} {\bf B421} (1994) 530--544,
  [\href{http://xxx.lanl.gov/abs/hep-ph/9311260}{{\tt hep-ph/9311260}}].

\bibitem{ADILVITEV}
A.~Adil and I.~Vitev, {\it {Collisional dissociation of heavy mesons in dense
  QCD matter}},  {\em Phys.Lett.} {\bf B649} (2007) 139--146,
  [\href{http://xxx.lanl.gov/abs/hep-ph/0611109}{{\tt hep-ph/0611109}}].

\bibitem{Vitev:2007jj}
I.~Vitev, A.~Adil, and H.~van Hees, {\it {Novel heavy flavor suppression
  mechanisms in the QGP}},  {\em J.Phys.} {\bf G34} (2007) S769--774,
  [\href{http://xxx.lanl.gov/abs/hep-ph/0701188}{{\tt hep-ph/0701188}}].

\bibitem{Wong:2004zr}
C.-Y. Wong, {\it {Heavy quarkonia in quark-gluon plasma}},  {\em Phys.Rev.}
  {\bf C72} (2005) 034906, [\href{http://xxx.lanl.gov/abs/hep-ph/0408020}{{\tt
  hep-ph/0408020}}].

\bibitem{Petreczky:2009cr}
{\bf RBC-Bielefeld Collaboration} Collaboration, P.~Petreczky, P.~Hegde, and
  A.~Velytsky, {\it {Quark number fluctuations at high temperatures}},  {\em
  PoS} {\bf LAT2009} (2009) 159, [\href{http://xxx.lanl.gov/abs/0911.0196}{{\tt
  arXiv:0911.0196}}].

\bibitem{KKP}
B.~A. Kniehl, G.~Kramer, and B.~Potter, {\it {Fragmentation functions for
  pions, kaons, and protons at next-to-leading order}},  {\em Nucl.Phys.} {\bf
  B582} (2000) 514--536, [\href{http://xxx.lanl.gov/abs/hep-ph/0010289}{{\tt
  hep-ph/0010289}}].

\bibitem{PETERSON}
C.~Peterson, D.~Schlatter, I.~Schmitt, and P.~M. Zerwas, {\it {Scaling
  Violations in Inclusive e+ e- Annihilation Spectra}},  {\em Phys.Rev.} {\bf
  D27} (1983) 105.

\bibitem{DGVW}
M.~Djordjevic, M.~Gyulassy, R.~Vogt, and S.~Wicks, {\it {Influence of bottom
  quark jet quenching on single electron tomography of Au + Au}},  {\em
  Phys.Lett.} {\bf B632} (2006) 81--86,
  [\href{http://xxx.lanl.gov/abs/nucl-th/0507019}{{\tt nucl-th/0507019}}].

\bibitem{VOGT}
M.~Cacciari, P.~Nason, and R.~Vogt, {\it {QCD predictions for charm and bottom
  production at RHIC}},  {\em Phys.Rev.Lett.} {\bf 95} (2005) 122001,
  [\href{http://xxx.lanl.gov/abs/hep-ph/0502203}{{\tt hep-ph/0502203}}].

\bibitem{Horowitz:2009eb}
W.~Horowitz and B.~Cole, {\it {Systematic theoretical uncertainties in jet
  quenching due to gluon kinematics}},  {\em Phys.Rev.} {\bf C81} (2010)
  024909, [\href{http://xxx.lanl.gov/abs/0910.1823}{{\tt arXiv:0910.1823}}].

\bibitem{Bjorken:1982tu}
J.~Bjorken, {\it {Energy Loss of Energetic Partons in Quark - Gluon Plasma:
  Possible Extinction of High p(t) Jets in Hadron - Hadron Collisions}}, .

\end{thebibliography}\endgroup

\end{document}